\documentclass[preprint,3p,times]{elsarticle}

\pdfoutput=1

\usepackage{amssymb}
\usepackage{amsthm}
\usepackage{amsmath}
\usepackage{epsfig}
\usepackage{subfigure}
\usepackage[usenames]{color}
\usepackage{algorithm}
\usepackage{algorithmic}
\usepackage{nicefrac}
\epsfverbosetrue
\graphicspath{{fig/}{../fig/}}
\usepackage{dutchcal}

\usepackage{bm}
\usepackage{array}

\newcommand{\mb}[1]{\ensuremath{\mathbf{#1}}}

\newcommand{\mbd}[1]{\textsf{\textbf{#1}}}
\newcommand{\mbds}[1] {\pmb{\bm{#1}}}

\journal{Computer Methods in Applied Mechanics and Engineering}

\begin{document}
 \setlength{\parindent}{0.0ex}
 \setcounter{secnumdepth}{4}
 \setcounter{tocdepth}{4}
\begin{frontmatter}



\title{Geometrically Exact Finite Element Formulations for Curved Slender Beams: Kirchhoff-Love Theory vs. Simo-Reissner Theory}

\author[mechanosynth]{Christoph Meier\corref{cor1}}
\ead{chrmeier@mit.edu}
\author[lnm]{Wolfgang A. Wall}
\author[lnm]{Alexander Popp}

\address[mechanosynth]{Mechanosynthesis Group, Massachussets Institute of Technology, 77 Massachusetts Avenue, Cambridge, 02139, MA, USA}
\address[lnm]{Institute for Computational Mechanics, Technical University of Munich, Boltzmannstrasse 15, D--85748 Garching b. M\"unchen, Germany}

\cortext[cor1]{Corresponding author}

\begin{abstract}
The present work focuses on geometrically exact finite elements for highly slender beams. It aims at the proposal of novel formulations of Kirchhoff-Love type, a detailed review of existing formulations of Kirchhoff-Love and Simo-Reissner type as well as a careful evaluation and comparison of the proposed and existing formulations. Two different rotation interpolation schemes with strong or weak Kirchhoff constraint enforcement, respectively, as well as two different choices of nodal triad parametrizations in terms of rotation or tangent vectors are proposed. The combination of these schemes leads to four novel finite element variants, all of them based on a $C^1$-continuous Hermite interpolation of the beam centerline. Essential requirements such as representability of general 3D, large-deformation, dynamic problems involving slender beams with arbitrary initial curvatures and anisotropic cross-section shapes, preservation of objectivity and path-independence, consistent convergence orders, avoidance of locking effects as well as conservation of energy and momentum by the employed spatial discretization schemes, but also a range of practically relevant secondary aspects will be investigated analytically and verified numerically for the different formulations. It will be shown that the geometrically exact Kirchhoff-Love beam elements proposed in this work are the first ones of this type that fulfill all these essential requirements. On the contrary, Simo-Reissner type formulations fulfilling these requirements can be found in the literature very well. However, it will be argued that the shear-free Kirchhoff-Love formulations can provide considerable numerical advantages such as lower spatial discretization error level, improved performance of time integration schemes as well as linear and nonlinear solvers or smooth geometry representation as compared to shear-deformable Simo-Reissner formulations when applied to highly slender beams. Concretely, several representative numerical test cases confirm that the proposed Kirchhoff-Love formulations exhibit a lower discretization error level as well as a considerably improved nonlinear solver performance in the range of high beam slenderness ratios as compared to two representative Simo-Reissner element formulations from the literature.
\end{abstract}


\end{frontmatter}

%
\section{Introduction}
\label{sec:intro}
%

Highly slender fiber- or rod-like components represent essential constituents of mechanical systems in countless fields of application and scientific disciplines such as mechanical engineering,
biomedical engineering, material science and bio- or molecular physics. Examples are high-tensile industrial webbings, fiber-reinforced composite materials,  fibrous materials with tailored porosity, synthetic polymer materials or also cellulose fibers determining the characteristics of paper~\cite{durville2010,vu_and_durville2015,kulachenko2012,kulachenko2012b,gadot_and_orgeas2015,rodney_and_orgeas2016}. On entirely different time and length scales, such slender components are relevant when analyzing the supercoiling process of DNA strands, the characteristics of carbon nanotubes or the Brownian dynamics within the cytoskeleton of biological cells, a biopolymer network of highly slender filaments that crucially influences biologically relevant processes such as cell division and cell migration~\cite{cyron2012,mueller2015,wang2007,yang1993}. Often, these slender components can be modeled as 1D Cosserat continua based on a geometrically nonlinear beam theory. In all mentioned cases, mechanical contact interaction crucially influences the overall system behavior. While in the authors' recent works~\cite{meier2015b} and~\cite{meier2015c} a novel beam-to-beam contact formulation has been proposed that faces the challenges arising from high beam slenderness ratios and complex 3D contact configurations, the present contribution focuses on the development of geometrically exact beam element formulations suitable for an accurate, efficient and robust modeling of highly slender beams. While the vast majority of geometrically exact beam element formulations available in the literature is based on the Simo-Reissner theory of thick rods, the current work proposes geometrically exact finite elements based on the Kirchhoff-Love theory of thin rods that are taylored for high beam slenderness ratios.\\

Basically, two essential motivations for applying a beam theory instead of a 3D continuum mechanics theory for the modeling of slender bodies can be identified: In the early days of beam theories, it was 
the accessibility of analytic solutions as for example "Euler's Elastica", even for large deformation problems, that motivated the development and application of one-dimensional theories. Nowadays, 
it is the knowledge that the modeling of highly slender bodies via beam theories yields considerably more efficient, but also more well-posed, numerical formulations as it would be the case for 3D continuum theories. So-called \textit{induced} beam theories can be regarded as reduced 1D continuum theories consistently derived from the 3D theory of continuum mechanics under consideration of a basic kinematic constraint that reflects the deformation states expected for slender bodies in a reasonable manner. Such 1D beam theories typically allow to describe the motion and deformation of slender bodies in 3D space on the basis of proper kinematic, kinetic and constitutive resultant quantities. In the case of induced beam theories, these resultant quantities can for example be derived via integration of 3D stress measures over the beam cross-section. The 3D stress measures typically result from the constrained 3D displacement field as well as standard 3D strain measures and constitutive relations. In this context, the cross-section of a beam represents the collection of all material points sharing the same beam length coordinate in the stress-free configuration. On the contrary, so-called \textit{intrinsic} beam theories directly postulate the 1D resultant quantities. These theories are internally consistent in the sense that the resultant quantities as well as the 1D relations connecting these quantities still fulfill essential mechanical principles such as equilibrium of forces and moments, conservation of energy or rather existence of work conjugated stress-strain pairs, observer invariance or path-independence of conservative problems. Nevertheless, intrinsic beam theories are decoupled from the 3D continuum mechanics theory. Typically, the postulated constitutive constants relating stress and strain measures are determined experimentally, while the constitutive constants of induced beam theories follow directly from the corresponding 3D constitutive laws. Such postulated constitutive laws based on experimentally determined constants are favorable for applications where no 3D continuum foundation exists: Considering the low number of discrete molecules distributed over the thickness of macromolecules as occurring for example in biopolymer networks, DNA strands or carbon nanotubes - to come back to the applications mentioned above - no 3D continuum theory can be applied in a reasonable manner. Nevertheless, it is well-established that these slender components can be described in good approximation by 1D continuum theories and associated experimentally determined constitutive constants~\cite{schmidt2015}. A compromise between induced and intrinsic theories are so-called \textit{semi-induced} beam theories, where only the constitutive law is postulated. The remaining kinetic and kinematic relations are consistently derived from the 3D theory.\\

Based on the Bernoulli hypothesis of undeformed cross-sections and the work of Euler, the ``Kirchhoff beam''~\cite{kirchhoff1859}, proposed by Kirchhoff zu Heidelberg in 1859, was the first formulation allowing for arbitrary initial curvatures and large three-dimensional deformations including the states of bending and torsion. In 1944, this theory was enhanced by Love~\cite{love1944} to also account for small axial tension. A comprehensive historic overview of these early developments is given in the work of Dill~\cite{dill1992}. It was Reissner in 1972 for the two-dimensional case~\cite{reissner1972} and in 1981 for the general three-dimensional case~\cite{reissner1981}, who completed the theory by two additional deformation measures representing the shear deformation of the beam. While the 3D problem statement of Reissner was still based on some additional approximations, Simo~\cite{simo1985} extended the work of Reissner to yield a formulation that is truly consistent in the sense of a semi-induced beam theory. Thus, starting from a basic kinematic assumption, all kinetic and kinematic quantities and relations are consistently derived from the 3D continuum theory, while the constitutive law has been postulated. Originally, this theory has been denoted as geometrically exact beam theory. Nowadays, it is also referred to as Simo-Reissner beam theory. According to the definition of Simo~\cite{simo1985}, also in this work, a beam theory is denoted as geometrically exact, if ``the relationships between the configuration and the strain measures are consistent with the virtual work principle and the equilibrium equations at a deformed state \textit{regardless of the magnitude of displacements, rotations and strains}'' (\cite{crisfield1999}, p. 1126). For that reason, also the notation ``finite-strain beams'' has been applied in the original work~\cite{simo1985}. However, as later argued by several authors (see e.g.~\cite{crisfield1999}) and in accordance with derivations in the literature (see e.g.~\cite{antmann1995,antman1974,love1944}, but also Section~\ref{sec:reissnerrelation3d1d} of this work), a consistency of the geometrically exact beam theory and the 3D theory of continuum mechanics in the sense of a (fully) induced beam theory can only be assumed as long as small strains are considered. The fulfillment of the basic kinematic assumption of rigid cross-sections underlying the geometrically exact beam theory requires pointwise six (translational and rotational) degrees of freedom in order to uniquely describe the (centroid) position and orientation of the cross-sections. Consequently, this beam theory can be identified as 1D Cosserat continuum~\cite{cosserat1909}, derived from a 3D Boltzmann continuum with pointwise three (translational) degrees of freedom. While there exists a variety of beam theories that also consider in-plane as well as out-of-plane cross-section distortion, this contribution focuses on geometrically exact beam formulations on the basis of the rigid cross-section assumption as applied by Simo and Reissner. Furthermore, throughout this work, the notion Simo-Reissner theory will be preferred since the notion geometrically exact beam theory, when following the definition presented above, also applies to consistently derived shear-free formulations on the basis of the Kirchhoff-Love theory. In the remainder of this work, the notion ``shear-free'' represents the opposite of "shear-deformable" and thus is equivalent to ``vanishing shear strains'', but, of course, not to "vanishing shear stresses''. Unfortunately, no absolute consensus concerning naming of different beam models can be found in the literature. For that reason, the following nomenclature, trying to be consistent with the most important representatives in the literature, will be applied here:\\

\begin{table}[h!]
\centering
\begin{tabular}{| m{4.2cm}|>{\centering\arraybackslash}m{1.4cm}|>{\centering\arraybackslash}m{1.1cm}|>{\centering\arraybackslash}m{1.1cm}|>{\centering\arraybackslash}m{1.1cm}|>{\centering\arraybackslash}m{1.1cm}|>{\centering\arraybackslash}m{1.1cm}|>{\centering\arraybackslash}m{1.1cm}|} \hline
Name  & Theory & $\kappa_0 \neq 0 $  & $I_2 \neq I_3$ & Tension & Shear & Torsion & Bending  \\ \hline
Simo-Reissner  & Nonlinear & +  & + & + & + & + & +  \\ \hline
(Anisotropic) Kirchhoff-Love  & Nonlinear & +  & + & + & - & + & +  \\ \hline
Straight Kirchhoff-Love  & Nonlinear & -  & + & + & - & + & +  \\ \hline
Isotropic Kirchhoff-Love \newline Nonlinear Euler-Bernoulli  & Nonlinear & -  & - & + & - & + & +  \\ \hline
Torsion-Free Simo-Reissner & Nonlinear & -  & - & + & + & - & +  \\ \hline
Torsion-Free Kirchhoff-Love & Nonlinear & -  & - & + & - & - & +  \\ \hline
Timoshenko  & Linear & -  & + & + & + & + & +  \\ \hline
Euler-Bernoulli  & Linear & -  & + & + & - & + & +  \\ \hline
Inextensible ...   & ... & ...  & ... & - & - & ... & ...  \\ \hline
\end{tabular}
\caption{Nomenclature applied within the current contribution for different (geometrically exact) beam theories / models.}
\label{tab:beam_models}
\end{table}

Geometrically nonlinear beam models capturing the modes of axial tension, torsion and bending and being appropriate for initially straight beams with isotropic cross-section shapes (identical moments of inertia of area $I_2\!=\!I_3$) are denoted as nonlinear Euler-Bernoulli beams. The extension to anisotropic cross-section shapes (different moments of inertia of area $I_2\! \neq \!I_3$) and arbitrary initial curvatures $\kappa_0 \! \neq \!0$ is referred to as Kirchhoff-Love theory. In order to simplify the subsequent comparison of shear-free beam formulations from the literature the more refined nomenclature of i) \textit{isotropic}, ii) \textit{straight} and iii) \textit{anisotropic} Kirchhoff-Love formulations is introduced for theories capturing i) neither anisotropic cross-section shapes nor initial curvatures (=nonlinear Euler-Bernoulli), ii) anisotropic cross-section shapes but no initial curvatures or iii) both, anisotropic cross-section shapes and initial curvatures. The supplementation by shear deformation modes is covered by the Simo-Reissner theory. Euler-Bernoulli or Kirchhoff-Love formulations that neglect the mode of axial tension are denoted as inextensible Euler-Bernoulli or Kirchhoff-Love variants. Recently, shear-deformable and shear-free beam formulations have been proposed that also neglect the mode of torsion~\cite{romero2014,meier2015}. These formulations are capable of accurately modeling ropes and cables (providing a mechanically consistent bending stabilization), quasi-continua such as chains but also general 1D continua with full bending and torsional stiffness if some restrictions concerning external loads (no torsional moment loads) and initial geometry ($I_2\!=\!I_3$,$\kappa_0 \! = \!0$) are fulfilled. Eventually, the restriction of shear-free or shear-deformable theories to the geometrically linear regime yields the linear Euler-Bernoulli and Timoshenko beam models~\cite{timoshenko1921}. An overview of these beam models is given in Table~\ref{tab:beam_models}.\\

By identifying the configuration space underlying the geometrically exact beam theory as nonlinear, differentiable manifold with Lie group structure and by pointing out important algorithmic consequences resulting from the non-additivity and non-commutativity of the associated group elements, the original work by  Simo~\cite{simo1985} and the subsequent work by Simo and Vu-Quoc~\cite{simo1986} laid the foundation for abundant research work on this topic in the following years. The static beam theory~\cite{simo1985,simo1986} has been extended to dynamics by  Cardona and Geradin~\cite{cardona1988,cardona1991} and by Simo and Vu-Quoc~\cite{simo1988}. The contributions of Kondoh et al.~\cite{kondoh1986}, Dvorkin et al.~\cite{dvorkin1988} as well as Iura and Atluri~\cite{iura1988} can be regarded as further pioneering works in this field. These contributions mark the starting point for the development of a large variety of geometrically exact beam element formulations~\cite{crivelli1993,ibrahimbegovic1995a,ibrahimbegovic1995b,smolenski1998,petrov1998,jelenic1995,jelenic1998,sansour2003} which basically differ in the type of rotation interpolation (e.g. interpolation of incremental, iterative or total rotational vectors), the choice of nodal rotation parametrization (via rotation vectors, quaternions etc.), the type of iterative rotation updates (multiplicative or additive), or the time integration scheme applied to the rotational degrees of freedom (e.g. based on additive or multiplicative rotation increments). Also extensions of the geometrically exact beam theory to arbitrary cross-section shapes with shear centers differing from the cross-section centroid can be found~\cite{gruttmann1998}. An overview of the most important developments at that time is exemplarily given in the text books of \mbox{Crisfield}~\cite{crisfield1997a} as well as Geradin and Cardona~\cite{geradin2001}. An break in this development is given by the works of \mbox{Crisfield} and Jeleni\'{c}~\cite{crisfield1999,jelenic1999}, who have shown that none of the rotation field discretizations of the formulations existent at that time could preserve both of the important properties objectivity and path-independence (see also~\cite{ibrahimbegovic2002} for a discussion of this topic). Furthermore, in~\cite{crisfield1999} and~\cite{jelenic1999}, a new, objective and path-independent orthogonal interpolation scheme was proposed that directly acts on the rotation manifold and not on any of its rotation vector parametrizations as done in the works before. This formulation was the starting point for the development of many alternative rotation interpolation strategies for geometrically exact beams that also preserve these properties. Among others, orthogonal interpolations of relative rotation vectors (see e.g. \cite{ghosh2009,sander2010}) or quaternions (see e.g. \cite{ghosh2008,romero2004,zupan2013}),  non-orthogonal interpolation strategies in combination with modified beam models (see e.g. \cite{romero2002,betsch2002,eugster2013}) and non-orthogonal interpolation strategies with subsequent orthogonalization (see e.g. \cite{romero2004}) can be identified. As reported in the original work~\cite{crisfield1999,jelenic1999}, the rotation interpolation scheme proposed by Crisfield and Jeleni\'{c} can exactly represent the state of constant curvature. Thus, it can be interpreted as geodesic, i.e. shortest, connection between two configurations on the rotation manifold. Consequently, these geodesic rotation interpolation schemes represent the counterpart to linear interpolations of translational quantities. The works~\cite{borri1994,borri1994b} as well as the recent contributions~\cite{sander2010,cesarek2012,sonneville2014,sonneville2014b} can be identified as further geometrically exact beam element formulations based on geodesic interpolations of the rotational (and translational) primary variable fields. A 2D extension of these so-called helicoidal interpolations to higher-order elements is given in~\cite{dukic2014}. A formulation with smooth centerline representation based on an isogeometric collocation scheme is proposed in~\cite{weeger2016}. Besides these purely displacement-based elements, also interpolation schemes directly acting at the strain level combined with a subsequent derivation of the position and rotation field via integration (see e.g. \cite{zupan2003,zupan2006,cesarek2012}) as well as mixed formulations ~\cite{santos2011b} have been proposed. Furthermore, a variety of contributions considering time integration of rotational variables can be found in this context~\cite{simo1991,bottasso1998,gonzalez1996,bachau1999,betsch2001,ibrahimbegovic2002b,lens2008,romero2008b,kane2000,demoures2015}.\\

\begin{figure}[ht]
 \centering
  \includegraphics[width=0.9\textwidth]{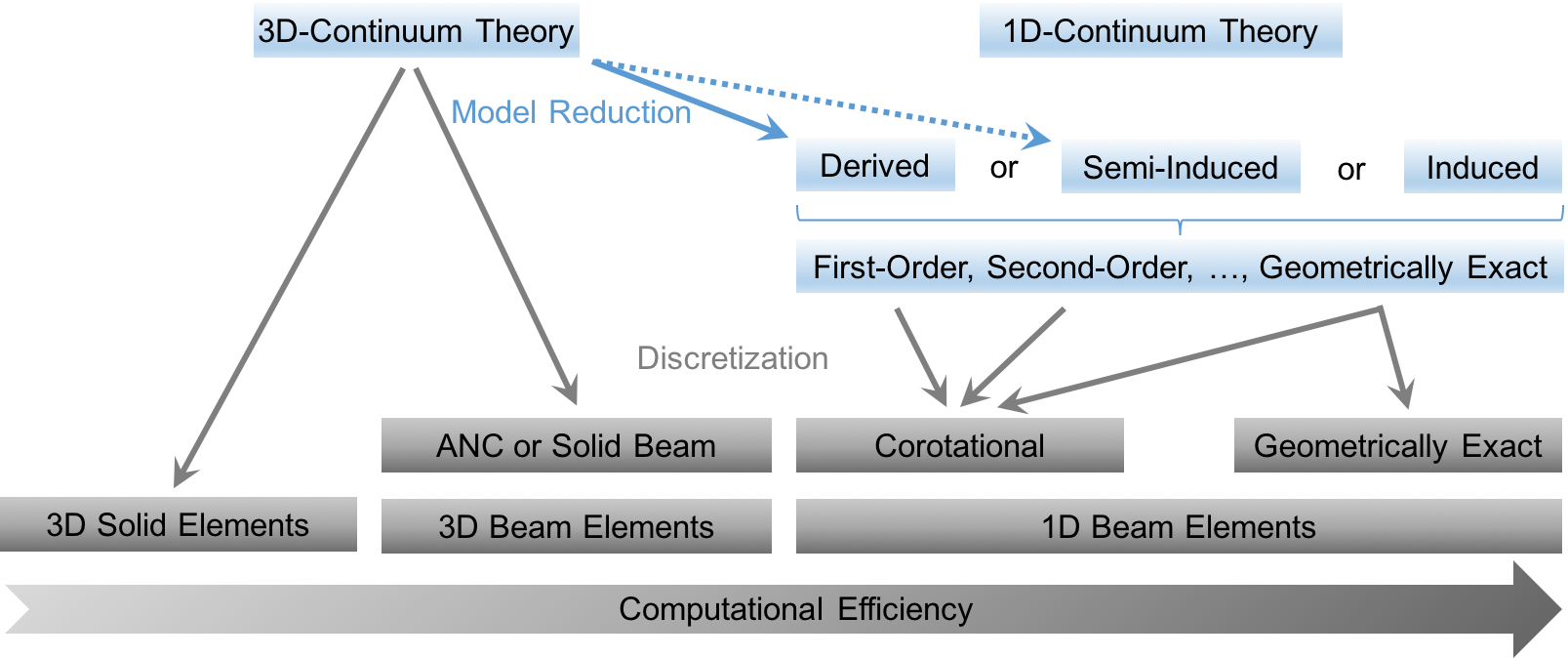}
  \caption{Overview of different discretization techniques for nonlinear problems of slender continua based on the finite element method.}
  \label{fig:intro_beamdiscretizations}
\end{figure}

While all the formulations presented above have been based on the finite element method (FEM), also discrete representatives of Simo-Reissner beam formulations based on finite difference schemes can be found in the literature~\cite{lang2011,lang2012,linn2013,jung2011}. These are often denoted as discrete elastic rods and based on the concept of discrete differential geometry (DDG). In the context of finite element formulations for geometrically nonlinear beam problems, again, a variety of alternatives to the geometrically exact formulations considered in the last two paragraphs can be found. The maybe most prominent representatives of these alternatives are the corotational method~\cite{crisfield1990,crisfield1997a,crisfield1997b,felippa2005,geradin1989} as well as Absolute Nodal Coordinate (ANC)~\cite{shabana2001,shabana1998} and solid beam element~\cite{bathe1979,frischkorn2013} formulations. The corotational technique was initially introduced by Wempner~\cite{wempner1969} as well as Belytschko et al.~\cite{belytschko1973,belytschko1979} and shows strong similarities to the "natural approach" of Argyris et al.~\cite{argyris1979b}. The basic idea is to split the overall non-linear deformation into a contribution stemming from large rotations and a part stemming from local deformations expressed in a local, "corotated" frame. Often, the local deformation can be modeled on the basis of first- (or second-) order theories such that the entire degree of nonlinearity is represented by the rotation of the local frame. The basic idea of ANC beam element formulations is to employ standard shape functions as known from solid finite element formulations in order to interpolate the 3D displacement field within the beam. Instead of introducing a kinematic constraint and deriving a resultant 1D model, different polynomial degrees are typically applied for the interpolation in beam length direction and in transverse directions. Numerical comparisons as performed e.g. by Romero~\cite{romero2004,romero2008} and Bauchau et al.~\cite{bachau2014} advocate geometrically exact beams in general, and orthogonal triad interpolation schemes (see e.g.~\cite{crisfield1999}) in particular, with regard to computational accuracy and efficiency. Especially the arguments I-III stated below in terms of possible benefits of shear-free as compared to shear-deformable beam element formulations in the range of high beam slenderness ratios are pronounced even stronger when comparing geometrically exact Kirchhoff-Love beam element formulations with ANC beam element formulations, with the latter additionally exhibiting highly stiff in-plane deformation modes when dealing with very slender beams. On the other hand, a clear distinction between corotational and geometrically exact formulations will not always be possible. For example, the interpolation scheme of the geometrically exact formulation proposed by \mbox{Crisfield} and Jeleni\'{c}~\cite{crisfield1999,jelenic1999} is also based on the definition of an element-local, corotated reference triad and can consequently be interpreted as corotational formulation at the same time. All in all, it can be stated that geometrically exact finite element formulations have become well-established in the meantime and can arguably be regarded as state-of-the-art methods for the computational treatment of geometrically nonlinear beam problems. An overview of the different FEM discretiation techniques is given in Figure~\ref{fig:intro_beamdiscretizations}.\\

In the context of the geometrically nonlinear Kirchhoff-Love beam theory, several discrete realizations based on finite difference schemes have recently been proposed~\cite{goyal2005,bergou2008,bertails2006,lazarus2013,ademir2014}. In contrast to the Simo-Reissner theory, also several works based on an analytic treatment of Kirchhoff-Love beam problems exist in the modern literature~\cite{shi1994,langer1996}. Interestingly, most approaches of these two categories can be found in the field of bio- or molecular physics. Although, the theoretical basis of shear-free Kirchhoff-Love beam formulations has a much longer tradition than the Simo-Reissner theory of shear-deformable beams, there are only a few geometrically nonlinear shear-free finite element representations, which have not reached the excellent properties of geometrically exact Simo-Reissner formulations so far. In his recent works~\cite{armero2012a} and~\cite{armero2012b}, Armero and Valverde gave a historic overview of existing Kirchhoff finite elements and pointed out their drawbacks. Accordingly, the first Kirchhoff type element formulations have applied different interpolations ($C^0$-continuous Lagrange polynomials for the axial displacements and $C^1$-continuous Hermite polynomials for the transversal displacements) to the different displacement-components, which led to a loss of objectivity~\cite{armero2012a}. In later works, the objectivity could be preserved by employing e.g. trigonometric shape functions, but the corresponding formulations were limited to the investigation of plane circular arches (see e.g.~\cite{ashwell1971a} or~\cite{ashwell1971b}). A different approach was the development of framework or corotational shear-free beams~\cite{crisfield2003,hsiao1999,hsiao1993,hsiao1994,le2014}, a category of formulations, which naturally preserves the objectivity of the continuous problem. As pointed out in~\cite{armero2012a}, these types of Kirchhoff type formulations often exhibit a comparatively poor accuracy; a fact, which can directly be traced back to the lack of an exact representation of the kinematic quantities. A further critical issue relevant in the context of thin Kirchhoff beams is membrane locking, a locking phenomenon given distinction to by Stolarski and Belytschko~\cite{stolarski1982}. In general, membrane locking denotes the inability of curved structural elements, e.g.~beams or shells, to represent the inextensibility constraint of vanishing membrane / axial strains. For thin Kirchhoff beams,~\cite{fried1973} was one of the first contributions in which this effect was investigated by relating the beam slenderness ratio to the condition number of the stiffness matrix, but without explicitly using the term locking. Diverse methods have been proposed in the literature in order to avoid membrane locking of Kirchhoff rods. Amongst others, these are the approaches of reduced / selective integration (see e.g.~\cite{noor1981,prathap1985,prathap1990}), assumed strains based on the  Hu-Washizu functional  (see e.g.~\cite{choi1995,karamanlidis1987,lee1993}), assumed stresses based on the Hellinger-Reissner functional (see e.g.~\cite{cannarozzi2008,noor1981}) or penalty relaxation / stabilization techniques in combination with membrane correction factors (see e.g.~\cite{lyly1993,tessler1986}). Most of these works are limited to 2D beam problems. A historic overview concerning the development of Kirchhoff beam elements in general and the key issues of objectivity and membrane locking in particular is given in the recent works of Armero and Valverde~\cite{armero2012a,armero2012b}.\\

The shear-free beam elements presented by now are typically based on additional kinematic assumptions, thus not being consistent with the concept of geometrically exact beams. Besides these contributions considered so far, also finite element and finite difference discretization approaches for Kirchhoff beams can be found in the literature which are in fact geometrically exact, but which rely on \textit{global}  interpolation strategies~\cite{yang1993,bertails2006,fan2016}. These are typically based on a rotation or curvature interpolation strategy and a subsequent integration of the rotation field along the entire beam length in order to yield an explicit beam centerline representation. Unfortunately, these global approaches typically rely on a serial finite element evaluation and yield dense system matrices and not the very desirable sparse system matrices with small bandwidths as typical for standard FEM approaches. Consequently, these approaches are not suitable for high-performance computing and will not further be considered in the following. The number of existing \textit{geometrically exact} (locally supported) finite element formulations on the basis of the Kirchhoff-Love theory is very limited. For example, the recent contribution of Sansour~\cite{sansour2015} proposes an energy-momentum method for 2D, initially straight geometrically exact elements based on the nonlinear Euler-Bernoulli theory. On the other hand, Armero and Valverde developed plane and three-dimensional locking-free Kirchhoff-Love beam element formulations accounting for arbitrarily curved initial geometries as well as anisotropic cross-section shapes that guarantee the fundamental properties of objectivity and geometrical exactness~\cite{armero2012a,armero2012b}. However, these beam elements only cover the geometrically linear case of infinitesimal deformations, but not the general large-deformation case. \\

Only a few geometrically exact, 3D, large-deformation, shear-free beam elements can be found in the literature. Arguably the first formulations of this kind have been proposed by Boyer~\cite{boyer2004} and Weiss~\cite{weiss2002a,weiss2002b}. In his recent work~\cite{boyer2011}, Boyer extended the original formulation~\cite{boyer2004} for the modeling of undersea cables. Also the essential requirement of objectivity and path-independence ar fulfilled by the approaches of Boyer and Weiss. However, these geometrically nonlinear Euler-Bernoulli formulations only treat the special case of beams with circular cross-sections and a straight initial configuration, i.e. the case of beams with rotationally symmetric reference geometry in the following denoted as \textit{isotropic} case (see first column of Table~\ref{tab:comparison_requirements}). This limitation simplifies the theory considerably, and already the modeling of simple piecewise straight frames is difficult since no variables are available that determine the cross-section orientation required for kinematic constraints at beam-to-beam joints. The more recent contributions of Zhang and Zhao~\cite{zhang2015,zhao2012} allow for anisotropic cross-sections but still focus on initially straight beams (denoted as \textit{straight} case in the second column of Table~\ref{tab:comparison_requirements}). However, later it will become clear that the first-order twist angle interpolation underlying these formulations might in general not allow for optimal spatial convergence rates considering the employed third-order centerline interpolation. In the very recent works of Greco et al.~\cite{greco2013,greco2016}, first steps towards geometrically nonlinear isogeometric Kirchhoff-Love beam elements accounting for initial curvature and anisotropic cross-sections have been made. However, the formulation has only been applied to geometrically linear examples. Bauer et al.~\cite{bauer2016} adapted the ideas of Greco and facilitated an application to geometrically nonlinear examples. These formulations are denoted as \textit{anisotropic} case in Table~\ref{tab:comparison_requirements}. Unfortunately, important properties such as objectivity and path-independence have not been considered in the works of Greco and Bauer. Also the question of membrane locking has not yet been consistently treated within the geometrically nonlinear realm by the textit{isotropic}, \textit{straight} and \textit{anisotropic} Kirchhoff-Love formulations considered in this paragraph. Nevertheless, they can be considered as the most elaborate geometrically exact Kirchhoff-Love beam element formulations from the literature, since being the only ones accounting for the numerical treatment of general 3D, large-deformation problems. Consequently, these formulations will be in the focus of detailed evaluations and comparisons performed throughout this work.\\

All in all, it can be concluded that none of the existing geometrically exact shear-free beam elements of Kirchhoff-Love type is comparable to the existing shear-deformable formulations of Simo-Reissner type in terms of generality and fulfillment of essential properties so far. Also detailed comparisons of shear-free and shear-deformable geometrically exact beam elements in the range of high slenderness ratios  are still missing. This backlog is the motivation for the development of novel geometrically exact Kirchhoff-Love beam elements fulfilling the following essential requirements:\\

\textbf{1) Geometrical exactness: } As already mentioned above, the proposed beam element formulations have to be geometrically exact in the sense that the derived deformation measures  are consistent with the virtual work principle and the equilibrium equations at any deformed state independent of the magnitude of displacements, rotations and strains.\\

\textbf{2) Representability of general geometries and external loads: } The most general case of 3D, large-deformation problems of thin beams with a) anisotropic cross-section shapes as well as b) arbitrary initial geometries and curvatures have to be represented. c) The proposed rotation interpolation schemes have to be capable of representing such general scenarios without exhibiting any singularities for practically relevant configurations.\\

\textbf{3) Fulfillment of essential mechanical principles: } The essential mechanical principles of objectivity, i.e. observer or frame invariance, as well as path-independence have to be preserved by the employed discretization schemes.\\

\textbf{4) Fulfillment of general requirements on spatial finite element discretization: } For all finite element formulations presented in this work, accuracy will be verified in terms of convergence towards proper reference solutions. Furthermore, the following criteria have to be fulfilled: a) Optimal spatial convergence rates measured in appropriate error norms have to be achieved. b) Furthermore, effective methods for the avoidance of membrane locking are required such that no remaining locking-related deterioration of the spatial convergence behavior will be observed for the resulting finite elements, even in the range of very high slenderness ratios. c) Further properties such as conservation of linear momentum, conservation of angular momentum and conservation of energy for arbitrarily rough spatial discretizations are very desirable for the proposed spatial interpolation schemes. d) It can also be beneficial for beam element formulations to fulfill simple patch tests, e.g. to exactly represent the states of constant axial tension, torsion and bending. e) It may be preferable that the finite element solution is independent from the chosen node numbering.\\

\textbf{5) Fulfillment of general requirements on temporal discretization: } The main focus of this work lies on the development and evaluation of spatial finite element discretizations for geometrically nonlinear beam problems. Nevertheless, it is at least required that a) dynamics can be represented in general, and that b) energy-stable time integration can be achieved. c) Often, energy-momentum conserving time integration schemes may be favorable.\\

\textbf{6) Simple realization of essential boundary conditions and joints: } Choices of nodal primary variables, especially with respect to the rotation parametrization, are demanded that enable the formulation of practically relevant Dirichlet boundary conditions but also of node-wise joints between several beams without the need of additional constraint equations. This is not standard for most of the existing geometrically exact shear-free beam element formulations.\\

\textbf{7) Avoidance of Lagrange multipliers and saddle point systems: } Also existing geometrically exact beam elements of Simo-Reissner type can be subjected to the Kirchhoff constraint via additional Lagrange multiplier fields. Such a procedure typically results in saddle point systems and the need of a special class of linear solvers or requires global condensation strategies. The shear-free beam elements considered here should neither rely on Lagrange multipliers nor should they yield in saddle point systems. All required calculations have to be feasible in an element-local manner.\\

\textbf{8) Suitability for high-performance computing: } In the literature, several finite difference and finite element discretizations of Kirchhoff-Love beams have been proposed that rely on global strategies for rotation field construction. Even though, these schemes show otherwise desirable properties, they typically suffer from two elementary drawbacks: Mostly, these schemes result in dense discrete system matrices and depend on a successive, i.e. serial, evaluation of the individual finite elements within a discretization. These two properties make such formulations virtually impossible for high-performance computing. From the finite elements considered here, it is required to result in sparse system matrices with small bandwidths and to be suitable for parallel computing element evaluation routines.\\ 

\begin{table}[h!]
\centering
\begin{tabular}{| p{4.8cm}|>{\centering\arraybackslash}p{1.4cm}|>{\centering\arraybackslash}p{1.3cm}|>{\centering\arraybackslash}p{1.5cm}|>{\centering\arraybackslash}p{1.5cm}|>{\centering\arraybackslash}p{1.3cm}|>{\centering\arraybackslash}p{1.4cm}|} \hline
Type  & \multicolumn{5}{c|}{Kirchhoff-Love}  & Simo-R. \\ \hline
Source  & \multicolumn{3}{c|}{Literature} & \multicolumn{2}{c|}{Authors}  & Literature \\ \hline
Formulation  & Isotropic & Straight & Anisotropic & \cite{meier2014,meier2015} & Current  & \cite{crisfield1999,jelenic1999} \\ \hline
1) \, Geometrically Exact   & + &  + &  + &  + & + & + \\ \hline
2a) Initial Curvature & - &  - &  + &   + &  + & + \\ \hline
2b) Anisotropic Cross-Sections& - &  + &  + &   + &  + & + \\ \hline
2c) Singularity-Free & + & + &  - &   + &  + & + \\ \hline
3) \, Objective \& Path-Independent   & + &  + &  - &  + & + & + \\ \hline
4a) Optimal Convergence Order & + &  - &  + &  + & + & + \\ \hline
4b) Locking Avoided & o &  ? &  o  &  + & + & + \\ \hline
4c) Conservation Properties (FE) & - & + & - & - & + & + \\ \hline
4d) Patch Test Passed & - &  - &  -  &  - & o & o \\ \hline
4e) Symmetric Discretization & + &  - &  - &  - & + & + \\ \hline
5a) Suitable for Dynamics & + &  + &  - &  - & + & + \\ \hline
5b) Energy-Stable (FD) & o &  ? &  ?  &  ?  & + & - \\ \hline
5c) Conservation Properties (FD) \!\!\! & o &  - &  ?  &   ? &  - & - \\ \hline
6) \, Realization of DBCs \& Joints &  - &  + &  - &  - &  + & + \\ \hline
7) \, Lagrange Multipliers Avoided & + &  + &  + &  + &  + & + \\ \hline
8) \, High-Performance Computing & + &  + &  + &  + &  + & + \\ \hline
\end{tabular}
\caption{Fulfillment of essential requirements on geometrically exact beam elements: Comparison of existing and proposed formulations.}
\label{tab:comparison_requirements}
\end{table}

In the authors' recent contribution~\cite{meier2014}, the first 3D large-deformation geometrically exact Kirchhoff-Love beam element formulation that fulfills the essential properties of objectivity and path-independence and that is capable of representing arbitrary initial curvatures and anisotropic cross-section shapes has been proposed. In the subsequent work~\cite{meier2015}, also the important question of membrane locking has successfully been addressed. The current contribution extends these methodologies by providing considerable improvements in terms of accuracy and practical applicability, a generalization to dynamic problems as well as theoretical and numerical comparisons to existing formulations. In Table~\ref{tab:comparison_requirements}, the fulfillment of the essential requirements stated above is verified for the existing categories of 3D, large-deformation geometrically exact Kirchhoff-Love beam elements identified so far, i.e. for the isotropic, straight and anisotropic formulations from the literature, the author's recent contributions~\cite{meier2014,meier2015} as well as the formulations proposed in the current work. Of course, not all of the stated requirements have explicitly been investigated in the mentioned references. However, the evaluation results presented in Table~\ref{tab:comparison_requirements} have been derived on the best of the authors' knowledge. Explanations and Derivations required for the evaluation of the individual categories can be found in Sections~\ref{sec:temporaldiscretization}-\ref{sec:eleweakkirchhoff} in form of paragraphs beinning with the expression "\textbf{Review:} ...". In general, the symbols "+", "o" and "-" in Table~\ref{tab:comparison_requirements} mean that a requirement is fully, partly or not fulfilled. The question mark~"?" indicates that the required information could not definitely be extracted from the literature. According to Table~\ref{tab:comparison_requirements}, the formulations newly proposed in the current work can indeed close most of the gaps left by existing geometrically exact Kirchhoff-Love formulations. However, several existing Simo-Reissner beam element formulations already fulfill most of the stated requirements, a fact that might indicate the increased complexity of consistently incorporating the additional constraint of vanishing shear strains in Kirchhoff-Love formulations. In order to underline this statement, also the well-known Simo-Reissner type formulation proposed by Crisfield and Jeleni\'c~\cite{crisfield1999,jelenic1999}, the first one of this type preserving objectivity and path-independence, is represented as one possible example in Table~\ref{tab:comparison_requirements} (right column).\\

Given the excellent properties of existing geometrically exact shear-deformable finite element formulations, the question arises which benefits can be gained from applying Kirchhoff-Love instead of Simo-Reissner beam element formulations. It is quite obvious, that the range of low beam slenderness ratios, where shear-deformation is not negligible, requires the application of beam element formulations taking this fact into account. Also the underlying continuum theory, based on an unconstrained 1D Cosserat continuum with pointwise six degrees of freedom, and the resulting discrete problem statement seem to be easier to be formulated for the Simo-Reissner case. However, in the range of high beam slenderness ratios, thus, exactly in scenarios prevalent in many of the practically relevant applications mentioned in the beginning of this contribution, shear-free beam element formulations of Kirchhoff-Love type may exhibit considerable advantages. In the following, some of these potential benefits shall be outlined:\\

\textbf{I) Lower discretization error level: } The most essential difference between the proposed Kirchhoff-Love and existing Simo-Reissner beam element formulations lies in the neglect of shear deformation in the beam theory underlying the former category. This property is independent from the chosen discretization strategy. Consequently, if otherwise comparable interpolation strategies are applied, it can be assumed that the Kirchhoff type formulations require less degrees of freedom in order to yield the same polynomial approximation quality, and eventually the same discretization error level, since no additional primary variables are required in order to represent shear deformation.\\

\textbf{II) Numerical benefits in the range of high slenderness ratios: } In the range of very high slenderness ratios, the influence of the shear modes on the overall beam deformation is not only negligible, it can also be very beneficial to abstain from these high stiffness contributions from a numerical point of view: Mechanical problems of slender beams typically lead to stiff differential equations and ill-conditioned numerical problems deteriorating the performance of time integration schemes, nonlinear solvers and linear solvers. The avoidance of the stiff shear mode contributions can considerably improve the situation. Concretely, detailed numerical investigations on several numerical test cases involving highly slender beams will be considered. These test cases will reveal a considerably improved performance of nonlinear solution schemes when Kirchhoff type instead of Reissner type discretizations are employed. Similar trends will be predicted - at least theoretically - for the behavior of linear solvers and time integration schemes.\\ 

\textbf{III) Smooth geometry representation: } The proposed Kirchhoff-Love beam elements will be based on $C^1$-continuous centerline interpolations. These interpolations will eventually result in smooth beam-to-beam contact kinematics, a property that is highly desirable in order to yield efficient and robust contact algorithms (see e.g.~\cite{meier2015b,meier2015c}).\\

\textbf{IV) Derivation of reduced models: } In~\cite{meier2015}, a special reduced model, denoted as torsion-free beam theory, has been derived from the general Kirchhoff-Love theory. This reduced model has been shown to only be valid for special problem classes concerning beam geometry and external loads, which are, however, present in many fields of application. The finite elements resulting from such a reduced model typically feature a simplified numerical implementation and an increased computational efficiency. Although this torsion-free theory does not lie in the focus of this work, it has to be emphasized1 that the proposed Kirchhoff-Love formulations lie the foundation for such reduced models.\\

These possible benefits will be further detailed in Section~\ref{sec:reissnermotivationshearfree} and verified numerically in Section~\ref{sec:elenumericalexamples}.
After having motivated the intention of the current work, its main scientific contributions shall briefly be highlighted:

\begin{enumerate}
\item[$\bullet$] Two novel rotation interpolation strategies for geometrically exact Kirchhoff-Love beam element formulations are proposed: i) The first represents an orthonormal interpolation scheme that fulfills the Kirchhoff constraint in a strong manner. It can be regarded as generalization of the scheme proposed in~\cite{meier2014} that allows for an exact conservation of energy and momentum and for alternative sets of nodal degrees of freedom that simplify the prescription of essential boundary conditions. ii) The space-continuous theory of the second variant is based on a weak enforcement of the Kirchhoff constraint. The discrete realization of the Kirchhoff constraint relies on a properly  chosen collocation strategy. In combination with the employed smooth centerline interpolation, also this strategy can completely avoid any additional Lagrange multipliers. For each of these two element formulations, two different sets of nodal rotation parametrizations are proposed. One is based on nodal rotation vectors and one on nodal tangent vectors. While these different choices are shown to yield identical FEM solutions, they differ in the resulting performance of nonlinear solvers and the effort required to prescribe essential boundary conditions and joints. The four finite element formulations resulting from a combination of the two interpolation schemes and the two choices of nodal primary variables are subject to detailed comparisons with respect to resulting discretization error levels and the performance of nonlinear solution schemes.\\
\item[$\bullet$] The resulting finite element formulations are combined with a finite difference time integration scheme for large rotations recently proposed in the literature by Br\"uls and Cardona~\cite{bruels2010}. This implicit scheme allows for energy-stable, second-order accurate time integration on the basis of optimized numerical dissipation and can be identified as a Lie-group extension of the well-known generalized-$\alpha$ scheme. Up to the best of the author's knowledge, the current work represents the first application of a Lie group time integration scheme based on optimized numerical dissipation to geometrically exact Kirchhoff-Love beam elements, and one of the first applications of such a scheme to geometrically nonlinear beam element formulations at all.\\
\item[$\bullet$] Furthermore, the current work intends to review and evaluate existing geometrically exact beam element formulations and to compare these with the formulations newly proposed in this work. Concretely, the fulfillment of the essential requirements according to Table~\ref{tab:comparison_requirements} will be studied \textit{analytically} for the three identified categories of isotropic, straight and anisotropic formulations, which can be identified as the most general and elaborated 3D, large-deformation geometrically exact Kirchhoff-Love formulations previously proposed in the literature, for one exemplary representative of Simo-Reissner formulations from the literature as well as for the Kirchhoff-Love formulations presented in the current and former works of the authors. A further original contribution of this work is given by the detailed and systematic \textit{numerical} comparison performed between the proposed geometrically exact Kirchhoff-Love and two representative Simo-Reissner beam element formulations from the literature. Specifically, resulting spatial convergence rates, discretization error levels as well as the performance of nonlinear solution schemes are compared for different beam slenderness ratios.
\end{enumerate}

Eventually, the organization of the remainder of this contribution shall briefly be given. One distinctive property of geometrically exact beam formulations is the presence of large rotations within the associated configuration space. In order to provide the theoretical basis for subsequent derivations, in the following Section~\ref{sec:largerotations}, the $S \! O(3)$ group of large rotations as well as possible parametrizations will be introduced. In Section~\ref{sec:simoreissner}, the most general type of geometrically exact beam formulations considered in this work, the Simo-Reissner theory of thick (shear-deformable) beams, will be presented. Subsequently, in Section~\ref{sec:kirchhoff}, the general theory will be restricted to the Kirchhoff-Love theory of thin (shear-free) beams. There, the different methodologies of imposing the Kirchhoff constraint of vanishing shear strains in a strong or weak sense will be investigated. Afterwards, the space-time-continuous beam problem will be discretized. In Section~\ref{sec:temporaldiscretization}, a recently proposed extension of the generalized-$\alpha$ method from vector spaces to Lie groups \cite{arnold2007,bruels2010,bruels2012}, will be applied to the beam formulations proposed in this work. Afterwards, in Section~\ref{sec:spatialdiscretization}, spatial discretization is performed, which represents the core topic in the development of geometrically exact Kirchhoff beam elements considered in this work. In Sections~\ref{sec:spatialdiscretization_beamcenterline} and~\ref{sec:spatialdiscretization_rotationfield}, specific finite element interpolations employed to the translational and rotational primary variable fields considered in this work will be proposed. In Section~\ref{sec:spatialdiscretization_requirements}, the already stated essential  requirements on the employed spatial discretizations and the resulting finite element formulations will be further detailed. Also detailed explanations and derivations for the evaluation results of Table~\ref{tab:comparison_requirements}
will be presented in Sections~\ref{sec:temporaldiscretization} and~\ref{sec:spatialdiscretization}. In the subsequent Sections~\ref{sec:elesimoreissner}-\ref{sec:eleweakkirchhoff}, different finite element realizations resulting from the proposed interpolation schemes and rotation parametrizations will be presented and the fulfillment of the basic requirements stated above will be confirmed analytically. Concretely, Section~\ref{sec:elesimoreissner} presents the basics of the geometrically exact Simo-Reissner type formulation proposed in~\cite{crisfield1999,jelenic1999}, which will serve as reference for several numerical comparisions performed in Section~\ref{sec:elenumericalexamples}. In the subsequent Section~\ref{sec:reissnermotivationshearfree}, the benefits of applying Kirchhoff-Love instead of Simo-Reissner element formulations in the range of high slenderness ratios will be further quantified. Afterwards, the two major developments of this work, a Kirchhoff-Love element formulation based on strong constraint enforcement as well as a Kirchhoff-Love element formulation based on weak constraint enforcement, will be proposed in Sections~\ref{sec:elestrongkirchhoff} and~\ref{sec:eleweakkirchhoff}. For each formulation, two different variants of nodal rotation parametrization, one based on nodal rotation vectors and one based nodal tangent vectors will be derived. Eventually, in Section~\ref{sec:elenumericalexamples}, all the proposed concepts and the resulting finite element formulations will be verified by means of proper numerical test cases. The reader who is primarily interested in the practical implementation of the newly proposed geometrically exact Kirchhoff-Love beam element formulations is referred to Sections~\ref{sec:temporaldiscretization},~\ref{sec:elestrongkirchhoff},~\ref{sec:eleweakkirchhoff} as well as to the appendices of this work.

\section{The Rotation Group SO(3)}
\label{sec:largerotations}

The category of beam theories considered in this contribution assumes the beam cross-sections to be rigid. Consequently, the cross-section 
kinematics are point-wise uniquely defined by six degrees of freedom, three translational ones representing the position vector of the cross-section 
centroid and three rotational ones describing the cross-section orientation. Thereto, an orthonormal triad consisting of the base vectors $\mb{g}_{1}, 
\mb{g}_{2}, \mb{g}_{3} \in \Re^3$ is attached on the beam cross-sections. Furthermore, a right-handed inertial Cartesian frame $\mb{E}_1, \mb{E}_2, \mb{E}_3\in \Re^3$ associated with the material configuration and a corresponding right-handed inertial Cartesian frame \mbox{$\mb{e}_1, \mb{e}_2, \mb{e}_3\in \Re^3$} of the spatial configuration are introduced. Nevertheless, for simplicity, it is assumed that both frames coincide, thus $\mb{e}_i=\mb{E}_i$ for $i=1,2,3$. Now, the rotation from the global frame $\mb{E}_{i}$ onto the local frame $\mb{g}_{i}$ is described via the orthogonal transformation $\mb{\Lambda} \in S\!O(3)$:
\begin{align}
\label{largerotations_lambda}
  \mb{g}_{i}=\mb{\Lambda} \mb{E}_{i}  \quad \text{with} \quad \mb{\Lambda} = \mb{g}_{j} \otimes \mb{E}_{j} = (\mb{g}_{1},\mb{g}_{2},\mb{g}_{3})_{\mb{E}_{j}} \quad \text{for} \quad  i,j=1,2,3.
\end{align}
Here and in the following, the summation convention over repeated indices holds. Throughout this work, the index near a matrix (for example the index $(.)_{\mb{E}_{i}}$ in equation~\eqref{largerotations_lambda}) denotes the basis in which the associated tensor is represented. In the context of geometrically exact beam theories, the two-point tensor $\mb{\Lambda}$ acts as push-forward operator (see e.g. \cite{marsden1994}) between material and spatial objects. From a rather mathematical point of view, the rotation tensor $\mb{\Lambda}$ can be identified as an element of the Special Orthogonal group $S \! O(3)$ of orthogonal transformations according to
\begin{align}
\label{largerotations_SO3}
  S \! O(3):= \{ \mb{\Lambda} \in \Re^{3\!\times\!3} | \mb{\Lambda}^T\!\mb{\Lambda} = \mb{I}_3, \text{det}(\mb{\Lambda})~=~1  \},
\end{align}
under the action of non-commutative multiplication $S \! O(3)\!\!\times\!\!S \! O(3) \!\!\rightarrow\!\! S \! O(3), \boldsymbol{\Lambda}_3\!\!=\!\!\boldsymbol{\Lambda}_1\boldsymbol{\Lambda}_2\!\!\neq\!\!\boldsymbol{\Lambda}_2\boldsymbol{\Lambda}_1$ with inverse element $\boldsymbol{\Lambda}^{-1}\!\!=\!\!\mb{\Lambda}^T\!\!$ and identity element $\mb{I}_3$. Here, $\text{det}(.)$, $(.)^T\!$ and $(.)^{-1}\!$ are the determinant, transpose and inverse. $\mb{I}_3$ is a $3\!\!\times\!\!3$ identity matrix. The nonlinear manifold $S \! O(3)$ can be classified as Lie group with tangent space
\begin{align}
\label{largerotations_tangentspace}
\delta \mb{\Lambda} \in T_{\mb{\Lambda}} S \! O(3):=\{ \mb{S}(\mb{a}) \mb{\Lambda} | \mb{S}(\mb{a}) \in  so(3)\}.
\end{align}
Here, $so(3)$ denotes the set of skew symmetric tensors with $\mb{S}(\mb{a})\mb{b} \!=\! \mb{a}\!\times\!\mb{b} \,\, \text{for} \,\, \mb{a},\mb{b}  \!\in\! \Re^3$. The isomorphism between $so(3)$ and $\Re^3$ enables a unique expression of $\mb{S}( \mb{a} ) \in so(3)$ by the vector $\mb{a} \in \Re^3$ denoted as axial vector. By inserting the special choice $\boldsymbol{\Lambda}\!=\!\mb{I}_3$ into~\eqref{largerotations_tangentspace} it can easily be verified that $so(3)$ represents the tangent space to $S \! O(3)$ at the identity:
\begin{align}
\label{largerotations_so3}
  so(3):= T_{\mb{I}} S \! O(3) = \{ \mb{S}( \mb{a} ) | \mb{S}( \mb{a} )=-\mb{S}( \mb{a} )^T \, \forall \, \mb{a} \in \Re^3  \}.
\end{align}
The Lie group $S \! O(3)$ and its Lie algebra $so(3)$ are related by the exponential map $\exp{(.)}\!\!: so(3)\! \rightarrow \!S \! O(3)$ according to:
\begin{align}
\label{largerotations_expmap}
\exp{ \! ( \mb{S}(  \mb{a} ) )  } := \mb{I}_3 \!+\! \mb{S}(\mb{a})
\!+\! \frac{\mb{S}(\mb{a})^2}{2!} \!+\! \frac{\mb{S}(\mb{a})^3}{3!}\!+\!... \,\,.
\end{align}
Up to now, the rotation group $S \! O(3)$ has been introduced without stating any specific parametrization of the rotation tensor $\mb{\Lambda}$. In the following sections, two possible parametrizations, which will be useful in the development of beam element formulations according to the Simo-Reissner and Kirchhoff-Love theory, will be presented.

\subsection{SO(3) parametrization via rotation vectors}
\label{sec:largerotations_rotvec}

There exist various parametrizations of the rotation tensor such as rotation (pseudo-) vectors, Euler angles or Rodrigues parameters that are based on a minimal set of three parameters. Also four-parametric representations of the rotation tensor such as quaternions have proven to be very useful for practical purposes. Within this work, two different parameterizations will be employed: The one presented in this section is based on rotation vectors $\boldsymbol{\psi} \!\in\! \Re^3$. In Section~\ref{sec:largerotations_sr}, an alternative parametrization especially suited for Kirchhoff formulations is presented. The rotation vector parametrization can explicitly be given via the well-known Rodrigues formula:
\begin{align}
\label{largerotations_rotrigues}
  \mb{\Lambda}(\boldsymbol{\psi}) = \exp{  \! ( \mb{S}(   \boldsymbol{\psi}  ) )   } = 
  \left[
  \mb{I}  +  \sin{\psi} \mb{S}(\boldsymbol{e_{\boldsymbol{\psi}}}) +  (1-\cos{\psi}) \mb{S}(\boldsymbol{e_{\boldsymbol{\psi}}}) \mb{S}(\boldsymbol{e_{\boldsymbol{\psi}} })
  \right].
\end{align}
Here, $\psi \!=\! ||\boldsymbol{\psi}||$ represents the scalar rotation angle and $\mb{e}_{\boldsymbol{\psi}} \!=\! \boldsymbol{\psi}/||\boldsymbol{\psi}||$ the axis of rotation. Throughout this work, $||(.)||$ denotes the Euclidean norm in $\Re^3$. As indicated by the notation $\exp{  \! ( \mb{S}(   \boldsymbol{\psi}  ) )   }$, equation~\eqref{largerotations_rotrigues} represents a closed-form representation of the exponential map initially introduced in~\eqref{largerotations_expmap} of the last section. The rotation vector of a given rotation tensor can for example be extracted by employing Spurrier's algorithm~\cite{spurrier1978}. The infinitesimal spatial quantity $\delta \boldsymbol{\theta}$, denoted as multiplicative rotation vector variation or spin vector, allows to express the variation of $\mb{\Lambda}$:
\begin{align}
\label{largerotations_deltalambdaspatial}
  \delta \mb{\Lambda} := \frac{d}{d \epsilon} \Big|_{\epsilon=0}\exp{(\epsilon \mb{S}(\delta \boldsymbol{\theta}))} \mb{\Lambda} = \mb{S}(\delta \boldsymbol{\theta}) \mb{\Lambda} \quad \text{or} \quad \delta \mb{g}_i = \delta \boldsymbol{\theta} \!\times\! \mb{g}_i.
\end{align}
Alternatively, the variation $\delta \mb{\Lambda}$ can be expressed by the material spin vector $\delta \boldsymbol{\Theta}=\mb{\Lambda}^T \delta \boldsymbol{\theta}$ via right translation:
\begin{align}
\label{largerotations_deltalambdamaterial}
  \delta \mb{\Lambda} := \frac{d}{d \epsilon} \Big|_{\epsilon=0} \mb{\Lambda} \exp{(\epsilon \delta \boldsymbol{\Theta})} =\mb{\Lambda}  \mb{S}(\delta \boldsymbol{\Theta}).
\end{align}
Expressing the spin vectors $\delta \boldsymbol{\Theta}$ and $\delta \boldsymbol{\theta}$ in the associated frames $\mb{E}_i$ and $\mb{g}_i$, respectively, yields:
\begin{align}
\label{largerotations_spincomponents}
  \delta \boldsymbol{\Theta}=\delta \Theta_i \mb{E}_i, \quad \delta \boldsymbol{\theta}=\delta \Theta_i \mb{g}_i.
\end{align}
Thus, the components of the spatial spin vector expressed in the local basis $\mb{g}_{i}$ are
identical to the components of the material spin vector expressed in the global basis $\mb{E}_{i}$. This relation also holds for all the other pairs of spatial / material quantities considered in this work (see e.g. \cite{simo1985}). Based on the defining equation~\eqref{largerotations_deltalambdamaterial} and the representation of skew-symmetric tensors by means of the Levi-Civita-Symbol $\epsilon_{ijk}$, the components of $\mb{S}(\delta \boldsymbol{\Theta})$ can be determined:
\begin{align}
 \label{largerotations_SdeltaTheta}
 \mb{S}(\delta \boldsymbol{\Theta}) \!=\!  \mb{\Lambda}^T \! \delta \mb{\Lambda} \!=\! (\mb{E}_{i} \!\otimes\! \mb{g}_{i}) \!\cdot\! (\delta \mb{g}_{j} \!\otimes\! \mb{E}_{j})  \!=\!\! \underbrace{\mb{g}_{i}^T \! \delta \mb{g}_{j}}_{=:S_{ij}(\delta \boldsymbol{\Theta})} \!\!\!
\mb{E}_{i} \!\otimes\! \mb{E}_{j} \,\,\,\, \text{with} \,\,\,\, S_{ij}(\delta \boldsymbol{\Theta})=-\epsilon_{ijk} \delta \Theta_k.
\end{align}
For later use, the components $\delta \Theta_i$ shall be expressed by $\mb{g}_{i}$ and $\delta \mb{g}_{i}$. From the previous equation~\eqref{largerotations_SdeltaTheta}, it follows:  
\begin{align}
\label{largerotations_deltaTheta}
 \delta \Theta_1 = \mb{g}_{3}^T \! \delta \mb{g}_{2} \!=\!-\mb{g}_{2}^T \! \delta \mb{g}_{3}, \quad \delta \Theta_2 = \mb{g}_{1}^T \! \delta \mb{g}_{3} \!=\!-\mb{g}_{3}^T \! \delta \mb{g}_{1}, \quad \delta \Theta_3 = \mb{g}_{2}^T \! \delta \mb{g}_{1} \!=\!-\mb{g}_{1}^T\!\delta \mb{g}_{2}.
\end{align}
A relation between infinitesimal additive and multiplicative increments is given by the tangent operator $\mb{T}$ defined as
\begin{align}
\label{largerotations_tmatrix}
  \delta \boldsymbol{\psi}=:\mb{T}\delta \boldsymbol{\theta}, \quad
  \mb{T}=\frac{1}{\psi^2} \mb{S}( \boldsymbol{\psi} ) \mb{S}( \boldsymbol{\psi} )^T
  + \frac{\psi/2}{\tan{(\psi/2)}} \left( \mb{I} - \frac{1}{\psi^2} \mb{S}( \boldsymbol{\psi} ) \mb{S}( \boldsymbol{\psi} )^T\right) -\frac{1}{2}\mb{S}( \boldsymbol{\psi} ).
\end{align}
The inversion of the previous equation~\eqref{largerotations_tmatrix}, expressing multiplicative by means of additive increments, is given by:
 \begin{align}
\label{largerotations_tmatrix_inv}
  \delta \boldsymbol{\theta}=\mb{T}^{-1}\delta \boldsymbol{\psi}, \,\,
  \mb{T}^{-1}\!=\!\frac{1}{\psi^2} \left(1\!-\!\frac{\sin{(\psi)}}{\psi} \right) \mb{S}( \boldsymbol{\psi} ) \mb{S}( \boldsymbol{\psi} )^T \!+\! \frac{\sin{(\psi)}}{\psi} \mb{I} \!+\! \frac{1\!-\!\cos{(\psi)}}{\psi^2}\mb{S}( \boldsymbol{\psi} ).
\end{align}
For details on the derivation of the transformations $\mb{T}$ and $\mb{T}^{-1}$, the interested reader is e.g.  referred to~\cite{simo1988,cardona1988,ibrahimbegovic1995b,crisfield1997a}. While the rotation vector parametrization presented so far represents a well-known tool in the formulation of geometrically exact beam elements of Simo-Reissner type, in the following section, an alternative parametrization of large rotations will be proposed which offers some advantages in the description of Kirchhoff type beam elements.

\subsection{SO(3) parametrization via "smallest rotation" triads}
\label{sec:largerotations_sr}

The alternative parametrization considered in this section consists of four degrees of freedom ($\mb{t}, \varphi$) with $\mb{t} \in \Re^3$ and $\varphi \in \Re$. In the context of Kirchhoff beam elements presented later in this work, $\mb{t}$ will be the \textit{non-unit }tangent vector aligned to the space curve representing the beam centerline. Due to the Kirchhoff constraint of vanishing shear strains, the first base vector $\mb{g_1}$ of the cross-section triad $\mb{\Lambda}\!=\!(\mb{g}_1, \mb{g}_2, \mb{g}_3)_{\mb{E}_i}$, can be expressed by this tangent vector:
 \begin{align}
\label{largerotations_unittangent}
  \mb{g_1}=\frac{\mb{t}}{||\mb{t}||}.
\end{align}
Based on the tangent vector $\mb{t}$ and an arbitrary given triad $\bar{\mb{g}}_{i}$, one can determine a triad $\mb{g}_{Mi}$, in the following denoted as interMediate or Medium triad (index M), that results when the triad $\bar{\mb{g}}_{i}$ is rotated onto the tangent vector $\mb{t}$ via the ``Smallest Rotation'' (SR) (see e.g. \cite{crisfield1997a,meier2014}. The resulting base vectors can be represented by the expressions
\begin{align}
\begin{split}
\label{largerotations_smallestrotation}
\mb{g}_{M1}\!=\!\mb{g}_{1}\!=\!\frac{\mb{t}}{||\mb{t}||},\,\,\,\,
\mb{g}_{M2}\!=\! \bar{\mb{g}}_{2} \!-\! \frac{\bar{\mb{g}}_{2}^T\mb{g}_{1}}{1 \!+\! \bar{\mb{g}}_{1}^T \mb{g}_{1}}\left(
\mb{g}_{1} \!+\! \bar{\mb{g}}_{1} \right), \,\,\,\,
\mb{g}_{M3}\!=\! \bar{\mb{g}}_{3} \!-\! \frac{\bar{\mb{g}}_{3}^T\mb{g}_{1}}{1 \!+\! \bar{\mb{g}}_{1}^T \mb{g}_{1}}\left(
\mb{g}_{1} \!+\! \bar{\mb{g}}_{1} \right) \quad \rightarrow \quad
\boldsymbol{\Lambda}_{M}=:\text{sr}(\bar{\boldsymbol{\Lambda}},\mb{g}_{1}).
\end{split}
\end{align}
In order to shorten notation, the abbreviation $\text{sr}(.)$ has been introduced for the SR mapping of~\eqref{largerotations_smallestrotation}. Practical choices for the triad $\bar{\boldsymbol{\Lambda}}$ will be presented in Section~\ref{sec:spatialdiscretization_rotationfield} (see also~\cite{meier2015} for a discussion of possible singularities of the SR mapping). Subsequently to the definition of an intermediate triad $\boldsymbol{\Lambda}_{M}$ according to~\eqref{largerotations_smallestrotation}, the cross-section triad $\mb{g}_i$ can be defined based on a relative rotation of the intermediate triad $\boldsymbol{\Lambda}_{M}$ with respect to the tangent $\mb{t}$ by an angle of $\varphi$:
\begin{align}
 \label{largerotations_2drot}
\boldsymbol{\Lambda}=\exp{ \! ( \mb{S}( \varphi \mb{g}_1) )} \boldsymbol{\Lambda}_M \quad  \Leftrightarrow  \quad
 \mb{g}_1 = \mb{g}_{M1}, \, \, \mb{g}_2 = &\mb{g}_{M2} \cos{\varphi} + \mb{g}_{M3} \sin{\varphi}, \, \, \mb{g}_3 = \mb{g}_{M3} \cos{\varphi} - \mb{g}_{M2} \sin{\varphi}.
\end{align}
Equations~\eqref{largerotations_smallestrotation} and~\eqref{largerotations_2drot} uniquely define a triad $\boldsymbol{\Lambda}$ parametrized by the four degrees of freedom $(\mb{t},\varphi)$. Evidently, one of these four degrees of freedom, namely the norm $||\mb{t}||$ of the tangent vector, will not influence the triad orientation. However, as it will turn out in the next sections, the non-unit tangent vector $\mb{t}$ is a quantity that directly results from the beam centerline description. Thus, the only additional degree of freedom introduced in order to describe the triad orientation is the relative angle $\varphi$ and consequently, the proposed type of triad parametrization is not redundant. For later use, the spatial spin vector $\delta \boldsymbol{\theta}$ shall be expressed by means of additive increments $(\delta \mb{t},\delta\varphi)$ of the four parameters $(\mb{t},\varphi)$. Therefore, it is split into a component $\delta
\boldsymbol{\theta}_{\parallel}$ parallel to $\mb{g}_1$ and a component $\delta \boldsymbol{\theta}_{\perp}$ perpendicular to $\mb{g}_1$ as follows:
\begin{align}
\label{largerotations_deltathetasplit}
  \delta \boldsymbol{\theta} = \delta \boldsymbol{\theta}_{\parallel} + \delta \boldsymbol{\theta}_{\perp} 
  = \delta \Theta_{1} \mb{g}_1 + \delta \boldsymbol{\theta}_{\perp}.
\end{align}
Throughout this work, the indices $(.)_{\parallel}$ and $(.)_{\perp}$ of a vector will denote the components of the vector which are parallel or perpendicular to the vector $\mb{g}_1$, respectively. Taking advantage of~\eqref{largerotations_deltalambdaspatial}, $\delta \boldsymbol{\theta}_{\perp}$ can be easily derived (see also~\cite{meier2015}):
\begin{align}
\label{largerotations_deltaThetaperp}
\delta \boldsymbol{\theta}_{\perp} = \mb{g}_1 \times \delta \mb{g}_1 = \frac{\mb{t}}{||\mb{t}||} \times \delta \left(\frac{\mb{t}}{||\mb{t}||}\right) = 
\frac{\mb{t} \times \delta \mb{t}}{||\mb{t}||^2}=\frac{\mb{g}_1 \times \delta \mb{t}}{||\mb{t}||}
 \quad \text{with} \quad \delta \mb{t} \in \Re^3.
\end{align}
In a next step,~\eqref{largerotations_deltaTheta} and~\eqref{largerotations_2drot} are exploited in order to formulate the tangential component $\delta \Theta_1$ of the spin vector:
\begin{align}
\label{largerotations_deltaTheta1}
\delta \Theta_1 = \mb{g}_{3}^T \! \delta \mb{g}_{2} = \mb{g}_{M3}^T \delta \mb{g}_{M2} + \delta \varphi =:\delta \Theta_{M1} + \delta \varphi
\end{align}
By variation of the basis vector $\delta \mb{g}_{M2}$ defined in~\eqref{largerotations_smallestrotation}, the tangential component $\delta \Theta_{M1}$ can be determined to:
\begin{align}
\label{largerotations_deltaThetaSR1}
\!\!\!\!\!\!
\delta \Theta_{M1} \!=\! \frac{ (\mb{g}_{1} \!\times\! \bar{\mb{g} }_{1})^T }{1\!+\! \mb{g}_{1}^T \bar{\mb{g}}_{1}} \delta \mb{g}_{1}
\!=\! -\frac{ \bar{\mb{g} }_{1}^T \mb{S}(\mb{g}_{1})}{1 \!+\! \mb{g}_{1}^T \bar{\mb{g}}_{1}}  \frac{\delta \mb{t} }{||\mb{t}||} \!=:\! \mb{T}_{\Theta_{M1} \! \mb{t}}\delta \mb{t} \quad \rightarrow \quad
\delta \Theta_1 \!=\! \mb{T}_{\Theta_{M1} \! \mb{t}}\delta \mb{t} \!+\! \delta \varphi
.\!\!\!\!\!\!
\end{align}
Inserting equations~\eqref{largerotations_deltaThetaperp}-\eqref{largerotations_deltaThetaSR1} into the split relation~\eqref{largerotations_deltathetasplit} yields the following expression for the spatial spin vector:
\begin{align}
\label{largerotations_deltathetasplitperp}
 \delta \boldsymbol{\theta} =  
\mb{g}_{1}  \delta \varphi + 
\underbrace{
\frac{1}{t}  \left( 
\mb{I}-\frac{ \mb{g}_{1} \otimes \bar{\mb{g} }_{1}^T }{1+ \mb{g}_{1}^T \bar{\mb{g}}_{1}} 
\right) \mb{S}(\mb{g}_{1})}_{=:\mb{T}_{\boldsymbol{\theta} \mb{t}}}
\delta \mb{t} \quad \text{with} \quad \mb{g}_1 = \frac{\mb{t}}{t}, \quad t:=||\mb{t}||.
\end{align}
So far, the four degrees of freedom $(\mb{t},\varphi)$ have been applied in order to uniquely describe a tangent vector $\mb{t}$, defined by its orientation and its length, as well as the orientation of a triad $\boldsymbol{\Lambda}\!=\!(\mb{g}_1,\mb{g}_2,\mb{g}_3)$ aligned parallel to this tangent vector. Next, the non-unit tangent vector $\mb{t}$ and the triad $\boldsymbol{\Lambda}$ shall be described by the alternative set of four degrees of freedom $(\boldsymbol{\psi},t)$. Here, $\boldsymbol{\psi}$ represents the rotation vector associated with the triad $\boldsymbol{\Lambda}$ via the Rodrigues formula~\eqref{largerotations_rotrigues} and $t:=||\mb{t}||$ is the norm of the tangent vector. The following transformations hold between the two sets $(\mb{t},\varphi)$ and $(\boldsymbol{\psi},t)$:
 \begin{align}
\label{largerotations_transformation_tandpsi}
 \boldsymbol{\Lambda}=\exp{ \! ( \mb{S}( \boldsymbol{\psi}) )}, \,\, \mb{g}_1 = \boldsymbol{\Lambda}  \mb{E}_1, \,\, \mb{t}=t \mb{g}_1,  \,\,
 \exp{ \! ( \mb{S}( \varphi \mb{g}_1) )}=\boldsymbol{\Lambda}\text{sr}(\bar{\boldsymbol{\Lambda}},\mb{g}_1)^{-1}.
\end{align}
Based on~\eqref{largerotations_transformation_tandpsi}, the set $(\mb{t},\varphi)$ can be calculated from $(\boldsymbol{\psi},t)$ and the other way round. Next, also a transformation rule between the variations $(\delta \mb{t},\delta\varphi)$ and the variations $(\delta \boldsymbol{\theta}, \delta t)$ associated with the set $(\boldsymbol{\psi},t)$ shall be derived:
\begin{align}
\label{largerotations_deltat}
 \delta t = \delta || \mb{t} || = \frac{\mb{t}^T}{t} \delta \mb{t} = \mb{g}_1^T \delta \mb{t}.
\end{align}
Combining equations~\eqref{largerotations_deltathetasplitperp} and~\eqref{largerotations_deltat} yields a transformation rule between the two sets $(\delta \boldsymbol{\theta}, \delta t)$ and $(\delta \mb{t}, \delta \varphi)$:
\begin{align}
\label{largerotations_tinvmatrixsr}
   \left(
   \begin{array}{c}
   \delta \boldsymbol{\theta} \\
   \delta t
   \end{array}
   \right)=
   \underbrace{
   \left(
   \begin{array}{cc}
   \mb{T}_{\boldsymbol{\theta} \mb{t}} & \mb{g}_{1}\\
   \mb{g}_{1}^T & 0
   \end{array}
   \right)
   }_{=:\mb{T}_{M}^{-1}}
   \left(
   \begin{array}{c}
   \delta \mb{t} \\
   \delta \varphi
   \end{array}
   \right) \quad  \Leftrightarrow  \quad
      \left(
   \begin{array}{c}
   \delta \mb{t} \\
   \delta \varphi
   \end{array}
   \right)=
   \underbrace{
   \left(
   \begin{array}{cc}
   - t \mb{S}(\mb{g}_1) & \mb{g}_{1}\\
    \mb{T}_{\varphi \boldsymbol{\theta}} & 0
   \end{array}
   \right)
   }_{=:\mb{T}_{M}}
      \left(
   \begin{array}{c}
   \delta \boldsymbol{\theta} \\
   \delta t
   \end{array}
   \right) \quad \text{with} \quad 
   \mb{T}_{\varphi \boldsymbol{\theta}}=\frac{(\mb{g}_{1} \!+\! \bar{\mb{g}}_{1})^T}{1+ \mb{g}_{1}^T \bar{\mb{g}}_{1}}.
\end{align}
The inverse transformation in~\eqref{largerotations_tinvmatrixsr} has been derived in a similar manner. The mappings $\mb{T}_{M}$ and $\mb{T}_{M}^{-1}$ transform between multiplicative rotation increments and additive increments of the chosen parametrization. Thus, they represent the analogon to the transformations $\mb{T}$ and $\mb{T}^{-1}$ in case of a rotation vector parametrization (see Section~\ref{sec:largerotations_rotvec}). Since the Kirchhoff constraint of vanishing shear deformation solely influences the component $\delta \boldsymbol{\theta}_{\perp}$ of the spin vector, it will in the following sections often be useful to express only this component by additive increments $\delta \mb{t}$, while the tangential spin vector component $\delta \Theta_{1}$ instead of the additive increment $\delta \varphi$ is regarded as independent primary variable:
\begin{align}
\label{largerotations_tildetinvmatrixsr}
   \left(
   \begin{array}{c}
   \delta \boldsymbol{\theta} \\
   \delta t
   \end{array}
   \right)=
   \underbrace{
   \left(
   \begin{array}{cc}
   \frac{1}{t} \mb{S}(\mb{g}_1) & \mb{g}_{1}\\
   \mb{g}_{1}^T & 0
   \end{array}
   \right)
   }_{=:\tilde{\mb{T}}^{-1}}
   \left(
   \begin{array}{c}
   \delta \mb{t} \\
   \delta \Theta_{1}
   \end{array}
   \right), 
   \quad
   \left(
   \begin{array}{c}
   \delta \mb{t} \\
   \delta \Theta_{1}
   \end{array}
   \right)=
   \underbrace{
   \left(
   \begin{array}{cc}
   -t \mb{S}(\mb{g}_1) & \mb{g}_{1}\\
   \mb{g}_{1}^T & 0
   \end{array}
   \right)
   }_{=:\tilde{\mb{T}}}
      \left(
   \begin{array}{c}
   \delta \boldsymbol{\theta} \\
   \delta t
   \end{array}
   \right),
\end{align}
The transformation~\eqref{largerotations_tildetinvmatrixsr} basically represents a reformulation of~\eqref{largerotations_deltathetasplit} and~\eqref{largerotations_deltaThetaperp}. Again, $\tilde{\mb{T}}$ and $\tilde{\mb{T}}^{-1}$ represent the corresponding mappings. Since these mappings solely transform the component $\delta \boldsymbol{\theta}_{\perp}$, they are independent from the actual definition of the triad $\boldsymbol{\Lambda}_M$. Consequently, the index $M$ has been omitted for the transformation matrices $\tilde{\mb{T}}$ and $\tilde{\mb{T}}^{-1}$.

\section{Simo-Reissner Beam Theory}
\label{sec:simoreissner}

In this section, the fundamentals of the geometrically exact Simo-Reissner beam theory based on the work of Reissner~\cite{reissner1972,reissner1981} as well as Simo and Vu-Quoc~\cite{simo1985,simo1986} will be briefly summarized. The results of this section will provide an essential basis for the subsequent derivation of Kirchhoff type beam formulations.

\begin{figure}[ht]
 \centering
  \includegraphics[width=1.0\textwidth]{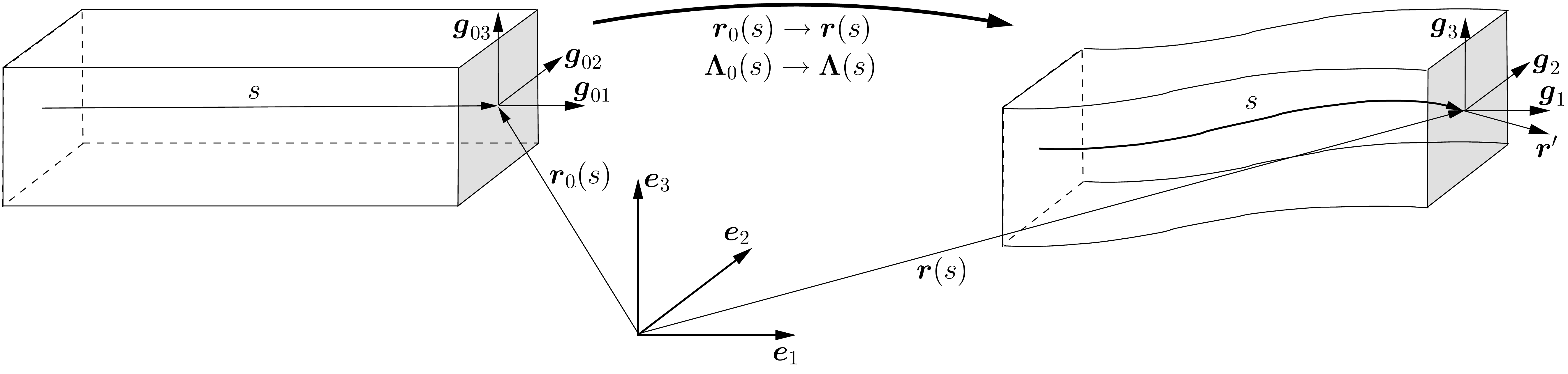}
  \caption{Kinematic quantities defining the initial and deformed configuration of the geometrically exact beam.}
  \label{fig:cpt1_beamkinematics1}
\end{figure}

\subsection{Basic kinematic assumptions}
\label{sec:simoreissner_kinematicassumptions}
%

Throughout this work, prismatic beams with anisotropic cross-section shape are considered. In the initial (unstressed) configuration, the beam centerline, which is defined as the line connecting the cross-section centroids, is described by the space curve $s \rightarrow \mb{r}_0(s) \in \Re^3$. Here and in the following, the index $0$ of a quantity refers to the unstressed, initial configuration. Furthermore, $s \in [0,l]=:\Omega_l \subset \Re$ is an arc-length parametrization of the curve and $l \in \Re$ the beam length in the initial configuration. The description of the initial configuration is completed by a field of right-handed orthonormal triads $s \rightarrow \mb{g}_{01}(s), \mb{g}_{02}(s), \mb{g}_{03}(s) \in \Re^3$, also denoted as material triads in the following. These are attached to the beam cross-sections, which are assumed to be undeformable according to the Bernoulli hypothesis. In this context, $\mb{g}_{01}(s)\!=\!\mb{r}_0^{\prime}(s)$ represents the unit tangential vector to the initial centerline and the base vectors $\mb{g}_{02}(s)$ and $\mb{g}_{03}(s)$ coincide with the principal axes of inertia of the cross-section at $s$. Throughout this work, the prime $(.)^{\prime}=\frac{d}{ds}(.)$ denotes the derivative with respect to the arc-length parameter~$s$. The rotation from the global frame $\mb{E}_{i}$ onto the initial local frame $\mb{g}_{0i}(s)$ is described via the orthogonal transformation $s \rightarrow \mb{\Lambda}_0(s) \in S\!O(3)$ as introduced in~\eqref{largerotations_lambda} leading to the following definition $s \!\rightarrow\! C_0 \!:=\! (\mb{r}_0(s), \mb{\Lambda}_0(\boldsymbol{\psi}_0(s))) \in \Re^3 \!\times\! S\!O(3)$ of the initial configuration. Correspondingly, the deformed configuration of the beam at time $t \!\in\! \Re$ is given by $s,t \rightarrow C := (\mb{r}(s,t), \mb{\Lambda}(\boldsymbol{\psi}(s,t))) \in \Re^3 \times S\!O(3)$, where the orthogonal transformation $\mb{\Lambda}(\boldsymbol{\psi}(s,t))$ maps from the global frame $\mb{E}_{i}$ onto the current local frame $\mb{g}_{i}(s)$ and the base vector $\mb{g}_{1}(s,t):= \mb{g}_{2}(s,t) \!\times\! \mb{g}_{3}(s,t)$ is in general not tangential to the deformed centerline anymore due to shear deformation. According to Section~\ref{sec:largerotations}, $\mb{\Lambda}(s,t)$ can be represented by three rotation parameters (e.g. by a rotation vector $\boldsymbol{\psi}(s,t)$), leading to pointwise six, three translational and three rotational, degrees of freedom. The basic kinematic assumption of the geometrically exact Simo-Reissner theory considered so far can easily be summarized by the following constrained position vector field associated with the initial and deformed configuration
\begin{align}
\label{simoreissner_x0}
 \!\!\!\!\!\!\mathbcal{X}(\mathbcal{s})=\mb{r}_{0}(s) + s_2 \mb{g}_{02}(s) + s_3 \mb{g}_{03}(s), \quad \mathbcal{x}(\mathbcal{s},t)=\mb{r}(s,t) + s_2 \mb{g}_{2}(s,t) + s_3 \mb{g}_{3}(s,t), \quad \mathbcal{s}\!:=\!(s,s_2,s_3)^T\!=:\!(s_1,s_2,s_3)^T,\!\!\!\!\!\!
\end{align}
where $(s_2,s_3)$ represent convective coordinates describing the position of an arbitrary material point within the (rigid) cross-section. In order to simplify the notation for subsequent derivations, the convective coordinate vector given by $\mathbcal{s}\!=\!(s_1,s_2,s_3)^T$ as well as the redundant name $s_1\!:=\!s$ for the arc-length $s$ has been introduced. The kinematic quantities defining the initial and deformed configuration as introduced so far are illustrated in Figure \ref{fig:cpt1_beamkinematics1}. In order to simplify notation, the time argument $t$ will be dropped in the following.  Later in this section, the arc-length derivative of the base vectors $\mb{g}_{i}(s)$ will be required. Similar to equation~\eqref{largerotations_deltalambdaspatial}, this derivative can be formulated as
\begin{align}
\label{simoreissner_spatialcurvature}
  \mb{\Lambda}^{\prime}(s) =  \mb{S}(
\mb{k}(s))\mb{\Lambda}(s) \quad \text{or} \quad  
  \mb{\Lambda}^{\prime}(s) = \mb{\Lambda}(s)\mb{S}(\mb{K}(s)),
\end{align}
where $\mb{k}(s)$ and $\mb{K}(s) \!:=\! \mb{\Lambda}^T(s) \mb{k}(s)$ are referred to as spatial and material curvature vector, which can be derived as:
\begin{align}
\label{simoreissner_explicitcurvatures}
\mb{S}(\mb{k}(s)) \!=\! \mb{\Lambda}^{\prime}(s) \mb{\Lambda}^T(s), \quad \mb{S}(\mb{K}(s)) \!=\! \mb{\Lambda}^T(s) \mb{\Lambda}^{\prime}(s)   \quad \quad
\mb{K}=K_i \mb{E}_i, \,\, \mb{k}=K_i \mb{g}_i.
\end{align}
In a similar manner, the spatial and material angular velocity vectors $\mb{w}$ and $\mb{W}$ are defined according to:
\begin{align}
\label{simoreissner_explicitangularvelocities}
\mb{S}(\mb{w}(s)) \!=\! \dot{\mb{\Lambda}}(s) \mb{\Lambda}^T(s), \quad \mb{S}(\mb{W}(s)) \!=\! \mb{\Lambda}^T(s) \dot{\mb{\Lambda}}(s)   \quad \quad
\mb{W}=W_i \mb{E}_i, \,\, \mb{w}=W_i \mb{g}_i.
\end{align}
For completeness, also the spin vectors, which have been introduced in Section~\ref{sec:largerotations}, are repeated here:
\begin{align}
\label{simoreissner_explicitspinvectors}
\mb{S}(\delta \boldsymbol{\theta}(s)) \!=\! \delta \mb{\Lambda}(s) \mb{\Lambda}^T(s), \quad \mb{S}(\delta \boldsymbol{\Theta}(s)) \!=\! \mb{\Lambda}^T(s) \delta \mb{\Lambda}(s)   \quad \quad
\delta \boldsymbol{\Theta}=\delta \Theta_i \mb{E}_i, \,\, \delta \boldsymbol{\theta}=\delta \Theta_i \mb{g}_i.
\end{align}
Throughout this work, the dot $\dot{(.)}\!=\!\frac{d}{dt}(.)$ denotes the derivative with respect to time~$t$. By applying the Young theorem $\dot{(\mb{\Lambda}^{\prime})}\!=\!(\dot{\mb{\Lambda}})^{\prime}$ and making use of $\mb{S}(\mb{a})\mb{S}(\mb{b})\!-\!\mb{S}(\mb{b})\mb{S}(\mb{a})\!=\!\mb{S}(\mb{S}(\mb{a})\mb{b})$, the following compatibility relations can be shown:
\begin{align}
\begin{split}
\label{simoreissner_relationskwtheta}
\!\!\!\! 
\dot{\mb{k}} & = \mb{w}^{\prime} + \mb{w} \times \mb{k}, 
\hspace{0.5cm} \quad
\delta \mb{k} = \delta \boldsymbol{\theta}^{\prime} + \delta \boldsymbol{\theta} \times \mb{k}, \hspace{0.5cm}  \quad
 \delta \mb{w} = \delta \dot{\boldsymbol{\theta}} + \delta \boldsymbol{\theta} \times \mb{w}, \\
\dot{\mb{K}} & = \mb{W}^{\prime} - \mb{W} \times \mb{K}, 
\quad
\delta \mb{K} = \delta \boldsymbol{\Theta}^{\prime} - \delta \boldsymbol{\Theta} \times \mb{K}, \quad
\delta \mb{W} = \delta \dot{\boldsymbol{\Theta}} - \delta \boldsymbol{\Theta} \times \mb{W}.
\end{split}
\end{align}
In the following sections, stress resultants mechanical equilibrium and proper constitutive relations will be presented.

\subsection{Stress resultants, mechanical equilibrium and objective deformation measures}
\label{sec:simoreissner_strongform}
%

With $\mb{\tilde{f}}$ and $\mb{\tilde{m}}$ denoting distributed external forces and moments per unit length and $\mb{f}_{\rho}$ and $\mb{m}_{\rho}$ representing the force and moment contributions due to inertia effects, the strong form of equilibrium reads (see e.g. \cite{antmann1995,reissner1981,simo1985}):
\begin{align}
\label{simoreissner_equilibriumspatial}
\begin{split}
\mb{f}^{\prime} + \mb{\tilde{f}} + \mb{f}_{\rho} & = \mb{0}, \\
\mb{m}^{\prime} + \mb{r}^{\prime} \times \mb{f} + \mb{\tilde{m}} + \mb{m}_{\rho}& = \mb{0}.
\end{split}
\end{align}
In~\eqref{simoreissner_equilibriumspatial}, $\mb{f}$ and $\mb{m}$ are the force and moment stress resultants acting on the beam cross-section area $A$. A material form of the 1D equilibrium equations can be derived by inserting the material stress resultants $\mb{F}\!\!:=\!\!\mb{\Lambda}^T \!\mb{f}$ and $\mb{M}\!\!:=\!\!\mb{\Lambda}^T\! \mb{m}$ into the balance equations~\eqref{simoreissner_equilibriumspatial}. Following the principle of virtual work, the admissible variations, i.e. infinitesimal small and arbitrary (additive or multiplicative) changes of the current configuration $s \rightarrow \delta C := (\delta \mb{r}(s), \delta \boldsymbol{\theta}(s)) \in \Re^3 \!\times\! \Re^3$ that are compatible with the employed boundary conditions are introduced in a next step. Here, $\delta \boldsymbol{r}(s) \!\in\! \Re^3$ represents the vector of (additive) virtual displacements and $\delta \boldsymbol{\theta}(s) \!\in\! \Re^3$ the vector of (multiplicative) virtual rotations, also denoted as spin vector. By multiplication of \eqref{simoreissner_equilibriumspatial} with $\delta \boldsymbol{r}$ and $\delta \boldsymbol{\theta}$ and integration by parts, the spatial weak form is derived:
\begin{align}
\label{weakformspatial}
\!\!\!\!\!\!G=  \int \limits_0^l \bigg( \underbrace{\delta \boldsymbol{\theta}^{\prime T}  \mb{m}}_{\delta_{o} \boldsymbol{\omega}^T  \mb{m}}  +  \underbrace{( \delta \mb{r}^{\prime} - \delta \boldsymbol{\theta} \times
\mb{r}^{\prime} )^T\mb{f}}_{\delta_{o} \boldsymbol{\gamma}^T\mb{f}} - \delta \boldsymbol{\theta}^T  (\tilde{\mb{m}}+\mb{m}_{\rho}) - \delta \mb{r}^T (\tilde{\mb{f}} + \mb{f}_{\rho}) \bigg) ds
-\Big[\delta \mb{r}^T  \mb{f}_{\sigma} \Big]_{\varGamma_{\sigma}} -  \Big[\delta \boldsymbol{\theta}^T  \mb{m}_{\sigma} \Big]_{\varGamma_{\sigma}}\dot{=}\,0.\!\!\!\!\!\!
\end{align}
Here, $\mb{f}_{\sigma}$ and $\mb{m}_{\sigma}$ denote external forces and moments at the Neumann boundary $\varGamma_{\!\sigma}$ of the considered beam. Based on the principle of virtual work, the following objective spatial deformation measures can be identified:
\begin{align}
\label{spatialdeformationmeasures}
\!\!\! \boldsymbol{\gamma} \!=\! \mb{r}^{\prime} \!-\! \mb{g}_1, \,\,\,\, 
& \mb{S}(\boldsymbol{\omega}) \!=\! \mb{\Lambda}^{\prime}\mb{\Lambda}^T\!
\!-\! \mb{\Lambda}\mb{\Lambda}_0^T \!\mb{\Lambda}_0^{\prime}\mb{\Lambda}^T \!=\! \mb{S}(\boldsymbol{k}) \!-\! \mb{S}(\mb{\Lambda} \mb{\Lambda}_0^T \! \boldsymbol{k_0}) 
\,\, \rightarrow \,\, \boldsymbol{\omega} \!=\! \boldsymbol{k} \!-\! \mb{\Lambda} \mb{\Lambda}_0^T \boldsymbol{k_0}. \!
\end{align}
If the material counterpart of equation~\eqref{simoreissner_equilibriumspatial} is chosen as starting point (see e.g \cite{crisfield1999}), the material deformation measures $\boldsymbol{\Gamma}$ and $\boldsymbol{\Omega}$ being work-conjugated to the material stress resultants $\mb{F}$ and $\mb{M}$ can be determined in an analogous manner:
\begin{align}
\label{materialdeformationmeasures}
\boldsymbol{\Gamma} \!=\! \mb{\Lambda}^T\!\mb{r}^{\prime}  \!-\! \mb{E}_1, \quad
& \mb{S}(\boldsymbol{\Omega}) \!=\! \mb{\Lambda}^T \!\mb{\Lambda}^{\prime} \!-\! \mb{\Lambda}_0^T\! \mb{\Lambda}_0^{\prime} \!=\! \mb{S}(\boldsymbol{K}) \!-\!
\mb{S}(\boldsymbol{K_0})
\,\, \rightarrow \,\, \boldsymbol{\Omega} \!=\! \boldsymbol{K} \!-\! \boldsymbol{K}_0.
\end{align}
The ``objective variation'' $\delta_{o}$ of an arbitrary vector $\mb{a} \! \in \! \Re^3$ appearing in~\eqref{weakformspatial} is defined as $\delta_{o} \mb{a}\!:=\!\delta \mb{a} \!-\! \delta \boldsymbol{\theta} \!\times\! \mb{a}$ (see e.g. \cite{simo1985}). It is easy to verify that all of these deformation measures vanish for the stress-free initial configuration, i.e. when the relations
$\mb{\Lambda}\!=\!\mb{\Lambda}_0$, $\mb{r}^{\prime}\!=\!\mb{r}_0^{\prime}$ and $\mb{g}_1\!=\!\mb{g}_{01}\!=\!\mb{r}_0^{\prime}\!=\!\mb{\Lambda}_0 \mb{E}_1$ are valid, and that the following variations can be derived:
\begin{align}
\label{variationspatialandmaterialdeformationmeasures}
\delta_{o} \boldsymbol{\gamma} = \delta \mb{r}^{\prime} - \delta \boldsymbol{\theta} \times
\mb{r}^{\prime}, \quad 
\delta_{o} \boldsymbol{\omega} =  \delta \boldsymbol{\theta}^{\prime}, \quad
\delta \boldsymbol{\Gamma} = \mb{\Lambda}^T \! \left(\delta \mb{r}^{\prime} + \mb{r}^{\prime} \times \delta \boldsymbol{\theta} \right), \quad 
\delta \boldsymbol{\Omega} = \mb{\Lambda}^T \delta \boldsymbol{\theta}^{\prime}. 
\end{align}
The components of $\boldsymbol{\Gamma}$ represent axial tension and 
shear, the components of $\boldsymbol{\Omega}$ represent torsion as well as bending. 

%
\subsection{Constitutive relations}
\label{sec:reissnerconstitutiverelations}
%

Finally, constitutive relations between the stress resultants $\mb{M}$ and $\mb{F}$ and the deformation measures $\mb{\Omega}$ and $\mb{\Gamma}$ are required. The simplest constitutive law of this type is given by the length-specific hyperelastic stored energy function:
\begin{align}
\label{storedenergyfunction}
\tilde{\Pi}_{int}(\mb{\Omega},\mb{\Gamma}) \!=\! \frac{1}{2} \mb{\Omega}^T \! \mb{C}_M \mb{\Omega} \!+\! \frac{1}{2} \mb{\Gamma}^T \mb{C}_F \mb{\Gamma},
\quad \mb{M}\!=\!\dfrac{\partial \tilde{\Pi}_{int}}{\partial \mb{\Omega}} \!=\! \mb{C}_M \!\cdot\! \mb{\Omega}, \quad
\mb{F}\!=\!\dfrac{\partial \tilde{\Pi}_{int}}{\partial \mb{\Gamma}} \!=\! \mb{C}_F \!\cdot\! \mb{\Gamma}.
\end{align}
Here, the material constitutive tensors $\mb{C}_M$ and $\mb{C}_F$ have the following diagonal structure:
\begin{align}
\label{materialconstitutive}
\mb{C}_M  = \textbf{diag}\big[GI_T,EI_2,EI_3\big]_{\mb{E}_{i}}   
\quad \text{and} \quad  
\mb{C}_F = \textbf{diag}\big[EA,G\bar{A}_2,G\bar{A}_3\big]_{\mb{E}_{i}}.  
\end{align}
Here, $E$ and $G$ are the Young's modulus and the shear modulus, $A$, $\bar{A}_2$ and $\bar{A}_3$ are the cross-section and the two reduced cross-sections, $I_2$ and $I_3$ are the two principal moments of inertia and $I_T$ is the torsional moment of inertia. In a similar manner to~\eqref{storedenergyfunction}, the length-specific kinetic energy $\tilde{\Pi}_{kin}$ of the beam can be formulated:
\begin{align}
\label{kineticenergy}
\!\!\!\!\tilde{\Pi}_{kin}(\mb{w},\dot{\mb{r}}) \!=\! \frac{1}{2} \mb{w}^T \! \mb{c}_{\rho} \mb{w} \!+\! \frac{1}{2} \rho A \dot{\mb{r}}^T \! \dot{\mb{r}}, \quad \mb{c}_{\rho}=\mb{\Lambda}\mb{C}_{\rho}\mb{\Lambda}^T, \quad \mb{C}_{\rho}\!=\!\textbf{diag}\big[\rho (\underbrace{I_2\!+\!I_3}_{=:I_P}),\rho I_2,\rho I_3 \big]_{\mb{E}_{i}}.\!\!\!\!
\end{align}
Here, $\rho $ is the mass density, $\mb{C}_{\rho}$ the material inertia tensor and $\mb{w}$ the spatial angular velocity vector, which has already been introduced in~\eqref{simoreissner_explicitangularvelocities}. From~\eqref{kineticenergy}, the length-specific linear momentum $\tilde{\mb{l}}$ and angular momentum $\tilde{\mb{h}}$ can be derived:
\begin{align}
\label{momentsofinertia}
\tilde{\mb{l}}\!:=\!\dfrac{\partial \tilde{\Pi}_{kin}}{\partial \dot{\mb{r}}}=\rho A\dot{\mb{r}}, \quad
\tilde{\mb{h}}\!:=\!\dfrac{\partial \tilde{\Pi}_{kin}}{\partial \mb{w}}= \mb{c}_{\rho} \mb{w}\!=\!\mb{\Lambda}\mb{C}_{\rho}\mb{\Lambda}^T \! \mb{w}\!=\!\mb{\Lambda}\mb{C}_{\rho} \mb{W}.
\end{align}
Similar to the length-specific external forces $\tilde{\mb{f}}$ and $\tilde{\mb{m}}$, also the length-specific energies $\tilde{\Pi}_{int},\tilde{\Pi}_{kin}$ and momenta $\tilde{\mb{l}},\tilde{\mb{h}}$ have been furnished with the $\tilde{(.)}$-symbol. The total counterparts $\mb{f}_{ext}$, $\mb{m}_{ext}$, $\Pi_{int}$, $\Pi_{kin}$, $\mb{l}$ and $\mb{h}$ are obtained by integration: 
\begin{align}
\label{integrated_quantities}
\begin{split}
\mb{f}_{ext}\!&:=\!\!\int \limits_0^l \! \mb{\tilde{f}} ds
\!+\! \Big[ \mb{f}_{\sigma} \Big]_{\Gamma_{\sigma}} \hspace{-0.1cm}, \quad 
\mb{m}_{ext}\!:=\!\!\int \limits_0^l \!\! \left(\mb{r} \times \mb{\tilde{f}} \!+\! \mb{\tilde{m}}\right) ds 
\!+\! \Big[\mb{r} \times \mb{f}_{\sigma} \!+\! \mb{m}_{\sigma}\Big]_{\Gamma_{\sigma}} \hspace{-0.1cm},\\
\Pi_{int}\!&:=\! \!\int \limits_0^l \!\tilde{\Pi}_{int} ds, \quad
\Pi_{kin}\!:=\! \!\int \limits_0^l \!\tilde{\Pi}_{kin} ds, \quad
\mb{l} \!:=\! \!\int \limits_0^l \!\tilde{\mb{l}} ds, \quad
\mb{h} \!:=\! \!\int \limits_0^l \!(\tilde{\mb{h}} \!+\! \mb{r} \!\times\! \tilde{\mb{l}}) ds.
\end{split}
\end{align}
These definitions will be required subsequently. Based on~\eqref{momentsofinertia}, the inertia forces $\mb{f}_{\rho}$ and moments $\mb{m}_{\rho}$ yield:
\begin{align}
\label{simoreissner_inertia}
\begin{split}
\!\!\!\!\!\!-\mb{f}_{\rho}  \! =\! \dot{\tilde{\mb{l}}} \!=\! \rho A \ddot{\mb{r}}, \,\,\, -\mb{m}_{\rho}\!=\!\dot{\tilde{\mb{h}}} 
\!=\! \mb{\Lambda} \left[\mb{S}(\mb{W})\mb{C}_{\rho} \mb{W} \!+\! \mb{C}_{\rho} \mb{A}\right]
\!=\! \mb{S}(\mb{w}) \mb{c}_{\rho} \mb{w} \!+\! \mb{c}_{\rho} \mb{a},
\,\,\,  \mb{a}\!:=\!\dot{\mb{w}}, \,\,\, \mb{A}\!:=\!\dot{\mb{W}}.\!\!\!\!\!\!
\end{split}
\end{align}
Similar to $\mb{w}\!=\!\mb{\Lambda}\mb{W}$, also the angular accelerations are related via the push-forward operator:
\begin{align}
\label{simoreissner_Avector}
\mb{a} \!:=\! \dot{\mb{w}} \!=\! \frac{d}{dt} \left(\mb{\Lambda}\mb{W} \right) \!=\! \mb{\Lambda}\mb{S}(\mb{W})\mb{W} + \mb{\Lambda}\dot{\mb{W}} \!=\! \mb{\Lambda}\dot{\mb{W}} \!=:\! \mb{\Lambda} \mb{A}, \,\,\,\, \text{since} \,\,\,\, \mb{S}(\mb{W})\mb{W} \!=\! \mb{0}.
\end{align}
When the beam problem has to be discretized in time (see Section~\ref{sec:temporaldiscretization} for details), the vectors $\mb{w}$ and $\mb{a}$ can either be directly  employed in a time integration scheme~\cite{simo1988, simo1991,jelenic1999} or, alternatively, they can be expressed via the (additive) rate of the primary variable $\boldsymbol{\psi}$ \cite{cardona1988}. Similar to~\eqref{largerotations_tmatrix}, the following relations between $\mb{w}$ and $\mb{a}$ can be formulated:
 \begin{align}
\label{simoreissner_additiveacceleration}
  \mb{w}=\mb{T}^{-1} \dot{\boldsymbol{\psi}}, \,\,\,\,
  \mb{a}=\dot{\mb{T}}^{-1} \dot{\boldsymbol{\psi}}+\mb{T}^{-1} \ddot{\boldsymbol{\psi}}.
\end{align}
Finally, the problem setup presented in Sections~\ref{sec:simoreissner_kinematicassumptions}-\ref{sec:reissnerconstitutiverelations} has to be completed by boundary and initial conditions in order to end up with a well-defined initial boundary value problem:
\begin{align}
\label{reissner_boundaryconditions}
\begin{split}
\mb{r} \!=\! \mb{r}_{u}, \, \boldsymbol{\psi} \!=\! \boldsymbol{\psi}_{u}  \,\, \text{on} \,\, \varGamma_{u},  \quad 
&\mb{f} \!=\! \mb{f}_{\sigma}, \, \mb{m} \!=\! \mb{m}_{\sigma} \, \, \text{on} \,\, \varGamma_{\sigma}, \quad \varGamma_{\sigma} \cap \varGamma_{u} \!=\! \varnothing, \, \varGamma_{\sigma} \cup \varGamma_{u} \!=\! \{0,L\}\\
&\mb{r}\!=\!\mb{r}_0, \, \dot{\mb{r}}\!=\!\mb{v}_0, \, 
\boldsymbol{\Lambda}\!=\!\boldsymbol{\Lambda}_0, \, \mb{w}\!=\!\mb{w}_0 \,\,
\text{at} \,\, t\!=\!0.
\end{split}
\end{align}
Based on a trial space $(\mb{r}(s,t),\boldsymbol{\Lambda}(\boldsymbol{\psi}(s,t))) \!\in\! \mathcal{U}$ of functions with square-integrable first derivatives satisfying~\eqref{reissner_boundaryconditions} and an weighting space $(\delta \mb{r}(s), \delta \boldsymbol{\theta}(s)) \!\in\! \mathcal{V}$ of functions with square-integrable first derivatives satisfying $\delta \mb{r}\!=\!\mb{0}, \, \delta \boldsymbol{\theta} \!=\! \mb{0} \, \text{on} \, \varGamma_{u}$, the weak form~\eqref{weakformspatial} is equivalent to the strong form~\eqref{simoreissner_equilibriumspatial} supplemented by the boundary conditions~\eqref{reissner_boundaryconditions}. The following two Sections~\ref{sec:reissnerhamilton} and~\ref{sec:reissnerrelation3d1d} are intended to supplement the geometrically exact beam theory sdsd so far. However, the provided information is not necessarily required  for the derivation of the space-continuous problem statement of the Simo-Reissner and Kirchhoff-Love beam theory and the subsequent discretization procedures. Thus, the reader may alternatively skip these content and proceed with Section~\ref{sec:kirchhoff}. Concretely, in Section~\ref{sec:reissnerhamilton}, the alternative procedure of deriving the weak and strong form of the balance equations presented in Section~\ref{sec:simoreissner_strongform} on the basis of a variational problem statement and properly defined Lagrangian is undertaken. Section~\ref{sec:reissnerrelation3d1d} is meant to confirm the constitutive relations presented above based on a derivation from 3D continuum mechanics. There, the consistency of the 1D and 3D continuum model in the sense of a fully induced beam theory will be verified for the case of locally small strains. \\

\hspace{0.2 cm}
\begin{minipage}{15.0 cm}
\textbf{Remark: } Two possible time integration schemes can be derived from the variants of either employing $\mb{W}$ and $\mb{A}$ directly or expressing them 
via additive rates given by~\eqref{simoreissner_additiveacceleration}:
\begin{align}
\begin{split}
\label{reissnerkinematics_timeintegrations}
\text{1)} \hspace{0.72cm} (\dot{\mb{r}},\ddot{\mb{r}},\mb{w},\mb{a})_{n+1}&=\mb{f}((\mb{r},\boldsymbol{\Lambda})_{n+1},(\mb{r},\boldsymbol{\Lambda})_{n},(\dot{\mb{r}},\mb{w})_{n},(\ddot{\mb{r}},\mb{a})_{n}), \\
\text{2)} \hspace{0.65cm} (\dot{\mb{r}},\ddot{\mb{r}},\dot{\boldsymbol{\psi}},\ddot{\boldsymbol{\psi}})_{n+1}&=\mb{f}((\mb{r},\boldsymbol{\psi})_{n+1},(\mb{r},\boldsymbol{\psi})_{n},(\dot{\mb{r}},\dot{\boldsymbol{\psi}})_{n},(\ddot{\mb{r}},\ddot{\boldsymbol{\psi}})_{n}).
\end{split}
\end{align}
Here, the indices $(.)_{n}$ and $(.)_{n+1}$ refer to two successive time steps of the time-discrete problem and $\mb{f}(.)$ represents a typical finite difference time integration scheme (e.g. a Newmark scheme). The first variant (see Section~\ref{sec:temporaldiscretization}) can be considered as being more flexible since it does not require any specific rotation parametrization. It can directly be applied to Reissner type beam formulations as well as to Kirchhoff type beam formulations with strong or weak Kirchhoff constraint enforcement without the need for further adaptions. Due to this flexibility and the very simple and compact time integrator resulting from this procedure, this will be the method of choice employed throughout this work.\\ 
\end{minipage}

%
\subsection{Interlude: Variational problem formulation of Simo-Reissner beam theory}
\label{sec:reissnerhamilton}
%

For the case that no external forces act on the beam, i.e. $\mb{f}_{\sigma} \!=\! \tilde{\mb{f}} \!=\!\mb{m}_{\sigma} \!=\!\tilde{\mb{m}} \!=\! \mb{0}$, the strong and weak forms~\eqref{simoreissner_equilibriumspatial} and~\eqref{weakformspatial} shall equivalently be formulated on the basis of the well-known Hamilton principle:
\begin{align}
\label{hamilton_principle}
  \delta \int \limits_{t=0}^T \! \! \mathcal{L} \, dt =0 \quad \text{with} \quad \left[\delta \mb{r}\!=\!\delta \boldsymbol{\theta}\!=\!\mb{0}\right]_{t=0}^{t=T}.
\end{align}
The Lagrangian $\mathcal{L}$ occurring in~\eqref{hamilton_principle} is defined as the difference of kinetic and potential energy:
\begin{align}
\label{hamilton_lagrangian}
  \mathcal{L}\!=\! \int \limits_{s=0}^l \! (\tilde{\Pi}_{kin}\!-\! \tilde{\Pi}_{int}) \, ds.
\end{align}
Based on the kinetic energy~\eqref{kineticenergy} and the hyper-elastic energy~\eqref{storedenergyfunction}, the variation~\eqref{hamilton_principle} reads:
\begin{align}
\begin{split}
\label{hamilton_principle2}
  \delta \int \limits_{t=0}^T \! \! \mathcal{L} \, dt & = 
  \int \limits_{t=0}^T \int \limits_{s=0}^l \left( \delta_o \mb{w}^T \! \mb{c}_{\rho} \mb{w}  +  \delta_o \dot{\mb{r}}^T \! \rho A \dot{\mb{r}} -  \delta_o \boldsymbol{\omega}^T \! \mb{c}_m \boldsymbol{\omega} - \delta_o \boldsymbol{\gamma}^T \! \mb{c}_f \boldsymbol{\gamma}   \right) ds \, dt \\
  & = \int \limits_{t=0}^T \int \limits_{s=0}^l \left( \delta \dot{\boldsymbol{\theta}}^T \tilde{\mb{h}}  +  \delta \dot{\mb{r}}^T \tilde{\mb{l}} -  \delta \boldsymbol{\theta}^{\prime T} \! \mb{m} -  
\left[\delta \mb{r}^{\prime} \!-\! \delta \boldsymbol{\theta} \!\times\! \mb{r}^{\prime}  \right]^T \!\! \mb{f}  \right) ds \, dt \\
  & = -\int \limits_{t=0}^T \int \limits_{s=0}^l \left( -\delta \boldsymbol{\theta}^T \! \mb{m}_{\rho}  -  \delta\mb{r}^T \! \mb{f}_{\rho}  +  \delta \boldsymbol{\theta}^{\prime T} \! \mb{m} +  
\left[\delta \mb{r}^{\prime} \!-\! \delta \boldsymbol{\theta} \!\times\! \mb{r}^{\prime} \right]^T \!\! \mb{f}  \right) ds \, dt = 0.
\end{split}
\end{align}
From the first to second line, the objective variations $\delta_o \mb{w} \!=\! \delta \dot{\boldsymbol{\theta}}, \delta_o \boldsymbol{\omega} \!=\! \delta \boldsymbol{\theta}^{\prime}$ and $\delta_o \boldsymbol{\gamma} \!=\! \left[\delta \mb{r}^{\prime} \!-\! \delta \boldsymbol{\theta} \!\times\! \mb{r}^{\prime}  \right]$ (see~\eqref{simoreissner_relationskwtheta} and~\eqref{variationspatialandmaterialdeformationmeasures}) have been inserted. Furthermore, the additional relation 
$\delta_o \dot{\mb{r}}^T \dot{\mb{r}} \!=\! (\delta \dot{\mb{r}} \!-\! \delta \boldsymbol{\theta} \!\times\! \dot{\mb{r}})^T \dot{\mb{r}} \!=\!\delta \dot{\mb{r}}^T \dot{\mb{r}}$
 has been applied. From the second to the third line, partial integration of the inertia terms together with the boundary conditions in~\eqref{hamilton_principle} have been applied. Since no external forces and moments are considered, i.e. $\mb{f}_{\sigma} \!=\! \mb{m}_{\sigma} \!=\! \mb{0}$, partial integration of~\eqref{hamilton_principle2} yields:
\begin{align}
\begin{split}
\label{hamilton_strongform}
  \delta \int \limits_{t=0}^T \! \! \mathcal{L} \, dt = 
  \int \limits_{t=0}^T \int \limits_{s=0}^l \big( \delta \boldsymbol{\theta}^T \! \big[  \underbrace{\mb{m}^{\prime} + \mb{r}^{\prime} \times \mb{f} + \mb{m}_{\rho}}_{\dot{=}0} \big] + \delta\mb{r}^T \big[ \underbrace{\mb{f}^{\prime} + \mb{f}_{\rho}}_{\dot{=}0} \big]   \big) ds \, dt = 0.
\end{split}
\end{align}
The arbitrariness of $\delta \boldsymbol{\theta}(s,t)$ and $ \delta \mb{r}(s,t)$ directly yields the strong form~\eqref{simoreissner_equilibriumspatial}. Inverting the last step from~\eqref{hamilton_principle2} to~\eqref{hamilton_strongform}, by partial integration of the weighted strong form~\eqref{simoreissner_equilibriumspatial}, this time only along the beam length l, yields the weak form of the balance equations~\eqref{weakformspatial}. The terms occurring in this weak form for an unloaded beam can already be identified in the third line of~\eqref{hamilton_principle2}. While conservative external forces could also be included into the Lagrangian~\eqref{hamilton_lagrangian}, for a consideration of non-conservative external forces and 3D external moments, which are known to be non-conservative (see e.g.~\cite{simo1986}), the starting point has to be the strong form of the mechanical balance equations~\eqref{simoreissner_equilibriumspatial}.

%
\subsection{Interlude: Relation between 1D and 3D constitutive laws}
\label{sec:reissnerrelation3d1d}
%

The aim of the following considerations is to derive the constitutive laws~\eqref{storedenergyfunction} and~\eqref{materialconstitutive} in a consistent manner from the 3D continuum theory. Thereto, the deformation gradient $\boldsymbol{\mathcal{F}}$ of the 3D position field subject to the kinematic constraints~\eqref{simoreissner_x0} of the geometrically exact beam theory shall be derived. Subsequently, also the Cauchy-Green deformation tensor $\boldsymbol{\mathcal{E}}$ is required. These two objects can be formulated based on the following definitions:
\begin{align}
\label{deformationgradient}
\boldsymbol{\mathcal{E}}:=\dfrac{1}{2}\!\left( \boldsymbol{\mathcal{F}}^T \! \boldsymbol{\mathcal{F}} - \mb{I}_3 \right) \quad \text{with} \quad \boldsymbol{\mathcal{F}}:=\frac{\partial \mathbcal{x}}{\partial \mathbcal{X}}=\frac{\partial \mathbcal{x}}{\partial \mathbcal{s}}\frac{\partial \mathbcal{s}}{\partial \mathbcal{X}}=\mathbcal{g}_i \otimes \mathbcal{G}^i.
\end{align}
The (non-orthonormal) covariant basis vectors $\mathbcal{g}_i\!=\! \partial \mathbcal{x}(\mathbcal{s})/\partial s_i$ and $\mathbcal{G}_{i}\!=\! \partial \mathbcal{X}(\mathbcal{s})/\partial s_i$ have to be determined from~\eqref{simoreissner_x0}:
\begin{align}
\label{covariantbasis}
\begin{split}
\mathbcal{g}_1\!&=\! \mb{r}^{\prime} + \mb{k} \times \left( s_2 \mb{g}_{2} + s_3 \mb{g}_{3} \right), \,\, \mathbcal{g}_2 \!=\! \mb{g}_{2}, \,\, \mathbcal{g}_3 \!=\! \mb{g}_{3}, \\
\mathbcal{G}_{1}\!&=\!\mb{g}_{01} + \mb{k}_0 \times \left( s_2 \mb{g}_{02} + s_3 \mb{g}_{03} \right), \,\, \mathbcal{G}_{2} \!=\! \mb{g}_{02}, \,\, \mathbcal{G}_{3} \!=\! \mb{g}_{03}.
\end{split}
\end{align}
Also the contravariant base vectors $\mathbcal{G}^i$ can be determined from the second line of~\eqref{covariantbasis} via the definition $\mathbcal{G}_{i}^T \mathbcal{G}^j\!=\!\delta_i^{\,\,j}$:
\begin{align}
\label{contravariantbasis}
\!\!\!\!\!\! \mathbcal{G}^1\!&=\! \frac{1}{C} \mb{g}_{01}, \,\,\,\, 
\mathbcal{G}^2 \!=\! \mb{g}_{02} \!+\! \frac{K_{01}s_3}{C}\mb{g}_{01}, \,\,\,\,
 \mathbcal{G}^3 \!=\! \mb{g}_{03} \!-\! \frac{K_{01} s_2}{C}\mb{g}_{01}, \,\,\,\,
C\!:=\!1\!-\!K_{03}s_2\!+\!K_{02}s_3.\!\!\!\!\!\!
\end{align}
Making use of the relation $\mb{r}^{\prime}\!=\!\mb{g}_1\!+\!\boldsymbol{\gamma}$~\eqref{spatialdeformationmeasures} and inserting the first line of~\eqref{covariantbasis} as well as~\eqref{contravariantbasis} into~\eqref{deformationgradient} yields:
\begin{align}
\label{deformationgradient2}
\!\!\!\!\!\!\boldsymbol{\mathcal{F}}\!=\! 
\frac{1}{C}\big[ \mb{g}_1\!+\!\boldsymbol{\gamma} \!+\! \mb{k} \!\times\! \left( s_2 \mb{g}_{2} \!+\! s_3 \mb{g}_{3} \right)\!\big] \!\otimes\! \mb{g}_{01}
\!+\!\mb{g}_2 \!\otimes\! \big[\mb{g}_{02} \!+\! \frac{K_{01}s_3}{C}\mb{g}_{01} \!\big]
\!+\!\mb{g}_3 \!\otimes\! \big[\mb{g}_{03} \!-\! \frac{K_{01} s_2}{C}\mb{g}_{01} \!\big].\!\!\!\!\!\!
\end{align}
Finally, by inserting~\eqref{deformationgradient2} into~\eqref{deformationgradient}, the individual components of the Cauchy-Green deformation tensor can be determined. However, in order to gain further insight into the underlying structure of the deformation gradient, the very ellegant procedure suggested by Geradin and Cardona~\cite{geradin2001} (for initially straight beams) as well as Linn et al.~\cite{linn2013} (for initially curved beams)  is employed by slightly reformulating the expression~\eqref{deformationgradient2} on the basis of the relative rotation tensor $\tilde{\mb{\Lambda}} \!:=\! \mb{\Lambda} \mb{\Lambda}_0^T$. By applying the auxiliary relation $1/C\!=\!1\!+\!(1\!-\!C)/C$ to the pre-factor of $\mb{g}_1$ and solving all the products in~\eqref{deformationgradient2}, the deformation gradient can eventually be reformulated according to:
\begin{align}
\label{deformationgradient3}
\!\!\!\!\!\!\boldsymbol{\mathcal{F}}\!=\!\tilde{\mb{\Lambda}}
\left(
\mb{I}_3\!+\!\boldsymbol{H}\!\otimes\!\mb{g}_{01}
\right) \quad \text{with} \quad \boldsymbol{H}\!:=\!H_i \mb{g}_{0i}, \,\,\,\, \mb{I}_3\!=\!\mb{g}_{0i}\!\otimes\!\mb{g}_{0i}, \,\,\,\, \tilde{\mb{\Lambda}} \!=\! \mb{g}_{i} \!\otimes\! \mb{g}_{0i}.
\end{align}
The components of the vector $\boldsymbol{H}$ identified in~\eqref{deformationgradient3}, denoted as \textit{material strain vector} in~\cite{geradin2001} and~\cite{linn2013}, read:
\begin{align}
\begin{split}
\label{materialstrainvector}
\!\!\!\!\!\!
H_1\!&=\!\frac{1}{C}\left[ \Gamma_1\!+\!(K_2\!-\!K_{02})s_3\!-\!(K_3\!-\!K_{03})s_2 \right], \\
H_2\!&=\!\frac{1}{C}\left[ \Gamma_2\!-\!(K_1\!-\!K_{01})s_3 \right], \,\,\,\,
H_3\!=\!\frac{1}{C}\left[ \Gamma_3\!-\!(K_1\!-\!K_{01})s_2 \right].
\!\!\!\!\!\!
\end{split}
\end{align}
Based on the deformation gradient~\eqref{deformationgradient3}, the Cauchy-Green deformation tensor can be derived according to:
\begin{align}
\label{deformationgradient4}
\boldsymbol{\mathcal{E}}=\frac{1}{2C}
\left[
\boldsymbol{H}\!\otimes \mb{g}_{01} +  \mb{g}_{01}\otimes \boldsymbol{H} 
\right]
+
\frac{\boldsymbol{H}^T\!\boldsymbol{H}}{2C^2} \mb{g}_{01} \otimes \mb{g}_{01}.
\end{align}
The result~\eqref{deformationgradient4} has been consistently derived from the basic kinematic assumptions~\eqref{simoreissner_x0} without any additional approximations. However, in order to finally end up with the simple constitutive laws of the geometrically exact beam theory based on the quadratic form~\eqref{storedenergyfunction}, the following well-known assumption of small local strains is made:
\begin{align}
\label{smallstrainassumption}
\Gamma_i \ll 1 \quad \text{and} \quad R \cdot K_i \ll 1, \,\, R \cdot K_{0i} \ll 1 \,\, \text{for} \,\, i\!=\!1,2,3.
\end{align}
The assumptions~\eqref{smallstrainassumption} state that small local axial and shear strains are considered and that the radii of initial and deformed centerline curvature have to be small as compared to the cross-section radius $R$. In the following, a first-order approximation in these small quantities is considered by setting $C \approx 1$ and neglecting the last, quadratic term in~\eqref{deformationgradient4}. Based on these small-strain assumptions, the approximated Cauchy-Green deformation tensor reads
\begin{align}
\label{deformationgradient5}
\boldsymbol{\mathcal{E}} \approx \bar{\boldsymbol{\mathcal{E}}}=\frac{1}{2}
\left[
\boldsymbol{H}\!\otimes \mb{g}_{01} +  \mb{g}_{01}\otimes \boldsymbol{H}
\right]
 =: \bar{\mathcal{E}}_{ij} \mb{g}_{0i} \otimes \mb{g}_{0j}.
\end{align}
The components $\bar{\mathcal{E}}_{ij}$ of the approximated Cauchy-Green deformation tensor $\bar{\boldsymbol{\mathcal{E}}}$ resulting from this procedure yield:
\begin{align}
\begin{split}
\label{deformationgradient6}
& \bar{\mathcal{E}}_{11} = \Gamma_1 + (K_2 - K_{02}) s_3 - (K_3 - K_{03}) s_2,\\
& \bar{\mathcal{E}}_{12} = \bar{\mathcal{E}}_{21} = \dfrac{1}{2}\left( \Gamma_2 - (K_1- K_{01}) s_3 \right), \\
& \bar{\mathcal{E}}_{13} = \bar{\mathcal{E}}_{31} = \dfrac{1}{2}\left( \Gamma_3 + (K_1- K_{01}) s_2 \right), \\
& \bar{\mathcal{E}}_{22} = \bar{\mathcal{E}}_{33} =  \bar{\mathcal{E}}_{23} = \bar{\mathcal{E}}_{32} = 0.
\end{split}
\end{align}
By applying a Saint-Venant-Kirchhoff material with constitutive tensor $\mathbcal{C}=\mathcal{C}^{ijkl} \mb{g}_{0i} \otimes \mb{g}_{0j} \otimes \mb{g}_{0k} \otimes \mb{g}_{0l}$, the 2. Piola-Kirchhoff stress tensor $\bar{\mb{\mathcal{S}}}\!=\!\bar{\mathcal{S}}^{ij}\mb{g}_{0i} \otimes \mb{g}_{0j}$ can be formulated based on the approximated Cauchy-Green tensor $\bar{\boldsymbol{\mathcal{E}}}$:
\begin{align}
\begin{split}
\label{secondpiolakirchhoff}
& \bar{\mathcal{S}}^{11} = \bar{E} (\Gamma_1 + (K_2 - K_{02}) s_3 - (K_3 - K_{03}) S_2), \\
& \bar{\mathcal{S}}^{22} = \bar{\mathcal{S}}^{33} = -\nu \mathcal{E}_{11}, \\
& \bar{\mathcal{S}}^{12} = \bar{\mathcal{S}}^{21} = G \left( \Gamma_2 - (K_1- K_{01}) s_3 \right), \\
& \bar{\mathcal{S}}^{13} = \bar{\mathcal{S}}^{31} = G \left( \Gamma_3 + (K_1- K_{01}) s_2 \right), \\
& \bar{\mathcal{S}}^{23} = \bar{\mathcal{S}}^{32} = 0.
\end{split}
\end{align}
Here, the scaled Young's modulus $\bar{E}$ and the shear modulus $G$ have been introduced in~\eqref{secondpiolakirchhoff} according to
\begin{align}
\begin{split}
\label{secondpiolakirchhoff_abbreviations}
\bar{E}:=\dfrac{(1-\nu)E}{(1+\nu)(1-2\nu)}, \quad G:=\dfrac{E}{2(1+\nu)},
\end{split}
\end{align}
where $E$ is Young's modulus and $\nu$ is Poisson's ratio. From~\eqref{secondpiolakirchhoff}, it gets obvious that the standard relations known in the geometrically exact beam theory in terms of vanishing in-plane stress components $\bar{\mathcal{S}}^{22} \!=\! \bar{\mathcal{S}}^{33} \!=\!0$ and of a constitutive parameter $E$ in front of the normal stress $\bar{\mathcal{S}}^{11}$, only holds for the special case $\nu\!=\!0$. This is a consequence of the kinematic assumption of rigid cross-sections, which requires the existence of in-plane reaction forces in general. In order to resolve these two putative contradictions for general cases $\nu\!\neq\!0$, the constraint of rigid cross-sections can be weakened by allowing for a uniform lateral contraction of the cross-section with in-plane strain components $\bar{\mathcal{E}}^{22} \!=\!\bar{\mathcal{E}}^{33} \!=\! -\nu E_{11}$ and for a proper in-plane warping field (see~\cite{linn2013} or~\cite{weiss2002a} for further details). Alternatively, the approximation $\nu\!=\!0$ can be employed in the first two lines of~\eqref{secondpiolakirchhoff}. In praxis, this slight inconsistency is often taken into account, which is not unusual in the field of structural theories (see e.g.~\cite{antmann1995} or \cite{koiter1966}). In a last step, the material force and moment stress resultants are determined by integration of the stress vector $\mathbcal{t}_1\!=\!\mathbcal{S} \!\cdot\! \mb{g}_{01}$ acting on a beam cross-section with material normal vector $\mb{g}_{01}$ according to the definition proposed in the original work~\cite{simo1985}:
\begin{align}
\label{forcefromcontinuum}
\begin{split}
\tilde{\mb{F}} \!:=\! \int_{A} \mathbcal{t}_1 dA \!&=\! \int_{A} \big[ E (\Gamma_1 \!+\! (K_2 \!-\! K_{02}) s_3 \!-\! (K_3 \!-\! K_{03}) s_2) \mb{g}_{01} \\ 
&\hspace{0.5cm} + G \left( \Gamma_2 \!-\! (K_1\!-\! K_{01}) s_3 \right) \mb{g}_{02} \!+\! G ( \Gamma_3  \! +\! (K_1\!-\! K_{01}) s_2 )  \mb{g}_{03} \big]dA \\
& =
\left[
\begin{array}{ccc}
EA & 0 & 0 \\
0 & G A & \\ 
0 & 0 & G A
\end{array}
\right]_{\mb{g}_{0i}}
\!\! \cdot
\left[
\begin{array}{c}
\Gamma_1 \\
\Gamma_2 \\ 
\Gamma_3
\end{array}
\right]_{\mb{g}_{0i}}
=\tilde{\mb{C}}_F \tilde{\boldsymbol{\Gamma}}.
\end{split}
\end{align}
\begin{align}
\label{momentfromcontinuum}
\begin{split}
\!\!\!\!\!\! \tilde{\mb{M}} \!:=\!\! \int_{A} \mathbcal{x}_A \! \times \mathbcal{t}_1 dA \!&=\! \int_{A} \left(s_2 \mb{g}_{02} \!+\! s_3\mb{g}_{03} \right) 
\!\times\! \big[ E (\Gamma_1 \!+\! (K_2 \!-\! K_{02}) s_3 \!-\! (K_3 \!-\! K_{03}) s_2) \mb{g}_{01} \!\!\!\!\!\!\\
\!&\hspace{0.5cm}+\! G \left( \Gamma_2 \!-\!(K_1\!-\!K_{01}) s_3 \right) \mb{g}_{02} + G \left( \Gamma_3 \!+\! (K_1\!-\!K_{01}) s_2 \right)  \mb{g}_{03} \big]dA \!\!\!\!\!\!\\
& =
\left[
\begin{array}{ccc}
GI_P & 0 & 0 \\
0 & EI_2 & \\ 
0 & 0 & EI_3
\end{array}
\right]_{\mb{g}_{0i}} \!\!
\cdot
\left[
\begin{array}{c}
K_1- K_{01} \\
K_2- K_{02} \\ 
K_3- K_{03}
\end{array}
\right]_{\mb{g}_{0i}}
=\tilde{\mb{C}}_M \tilde{\boldsymbol{\Omega}}.\!\!\!\!\!\!
\end{split}
\end{align}
Here, the defintions of the moments of inertia of area $I_2\!:=\!\int_A\! s_3^2 dA$ and $I_3\!:=\!\int_A\! s_2^2 dA$ as well as $\int_A \!s_3 dA\!=\!\int_A \!s_2 dA\!=\!\int_A \! s_2 s_3 dA\!=\!0$ have been applied. As expected,~\eqref{forcefromcontinuum} and~\eqref{momentfromcontinuum} yield a constitutive law that is identical to the one postulated in~\eqref{storedenergyfunction}. Comparable derivations based on similar small-strain assumptions can e.g. be found in the original works of Kirchhoff~\cite{kirchhoff1859} and Love~\cite{love1944} in the context of shear-free beam formulations as well as in the current contributions~\cite{geradin2001,kapania2003,linn2013} in the context of geometrically exact Simo-Reissner type formulations. It has to be mentioned that the presented derivation yields alternative material objects $\tilde{\mb{F}}$, $\tilde{\mb{C}}_F$ and $\tilde{\boldsymbol{\Gamma}}$ as well as $ \tilde{\mb{M}}$, $\tilde{\mb{C}}_M$ and $\tilde{\boldsymbol{\Omega}}$ that are pulled-back to the curved, initial reference configuration and not to the straight reference configuration as it was the case for the material objects considered so far (see also the remark in~\cite{meier2014}, page 452). However, the components of the alternative material objects, e.g. $\tilde{\mb{F}}\!=\!F_i\mb{g}_{0i}$, when expressed with respect to the "curved", local basis $\mb{g}_{0i}$ are identical to the components of the original material objects, e.g. $\mb{F}\!=\!F_i\mb{E}_i$, when expressed with respect to the "straight", global basis $\mb{E}_i$. Via push-forward, also the 1. Piola-Kirchhoff stress tensor $\mathbcal{P}$  and the Cauchy stress tensor $\boldsymbol{\sigma}$ can be determined.
\begin{align}
\mathbcal{P}=\mathbcal{F} \mathbcal{S}, \quad \boldsymbol{\sigma}=\frac{1}{\text{det}{\mathbcal{F}}}   \mathbcal{F} \mathbcal{S}\mathbcal{F}^T.
\end{align}
Starting with the 1. Piola-Kirchhoff stress tensor or with the Cauchy stress tensor, the spatial stress resultants, e.g. $\mb{f}\!=\!F_i\mb{g}_{i}$, can be derived similarly to~\eqref{forcefromcontinuum} and~\eqref{momentfromcontinuum}. Since again only first-order terms of the small strains~\eqref{smallstrainassumption} are relevant, it is sufficient to approximate the deformation gradient, required for the push-forward, according to:
\begin{align}
\mathbcal{F} \approx \tilde{\mb{\Lambda}} = \mb{g}_{i} \otimes \mb{g}_{0i}, \quad \text{det}\mathbcal{F} \approx  1.
\end{align}
Consequently, as already postulated in the sections before, the relevant pull-back / push-forward operator is given by a rotation tensor $ \tilde{\mb{\Lambda}}$. If for the derivations above a 3D continuum formulation with material strain and stress measures based on a straight reference configuration had been applied, the resulting 1D material objects would be based on the global basis $\mb{E}_i$ and the total rotation tensor $\mb{\Lambda}$ could be identified as the relevant pull-back / push-forward operator. \\

\section{Kirchhoff-Love Beam Theory}
\label{sec:kirchhoff}

The configuration space of Reissner type beams is described by pointwise six degrees of freedom, namely the three translational components of $\mb{r}(s)$ and three rotational degrees of freedom $\boldsymbol{\psi}(s)$ parametrizing $\mb{\Lambda}(s)$. In this section, the assumption of vanishing shear strains is made, which can be assumed as a sensible approximation for highly slender beams (see e.g. \cite{love1944}). Thus, the cross-sections spanned by $\mb{g}_2$ and $\mb{g}_3$ have to remain perpendicular to $\mb{t}(s):=\mb{r}^{\prime}(s)$:
\begin{align}
\label{kirchhoffconstraints}
\mb{g}_2(s) \cdot \mb{t}(s) \equiv 0 \quad \text{and} \quad \mb{g}_3(s) \cdot \mb{t}(s) \equiv 0 \quad \text{or} \quad \mb{g}_1(s) \equiv \frac{\mb{t}(s)}{||\mb{t}(s)||}.
\end{align}
Principally, this so-called Kirchhoff constraint of vanishing shear deformations can be enforced in a strong or in a weak manner. If the same parametrization ($\mb{r}(s), \boldsymbol{\psi}(s)$) as in the Reissner case is chosen, additional fields of Lagrange multipliers $\lambda_2(s)$ and $\lambda_3(s)$ are necessary, in order to integrate \eqref{kirchhoffconstraints} into a constrained variational problem in a weak sense (see Section~\ref{sec:weakkirchhoff}). In the following Sections~\ref{sec:kirchhoffkinematics}-\ref{sec:kirchhoffweakform}, the concept of developing a parametrization with 4 degrees of freedom ($\mb{r}(s), \varphi(s)$) that fulfills the Kirchhoff constraint in a strong manner as already derived in~\cite{meier2014} will be briefly repeated and extended to the dynamic case. As already introduced in Section~\ref{sec:largerotations_sr}, the scalar-valued quantity $\varphi(s)$ will describe the relative rotation between the material frame $\mb{g}_i(s)$ and an intermediate frame $\mb{g}_{Mi}(s)$ with respect to the tangent vector $\mb{t}(s)$ according to~\eqref{largerotations_2drot}. By means of~\eqref{largerotations_smallestrotation}, one example for a suitable  intermediate frame $\mb{g}_{Mi}(s)$ has already been given, the "Smallest Rotation" intermediate frame. Nevertheless, the following derivations are presented in a rather  general manner, which allows to insert arbitrary alternative intermediate frame definitions $\mb{g}_{Mi}(s)$.

%
\subsection{Kinematics}
\label{sec:kirchhoffkinematics}
%

Within the following three sections, the Kirchhoff constraint~\eqref{kirchhoffconstraints} is strongly enforced (see also~\eqref{largerotations_2drot}) according to 
\begin{align}
\begin{split}
 \label{kirchhoff_strongbasevectors}
 \!\!\!\!\!\!
\mb{g}_1(s) \!= &\mb{g}_{M1}(s) = \frac{\mb{t}(s)}{||\mb{t}(s)||} \quad \text{with} \quad \mb{t}(s)=\mb{r}^{\prime}(s)
\quad \text{and} \quad
\boldsymbol{\Lambda}=\exp{ \! ( \mb{S}( \varphi \mb{g}_1) )} \boldsymbol{\Lambda}_M \quad \text{with} \quad 
\boldsymbol{\Lambda}_M=(\mb{g}_{M1},\mb{g}_{M2},\mb{g}_{M3})_{\mb{E}_i},\!\!\!\!\!\!
\end{split}
\end{align}
where the centerline-aligned intermediate triad base vectors $\mb{g}_{Mi}(s)\!=\!\mb{g}_{Mi}(\mb{r}^{\prime}(s))$ are completely defined by the centerline field $\mb{r}(s)$ but not further specified for now (a possible example is given by~\eqref{largerotations_smallestrotation}). Now, the Kirchhoff constraint is incorporated by expressing the current configuration 
$s,t \rightarrow C := (\mb{r}(s,t), \mb{\Lambda}(\mb{r}(s,t), \varphi(s,t))) \in \Re^3 \times S\!O(3)$ via the new set of primary variables ($\mb{r}(s,t), \varphi(s,t)$). Inserting~\eqref{kirchhoff_strongbasevectors} in the definition of the curvature vectors~\eqref{simoreissner_explicitcurvatures} yields (see~\cite{meier2014}):
\begin{align}
\label{kirchhoffkinematics_ksplit}
   \mb{k} = \mb{k} _{\parallel} + \mb{k} _{\perp} 
  = K_{1} \mb{g}_1 + \mb{S}(\mb{g}_1) \mb{g}_1^{\prime}  = 
  (K_{M1} + \varphi^{\prime}) \mb{g}_1 +\underbrace{\frac{\mb{S}(\mb{r}^{\prime}) \mb{r}^{\prime \prime}}{|| \mb{r}^{\prime} ||^2}}_{=:\boldsymbol{\kappa}} \quad \text{with} \quad K_{M1}:= \mb{g}_{M3}^T \mb{g}_{M2}^{\prime}.
\end{align}
Here, $K_{M1}$ represents the torsion of the (arbitrary) intermediate triad field and $\boldsymbol{\kappa}$ is the Frenet-Serret curvature of the beam centerline $\mb{r}(s)$. In components, the spatial as well as the material curvature vector read:
\begin{align}
\label{curvaturekirchhoffexplicit}
   \boldsymbol{k}=\left(K_{M1} \!+\! \varphi^{\prime}, \, \mb{g}_{2}^T \boldsymbol{\kappa}, \, \mb{g}_{3}^T \boldsymbol{\kappa}\right)^T_{\mb{g}_{i}} \quad \text{and} \quad 
  \boldsymbol{K}=\left(K_{M1} \!+\! \varphi^{\prime}, \, \mb{g}_{2}^T \boldsymbol{\kappa}, \, \mb{g}_{3}^T \boldsymbol{\kappa}\right)^T_{\mb{E}_{i}}.
\end{align}
The intermediate torsion $K_{M1}$ is the only term in~\eqref{curvaturekirchhoffexplicit} that depends on the specific choice of the intermediate triad. Besides the curvature vectors, also the spin vector $\delta \boldsymbol{\theta}$ has to be adapted to the Kirchhoff constraint (see also~\eqref{largerotations_tildetinvmatrixsr}): 
\begin{align}
\label{kirchhoffkinematics_spinvector}
   \delta \boldsymbol{\theta} = \delta \boldsymbol{\theta}_{\parallel} + \delta \boldsymbol{\theta}_{\perp}
  = \delta \Theta_1 \mb{g}_1 + \mb{S}(\mb{g}_1) \delta \mb{g}_1
  = (\delta \Theta_{M1} + \delta \varphi) \mb{g}_1 \!+\! \frac{\mb{S}(\mb{r}^{\prime}) \delta \mb{r}^{\prime}}{|| \mb{r}^{\prime} ||^2}.
\end{align}
In analogy to Reissner beam formulations, the first component $\delta \Theta_1$ of the spin vector, representing a multiplicative increment, will directly be employed in the weak form and not further expressed via additive increments according to $\delta \Theta_1\!=\!\delta \Theta_{M1} \!+\! \delta \varphi$. Consequently, the admissible variations are $s \rightarrow \delta C \!:=\! (\delta \mb{r}(s),\! \delta \boldsymbol{\theta} (\delta \mb{r}(s),\delta \Theta_1(s),\mb{r}(s,t))) \! \in \! \Re^3 \! \times \! \Re^3$ with the new set of variational primary variables $(\delta \mb{r}(s),\delta \Theta_1(s))$ defining the Kirchhoff case. Again, compatibility conditions similar to~\eqref{simoreissner_relationskwtheta} can be stated. Due to the Kirchhoff constraint, these conditions are only required for the tangential vector components. Left-multiplication of the relations in the second line of~\eqref{simoreissner_relationskwtheta} with $\mb{E}_1^T\!=\! (\mb{\Lambda}^T\mb{g}_1)^T$ yields:
\begin{align}
\begin{split}
\label{torsionfree_K1andW1_3}
\!\!\!\!\! \dot{K}_1 \!=\! W_1^{\prime} \!+\! \dot{\mb{g}}_1^T\!(\mb{g}_1 \!\times\! \mb{g}_1^\prime), \,\,\,
\delta {K}_1 \!=\! \delta \Theta_1^{\prime} \!+\! \delta {\mb{g}}_1^T\!(\mb{g}_1 \!\times\! \mb{g}_1^\prime), \,\,\,
\delta {W}_1 \!=\! \delta \dot{\Theta}_1 \!+\! \delta {\mb{g}}_1^T\!(\mb{g}_1 \!\times\! \dot{\mb{g}}_1). \!\!\!\!\!
\end{split}
\end{align}

%
\subsection{Deformation measures and stress resultants}
\label{sec:kirchhoffresultants}
%

Having defined kinematics that are compatible with the Kirchhoff constraint according to~\eqref{kirchhoffconstraints}, the deformation measures, constitutive relations and stress resultants presented in Section~\ref{sec:simoreissner} can be adapted to the shear-free case. Inserting the constrained curvature vectors from \eqref{curvaturekirchhoffexplicit} into the deformation measures $\boldsymbol{\Omega}$ and $\boldsymbol{\omega}$ according to~\eqref{spatialdeformationmeasures} and~\eqref{materialdeformationmeasures} yields
\begin{align}
\label{kirchhoffdeformationmeasures}
   \!\!\!\!\!\! \boldsymbol{\Omega}=
   \left(
   \!K_{M1} \!+\! \varphi^{\prime} \!-\! K_{M01} \!-\! \varphi_0^{\prime},
   \,\,\mb{g}_{2}^T \boldsymbol{\kappa} \!-\! \mb{g}_{02}^T \boldsymbol{\kappa}_0,
   \,\,\mb{g}_{3}^T \boldsymbol{\kappa} \!-\! \mb{g}_{03}^T \boldsymbol{\kappa}_0
   \right)^T_{\mb{E}_{i}}\!,
   \quad
   \boldsymbol{\omega}=
   \left(
   \!K_{M1} \!+\! \varphi^{\prime} \!-\! K_{M01} \!-\! \varphi_0^{\prime},
   \,\,\mb{g}_{2}^T \boldsymbol{\kappa} \!-\! \mb{g}_{02}^T \boldsymbol{\kappa}_0,
   \,\,\mb{g}_{3}^T \boldsymbol{\kappa} \!-\! \mb{g}_{03}^T \boldsymbol{\kappa}_0
   \right)^T_{\mb{g}_{i}}\!.\!\!\!\!\!\!
\end{align}
The objective variation of~\eqref{kirchhoffdeformationmeasures} is still given by $\delta \boldsymbol{\Omega}\!=\!\boldsymbol{\Lambda}^T \! \delta \boldsymbol{\theta}^{\prime}, \, \delta_o \boldsymbol{\omega}\!=\!\delta \boldsymbol{\theta}^{\prime},$ but now the spin vector is constrained according to~\eqref{kirchhoffkinematics_spinvector}. By construction, the shear components of $\boldsymbol{\Gamma}$ and $\boldsymbol{\gamma}$ in \eqref{spatialdeformationmeasures} and
\eqref{materialdeformationmeasures}, vanish due to the Kirchhoff constraint~\eqref{kirchhoffconstraints}:
\begin{align}
\label{kirchhoffdeformationmeasures2}
\boldsymbol{\Gamma} = \epsilon \mb{E}_1, \,\,\,\, \boldsymbol{\gamma} = \epsilon \mb{g}_1 \,\, \rightarrow \,\,
\delta \boldsymbol{\Gamma} = \delta \epsilon \mb{E}_1, \,\,\,\, \delta_o \boldsymbol{\gamma} = \delta \epsilon \mb{g}_1
\quad \text{with} \quad
\epsilon := ||\mb{r}^{\prime}|| - 1, \,\,\,\, \delta \epsilon = \delta \mb{r}^{\prime T} \mb{g}_1.
\end{align}
Here, the abbreviation $\epsilon$ has been introduced for the remaining component representing the axial tension.
Based on the force split $\mb{f}\!=\!\mb{f}_{\parallel}\!+\!\mb{f}_{\perp}\!=\!\mb{F}_1 \mb{g}_1\!+\!\mb{f}_{\perp}$, also the constitutive relations simply as consequence of the Kirchhoff constraint :
\begin{align}
\label{storedenergyfunctionkirchhoff}
\!\!\!\!\!\!\tilde{\Pi}_{int}(\mb{\Omega},\epsilon) \!=\! \frac{1}{2} \mb{\Omega}^T \! \mb{C}_M \mb{\Omega} \!+\! \frac{1}{2} EA \epsilon^2, \,\,
\mb{M}\!=\!\dfrac{\partial \tilde{\Pi}_{int}}{\partial \mb{\Omega}} \!=\! \mb{C}_M \mb{\Omega}\!=\! \mb{\Lambda}^T\!
\mb{m}, \,\,
F_1 \!=\! \frac{\partial \tilde{\Pi}_{int}}{\partial \epsilon} \!=\! EA \epsilon.\!\!\!\!\!\!
\end{align}
The inertia forces $\mb{f}_{\rho}$ as well as the inertia moments $\mb{m}_{\rho}$ are identical to~\eqref{simoreissner_inertia} and~\eqref{simoreissner_Avector}. Like in the Reissner case, the spatial or material angular velocities $\mb{w}$ and $\mb{W}$ as well as the spatial or material angular accelerations $\mb{a}$ and $\mb{A}$ can either be directly used in the employed time integration scheme or they can be expressed via the (additive) rate of the primary variables $(\mb{r}(s),\varphi(s))$. For the latter approach, transformation matrices depending on the definition of the employed intermediate triad (see e.g. $\mb{T}_{M}^{-1}$ in~\eqref{largerotations_tinvmatrixsr} in the case of the SR intermediate triad) as well as their time-derivatives are required in order to formulate relations similar to~\eqref{simoreissner_additiveacceleration}.
\begin{align}
\label{kirchhoffkinematics_angularvelocity0}
\mb{w}=\mb{T}_{\boldsymbol{\theta} \mb{t}} \dot{\mb{t}} + \dot{\varphi} \mb{g}_{1}, \quad \mb{a}=\dot{\mb{T}}_{\boldsymbol{\theta} \mb{t}} \dot{\mb{t}} +  \mb{T}_{\boldsymbol{\theta} \mb{t}} \ddot{\mb{t}} + \dot{\varphi} \dot{\mb{g}}_{1}+ \ddot{\varphi} \mb{g}_{1}.
\end{align}
In the Kirchhoff case, a third variant can be advantageous: Similar to the curvature vector $\mb{k}$ (see \eqref{kirchhoffkinematics_ksplit}) and the spin vector $\delta \boldsymbol{\theta}$ (see \eqref{kirchhoffkinematics_spinvector} or \eqref{largerotations_tildetinvmatrixsr}), also the angular velocity $\mb{w}$ can be split into the following two components
\begin{align}
\label{kirchhoffkinematics_angularvelocity}
  \!\!\!\!\!\! \mb{w} \!=\! \mb{w}_{\parallel}\!+\!\mb{w}_{\perp} \!=\!W_1 \mb{g}_1 \!+\! \mb{S}(\mb{g}_1) \dot{\mb{g}}_1 \!=\!   
  W_1 \mb{g}_1 \!+\! \frac{\mb{S}(\mb{r}^{\prime}) \dot{\mb{r}}^{\prime}}{|| \mb{r}^{\prime} ||^2}, \quad
  \mb{a}\!=\!\dot{W}_1 \mb{g}_1\!+\!W_1 \dot{\mb{g}}_1 \!+\! \mb{S}(\mb{g}_1) \ddot{\mb{g}}_1.\!\!\!\!\!\!
\end{align}
While $\mb{w}_{\perp}$ is determined by $\mb{r}(s)$, $W_1$ is not specified any further, i.e. it is not expressed via $\mb{r}(s)$ and $\varphi(s)$ as done in~\eqref{kirchhoffkinematics_angularvelocity0}.\\

\hspace{0.2 cm}
\begin{minipage}{15.0 cm}
\textbf{Remark: } Three possible time integration schemes result from the different variants given above:
\begin{align}
\begin{split}
\label{kirchhoffkinematics_timeintegrations}
\text{1)} \hspace{0.72cm} (\dot{\mb{r}},\ddot{\mb{r}},\mb{w},\mb{a})_{n+1}&=\mb{f}((\mb{r},\boldsymbol{\Lambda})_{n+1},(\mb{r},\boldsymbol{\Lambda})_{n},(\dot{\mb{r}},\mb{w})_{n},(\ddot{\mb{r}},\mb{a})_{n}), \\
\text{2)} \hspace{0.72cm} (\dot{\mb{r}},\ddot{\mb{r}},\dot{\varphi},\ddot{\varphi})_{n+1}&=\mb{f}((\mb{r},\varphi)_{n+1},(\mb{r},\varphi)_{n},(\dot{\mb{r}},\dot{\varphi})_{n},(\ddot{\mb{r}},\ddot{\varphi})_{n}), \\
\text{3)} \,\,\,\, (\dot{\mb{r}},\ddot{\mb{r}},W_1,\dot{W}_1)_{n+1}&=\mb{f}((\mb{r},\boldsymbol{\Lambda})_{n+1},(\mb{r},\boldsymbol{\Lambda})_{n},(\dot{\mb{r}},W_1)_{n},(\ddot{\mb{r}},\dot{W}_1)_{n}).
\end{split}
\end{align}
sd\end{minipage}

\hspace{0.2 cm}
\begin{minipage}{15.0 cm}
\textbf{Remark: } From the tangent transformations~\eqref{kirchhoffkinematics_ksplit},~\eqref{kirchhoffkinematics_spinvector} and~\eqref{kirchhoffkinematics_angularvelocity}, the following similarities become obvious:
\begin{align}sd
\label{kirchhoff_tildetinvmatrixsr}
   \delta \boldsymbol{\theta}
   =
   \left(
   \begin{array}{cc}
   \frac{1}{|| \mb{t} ||} \mb{S}(\mb{g}_1) & \mb{g}_{1}
   \end{array}
   \right)
   \left(
   \begin{array}{c}
   \delta \mb{t}\\
   \delta \Theta_1
   \end{array}
   \right), 
   \,\,\,\,
   \mb{k}
     =
   \left(
   \begin{array}{cc}
   \frac{1}{|| \mb{t} ||} \mb{S}(\mb{g}_1) & \mb{g}_{1}
   \end{array}
   \right)
   \left(
   \begin{array}{c}
   \mb{t}^{\prime} \\
   K_1
   \end{array}
   \right),
   \,\,\,\,
   \mb{w}
     =
   \left(
   \begin{array}{cc}
   \frac{1}{|| \mb{t} ||} \mb{S}(\mb{g}_1) & \mb{g}_{1}
   \end{array}
   \right)
   \left(
   \begin{array}{c}
   \dot{\mb{t}} \\
   W_1
   \end{array}
   \right).
  \end{align}
\end{minipage}

%
\subsection{Strong and weak form}
\label{sec:kirchhoffweakform}
%

In this section, the spatial representation of mechanical equilibrium will be considered. In the following, the notation will be simplified by summarizing external forces and moments as well as inertia forces and moments according to $\mb{\tilde{f}}_{\rho}\!:=\!\mb{\tilde{f}}\!+\!\mb{f}_{\rho}$ and $\mb{\tilde{m}}_{\rho}\!:=\!\mb{\tilde{m}}\!+\!\mb{m}_{\rho}$. The shear forces $\mb{f}_{\perp}$, which provide no work contribution in the Kirchhoff case, can be eliminated from the strong form of mechanical equilibrium yielding the following set of four differential equations:
\begin{align}
\label{condensedequilibrium}
\begin{split}
 \mb{g}_1^T \left( \mb{m}^{\prime} + \mb{\tilde{m}}_{\rho} \right) & = 0, \\
 ({F}_{1}\mb{g}_1)^{\prime} + \left[\frac{\mb{r}^{\prime}}{||\mb{r}^{\prime}||^2} \times \left( \mb{m}^{\prime} + \mb{\tilde{m}}_{\rho} \right) \right]^{\prime} +
\mb{\tilde{f}}_{\rho} & = \mb{0}.
\end{split}
\end{align}
The set~\eqref{condensedequilibrium} is sufficient in order to solve for the four primary variables~$\mb{r}$ and~$\varphi$ as soon as the stress resultants $\mb{f}_{\parallel}$ and $\mb{m}$ are expressed by kinematic and constitutive relations from Section~\ref{sec:kirchhoffresultants}. Multiplying~\eqref{condensedequilibrium} with the admissible translational and rotational variations $\delta \mb{r}$ and $\delta \Theta_1$ and integrating by parts gives the equivalent weak form of equilibrium:
\begin{align}
\label{weakform3}
  \!\!\!\!\!\!G\!=\!\!\int \limits_0^l \!
  \Big[\!
    \underbrace{\delta \boldsymbol{\theta}^{\prime T} \! \mb{m}}_{\delta_o \boldsymbol{\omega}^T \! \mb{m}} 
     +\underbrace{\delta \mb{r}^{\prime T} \! \mb{g}_1  F_1}_{\delta \epsilon F_1 } 
     -\delta \mb{r}^T \mb{\tilde{f}}_{\rho}
     -\delta \boldsymbol{\theta}^T \! \mb{\tilde{m}} _{\rho}  
  \Big] ds
 - \Big[ \delta \mb{r}^T \! \mb{f}_{\sigma} \!+\! \delta \boldsymbol{\theta}^T \! \mb{m}_{\sigma} \Big]_{\Gamma_{\sigma}} \hspace{-0.1cm}
 \dot{=} \, 0, \,\,\,\,\,\, \delta \boldsymbol{\theta} \!=\! \delta \Theta_1 \mb{g}_1 \!+\! \frac{\mb{r}^{\prime} \!\times\! \delta \mb{r}^{\prime}}{||\mb{r}^{\prime}||^2}.\!\!\!\!\!\!
\end{align}
In \eqref{weakform3}, the constrained spatial spin vector according to \eqref{kirchhoffkinematics_spinvector} has been identified and already substituted with the symbol $\delta \boldsymbol{\theta}$. As indicated by the curly brackets in~\eqref{weakform3}, the pre-factors of the stress resultants $\mb{m}$ and $F_{1}$ are represented by the objective variations $\delta_o \boldsymbol{\omega}$ and $\delta \epsilon$, underlining the geometrical exactness of the proposed Kirchhoff beam formulation. Finally, the problem setup has to be completed by proper boundary and initial conditions:
\begin{align}
\label{boundaryconditions}
\begin{split}
\!\!\!\!\!\!\mb{r} \!=\! \mb{r}_{u}, \, \mb{g}_1 \!=\! \mb{g}_{1u}, \, \varphi \!=\! \varphi_{u} \,\, \text{on} \,\, \varGamma_{u},  \,\, 
&\mb{f} \!=\! \mb{f}_{\sigma}, \, \mb{m} \!=\! \mb{m}_{\sigma} \, \, \text{on} \,\, \varGamma_{\sigma}, \,\,  \varGamma_{\sigma} \cap \varGamma_{u} \!=\! \varnothing, \, \varGamma_{\sigma} \cup \varGamma_{u} \!=\! \{0,l\}\!\!\!\!\!\!\\
\!\!\!\!\!\! \,\, &\mb{r}\!=\!\mb{r}_0, \, \dot{\mb{r}}\!=\!\mb{v}_0, \, 
\varphi \!=\! \varphi_0, \mb{w}\!=\!\mb{w}_0
\, \, \text{at} \,\, t\!=\!0.\!\!\!\!\!\!
\end{split}
\end{align}
Here $\mb{g}_{1u}$ prescribes the orientation of the tangent vector and $\varphi_u$ the orientation of the cross-section with respect to a rotation around the tangent vector. How these conditions can be modeled in practice is shown in~\ref{anhang:elestrongkirchhoff_dirichletconditions}. By introducing the trial space $(\mb{r},\boldsymbol{\Lambda}(\mb{r},\varphi)) \! \in \!  \mathcal{U} $ satisfying \eqref{boundaryconditions} and the weighting space $(\delta \mb{r},\delta \boldsymbol{\theta}(\delta \mb{r},\delta \Theta_1,\mb{r})) \!\in\! \mathcal{V}$, with $\delta \mb{r} \!= \mb{0}, \, \delta \Theta_1 \!=\! 0 \, \text{on} \, \varGamma_{u}$, the beam problem is fully defined. It should be emphasized that only the concrete analytic expressions for $\mb{g}_{M2}(\mb{r}(s))$ and $\mb{g}_{M3}(\mb{r}(s))$ in~\eqref{kirchhoff_strongbasevectors} depend on the specific choice of the intermediate triad definition.\\

\hspace{0.2 cm}
\begin{minipage}{15.0 cm}
\textbf{Remark: } In~\cite{meier2014}, a special choice of intermediate triad field in~\eqref{kirchhoff_strongbasevectors} has been proposed, the Frenet-Serret triad of the beam centerline. It has been argued in this work, that this choice is not favorable for numerical purposes as consequence of singularities occurring on straight beam segments. However, the Frenet-Serret frame is very beneficial for the analytical treatment of Kirchhoff-Love beam problems and has e.g. been applied in order to derive analytical reference solutions for the numerical examples of Section~\ref{sec:elenumericalexamples}.\\
\end{minipage}

\hspace{0.2 cm}
\begin{minipage}{15.0 cm}
\textbf{Remark: } As shown in~\cite{meier2015}, the Kirchhoff-Love beam theory presented so far provides an ideal basis for the derivation of reduced beam models, i.e. beam models that are valid under certain restrictions concerning initial geometry and external loads and will eventually yield considerably simplified finite element formulations. On this basis, a special torsion-free beam element formulation could be derived  that completely abstains from any rotational degrees of freedom, a distinct property that drastically simplifies many standard procedures in the numerical treatment of geometrically exact beam element formulations.
\end{minipage}

%
\subsection{Weak enforcement of Kirchhoff constraint}
\label{sec:weakkirchhoff}
%

In the last sections, the set of primary variable fields ($\mb{r}(s), \varphi(s)$) has been chosen in a way such that the Kirchhoff constraint~\eqref{kirchhoffconstraints} of vanishing shear strains is strongly fulfilled by construction. However, more flexibility in the subsequent discretization process (see e.g. Section~\ref{sec:spatialdiscretization}) could be gained by formulating a Reissner type beam problem, which allows for two independent interpolations for the centerline field $\mb{r}(s)$ as well as the triad field $\boldsymbol{\Lambda}(s)$, and by weakly enforcing the Kirchhoff constraint of vanishing shear strains by means of additional constraint equations:
\begin{align}
\label{weakkirchhoff_strongconstraint}
\Gamma_{j}(s) \equiv \mb{g}_j^T(s)\mb{r}^{\prime}(s) \equiv 0 \quad \text{for} \quad j\!=\!2,3.
\end{align}
In order to integrate these constraints into a variational framework, an additional Lagrange multiplier potential
\begin{align}
\label{weakkirchhoff_weakconstraintpotential0}
\Pi_{\lambda_{\Gamma{23}}}\!=\! \int \limits_{0}^l (\lambda_{\Gamma2}(s) \Gamma_2 (s)+\lambda_{\Gamma3}(s) \Gamma_3 (s))ds
\end{align}
is required. The Lagrange multiplier fields $\lambda_{\Gamma2}(s)$ and $\lambda_{\Gamma3}(s)$ in~\eqref{weakkirchhoff_weakconstraintpotential0} can be interpreted as the shear force components $F_2(s)$ and $F_3(s)$, i.e. reaction forces which enforce the constraint of vanishing shear strains. Variation of the Lagrange multiplier potential~\eqref{weakkirchhoff_weakconstraintpotential0} leads to the contribution of the Kirchhoff constraint to the weak form:
\begin{align}
\label{weakkirchhoff_weakconstraintpotential}
\!\!\!\!\!\!\delta \Pi_{\lambda_{\Gamma{23}}}\!=\! \int \limits_{0}^l (\delta \lambda_{\Gamma2}(s) \Gamma_2 (s)\!+\!\delta \lambda_{\Gamma3}(s) \Gamma_3 (s))ds\!+\!
\int \limits_{0}^l (\lambda_{\Gamma2}(s) \delta \Gamma_2 (s)\!+\!\lambda_{\Gamma3}(s) \delta \Gamma_3 (s))ds.\!\!\!\!\!\!
\end{align}
The first term in~\eqref{weakkirchhoff_weakconstraintpotential} represents the weak statement of the Kirchhoff constraint~\eqref{weakkirchhoff_strongconstraint} while the second term can be interpreted as the work contribution of the shear reaction forces. Similar to the displacement primary fields, a proper trial space $(\lambda_{\Gamma2},\lambda_{\Gamma3}) \in \mathcal{U}_{\lambda_{\Gamma{23}}}$ and a proper weighting space $(\delta \lambda_{\Gamma2},\delta \lambda_{\Gamma3}) \in \mathcal{V}_{\lambda_{\Gamma{23}}}$ have to be introduced which uniquely define the resulting mixed beam formulation. The discrete realization of~\eqref{weakkirchhoff_weakconstraintpotential} will be presented in Section~\ref{sec:eleweakkirchhoff}.

\section{Temporal Discretization of Primary Fields}
\label{sec:temporaldiscretization}
In this and the next section, the space-time-continuous beam problems presented so far will be discretized. While spatial discretization as discussed in Section~\ref{sec:spatialdiscretization} will be based on the finite element method (FEM), a recently proposed extension of the well-known generalized-$\alpha$ method~\cite{chung1993} from vector spaces to Lie groups~\cite{arnold2007,bruels2010,bruels2012}, which is directly applicable to the beam element formulations proposed in this work, will be presented in this section. This time integrator represents an implicit, one-step finite difference scheme that inherits most of the desirable properties of the standard generalized-$\alpha$ variant. In the context of finite element methods for solid mechanics, it is often more convenient to perform time discretization on the semi-discrete problem setting resulting from spatial discretization. Here, just the opposite succession, i.e. the initial time discretization is followed by a subsequent spatial discretization, is chosen. This second variant is often applied in the development of geometrically exact beam finite element formulations and will lead to simpler discrete expressions. In the following, the considered total time interval $t \in [0,T]$ is subdivided into equidistant subintervals $[t_n,t_{n+1}]$ with constant time step size $\Delta t$, where $n \in \mathbb{N}_0$ is the time step index. Consequently, the solution for the primary variable fields describing the current configuration $C(s,t) \!:=\! (\mb{r}(s,t), \mb{\Lambda}(s,t))$ is computed at a series of discrete points in time with associated configurations $C(s,t_n) \!:=\! (\mb{r}(s,t_n), \mb{\Lambda}(s,t_n))\!=:\! (\mb{r}_n(s), \mb{\Lambda}_n(s))$. In order to simplify notation required for subsequent derivations, the weak form $G$ (see e.g.~\eqref{weakformspatial} or~\eqref{weakform3}) is split into the contributions $G_{int}$ of internal forces, $G_{kin}$ of kinetic forces and $G_{ext}$ of external forces:
\begin{align}
\label{temporaldiscretization_Gsplit}
G\!=\!G_{int}\!+\!G_{kin}\!-\!G_{ext}.
\end{align}
In a next step, the basics of the Lie group generalized-$\alpha$ method originally proposed by \cite{arnold2007,bruels2010,bruels2012} will be presented. This method will be applied for time discretization of the Reissner and Kirchhoff type beam element formulations presented in subsequent sections, whose configuration space $C(s,t) \!=\! (\mb{r}(s,t), \mb{\Lambda}(s,t)) \in \Re^3 \times S\!O(3)$ is defined by the position field $\mb{r}(s,t)$ and the rotation field $\mb{\Lambda}(s,t)$. It is emphasized that the following procedure is independent from the rotation parametrization of $\mb{\Lambda}(s)$ employed to these different beam element formulations. In order to express the (material) angular velocities and accelerations $\mb{W}_{n+1}$ and $\mb{A}_{n+1}$ at the end of a time interval $[t_n,t_{n+1}]$ in terms of known quantities at time $t_n$ and the unknown rotation field $\mb{\Lambda}_{n+1}(s)$, the vectors $\tilde{\boldsymbol{\theta}}_{n+1}$ and $\boldsymbol{\tilde{\Theta}}_{n+1}$, representing the spatial and material multiplicative rotation increment between the time steps $t_n$ and $t_{n+1}$, are introduced:
\begin{align}
\label{liegenalpha_rotationincrement_timestep}
\!\!\!\!\!\!\exp{  \! ( \mb{S}(   \boldsymbol{\tilde{\Theta}}_{n+1}  ) )   }
\!=\! \boldsymbol{\Lambda}_{n}^T \boldsymbol{\Lambda}_{n+1}, \,\,\,\,\,\, \boldsymbol{\tilde{\Theta}}_{n+1}\!=\! \boldsymbol{\Lambda}^T_n \boldsymbol{\tilde{\theta}}_{n+1}
\!=\! \boldsymbol{\Lambda}^T_{n+1} \boldsymbol{\tilde{\theta}}_{n+1}.\!\!\!\!\!\!
\end{align}
Besides the distinctions that can be made for vector space time integrators (e.g. implicit or explicit scheme, one-step or multi-step scheme, employed methodology in order to guarantee stability and/or conservation properties) two further classifications can be made for time integration schemes applied to rotational variables: First, depending on the type of spatial rotation interpolation, the succession of spatial and temporal discretization will in most cases influence the resulting discrete solution. Secondly, it can be distinguished between approaches that apply a time integration scheme directly to the vectors $(\boldsymbol{\tilde{\Theta}},\mb{W},\mb{A})$ and approaches that express angular velocities and accelerations by means of (additive) rates of the primary variables, i.e. $(\boldsymbol{\psi},\dot{\boldsymbol{\psi}},\ddot{\boldsymbol{\psi}})$ (see also~\eqref{simoreissner_additiveacceleration}). The axial vectors $\boldsymbol{\tilde{\Theta}},\mb{W}$ and $\mb{A}$ are associated with elements of the Lie algebra $so(3)$. Consequently, time integration schemes of the former type are commonly denoted as Lie group time integration schemes. In this context, it can be further distinguished between Lie group schemes that are based on the material vectors $(\boldsymbol{\tilde{\Theta}},\mb{W},\mb{A})$ and schemes that are based on their spatial counterparts $(\boldsymbol{\tilde{\theta}},\mb{w},\mb{a})$. Only the former variant will be considered in this work. Arguably, one of the first Lie group time integration schemes, at least in the context of geometrically exact beam formulations, has been proposed by Simo and Vu-Quoc~\cite{simo1988} and represents the Lie group extension of the classical Newmark scheme. On the contrary, the scheme of~\cite{arnold2007,bruels2010,bruels2012}, which will be presented in the following, is the Lie group extension of the standard generalized-$\alpha$ method. Also this scheme is based on the four parameters $\beta,\gamma,\alpha_m$ and $\alpha_f$ and simplifies to the variant of Simo and Vu-Quoc~\cite{simo1988} for the special choice $\alpha_m\!=\!\alpha_f\!=\!0$. A distinctive feature of the Lie group generalized-$\alpha$ scheme lies in the fact that all terms of the weak form are evaluated at the end point $t_{n+1}$ of the considered time interval:
\begin{align}
\label{liegenalpha_finalweakform}
\!\!\!\!\!\!G_{n+1}\!=\!G(\mb{r}_{n+1},\dot{\mb{r}}_{n+1},\ddot{\mb{r}}_{n+1},\boldsymbol{\Lambda}_{n+1},\mb{W}_{n+1},\mb{A}_{n+1},\mb{\tilde{f}}_{n+1},\mb{\tilde{m}}_{n+1},\mb{f}_{\sigma,n+1},\mb{m}_{\sigma,n+1})\dot{=}0.\!\!\!\!\!\!
\end{align}
In a next step, the update formulas for translational quantities are given by a standard Newmark scheme
\begin{align}
\label{liegenalpha_newmark}
\begin{split}
 \mb{\tilde{u}}_{n+1} &:=\mb{r}_{n+1}-\mb{r}_{n},\\
\mb{\tilde{u}}_{n+1} &=\Delta t \dot{\mb{r}}_{n} + + \Delta t^2[(0.5-\beta)\ddot{\mb{r}}_{mod,n} + \beta\ddot{\mb{r}}_{mod,n+1}], \\
\dot{\mb{r}}_{n+1}&= \dot{\mb{r}}_{n} + \Delta t[(1-\gamma)\ddot{\mb{r}}_{mod,n} + \gamma \ddot{\mb{r}}_{mod,n+1}].
\end{split}
\end{align}
This update scheme is slightly changed in form of a multiplicative configuration update for the rotations:
\begin{align}
\label{liegenalpha_newmark2}
\begin{split}
\exp{  \! ( \mb{S}(   \boldsymbol{\tilde{\Theta}}_{n+1}  ) )   } & = \boldsymbol{\Lambda}_{n}^T\boldsymbol{\Lambda}_{n+1}, \\
\boldsymbol{\tilde{\Theta}}_{n+1}&=\Delta t \mb{W}_{n} + + \Delta t^2[(0.5-\beta)\mb{A}_{mod,n} + \beta \mb{A}_{mod,n+1}], \\
 \mb{W}_{n+1}&= \mb{W}_{n} + \Delta t[(1-\gamma)\mb{A}_{mod,n} + \gamma \mb{A}_{mod,n+1}].
\end{split}
\end{align}
The modified acceleration vectors $\ddot{\mb{r}}_{mod}$ occuring in~\eqref{liegenalpha_newmark} are related to the physical acceleration vectors $\ddot{\mb{r}}$ according to:
\begin{align}
\label{liegenalpha_modaccel}
\begin{split}
(1-\alpha_m)\ddot{\mb{r}}_{mod,n+1}+\alpha_m\ddot{\mb{r}}_{mod,n}=
(1-\alpha_f)\ddot{\mb{r}}_{n+1}+\alpha_f\ddot{\mb{r}}_{n} \quad \text{with} \quad \ddot{\mb{r}}_{mod,0}\!=\!\ddot{\mb{r}}_{0}.
\end{split}
\end{align}
In a similar manner, the modified as well as the real / physical angular accelerations are related according to:
\begin{align}
\label{liegenalpha_modaccel2}
\begin{split}
(1-\alpha_m)\mb{A}_{mod,n+1}+\alpha_m\mb{A}_{mod,n}=
(1-\alpha_f)\mb{A}_{n+1}+\alpha_f\mb{A}_{n} \quad \text{with} \quad \mb{A}_{mod,0}\!=\!\mb{A}_{0}.
\end{split}
\end{align}
For later use, it is favorable to express the vectors $\dot{\mb{r}}_{n+1}$ and $\ddot{\mb{r}}_{n+1}$ in terms of the primary unknown $\mb{r}_{n+1}$:
\begin{align}
\label{liegenalpha_finalvelacc}
\begin{split}
\dot{\mb{r}}_{n+1}&=\frac{\gamma}{\beta \Delta t} \mb{\tilde{u}}_{n+1}+\left(1-\frac{\gamma}{\beta}\right)\dot{\mb{r}}_{n}+\Delta t\left(1-\frac{\gamma}{2\beta}\right)\ddot{\mb{r}}_{mod,n},\\
\ddot{\mb{r}}_{n+1}&=\frac{1-\alpha_m}{\beta \Delta t^2(1-\alpha_f)}\mb{\tilde{u}}_{n+1}-\frac{1-\alpha_m}{\beta \Delta t(1-\alpha_f)}\dot{\mb{r}}_{n} \\
&+\left[-\frac{(1-\alpha_m)(0.5-\beta)}{\beta(1-\alpha_f)}+\frac{\alpha_m}{1-\alpha_f}\right]\ddot{\mb{r}}_{mod,n}-\frac{\alpha_f}{1-\alpha_f}\ddot{\mb{r}}_{n}.
\end{split}
\end{align}
A similar relation can be formulated for the material vectors of angular velocity $\mb{W}_{n+1}$ and acceleration $\mb{A}_{n+1}$:
\begin{align}
\label{liegenalpha_finalvelacc2}
\begin{split}
\mb{W}_{n+1}&=\frac{\gamma}{\beta \Delta t} \boldsymbol{\tilde{\Theta}}_{n+1}+\left(1-\frac{\gamma}{\beta}\right)\mb{W}_{n}+\Delta t\left(1-\frac{\gamma}{2\beta}\right)\mb{A}_{mod,n},\\
\mb{A}_{n+1}&=\frac{1-\alpha_m}{\beta \Delta t^2(1-\alpha_f)}\boldsymbol{\tilde{\Theta}}_{n+1}-\frac{1-\alpha_m}{\beta \Delta t(1-\alpha_f)}\mb{W}_{n} \\
&+\left[-\frac{(1-\alpha_m)(0.5-\beta)}{\beta(1-\alpha_f)}+\frac{\alpha_m}{1-\alpha_f}\right]\mb{A}_{mod,n}-\frac{\alpha_f}{1-\alpha_f}\mb{A}_{n}.
\end{split}
\end{align}
In~\cite{bruels2010} and~\cite{bruels2012}, it has been proven that the integration scheme given by equations~\eqref{liegenalpha_finalweakform}-\eqref{liegenalpha_finalvelacc2} yields the same favorable properties as the standard generalized-$\alpha$ method, which are second-order accuracy, unconditional stability (within the linear regime), controllable damping of the high-frequency modes and minimized damping of the low-frequency modes. Remarkably, the parameter choice leading to this optimal behavior is identical to that of the standard generalized-$\alpha$ scheme. Furthermore, it has been shown that this scheme can consistently treat non-constant mass matrix contributions, such as the term $\mb{\Lambda} \mb{C}_{\rho} \mb{A}$ occurring in geometrically exact Reissner and Kirchhoff type beam formulations. It has to be stated that the extended generalized-$\alpha$ scheme cannot guarantee for exact conservation of energy, linear and angular momentum. In the field of Lie group time integration schemes, a large variety of methods aiming to guarantee these conservation properties has been proposed \cite{simo1991,bottasso1998,gonzalez1996,bachau1999,betsch2001,ibrahimbegovic2002b,lens2008,romero2008b,kane2000,demoures2015}. However, the perhaps most essential advantage of the extended generalized-$\alpha$ scheme as compared to these alternatives lies again in its simplicity and flexibility. Independent of the beam theory (Reissner or Kirchhoff type), the employed spatial interpolation schemes as well as the chosen set of nodal primary variables (e.g. in terms of rotation parametrization), this time integration scheme can directly be applied without the need for any further adaptions.\\

\hspace{0.2 cm}
\begin{minipage}{15.0 cm}
\textbf{Review:} As mentioned in Section~\ref{sec:temporaldiscretization}, a large number of scientific contributions considers the development of energy-momentum schemes for Simo-Reissner type formulations. However, this does not hold true for Kirchhoff-Love beam element formulations. For example, the formulations of anisotropic type (see requirements 5a)-5c) in Table~\ref{tab:comparison_requirements}) haven't considered dynamics so far. In~\cite{weiss2002b}, a energy-momentum scheme is proposed, but only for temporal discretization of the beam centerline. The torsional problem associated with the rotational degree of freedom is only considered in a static manner, i.e. arbitrary dynamic problems containing also rigid body rotations with respect to the beam axis are not accessible by this approach. The temporal discretization applied in the present work encompasses both, the translational and the rotational fields. As verified numerically in Section~\ref{sec:elenumericalexamples}, the numerical dissipation provided by the extended generalized-$\alpha$ scheme will enable energy-stable time integration for highly nonlinear problems. Concretely, in the well-known ellbow-cantilever example of Section~\ref{sec:examples_freeoscillationselbow}, a considerably improved energy stability can be observed as compared to~\cite{jelenic1999}, where a Lie-group extension of the standard Newmark scheme~\cite{simo1988} has been employed. Nevertheless, a supplementation of the proposed Kirchhoff-Love beam elements by energy-momentum schemes for temporal discretization seems to be a interesting direction of future research.
\end{minipage}

\section{Spatial Discretization Methods for Primary Fields}
\label{sec:spatialdiscretization}

Spatial discretization is exclusively considered in the context of finite element methods within this work. It represents the core topic in the development of geometrically exact Kirchhoff beam elements considered in this work.

%
\subsection{Discretization of beam centerline}
\label{sec:spatialdiscretization_beamcenterline}
%

First, the spatial discretization of the beam centerline $\mb{r}(s)$ will be conducted. Thereto, an elementwise parameter space $\xi \in [-1;1]$. is introduced with the element Jacobian $J(\xi)\!:=||\mb{r}_{0h,\xi}(\xi)||$ mapping between infinitesimal increments in the parameter space and the arc-length space according to $ds\!=\!J(\xi) d\xi$. In the following two sections, two different interpolation schemes based on Lagrange polynomials or Hermite polynomials respectively will be presented.

%
\subsubsection{Discretization of beam centerline based on Lagrange polynomials}
\label{sec:spatialdiscretization_lagrange}
%

The highest derivative of the primary variable $\mb{r}(s)$ occurring in the weak form~\eqref{weakformspatial} of the Simo-Reissner beam theory is the first derivative $\mb{r}^{\prime}(s)$ of the centerline curve. Consequently, a $C^0$-continuous interpolation of the beam centerline $\mb{r}(s)$ is sufficient in this case. Thus, the standard choice of trial functions is given by Lagrange polynomials, yielding
\begin{align}
\label{lagrange_interpolation}
\mb{r}_h(\xi) = \sum_{i=1}^{n_r} L^i(\xi) \hat{\mb{d}}^i =: \mbd{L} \hat{\mbd{d}} \quad \text{and} \quad 
\mb{r}_{0h}(\xi) = \sum_{i=1}^{n_r} L^i(\xi) \hat{\mb{d}}_0^i =: \mbd{L} \hat{\mbd{d}}_0.
\end{align}
In~\eqref{lagrange_interpolation}, the vectors $\hat{\mb{d}}^i, \hat{\mb{d}}_0^i \in \Re^3$ represent the nodal positions, whereas $L^i(\xi)$ are standard Lagrange polynomials satisfying the interpolation property $L^i(\xi^j)\!=\!\delta^{ij}$ at the element node coordinates $\xi^j$ as well as proper completeness conditions. Here, $\delta^{ij}$ represents the Kronecker delta symbol. The matrix $\mbd{L}$ and the vector $\hat{\mbd{d}}$ represent element-wise assemblies of the shape functions $L^i$ and the position vectors $\hat{\mb{d}}^i$ for $i\!=\!1,...,n_r$, which are given by $\mbd{L}\!:=\!(L^1(\xi)\mb{I}_3,...,L^{n_r}(\xi)\mb{I}_3)$ as well as $\hat{\mbd{d}}\!:=\!(\hat{\mb{d}}^{1 T},...,\hat{\mb{d}}^{n_r T})^T$. Following a Bubnov-Galerkin approach, the interpolation of the trial functions $\delta \mb{r}(s)$ reads:
\begin{align}
\label{lagrange_testfunctions}
\delta \mb{r}_h(\xi) = \sum_{i=1}^{n_r} L^i(\xi) \delta \hat{\mb{d}}^i=: \mbd{L} \delta \hat{\mbd{d}}.
\end{align}

%
\subsubsection{Discretization of beam centerline based on Hermite polynomials}
\label{sec:spatialdiscretization_hermite}
%

The highest derivative of the primary variable $\mb{r}(s)$ occurring in the weak form~\eqref{weakform3} of the Kirchhoff beam is given by the second derivative $\mb{r}^{\prime \prime}(s)$. Consequently, for the interpolation of the centerline $\mb{r}(s)$ shape functions are required that fulfill $C^1$-continuity at the element boundaries. Besides this requirement, a $C^1$-continuous centerline representation can be very beneficial for problem class requiring a smooth geometry representation such as beam-to-beam contact problems (see e.g. \cite{meier2015b,meier2015c}). In order to guarantee for $C^1$-continuity, Hermite shape functions are employed:
\begin{align}
\label{hermite_interpolation}
\!\!\!\!\!\!\mb{r}_h(\xi) \!=\!\!\! \sum_{i=1}^{2} H^i_{d}(\xi) \hat{\mb{d}}^i \!+\! \frac{c}{2} \! \sum_{i=1}^{2} H^i_{t}(\xi) \hat{\mb{t}}^i \!=:\! \mbd{H} \hat{\mbd{d}}, \,\, 
\mb{r}_{0h}(\xi) \!=\!\!\! \sum_{i=1}^{2} H^i_{d}(\xi) \hat{\mb{d}}_0^i \!+\! \frac{c}{2} \! \sum_{i=1}^{2} H^i_{t}(\xi) \hat{\mb{t}}_0^i\!=:\! \mbd{H} \hat{\mbd{d}}_0.\!\!\!\!\!\!
\end{align}
In \eqref{hermite_interpolation}, the vectors $\hat{\mb{d}}^i, \hat{\mb{d}}_0^i \in \Re^3$ and $\hat{\mb{t}}^i, \hat{\mb{t}}_0^i \in \Re^3$ represent nodal position and nodal tangent vectors at the two boundary nodes of the resulting finite elements. Again, the matrix $\mbd{H}$ and the vector $\hat{\mbd{d}}$ represent proper element-wise assemblies of the shape functions $H^i_{t}$ and $H^i_{d}$ as well as the nodal position and tangent vectors $\hat{\mb{d}}^i$ and $\hat{\mb{t}}^i$ for $i\!=\!1,2$. The explicit expressions are $\mbd{H}\!:=\!(H^1_d(\xi) \mb{I}_3,0.5cH^1_t(\xi)\mb{I}_3,H^2_d(\xi)\mb{I}_3,0.5cH^2_t(\xi)\mb{I}_3)$ as well as $\hat{\mbd{d}}\!:=\!(\hat{\mb{d}}^{1 T},\hat{\mb{t}}^{1 T},\hat{\mb{d}}^{2 T},\hat{\mb{t}}^{2 T})^T$. In~\cite{meier2014}, it has already been shown that the Hermite shape functions in~\eqref{hermite_interpolation} fulfill the interpolation property for the nodal positon and tangent vectors as well as proper completeness conditions for third-order polynomials and that the optimal choice for the constant $c\!=\!l_{ele,h}$ is given by the element length $l_{ele,h}$ of the discretized initial geometry, which can be determined in an iterative manner. On the basis of~\eqref{hermite_interpolation}, the first and second arc-length derivative can be determined as 
\begin{align}
\label{hermite_centerlinederivatives}
\mb{r}_{h}^{\prime}(\xi)=\frac{1}{J(\xi)}\mb{r}_{h,\xi}(\xi), \quad 
\mb{r}_{h}^{\prime \prime}(\xi)=\frac{1}{J(\xi)^2}\mb{r}_{h,\xi \xi}(\xi) - \frac{J_{,\xi}(\xi)}{J(\xi)^3} \mb{r}_{h,\xi}(\xi),
\end{align}
where the Jacobi factor $J(\xi)$ and its parameter derivative $J_{,\xi}(\xi)$ appearing in~\eqref{hermite_centerlinederivatives} are given by the following relations:
\begin{align}
\label{hermite_jacobian}
J(\xi):=||\mb{r}_{0h,\xi}(\xi)|| \quad \text{and} \quad J_{,\xi}(\xi)\!=\!\frac{d}{d\xi}||\mb{r}_{0h,\xi}||\!=\!\frac{\mb{r}^T_{0h,\xi}\mb{r}_{0h,\xi \xi}}{||\mb{r}_{0h,\xi}||}.
\end{align}
Finally, also the variations $\delta \mb{r}(s)$ have to be discretized by properly chosen test functions. Following a Bubnov-Galerkin approach, also the interpolation of $\delta \mb{r}(s)$ is based on Hermite polynomials and given by the expression:
\begin{align}
\label{testfunctions}
\begin{split}
\delta \mb{r}_h(\xi) \!=\! \sum_{i=1}^{2} H^i_{d}(\xi) \delta \hat{\mb{d}}^i \!+\! \frac{c}{2} \sum_{i=1}^{2} H^i_{t}(\xi) \delta \hat{\mb{t}}^i\!=:\! \mbd{H} \delta \hat{\mbd{d}}.
\end{split}
\end{align}
In an analogous manner, the procedure presented here can also be extended to Hermite polynomials of higher order.

%
\subsection{Discretization of rotation field}
\label{sec:spatialdiscretization_rotationfield}
%

In Section~\ref{sec:largerotations_rotvec}, two parametrizations of rotation tensors have been investigated: a parametrization via rotation vectors $\boldsymbol{\psi}$ and a parametrization via the SR mapping on the basis of the set $(\mb{t},\varphi)$. In the following two sections, these two variants will be employed in order to parametrize the rotation tensors $\mb{\Lambda}^i$ at the element nodes $i\!=\!1,...,n_{\Lambda}$. In the subsequent Sections~\ref{sec:spatialdiscretization_triadrelrot} and~\ref{sec:spatialdiscretization_triadsr}, also two possible approaches for the interpolation of these nodal triads in the elements interior, one based on rotation vectors and one based on the SR mapping, will be presented.

%
\subsubsection{Parametrization of nodal triads via rotation vectors}
\label{sec:spatialdiscretization_nodaltriads_rotationvector}
%

According to Section~\ref{sec:largerotations_rotvec}, rotation vectors $\hat{\boldsymbol{\psi}}^1_{n},...,\hat{\boldsymbol{\psi}}^{n_{\Lambda}}_{n}$ can be employed as primary variables describing the nodal triads $\mb{\Lambda}^i_{n}\!=\!\mb{\Lambda}^i_{n}(\hat{\boldsymbol{\psi}}^i_{n}), \,i\!=\!1,2,...,n_{\Lambda}$, at time step $t_{n}$. The update from iteration $k$ to $k+1$ of a nonlinear solution scheme might either be based on additive rotation increments $\Delta \hat{\boldsymbol{\psi}}^{i,k+1}_{n}$ or on multiplicative rotation increments $\Delta \hat{\boldsymbol{\theta}}^{i,k+1}_{n}$ given by:
\begin{align}
\label{newtoniteration_triadupdates}
\begin{split}
\!\!\!\!\!\!\mb{\Lambda}^{i,k+1}_{n}(\hat{\boldsymbol{\psi}}^{i,k+1}_{n})\!&=\!\mb{\Lambda}^{i,k}_{n}(\hat{\boldsymbol{\psi}}^{i,k}_{n}\!+\!\Delta \hat{\boldsymbol{\psi}}^{i,k+1}_{n})\\ \text{or} \,\, \mb{\Lambda}^{i,k+1}_{n}(\hat{\boldsymbol{\psi}}^{i,k+1}_{n})\!&=\!\exp{\!(\mb{S}[\Delta \hat{\boldsymbol{\theta}}^{i,k+1}_{n}])} \mb{\Lambda}^{i,k}_{n}(\hat{\boldsymbol{\psi}}^{i,k}_{n}).\!\!\!\!\!\!
\end{split}
\end{align}
Only for rotations with a magnitude smaller than $180^{\circ}$ a unique rotation vector can be extracted from a given triad (by applying e.g. Spurrier's algorithm, see \cite{spurrier1978}). Within this work, rotation vectors are always extracted in a manner such that $\boldsymbol{\psi} \in ]-\pi;\pi]$. Within this range, the transformation matrix $\mb{T}$ between additive and multiplicative rotation vector increments (see~\eqref{largerotations_tmatrix}) and its inverse, which do not exist at $\boldsymbol{\psi} \!=\! 2\pi$, are always well-defined.

%
\subsubsection{Parametrization of nodal triads via smallest rotation mapping}
\label{sec:spatialdiscretization_nodaltriadsrotation_smallestrotation}
%

Alternatively, the nodal triads can be defined via a relative rotation of nodal intermediate triads $\mb{\Lambda}_{M_{\hat{\varphi}}}^i$ by a relative angle $\hat{\varphi}^i$ with respect to the tangent (see~\eqref{largerotations_2drot}). The nodal intermediate triads are defined by the smallest rotation mapping of the nodal intermediate triad $\bar{\mb{\Lambda}}_{M_{\hat{\varphi}},n}^i\!:=\!\mb{\Lambda}_{M_{\hat{\varphi}},n-1}^i$ of the last time step onto the basis vector $\mb{g}_{1,n}^i$ of the current step $t_{n}$:
\begin{align}
\label{triadsr_nodalreftriads}
\mb{\Lambda}^i_{n}\!=\! \exp{\!(\mb{S}[\hat{\varphi}^i_{n} \mb{g}_{1,n}^i])}\mb{\Lambda}_{M_{\hat{\varphi}},n}^i, \quad \mb{\Lambda}_{M_{\hat{\varphi}},n}^i\!=\!\text{sr}(\bar{\mb{\Lambda}}_{M_{\hat{\varphi}},n}^i,\mb{g}_{1,n}^i) \quad \text{with} \quad
\bar{\mb{\Lambda}}_{M_{\hat{\varphi}},n}^i\!:=\!\mb{\Lambda}_{M_{\hat{\varphi}},n-1}^i.
\end{align}
This variant will be used for Kirchhoff type beam element formulations in combination with the Hermite centerline interpolation~\eqref{hermite_interpolation}, where the first base vector $\mb{g}_1^i$ is defined via the tangent vector to the beam centerline, i.e. $\mb{g}_1^i\!=\!\mb{t}(\xi^i)/||\mb{t}(\xi^i)||$. All in all, the nodal triad is defined by the nodal relative angle $\hat{\varphi}^i$ and the tangent vector $\mb{t}(\xi^i)$ at the node $i$, i.e. $\mb{\Lambda}^i=\mb{\Lambda}^i(\varphi^i,\mb{t}(\xi^i))$. However, it has to be emphasized that in contrary to $\hat{\varphi}^i$, the vector $\mb{t}(\xi^i)$ does not necessarily have to be a nodal primary variable. If the considered node $i$ coincides with one of the two element boundary nodes employed in the Hermite interpolation~\eqref{hermite_interpolation}, this tangent vector indeed represents a nodal primary variable, i.e. $\mb{t}(\xi^i)\!=\!\hat{\mb{t}}^i$ for $i\!=\!1,2$, otherwise the tangent vector simply represents the interpolated centerline derivative at this position, i.e. $\mb{t}(\xi^i)\!=\!\mb{r}^{\prime}(\xi^i)$. Based on (additive) increments $\Delta \hat{\varphi}^{i,k+1}_n$ of the nodal relative angles, the configuration update from iteration $k$ to iteration $k+1$ of the Newton-Raphson scheme at time step $t_n$ reads:
\begin{align}
\label{newtoniteration_srtriadupdates}
\!\!\!\!\!\!\mb{\Lambda}^{i,k+1}_{n}(\hat{\varphi}^{i,k+1}_{n}\!,\mb{g}_{1,n}^{i,k+1})\!=\! \exp{\!(\mb{S}[\hat{\varphi}^{i,k}_{n}\!+\!\Delta \hat{\varphi}^{i,k+1}_{n}\!, \mb{g}_{1,n}^{i,k+1}])}\mb{\Lambda}_{M_{\hat{\varphi}},n}^{i,k+1}, \,\,\, \mb{\Lambda}_{M_{\hat{\varphi}},n}^{i,k+1}\!=\!\text{sr}(\bar{\mb{\Lambda}}_{M_{\hat{\varphi}},n}^{i},\mb{g}_{1,n}^{i,k+1}).\!\!\!\!\!\!
\end{align}
The base vectors $\mb{g}_{1,n}^{i,k+1}\!=\!\mb{r}^{\prime\,k+1}_n(\xi^i)/||\mb{r}^{\prime\,k+1}_n(\xi^i)||$ are fully defined by the centerline $\mb{r}_n^{\prime\,k+1}\!=\!\mb{r}^{\prime}(\hat{\mb{d}}^{1,k+1}_n,\hat{\mb{d}}^{2,k+1}_n,\hat{\mb{t}}^{1,k+1}_n,\hat{\mb{t}}^{2,k+1}_n)$ based on the additive updates $\hat{\mb{d}}^{i,k+1}_n\!=\!\hat{\mb{d}}^{i,k}_n\!+\!\Delta \hat{\mb{d}}^{i,k+1}_n$ as well as $\hat{\mb{t}}^{i,k+1}_n\!=\!\hat{\mb{t}}^{i,k}_n\!+\!\Delta \hat{\mb{t}}^{i,k+1}_n$.\\

\hspace{0.2 cm}
\begin{minipage}{15.0 cm}
\textbf{Remark:} Within this work, intermediate triads $\mb{\Lambda}_M$ based on the SR mapping are used for two different purposes: Firstly, they are used for the definition of nodal material triads $\mb{\Lambda}^i$ based on nodal relative angles $\hat{\varphi}^i$ and associated nodal intermediate triads $\mb{\Lambda}_{M_{\hat{\varphi}}}^i$ (SR mapping "in time" from $\bar{\mb{\Lambda}}_{M_{\hat{\varphi}},n}^i\!=\!\mb{\Lambda}_{M_{\hat{\varphi}},n-1}^i$ to $\mb{\Lambda}_{M_{\hat{\varphi}},n}^i$, see~\eqref{triadsr_nodalreftriads}). Secondly, they are used for the definition of an interpolated material triad field $\mb{\Lambda}(\xi)$ based on a relative angle field $\varphi(\xi)$ and an associated intermediate triad field $\mb{\Lambda}_{M_{\varphi},n}(\xi)$ (SR mapping "in space" from a reference triad $\mb{\Lambda}_{r,n}\!=\!\mb{\Lambda}_{n}(\xi^r)$ to $\mb{\Lambda}_{M_{\varphi},n}(\xi)$, see~\eqref{triadsr_triadfield}). In order to distinguish these two applications, the additional index $\hat{\varphi}$ or $\varphi$ of $\mb{\Lambda}_{M}$ refers to the associated relative angle.\\
\end{minipage}

\hspace{0.2 cm}
\begin{minipage}{15.0 cm}
\textbf{Review:} In Appendix~\ref{anhang:elestrongkirchhoff_dirichletconditions} it is shown that a rotation vector-based parametrization of the nodal triads according to the last section will considerably simplify the modeling of complex Dirichlet boundary conditions and joints as compared to the tangent vector-based variant considered in this section. Only the straight formulations from the literature (see requirement 6) of Table~\ref{tab:comparison_requirements}) employ a rotation vector-based triad parametrization, which supports a simple prescription of such conditions. In Section~\ref{sec:elenumericalexamples}, it will turn out that the tangent vector-based parametrization of nodal triads on the other hand will lead to a better nonlinear solver performance. Thus, in practice one might combine the advantage of these to approaches by employing a rotation vector-based parametrization only at nodes where complex Dirichlet or constraint conditions have to be modeled and the tangent vector-based variant to all the remaining nodes.\\
\end{minipage}

%
\subsubsection{Triad interpolation based on local rotation vectors}
\label{sec:spatialdiscretization_triadrelrot}
%

In this section, a triad interpolation is presentented that has originally been proposed by Shoemake~\cite{shoemake1985} and for the first time employed to geomerically exact beam element formulations by Crisfield and Jeleni\'{c}~\cite{crisfield1999,jelenic1999}. On each of the considered $n_{\Lambda}$ nodes, a triad $\mb{\Lambda}^i$, with $i=1,2,...,n_{\Lambda}$, is defined by primary degrees of freedom either according to Section~\ref{sec:spatialdiscretization_nodaltriads_rotationvector} or to Section~\ref{sec:spatialdiscretization_nodaltriadsrotation_smallestrotation}. The interpolation strategy presented in this section is independent from the specific choice of nodal primary variables. First, a reference triad $\mb{\Lambda}_r$ based on the triads at nodes $I$ and $J$ is defined:
\begin{align}
\label{triadrelrot_lambdar}
\mb{\Lambda}_r \!=\! \mb{\Lambda}^I \exp{\!(\mb{S}(\boldsymbol{\Phi}^{IJ}/2))} \quad \text{with} \quad
\exp{\!(\mb{S}(\boldsymbol{\Phi}^{IJ}))}=\mb{\Lambda}^{IT} \! \mb{\Lambda}^J.
\end{align}
The nodes $I$ and $J$ are chosen as the \textit{two} middle triads for elements with an even number $n_{\Lambda}$ of nodes and as the \textit{one} middle triad (i.e. $I\!=\!J$) for elements with an odd number $n_{\Lambda}$ of nodes (see also equation (6.2) in~\cite{crisfield1999}, which is based on a sightly different node numbering). Based on the definition $\mb{\Lambda}_r$, the interpolated triad field is defined as follows:
\begin{align}
\label{triadrelrot_interpolation}
\mb{\Lambda}_h(\xi) \!=\! \mb{\Lambda}_r \exp{\!(\mb{S}(\boldsymbol{\Phi}_{lh}(\xi)))}, \quad
\boldsymbol{\Phi}_{lh}(\xi) \!=\! \sum_{i=1}^{n_{\Lambda}} L^i(\xi) \boldsymbol{\Phi}_{l}^i, \quad
\exp{\!(\mb{S}(\boldsymbol{\Phi}_{l}^i))}=\mb{\Lambda}_r^T \! \mb{\Lambda}^i.
\end{align}
Again, $L^i(\xi)$ represent the standard Lagrange polynomials of order $n_{\Lambda}\!-\!1$ and $\boldsymbol{\Phi}_{l}^i$ the (material) rotation vectors associated with the relative rotation between the triad $\mb{\Lambda}_i$ at node $i$ and the reference triad $\mb{\Lambda}_r$. The interpolation~\eqref{triadrelrot_interpolation} represents an orthonormal interpolation scheme. Thus, the interpolated triad field is still an element of the rotation group, i.e. $\mb{\Lambda}_h(\xi) \! \in \! S \! O(3) \,\forall \, \xi \in [-1;1]$. Furthermore, the interpolation scheme~\eqref{triadrelrot_interpolation} preserves the objectivity of the space-continuous deformation measures (see \cite{crisfield1999}). The curvature vector (see~\eqref{simoreissner_explicitcurvatures}) resulting from~\eqref{triadrelrot_interpolation} reads
\begin{align}
\label{triadrelrot_kappa}
\mb{K}_h(\xi)\!=\! \mb{T}^{-T}(  \boldsymbol{\Phi}_{lh}(\xi)  ) \boldsymbol{\Phi}_{lh}^{\prime}(\xi),
\end{align}
and can exactly represent the state of constant curvature $\mb{K}_h\!=\!\text{const\,}$. Thus, the two-noded variant of~\eqref{triadrelrot_interpolation} can be identified as a geodesic interpolation scheme, since it connects two points on the nonlinear manifold $S \! O(3)$ via the "shortest distance". Consequently, the two-noded variant of this interpolation represents the $S \! O(3)$-counterpart to the linear interpolation~\eqref{lagrange_interpolation} of quantities in $\Re^3$.  In contrast to the interpolations~\eqref{lagrange_interpolation} and~\eqref{hermite_interpolation} of the beam centerline, the rotation interpolation~\eqref{triadrelrot_interpolation} is nonlinear in the nodal degrees of freedom. Thus, if e.g. the field of rotation vectors $\boldsymbol{\psi}_h(\xi)$ with nodal values $\boldsymbol{\psi}_h(\xi^i)\!=\!\hat{\boldsymbol{\psi}}^i$ is employed for triad parametrization, the rotation vector interpolation resulting from~\eqref{triadrelrot_interpolation} can be written in an abstract manner in the form $\boldsymbol{\psi}_h(\xi)\!=\!\text{nl}(\hat{\boldsymbol{\psi}}^1,...,\hat{\boldsymbol{\psi}}^n_{\Lambda},\xi)$. While an explicit interpolation rule for the rotation vectors $\boldsymbol{\psi}_h(\xi)$ is not needed for practical purposes - the triad field is already given by~\eqref{triadrelrot_interpolation} - the discrete version of the spin vector field $\delta \boldsymbol{\theta}_h(\xi)$ and the field of (multiplicative) rotation vector increments $\Delta \boldsymbol{\theta}_h(\xi)$ will be required in the next sections for the spatially discretized weak form of the balance equations and its linearization. In~\cite{crisfield1999}, these discretized fields have consistently been derived from the triad interpolation~\eqref{triadrelrot_interpolation}, leading to:
\begin{align}
\label{crisfieldspininterpolation}
\Delta \boldsymbol{\theta}_h(\xi) = \sum_{i=1}^{n_{\Lambda}} \tilde{\mb{I}}^i(\xi) \Delta \hat{\boldsymbol{\theta}}^i 
=: \tilde{\mbd{I}} \Delta \hat{\mbds{\theta}}, \quad \delta \boldsymbol{\theta}_h(\xi) = \sum_{i=1}^{n_{\Lambda}} \tilde{\mb{I}}^i(\xi) \delta \hat{\boldsymbol{\theta}}^i =: \tilde{\mbd{I}} \delta \hat{\mbds{\theta}}.
\end{align}
The arc-length derivatives of the interpolations~\eqref{crisfieldspininterpolation} follow in a straightforward manner and are given by:
\begin{align}
\label{crisfieldspininterpolationderiv}
\Delta \boldsymbol{\theta}_h^{\prime}(\xi) = \sum_{i=1}^{n_{\Lambda}} \frac{1}{J(\xi)}\tilde{\mb{I}}_{,\xi}^i(\xi) \Delta \hat{\boldsymbol{\theta}}^i, \quad \delta \boldsymbol{\theta}_h^{\prime}(\xi) = \sum_{i=1}^{n_{\Lambda}} \frac{1}{J(\xi)} \tilde{\mb{I}}^i_{,\xi}(\xi) \delta \hat{\boldsymbol{\theta}}^i.
\end{align}
The generalized shape function matrices $\tilde{\mb{I}}^i(\xi) \!\in\! \Re^3 \!\times\! \Re^3$ as well as their derivatives $\tilde{\mb{I}}_{,\xi}^i(\xi)$ have been derived in the original work~\cite{crisfield1999} (see also~\ref{anhang:rotshapefunctions}). Again, assembly matrices and vectors $\tilde{\mbd{I}}\!:=\!(\tilde{\mb{I}}^{1},...,\tilde{\mb{I}}^{n_{\Lambda}})$ as well as $\Delta \hat{\mbds{\theta}}\!:=\!(\Delta \hat{\boldsymbol{\theta}}^{1 T},...,\Delta \hat{\boldsymbol{\theta}}^{n_{\Lambda} T})^T$ and $\delta \hat{\mbds{\theta}}\!:=\!(\delta \hat{\boldsymbol{\theta}}^{1 T},...,\delta \hat{\boldsymbol{\theta}}^{n_{\Lambda} T})^T$ have been introduced. These shape functions depend on the rotational primary variables in a nonlinear manner, e.g. $\tilde{\mb{I}}^i(\xi)\!=\!\text{nl}(\hat{\boldsymbol{\psi}}^1,...,\hat{\boldsymbol{\psi}}^n,\xi)$ if nodal rotation vectors according to Section~\ref{sec:spatialdiscretization_nodaltriads_rotationvector} are employed. Consequently, they have to be re-calculated for every new configuration and this dependency on the rotational primary variables would have to be considered within a consistent linearization procedure in case the spin vector interpolation $\delta \boldsymbol{\theta}_h(\xi)$ given in~\eqref{crisfieldspininterpolation} is used in the weak form according to a Bubnov-Galerkin procedure. In order to avoid this additional linearization, it can be sensible to follow a Petrov-Galerkin approach based on an interpolation of $\delta \boldsymbol{\theta}_h(\xi)$ via Lagrange polynomials. This strategy is also applied within this work and leads to:
\begin{align}
\label{petrovspininterpolation}
\delta \boldsymbol{\theta}_h(\xi) = \sum_{i=1}^{n_{\Lambda}} L^i(\xi) \delta \hat{\boldsymbol{\theta}}^i =: \mbd{L} \delta \hat{\mbds{\theta}}, \quad 
\delta \boldsymbol{\theta}_h^{\prime}(\xi) = \sum_{i=1}^{n_{\Lambda}} \frac{1}{J(\xi)} L^i_{,\xi}(\xi) \delta \hat{\boldsymbol{\theta}}^i=:\mbd{L}^{\!\! \prime} \delta \hat{\mbds{\theta}}.
\end{align}
Nevertheless, the interpolation $\Delta \boldsymbol{\theta}_h(\xi)$ is still based on~\eqref{crisfieldspininterpolation} in order to end up with a consistent linearization. As emphasized in~\cite{jelenic1999}, the generalized shape functions fulfill the following interpolation and completeness properties
\begin{align}
\label{generalizedshapefunction_completeness}
\tilde{\mb{I}}^i(\xi^j)=\delta^{ij} \mb{I}_3, \quad \sum_{i=1}^{n_{\Lambda}}\tilde{\mb{I}}^i(\xi) \equiv \mb{I}_3, \quad \sum_{i=1}^{n_{\Lambda}}\tilde{\mb{I}}_{,\xi}^{i}(\xi) \equiv \mb{0},
\end{align}
i.e. these shape functions can exactly represent constant rotation vector increment fields. Since these properties are also fulfilled by the Lagrange polynomials, both the Bubnov-Galerkin interpolation~\eqref{crisfieldspininterpolation} as well as the Petrov-Galerkin interpolation~\eqref{petrovspininterpolation} of the spin vector field $\delta \boldsymbol{\theta}_h(\xi)$ can exactly represent a constant distribution $\delta \boldsymbol{\theta}_h(\xi)\!=\!\text{const.}$ in an element. This property is important with respect to conservation of angular momentum (see e.g. Sections~\ref{sec:spatialdiscretization_conservation}).

%
\subsubsection{Triad interpolation based on "Smallest Rotation" mapping}
\label{sec:spatialdiscretization_triadsr}
%

\begin{figure}[h!!]
 \centering
   \subfigure[Node numbering applied to Simo-Reissner elements.]
   {
    \includegraphics[width=0.48\textwidth]{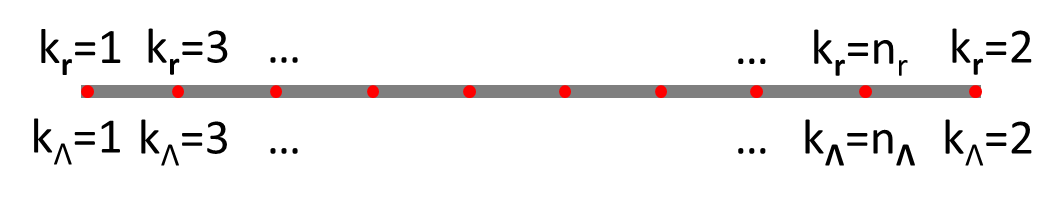}
    \label{fig:nodenum1}
   }
   \subfigure[Node numbering applied to Kirchhoff-Love elements.]
   {
    \includegraphics[width=0.48\textwidth]{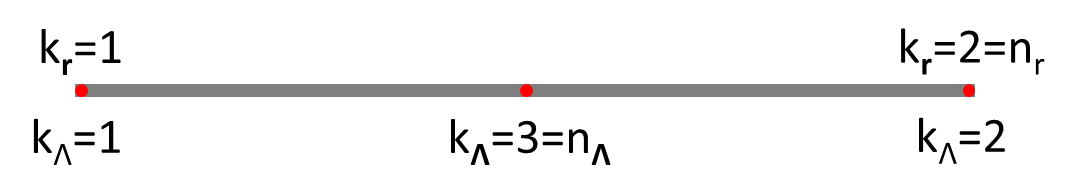}
    \label{fig:nodenum2}
   }
  \caption{Element-local node numbering of translational and rotational primary fields.}
  \label{fig:nodenum}
\end{figure}

In this section, again, a triad interpolation with $n_{\Lambda}$ nodes is considered. On each of these nodes, a triad $\mb{\Lambda}^i$, with $i=1,2,...,n_{\Lambda}$, is defined by primary degrees of freedom either according to Section~\ref{sec:spatialdiscretization_nodaltriads_rotationvector} or to Section~\ref{sec:spatialdiscretization_nodaltriadsrotation_smallestrotation}. Similar to the last section, the interpolation strategy presented in the following is independent from the specific choice of nodal primary variables. Concretely, a novel interpolation scheme is proposed that defines an orthonormal triad field $\mb{\Lambda}_h(\xi) \! \in \! S \! O(3) \,\forall \, \xi \in [-1;1]$ based on a given tangent vector field $\mb{t}(\xi)\!=\!\mb{r}^{\prime}(\xi)$ and nodal triads $\mb{\Lambda}^i\!=\!(\mb{g}_1^i,\mb{g}_2^i,\mb{g}_3^i)$ with $i=1,2,...,n_{\Lambda}$. In the following, this tangent vector field is defined by a Hermite interpolation of the beam centerline according to~\eqref{hermite_interpolation} based on two nodes at the element boundary with six degrees of freedom $\hat{\mb{d}}^j, \hat{\mb{t}}^j$ and $j\!=\!1,2$, respectively. It has to be emphasized that the number of nodes of the triad interpolation can in general differ from the number of nodes of the Hermite centerline interpolation, i.e. $n_{\Lambda} \neq 2$ (see Figure~\ref{fig:nodenum2}).\\

The nodal triads are oriented tangential to the beam centerline curve. Thus, the first base vectors yield:
\begin{align}
\label{triadsr_nodaltriads}
\mb{g}_1(\xi)\!=\!\frac{\mb{r}^{\prime}(\xi)}{||\mb{r}^{\prime}(\xi)||} \rightarrow \mb{g}_1^i\!=\!\frac{\mb{r}^{\prime}(\xi^i)}{||\mb{r}^{\prime}(\xi^i)||}.
\end{align}
Similarly to Section~\ref{sec:spatialdiscretization_triadrelrot}, one nodal triad $\mb{\Lambda}^{I}$ initially has to be chosen as reference triad of the interpolation scheme:
\begin{align}
\label{triadsr_nodalreftriadforinterpolation}
\mb{\Lambda}_r\!=\!\mb{\Lambda}^{I} \quad \text{with} \quad I \in \{1,2,...,n_{\Lambda}\}.
\end{align}
Based on the reference triad $\mb{\Lambda}_r$ and the nodal triads $\mb{\Lambda}^i$ the interpolation procedure is defined according to:
\begin{align}
\label{triadsr_triadfield}
\begin{split}
\mb{\Lambda}_h(\xi) \!&=\! \exp{\!(\mb{S}[\varphi_h(\xi) \mb{g}_1(\xi)])} \, \mb{\Lambda}_{M_{\varphi}}(\xi), \quad \quad \quad \quad \quad
\mb{\Lambda}_{M_{\varphi}}(\xi) \!=\! \text{sr}(\mb{\Lambda}_r,\mb{g}_1(\xi)), \\
\varphi_h(\xi)\!&=\! \sum_{i=1}^{n_{\Lambda}} L^i(\xi) \varphi^i, \hspace{4.0cm}
\exp{\!(\mb{S}[\varphi^i \mb{g}_1^i])} \!=\! \mb{\Lambda}^i \mb{\Lambda}_{M_{\varphi}}^T(\xi^i).
\end{split}
\end{align}
The general curvature vector $\mb{K}_h(\xi)$ for interpolations that fulfill the Kirchhoff constraint in a strong manner is given by~\eqref{curvaturekirchhoffexplicit}. The total 
torsion resulting from~\eqref{triadsr_triadfield} can be derived in a straight-forward manner to:
\begin{align}
\label{triadsr_totaltorsion}
\begin{split}
K_1 \!=\! K_{M_{\varphi}1} + \varphi_h^{\prime}, \quad K_{M_{\varphi}1}\!=\!  \mb{g}^T_{M_{\varphi}3} \mb{g}^{\prime}_{M_{\varphi}2} \!=\!-\frac{\boldsymbol{\kappa}^T \! \mb{g}_{1}^I}{1\!+\!\mb{g}_{1}^T\mb{g}_{1}^I}.
\end{split}
\end{align}

\hspace{0.2 cm}
\begin{minipage}{15.0 cm}
\textbf{Remark:} Ihe nodal relative angles $\varphi^i\!=\!\varphi_h(\xi^i)$ in~\eqref{triadsr_triadfield} are different from the nodal primary variables $\hat{\varphi}^i$ in~\eqref{triadsr_nodalreftriads} of Section~\ref{sec:spatialdiscretization_nodaltriadsrotation_smallestrotation} for parametrization of nodal triads. For both quantities, the symbol $\varphi$ has been chosen since in both cases the relative angle between the material triad $\mb{\Lambda}(\xi^i)$ and an intermediate triad $\mb{\Lambda}_{M_{\varphi}}(\xi^i)$ in case of $\varphi^i$ (stemming from a SR mapping "in space") or an intermediate triad $\mb{\Lambda}_{M_{\hat{\varphi}}}^i$ in case of $\hat{\varphi}^i$ (stemming from a SR mapping "in time") is measured. The difference becomes clear by realizing that the intermediate triad $\mb{\Lambda}_{M_{\varphi}}(\xi^i)$ resulting from the smallest rotation of $\mb{\Lambda}_r$ onto $\mb{g}_1(\xi^i)$ (see~\eqref{triadsr_triadfield}) will in general differ from the intermediate triad $\mb{\Lambda}_{M_{\hat{\varphi}}}^i$ resulting from the smallest rotation of $\bar{\mb{\Lambda}}_{M_{\hat{\varphi}}}^i$ onto $\mb{g}_1(\xi^i)$ (see~\eqref{triadsr_nodalreftriads}).\\
\end{minipage}

\hspace{0.2 cm}
\begin{minipage}{15.0 cm}
\textbf{Review:} In~\cite{meier2014}, it has been argued that the SR mapping exhibits a singularity for $\bar{\mb{g}}_{1}^T \mb{g}_{1}\!=\!-1$ in \eqref{largerotations_smallestrotation}. For the proposed finite element formulations this can only occur for rotation increments per time step that are larger than $180^{\circ}$ or large element deformations exhibiting relative rotations between the element center and boundary nodes that are larger than $180^{\circ}$ (see~\cite{meier2014}). Consequently, these singularities are not practically relevant for reasonable spatial and temporal discretizations. In~\cite{meier2014}, it has also been argued that these singularities occurring at relative rotations of $180^{\circ}$ represent the optimum that can be achieved for mappings on the tangent vector. The approach employed by the anisotropic formulations from the literature is sightly different (see requirement 2c) in Table~\ref{tab:comparison_requirements}). There, only one spatially fixed reference triad at the beam endpoint is used for the initial SR mapping in space, and only one temporarily fixed reference triad at every centerline position $s$ for the SR mapping in time. Consequently, for these formulations, singularities could occur for practically relevant configurations in case the relative rotation between the cross-sections at the beam end point and at an arbitrary centerline position $s$ exceeds $180^{\circ}$ or in case the total cross-section rotation from the initial to the current configuration exceeds $180^{\circ}$ at any centerline position $s$.\\
\end{minipage}

\hspace{0.2 cm}
\begin{minipage}{15.0 cm}
\textbf{Review: } In the literature, the smallest rotation mapping defined in~\eqref{largerotations_smallestrotation} of Section~\ref{sec:largerotations_sr} is often denoted as "rotation without twist". Thus, sometimes it is mistakenly assumed that an intermediate triad field as employed in~\eqref{triadsr_triadfield} would exhibit a vanishing torsion (see e.g.~\cite{greco2013}, where the slightly different interpolation scheme~\eqref{triadsr_triadfield_noneobjective} has been employed). However, according to~\eqref{triadsr_totaltorsion}, the torsion of the intermediate triad field~\eqref{triadsr_triadfield} constructed via the SR mapping does not vanish for general curved 3D configurations. It can easily be shown that the torsion vanishes in the limit of fine discretizations $l_{ele}\!:=\! s^{(e),2}\!-\!s^{(e),1} \rightarrow 0$. In this limit, the intermediate triad field becomes identical to an elementwise Bishop frame~\cite{bishop1975} and the relation
\begin{align}
\label{triadsr_intermediatetorsion_limit}
\begin{split}
\lim \limits_{l_{ele} \rightarrow 0} K_{M_{\varphi}1}(\xi)  \!=\!  - \lim \limits_{\xi^I \rightarrow \xi} \frac{\boldsymbol{\kappa}^T \! \mb{g}_{1}^I}{1\!+\!\mb{g}_{1}^T\mb{g}_{1}^I}
\!=\!  - \lim \limits_{\xi^I \rightarrow \xi} \frac{(\mb{g}_{1}(\xi)\times \mb{g}_{1}^{\prime}(\xi))^T \! \mb{g}_{1}(\xi^I)}{1\!+\!\mb{g}_{1}^T(\xi)\mb{g}_{1}(\xi^I)}\!=\! 0.
\end{split}
\end{align}
holds true. However, as verified numerically in Section~\ref{sec:examples_arcsegment}, a neglecting of $K_{M_{\varphi}1}$ in the range of finite element lengths will in general lead to a decline in the spatial convergence rate. On the contrary, the arc-length derivatives (and variations) derived in the alternative anisotropic Kirchhoff-Love formulation~\cite{bauer2016} (see requirement 4a) in Table~\ref{tab:comparison_requirements}), contain the required terms stemming from the SR mapping.\\ 
\end{minipage}

Again, the discrete version of the spin vector field $\delta \boldsymbol{\theta}_h(\xi)$ has to be determined. Following a Petrov-Galerkin approach, the spin vector of~\eqref{kirchhoffkinematics_spinvector} can be discretized as follows:
\begin{align}
\label{kirchhoffkinematics_spinvector_variant1}
   \delta \boldsymbol{\theta}_h(\xi) = \delta \Theta_{1h}(\xi) \mb{g}_1 + \frac{\mb{S}(\mb{r}^{\prime}) \delta \mb{r}^{\prime}}{|| \mb{r}^{\prime} ||^2} \quad \text{with} \quad \delta \Theta_{1h}(\xi)\!=\! \sum_{i=1}^{n_{\Lambda}} L^i(\xi) \delta \hat{\Theta}_1^i \!=:\!\mbd{L}_{\parallel} \delta \hat{\mbds{\Theta}}_1,
\end{align}
where the interpolation of $\mb{r}^{\prime}$ as well as $\delta \mb{r}^{\prime}$ follows~\eqref{hermite_interpolation} and~\eqref{testfunctions}. The matrix $\mbd{L}_{\parallel}\!:=\!(L^1,...,L^{n_{\Lambda}})$ as well as the vector $\delta \hat{\mbds{\Theta}}_1\!:=\!(\delta \hat{\Theta}_1^1,...,\delta \hat{\Theta}_1^{n_{\Lambda}})^T$ represent assemblies of the Lagrange shape functions $L^i$ and the nodal twist components $\delta \hat{\Theta}_1^i$. Alternatively, it can be discretized in a Bubnov-Galerkin manner based on~\eqref{triadsr_triadfield}. In this case, it reads:
\begin{align}
\label{kirchhoffkinematics_spinvector_variant2}
   \!\!\!\!\!\delta \boldsymbol{\theta}_h(\xi) \!=\! \delta \Theta_{1h}(\xi) \mb{g}_1 \!+\! \frac{\mb{S}(\mb{r}^{\prime}) \delta \mb{r}^{\prime}}{|| \mb{r}^{\prime} ||^2}, \,\, \delta \Theta_{1h}(\xi)\!=\! \delta \Theta_{M_{\varphi}1}(\xi) \!+\! \delta \varphi_h(\xi), \,\, \delta \varphi_h(\xi) \!&=\! \sum_{i=1}^{n_{\Lambda}} L^i(\xi) \delta \varphi^i.\!\!\!\!\!
\end{align}
The spin vector $\delta \boldsymbol{\theta}_h(\xi)$ in~\eqref{kirchhoffkinematics_spinvector_variant2} shall be completely expressed via nodal variations $\delta \hat{\mbds{\Theta}}_1$ and $\delta \hat{\mbd{d}}$, i.e. via assemblies of nodal primary variables. Thereto, $\delta \Theta_{M_{\varphi}1}(\xi)$ and $\delta \varphi^i$ are expressed as
\begin{align}
\begin{split}
\label{kirchhoffkinematics_deltaTheta1bar}
  \delta \Theta_{M_{\varphi}1}(\xi)\!& =\! \delta \Theta_1^I \!+\!\frac{\mb{g}_{1}^T(\mb{g}_{1}^{I} \! \times \! \delta \mb{g}_{1}^{I} ) \!-\! \mb{g}_{1}^{I \, T}(\mb{g}_{1} \! \times \! \delta \mb{g}_{1} ) }{1\!+\!\mb{g}_{1}^T\mb{g}_{1}^I}, \\
\delta \varphi^i \!& =\! \delta \Theta_1^i \!-\! \delta \Theta_{M_{\varphi}1}^i \!=\!
\delta \Theta_1^i \!-\! \delta \Theta_1^I \!-\!\frac{\mb{g}_{1}^{i \, T}(\mb{g}_{1}^{I} \! \times \! \delta \mb{g}_{1}^{I} ) \!-\! \mb{g}_{1}^{I \, T}(\mb{g}_{1}^i \! \times \! \delta \mb{g}_{1}^i ) }{1 \!+\!\mb{g}_{1}^{i \, T} \mb{g}_{1}^I},
\end{split}
\end{align}
which directly follows from interpolation~\eqref{triadsr_triadfield}. Inserting the relations~\eqref{kirchhoffkinematics_deltaTheta1bar} into the spin vector~\eqref{kirchhoffkinematics_spinvector_variant2} yields:
\begin{align}
\begin{split}
\label{kirchhoffkinematics_deltaTheta1final}
\!\!\!\!\!  \delta \Theta_{1h}(\xi) \!& =\! \mbd{L}_{\parallel} \delta \hat{\mbds{\Theta}}_1 \underbrace{\!-\! \sum_{i=1}^{n_{\Lambda}} \! L^i
  \frac{\mb{g}_{1}^{i T}\! (\mb{g}_{1}^{I} \! \times \! \delta \mb{g}_{1}^{I} ) \!-\! \mb{g}_{1}^{I T}\!(\mb{g}_{1}^i \! \times \! \delta \mb{g}_{1}^i ) }{1 \!+\!\mb{g}_{1}^{i T} \mb{g}_{1}^I}
 \!+\! \frac{\mb{g}_{1}^T\!(\mb{g}_{1}^{I} \! \times \! \delta \mb{g}_{1}^{I} ) \!-\! \mb{g}_{1}^{I T}\!(\mb{g}_{1} \! \times \! \delta \mb{g}_{1} ) }{1\!+\!\mb{g}_{1}^T\mb{g}_{1}^I}}_{=:\delta \Theta_{1,diff}(\xi)} \,\,. \!\!\!\!\!
\end{split}
\end{align}
In~\eqref{kirchhoffkinematics_deltaTheta1final}, the notion $\delta \Theta_{1,diff}(\xi)$ has been introduced for the term distinguishing the Petrov-Galerkin~\eqref{kirchhoffkinematics_spinvector_variant1} and the Bubnov-Galerkin variant~\eqref{kirchhoffkinematics_spinvector_variant2}. By making use of the abbreviations $\mb{x}_b=\mb{x}(\xi_b)$, $\mb{x}_a=\mb{x}(\xi_a)$ and the relations
\begin{align}
\label{kirchhoffkinematics_deltaTheta1_usefulrelations}
  \mb{g}_{1a}^{T} (\mb{g}_{1b} \! \times \! \delta \mb{g}_{1b} ) \!=\!  
  \mb{g}_{1a}^{T} (\tilde{\mb{t}}_b \! \times \! \delta \mb{r}_b^{\prime}) \!=\!
  (\mb{g}_{1a} \! \times \! \tilde{\mb{t}}_b)^T \mbd{H}^{\prime}(\xi_b) \delta \hat{\mbd{d}}, \quad
  \tilde{\mb{t}} \!:=\! \frac{\mb{g}_{1}}{||\mb{r}^{\prime}||}
  \!=\! \frac{\mb{r}^{\prime}}{||\mb{r}^{\prime}||^2},
\end{align}
the Bubnov-Galerkin interpolation of the spin vector given in equation~\eqref{kirchhoffkinematics_spinvector_variant2} can finally be formulated as:
\begin{align}
\begin{split}
\label{kirchhoffkinematics_spinvector_variant2final}
  \delta \boldsymbol{\theta}^T_h(\xi) \!& =\! \delta \hat{\mbds{\Theta}}_1^{\!T} \! \mbd{v}_{\theta_{\parallel \Theta}} \!+\!  
  \delta \hat{\mbd{d}}^T \! \mbd{v}_{\theta_{\perp}} \!+\! \delta \hat{\mbd{d}}^T \! \mbd{v}_{\theta_{\parallel d}} 
  \quad \mbd{v}_{\theta_{\parallel \Theta}} \!\!=\! \mbd{L}_{\parallel}^{\!T} \!\otimes\! \mb{g}_1^T, \,\,\,
  \mbd{v}_{\theta_{\perp}} \!\!=\! -\mbd{H}^{\prime T} \mb{S}(\tilde{\mb{t}}), \\
  \mbd{v}_{\theta_{\parallel d}} \!&=\! \left(\sum_{i=1}^{n_{\Lambda}} L^i \, \mbd{v}_{1i} \!-\! \mbd{v}_{1} \right) \! \otimes 
  \!\mb{g}_1^T, \,\, 
  \mbd{v}_{1} \!=\! \frac{\mbd{H}^{\prime T}(\xi)(\mb{g}_{1}^I \! \times \! \tilde{\mb{t}}) - 
  \mbd{H}^{\prime T}(\xi_I)(\mb{g}_{1} \! \times \! \tilde{\mb{t}}^I) }{1 \!+\!\mb{g}_{1}^{\, T} \mb{g}_{1}^I}, \\
  \mbd{v}_{1i} \!&=\! \mbd{v}_{1}(\xi_i)\!=\!\frac{ \mbd{H}^{\prime T}(\xi_i)(\mb{g}_{1}^I \! \times \! \tilde{\mb{t}}^i) - 
  \mbd{H}^{\prime T}(\xi_I) (\mb{g}_{1}^i \! \times \! \tilde{\mb{t}}^I) }{1 \!+\!\mb{g}_{1}^{i \, T} \mb{g}_{1}^I}.
\end{split}
\end{align}
Comparing equations~\eqref{kirchhoffkinematics_spinvector_variant1} and~\eqref{kirchhoffkinematics_spinvector_variant2final} leads to the conclusion that the difference between the Bubnov-Galerkin and the Petrov-Galerkin variant is expressed by one additional term involving~$\mbd{v}_{\theta_{\parallel d}}$. The arc-length derivative of~\eqref{kirchhoffkinematics_spinvector_variant2final} reads:
\begin{align}
\begin{split}
\label{kirchhoffkinematics_spinvector_variant2final2}
  \!\!\!\!\!
  \delta \boldsymbol{\theta}^{\prime T}_h(\xi) \!& =\! \delta \hat{\mbds{\Theta}}_1^{\!T} \mbd{v}_{\theta_{\parallel \Theta}}^{\prime} \!\!+\!  
  \delta \hat{\mbd{d}}^T \! \mbd{v}_{\theta_{\perp}}^{\prime} \!\!+\! \delta \hat{\mbd{d}}^T \! \mbd{v}_{\theta_{\parallel d}}^{\prime},\\ 
 \mbd{v}_{\theta_{\parallel \Theta}}^{\prime} \!\!& =\! \mbd{L}^{\! \prime T}_{\parallel} \!\otimes\! \mb{g}_1^T \!+\! \mbd{L}_{\! \parallel}^T \!\otimes\! \mb{g}_1^{\prime T}, \,\,
  \mbd{v}_{\theta_{\perp}}^{\prime} \!\!=\! -\mbd{H}^{\prime \prime T} \mb{S}(\tilde{\mb{t}}) \!-\! \mbd{H}^{\prime T} \mb{S}(\tilde{\mb{t}}^{\prime}),\,\,
  \tilde{\mb{t}}^{\prime}\!=\!\frac{\mb{r}^{\prime \prime}}{||\mb{r}^{\prime}||^2}
  \!-\!\frac{2(\mb{r}^{\prime T}\mb{r}^{\prime \prime})\mb{r}^{\prime}}{||\mb{r}^{\prime}||^4}
   \!\!\!\!\! \\ \!\!\!\!\!
  \mbd{v}_{\theta_{\parallel d}}^{\prime} \!&=\! \Big(\sum_{i=1}^{n_{\Lambda}} L^{i \prime} \, \mbd{v}_{1i} \!-\! \mbd{v}_{1}^{\prime} \Big) 
  \!\otimes \!\mb{g}_1^T \!+\! \Big(\sum_{i=1}^{n_{\Lambda}} L^{i} \, \mbd{v}_{1i} \!-\! \mbd{v}_{1} \Big) \! \otimes 
  \!\mb{g}_1^{\prime T}, \!\!\!\!\! \\ \!\!\!\!\!
  \mbd{v}_{1}^{\prime} \!&=\! \frac{
  \mbd{H}^{\prime T}(\xi) (\mb{g}_{1}^I \! \times \! \tilde{\mb{t}}^{\prime}) \!+\! 
  \mbd{H}^{\prime \prime T}(\xi) (\mb{g}_{1}^I \! \times \! \tilde{\mb{t}}) \!-\! 
  \mbd{H}^{\prime T}(\xi_I) (\mb{g}_{1}^{\prime} \! \times \! \tilde{\mb{t}}^I)
  }{1 \!+\!\mb{g}_{1}^{T} \! \mb{g}_{1}^I} \!-\! \frac{(\mb{g}_{1}^{\prime T} \mb{g}_{1}^I)\mb{v}_{1}}{1 \!+\!\mb{g}_{1}^{\, T} \mb{g}_{1}^I}.\!\!\!\!\!
\end{split}
\end{align}
It shall be investigated if the variants~\eqref{kirchhoffkinematics_spinvector_variant1} and~\eqref{kirchhoffkinematics_spinvector_variant2} can represent a constant distribution $\delta \boldsymbol{\theta}^T_h(\xi)\!=\!\mb{w}_0\!=\!\text{const.}$ and $\delta \boldsymbol{\theta}^{\prime T}_h(\xi)\!=\!\mb{0}$, as it is the case for their counterparts~\eqref{crisfieldspininterpolation} and~\eqref{petrovspininterpolation}. Thereto, the nodal variations are chosen as
\begin{align}
\label{elestrongkirchhoff_rigid_rotation}
\delta \hat{\mb{d}}^j = \mb{w}_0 \times \hat{\mb{d}}^j, \quad 
\delta \hat{\mb{t}}^j = \mb{w}_0 \times \hat{\mb{t}}^j, \quad
\delta \hat{\Theta}_1^i=\mb{g}_1^T(\xi^i) \mb{w}_0 \,\,\, \text{for} \,\,\, j=1,2; \,\,\,\, i\!=\!1,...,n_{\Lambda}.
\end{align}
Inserting $\delta \hat{\mb{d}}^j$ and $\delta \hat{\mb{t}}^j$ according to~\eqref{elestrongkirchhoff_rigid_rotation} into~\eqref{testfunctions} gives the desired result $\delta \mb{r}_h(\xi)\!=\!\mb{w}_0\!\times\!\mb{r}_h(\xi)$. With this result and by inserting~\eqref{elestrongkirchhoff_rigid_rotation} into the spin vector interpolations~\eqref{kirchhoffkinematics_spinvector_variant1} and~\eqref{kirchhoffkinematics_spinvector_variant2}, which both fulfill the interpolation property for the tangential spin vector components $\delta \hat{\Theta}_1^i$, it can be shown that $\mb{w}_0$ is at least represented correctly at the element nodes:
\begin{align}
\label{elestrongkirchhoff_rigid_rotation2}
\delta \boldsymbol{\theta}_h(\xi^i) \!=\! (\mb{g}_1^T(\xi^i) \mb{w}_0) \mb{g}_1(\xi^i) + \frac{\mb{r}^{\prime}(\xi^i) \!\times\! (\mb{w}_0\!\times\!\mb{r}^{\prime}(\xi^i))}{|| \mb{r}^{\prime}(\xi^i) ||^2} \!=\! \mb{w}_0 \,\,\, \text{for} \,\,\, i\!=\!1,...,n_{\Lambda}.
\end{align}
Here, the Grassmann identity for cross-products $\mb{a}\!\times\!(\mb{b}\!\times\!\mb{c})\!=\!(\mb{a}^T\mb{c})\mb{b}\!-\!(\mb{a}^T\mb{b})\mb{c}$ with $\mb{a},\mb{b},\mb{c} \in \Re^3$ as well as $\mb{g}_1\!=\!\mb{r}^{\prime}/||\mb{r}^{\prime}||$ have been employed. Next, it will be investigated for the variants~\eqref{kirchhoffkinematics_spinvector_variant1} and~\eqref{kirchhoffkinematics_spinvector_variant2}, if the choice~\eqref{elestrongkirchhoff_rigid_rotation} leads to a constant spin vector field along the beam element. Inserting~\eqref{elestrongkirchhoff_rigid_rotation} into~\eqref{kirchhoffkinematics_spinvector_variant1} and using the Grassmann identity yields:
\begin{align}
\label{elestrongkirchhoff_rigid_rotation3}
\begin{split}
\delta \boldsymbol{\theta}_h(\xi) \!&=\! \left(\sum_{i=1}^{n_{\Lambda}} L^i(\xi) \mb{g}_1^T(\xi^i) \mb{w}_0\right) \mb{g}_1(\xi) + \frac{\mb{r}^{\prime}(\xi) \!\times\! (\mb{w}_0\!\times\!\mb{r}^{\prime}(\xi))}{|| \mb{r}^{\prime}(\xi) ||^2} \\ \!&=\!
\Bigg[\mb{I}_3 - \mb{g}_1(\xi)\otimes \underbrace{\Bigg(\mb{g}_1^T(\xi) -\sum_{i=1}^{n_{\Lambda}} L^i(\xi) \mb{g}_1^T(\xi^i) \Bigg)}_{\neq \mb{0}} \Bigg] \mb{w}_0 \neq \mb{w}_0.
\end{split}
\end{align}
Thus, the Petrov-Galerkin variant~\eqref{kirchhoffkinematics_spinvector_variant1} cannot represent a constant spin vector field. Since~\eqref{kirchhoffkinematics_spinvector_variant1} and~\eqref{kirchhoffkinematics_spinvector_variant2} solely differ in $\delta \Theta_{1,diff}(\xi)$ (see~\eqref{kirchhoffkinematics_deltaTheta1final}), only this term has to be further investigated. Inserting~\eqref{elestrongkirchhoff_rigid_rotation} in~$\delta \Theta_{1,diff}(\xi)$ yields:
\begin{align}
\label{elestrongkirchhoff_deltatheta1diff}
\begin{split}
\!\!\!\!\!\! \delta \Theta_{1,diff}(\xi)\!&=\!-\!\!\sum_{i=1}^{n_{\Lambda}} \! L^i(\xi) (\mb{g}_1^T \!(\xi^i)\mb{w}_0) \!+\!\! \underbrace{\sum_{i=1}^{n_{\Lambda}} \! L^i(\xi)}_{=1} (\mb{g}_1^T\!(\xi^I)\mb{w}_0) \!+\! (\mb{g}_1^T(\xi)\mb{w}_0) \!-\! (\mb{g}_1^T\!(\xi^I)\mb{w}_0) \!\!\!\!\!\! \\ &
= \Bigg(\mb{g}_1^T(\xi) -\sum_{i=1}^{n_{\Lambda}} L^i(\xi) \mb{g}_1^T(\xi^i) \Bigg) \mb{w}_0.
\end{split}
\end{align}
Thus, adding the term $\delta \Theta_{1,diff}(\xi)\mb{g}_1(\xi)$ to the Petrov-Galerkin variant~\eqref{elestrongkirchhoff_rigid_rotation3} yields the desired result $\delta \boldsymbol{\theta}_h(\xi)\!=\!\mb{w}_0\!=\!\text{const.}$ for the Bubnov-Galerkin spin vector interpolation in case the nodal variations are given by~\eqref{elestrongkirchhoff_rigid_rotation}. Alternatively, this result can be obtained by considering that~\eqref{kirchhoffkinematics_spinvector_variant2} represents the consistent variation of the objective triad interpolation~\eqref{triadsr_triadfield} (see also Section~\ref{sec:spatialdiscretization_objectivity}). Since this interpolation is objective, the variation of the discrete internal energy has to vanish for infinitesimal rigid body rotations. For an arbitrary stress resultant $\mb{m}$, this is only possible if $\delta \boldsymbol{\theta}_h^{\prime}(\xi) \!\equiv\! \mb{0}$ and consequently $\delta \boldsymbol{\theta}_h(\xi) \!\equiv\! \text{const.}$ can be displayed exactly (see also the weak form~\eqref{weakform3}). Recapitulatory, the interpolation~\eqref{kirchhoffkinematics_spinvector_variant2} can represent arbitrary constant spin vector distributions~$\mb{w}_0$, while for the interpolation~\eqref{kirchhoffkinematics_spinvector_variant1} this is only possible for 2D problems or in the special case $\mb{w}_0\!=\!\mb{0}$. This result will be important in order to investigate the conservation properties of the resulting finite element formulations (see e.g. Sections~\ref{sec:spatialdiscretization_conservation} and~\ref{sec:elestrongkirchhoff_spatialdiscretization_conservation}). Finally, the field of (multiplicative) rotation vector increments $\Delta \boldsymbol{\theta}_h(\xi)$ and the derivative $\Delta \boldsymbol{\theta}^{\prime}_h(\xi)$, required for a consistent linearization of the discretized weak form, follow from equations~\eqref{kirchhoffkinematics_spinvector_variant2final} and~\eqref{kirchhoffkinematics_spinvector_variant2final2} by simply replacing the variations $\delta(.)$ by increments $\Delta(.)$. The spin vector considered above is expressed via multiplicative nodal increments $\delta \hat{\Theta}^i_1$ as nodal primary variables. In case a rotation parametrization of the nodal triads via nodal rotation vectors according to Section~\ref{sec:spatialdiscretization_nodaltriads_rotationvector} is employed, the nodal vector of multiplicative iterative rotation increments given by $\Delta \hat{\boldsymbol{\Theta}}^i\!=\!(\Delta \hat{\Theta}^i_1,\Delta \hat{\Theta}^i_2,\Delta \hat{\Theta}^i_3)^T_{\mb{g}^i}$ can be directly used for triad update as shown in~\eqref{newtoniteration_triadupdates}. However, if a rotation parametrization of the nodal triads via the SR mapping and nodal relative angles $\hat{\varphi}^i$ according to Section~\ref{sec:spatialdiscretization_nodaltriadsrotation_smallestrotation} is employed, the rotation vector increments shall be expressed by means of additive increments $\Delta \hat{\varphi}^i$ of the nodal primary variables $\hat{\varphi}^i$ as shown in~\eqref{newtoniteration_srtriadupdates}. A relation between $\Delta \hat{\Theta}^i_1$ and $\Delta \hat{\varphi}^i$ can easily be derived on the basis of equations~\eqref{largerotations_deltaTheta1} and~\eqref{largerotations_deltaThetaSR1}:
\begin{align}
\label{DeltaTheta1}
\Delta \hat{\Theta}_1^i = \Delta \Theta_{M_{\hat{\varphi}}1}^i + \Delta \hat{\varphi}^i = -\frac{ \bar{\mb{g} }_{1}^{i T} \mb{S}(\mb{g}_{1}^i)}{1+ \mb{g}_{1}^{i T} \bar{\mb{g}}_{1}^i}  \frac{\Delta \mb{t}^i}{|| \mb{t}^i ||} + \Delta \hat{\varphi}^i, \quad \Delta \mb{t}^i = \Delta \mb{r}^{\prime}(\xi^i) = \mbd{H}^{\prime}(\xi^i) \Delta \hat{\mbd{d}}.
\end{align}
At the element boundary nodes, the last term in~\eqref{DeltaTheta1} can be simplified: $\Delta \mb{t}^j \!=\! \Delta \mb{r}^{\prime}(\xi^j) \!=\! \mbd{H}^{\prime}(\xi^j) \Delta \hat{\mbd{d}} \!=\! \Delta \hat{\mb{t}}^j$ if $j\!=\!1$ or $j\!=\!2$.\\

\hspace{0.2 cm}
\begin{minipage}{15.0 cm}
\textbf{Review: } The triad interpolation scheme presented in this section is very similar to the approach proposed in the authors' earlier contributions~\cite{meier2014} (see Section 3.5.2). There, an intermediate triad field has been constructed in a manner similar to~\eqref{triadsr_triadfield}, but with the choice $\mb{\Lambda}_r\!=\!\mb{\Lambda}_{M_{\hat{\varphi}}}^1$. While the most essential properties of these two approaches are comparable, there are some slight advantages of the procedure presented here: Choosing a material triad $\mb{\Lambda}^i$ as reference triad $\mb{\Lambda}_r$ makes the interpolation scheme independent from the choice of the nodal primary variables (according to Section~\ref{sec:spatialdiscretization_nodaltriads_rotationvector} or according to Section~\ref{sec:spatialdiscretization_nodaltriadsrotation_smallestrotation}). Furthermore, locating the reference triad at the element middle node makes the element formulation symmetric and extends the maximal orientation difference of the material triads at the element boundary nodes that can be represented from $180^{\circ}$ to $360^{\circ}$. The latter property results from the maximal orientation difference of $180^{\circ}$ allowed for two tangent vectors in order to yield a unique SR mapping (see also Section~\ref{sec:largerotations_sr}). Besides the authors' earlier work~\cite{meier2014}, also the straight and anisotropic Kirchhoff-Love formulations from the literature exhibit the mentioned lack of symmetry (see requirement 4e) in Table~\ref{tab:comparison_requirements}).\\ 
\end{minipage}

\hspace{0.2 cm}
\begin{minipage}{15.0 cm}
\textbf{Review: } In~\cite{meier2014}, an alternative triad interpolation (see Section 3.5.1 of~\cite{meier2014}) has been investigated, which defines an intermediate triad field $\mb{\Lambda}_{M_{\varphi}}(\xi)$ directly via the smallest rotation from the intermediate triad field $\bar{\mb{\Lambda}}_{M_{\varphi}}(\xi)$ of the last time step onto the current tangent vector field $\mb{g}_1(\xi)$ according to:
\begin{align}
\label{triadsr_triadfield_noneobjective}
\begin{split}
\mb{\Lambda}_h(\xi) \!&=\! \exp{\!(\mb{S}[\varphi_h(\xi) \mb{g}_1(\xi)])} \, \mb{\Lambda}_{M_{\varphi}}(\xi), \quad \quad \quad \quad
\mb{\Lambda}_{M_{\varphi}}(\xi) \!=\! \text{sr}(\bar{\mb{\Lambda}}_{M_{\varphi}}(\xi),\mb{g}_1(\xi)), \\
\varphi_h(\xi)\!&=\! \sum_{i=1}^{n_{\Lambda}} L^i(\xi) \hat{\varphi}^i, \hspace{4.5cm} 
\bar{\mb{\Lambda}}_{M_{\varphi},n+1}(\xi) \!=\! \mb{\Lambda}_{M_{\varphi},n}(\xi).
\end{split}
\end{align}
At first glance, this interpolation seems to be more straightforward than~\eqref{triadsr_triadfield} since no nodal triads are required for constructing the intermediate triad field. However, as shown in~\cite{meier2014}, an interpolation of this kind is neither objective nor path-independent. This non-objective and path-dependent interpolation has e.g. been applied by the anisotropic formulations from the literature (see requirement 3) of Table~\ref{tab:comparison_requirements}).
\end{minipage}

\subsection{Requirements on spatial discretization methods}
\label{sec:spatialdiscretization_requirements}
%
%

In this section, essential requirements on the spatial discretizations of translational and rotational fields will be stated. Subsequently, different beam elements will be presented and the fulfillment of these requirements investigated.

%
\subsubsection{Differentiability of discrete fields}
\label{sec:spatialdiscretization_differentiability}
%

The first requirement for spatial discretization methods concerns differentiability. On the one hand, this requirement is related to the weak form of the balance equations: The highest arc-length derivative occurring in the weak form of the Simo-Reissner beam theory is of order one, leading to the requirement of at least $C^0$-continuous discrete centerline and triad fields. Such a continuity at the element boundaries is provided by the Lagrange centerline interpolation~\eqref{lagrange_interpolation} as well as by the two discussed approaches of triad interpolation according to Sections~\ref{sec:spatialdiscretization_triadrelrot} and~\ref{sec:spatialdiscretization_triadsr}. The second arc-length derivative of the beam centerline in the weak form of the balance equations is a distinctive property of the Kirchhoff-Love beam theory and requires the interpolation of the centerline to be at least $C^1$-continuous as guaranteed by~\eqref{hermite_interpolation}. On the other hand, certain applications such as beam-to-beam contact formulations (see e.g.~\cite{meier2015b,meier2015c}) benefit considerably from a smooth geometry representation and the existence of a well-defined tangent vector field along the entire beam centerline, which can conveniently be furnished by the Hermite interpolation~\eqref{hermite_interpolation}.

%
\subsubsection{Objectivity and path-independence}
\label{sec:spatialdiscretization_objectivity}
%

The properties of objectivity and path-independence play a central role in the development of (geometrically exact) beam finite element formulations. The importance of these properties can be traced back to the nonlinear nature of the configuration space (resulting from the occurrence of large rotations) which complicates the interpolation of rotational quantities. Furthermore, it can be explained by the historic background that none of the early geometrically exact beam formulations fulfilled both of these properties (see~\cite{crisfield1999}). As already explained in~\cite{crisfield1999}, the path-independence of the employed discretizations can directly be concluded from the fact that none of these interpolation schemes is based on any history values of interpolated quantities. Only the nodal primary variables depend on history values. However, the corresponding nodal displacements can always arise in a way such that the finite element solution is independent from the actual load path in case the considered physical problem is path-independent. In other words, the arising nodal displacements yield the path-independent solution to the discrete optimization problem (based on a proberly defined Lagrangian) which is associated with the path-independent physical problem. In the numerical investigations performed at the end of this chapter, this property will be verified. However, throughout this section, the fundamental property of objectivity, i.e. the invariance of the applied deformation measures under rigid body motions, will be investigated. Thereto, a rigid body translation $\mb{r}_R$ and a rigid body rotation $\mb{\Lambda}_R$ are superimposed onto the beam centerline curve $\mb{r}(\xi)$~and the triad field~$\mb{\Lambda}(\xi)$. A rigid body motion is characterized by constant fields $\mb{r}_R$ and $\mb{\Lambda}_R$ along the beam, thus $\mb{r}_R^{\prime} \!\equiv\! \mb{0}$ and $\mb{\Lambda}_R^{\prime} \!\equiv\! \mb{0}$. In the following, the subscript $(.)^*$ denotes quantities that result from the superimposed rigid body motion, thus $\mb{r}^*(\xi)\!=\!\mb{\Lambda}_R\left(\mb{r}(\xi)\!+\!\mb{r}_R\right)$ and $\mb{\Lambda}^*(\xi)\!=\!\mb{\Lambda}_R\mb{\Lambda}(\xi)$. A formulation is denoted as being objective if such a rigid body motion does not affect the material deformation measures $\mb{\Omega}$ and $\mb{\Gamma}$. It is straightforward to show that the space-continuous versions of the deformation measures $\mb{\Omega}$ and $\mb{\Gamma}$ are objective, i.e. $\mb{\Omega}^*\!=\!\mb{\Omega}$ and $\mb{\Gamma}^*\!=\!\mb{\Gamma}$ (see e.g.~\cite{crisfield1999}). The question of interest is if this objectivity is preserved by the employed spatial discretization schemes. In \cite{crisfield1999}, it is shown that the fulfillment of the following requirement guarantees for the objectivity of a geometrically exact beam formulation based on the Simo-Reissner theory as introduced in Section~\ref{sec:simoreissner}:
\begin{align}
\label{requirements}
\mb{r}^*_h(\xi) \dot{=} \mb{\Lambda}_R \! \left(\mb{r}_h(\xi)\!+\!\mb{r}_R\right), \quad \mb{\Lambda}^*_h(\xi) \dot{=} \mb{\Lambda}_R\mb{\Lambda}_h(\xi) \,\, \rightarrow \,\, 
\mb{\Omega}_h^*\!=\!\mb{\Omega}_h, \quad \mb{\Gamma}^*_h\!=\!\mb{\Gamma}_h.
\end{align}
The following investigations will exclusively be applied to discretized quantities. In order to shorten notation, the subscript $(.)_h$ will be omitted throughout this section. First, it shall be shown that the validity of~\eqref{requirements} is also sufficient for the invariance of the deformation measures \eqref{kirchhoffdeformationmeasures} and \eqref{kirchhoffdeformationmeasures2} of the Kirchhoff beam theory. If \eqref{requirements} is valid, it follows:
\begin{align}
\label{consequencesofrequirements}
\mb{r}^{* \prime}(\xi)=\mb{\Lambda}_R\mb{r}^{\prime}(\xi), \quad 
\mb{r}^{* \prime \prime}(\xi)=\mb{\Lambda}_R\mb{r}^{\prime \prime}(\xi), \quad
\mb{g}_{i}^*(\xi)=\mb{\Lambda}_R\mb{g}_{i}(\xi), \quad
\mb{g}_{i}^{* \prime}(\xi)=\mb{\Lambda}_R\mb{g}_{i}^{\prime}(\xi).
\end{align}
Based on~\eqref{consequencesofrequirements}, the axial tension, the Frenet-Serret curvature vector as well as the total torsion read:
\begin{align}
\label{consequencesofrequirements2}
\begin{split}
||\mb{r}^{* \prime}|| & =\left(\mb{r}^{\prime T}\mb{\Lambda}_R^T \mb{\Lambda}_R\mb{r}^{\prime}\right)^{0.5}\!=||\mb{r}^{\prime}|| \quad \rightarrow \quad \epsilon^*=||\mb{r}^{* \prime}||-1 =||\mb{r}^{\prime}||-1= \epsilon, \\
\boldsymbol{\kappa}^* & =   \frac{\mb{r}^{* \prime} \times \mb{r}^{* \prime \prime}}{||\mb{r}^{* \prime}||^2}
 = \frac{\mb{\Lambda}_R\!\left(\mb{r}^{\prime} \times \mb{r}^{\prime \prime} \right)}{||\mb{r}^{\prime}||^2}
 = \mb{\Lambda}_R \boldsymbol{\kappa} \,\, \rightarrow \,\, \mb{g}_{i}^{*T} \boldsymbol{\kappa}^*
 \!=\!\mb{g}_{i}^T\mb{\Lambda}_R^T \mb{\Lambda}_R\boldsymbol{\kappa} \!=\! \mb{g}_{i}^T\boldsymbol{\kappa}, \\
 K_{1}^* & = K_{M1}^* + \varphi^{* \prime} = \mb{g}_{2}^{* \prime T} \mb{g}_{3}^* = \mb{g}_{2}^{\prime T}\mb{\Lambda}_R^T\mb{\Lambda}_R
\mb{g}_{3} = \mb{g}_{2}^{\prime T} \mb{g}_{3} = K_{M1} + \varphi^{\prime}=K_{1}.
\end{split}
\end{align}
The identities $\mb{\Omega}^*\!=\!\mb{\Omega}$ and $\mb{\Gamma}^*\!=\!\mb{\Gamma}$ for the original and the rotated deformation measures~\eqref{kirchhoffdeformationmeasures} and \eqref{kirchhoffdeformationmeasures2} are a direct consequence of \eqref{consequencesofrequirements2}. Thus, also for Kirchhoff beam elements, the requirements~\eqref{requirements} are sufficient in order to ensure objectivity. In the following, the validity of \eqref{requirements} will be investigated for the interpolations in Sections~\ref{sec:spatialdiscretization_beamcenterline} and~\ref{sec:spatialdiscretization_rotationfield}.\\

\paragraph{Objectivity of centerline interpolations} Due to the linear dependence of the centerline interpolations~\eqref{lagrange_interpolation} or~\eqref{hermite_interpolation} on the nodal vectors, the proof of the first part of \eqref{requirements} is independent from the intermediate triad field:
\begin{align}
\label{objectivity_centerline}
\mb{r}^*(\xi) = \mb{\Lambda}_R \! \left(\mb{r}(\xi)\!+\!\mb{r}_R\right) \,\,\,\, \text{for} \,\,\,\, 
\hat{\mb{d}}^{*i} = \mb{\Lambda}_R \big(\hat{\mb{d}}^i \!+\!\mb{r}_R \big) \,\,\,\, \text{and} \,\,\,\,
\hat{\mb{t}}^{*i} = \mb{\Lambda}_R \hat{\mb{d}}^i. \,\,\,\, \square
\end{align}

\paragraph{Objectivity of triad interpolation based on local rotation vectors} The fulfillment of~\eqref{requirements}, i.e. the objectivity of interpolation~\eqref{triadrelrot_interpolation} has been shown in the original work~\cite{crisfield1999}. The interested reader is referred to this reference.\\

\paragraph{Objectivity of triad interpolation based on "Smallest Rotation" mapping} Based on the relations~\eqref{consequencesofrequirements}-\eqref{objectivity_centerline} and the strong fulfillment of the Kirchhoff constraint, the base vector $\mb{g}_{1}^*(\xi)$ resulting from the rigid body rotation yields:
\begin{align}
\label{objectivity_firstbasevector}
 \mb{g}_{1}^*(\xi)= \frac{\mb{r}^{*\prime}(\xi)}{||\mb{r}^{*\prime}(\xi)||} = 
 \frac{ \mb{\Lambda}_R \mb{r}^{\prime}(\xi)}{||\mb{r}^{\prime}(\xi)||}=\mb{\Lambda}_R\mb{g}_{1}(\xi).
\end{align}
In a next step, the nodal primary variables are chosen such that the nodal triads are also rigidly rotated:
\begin{align}
\begin{split}
\label{objectivity_smallestrotation3}
\mb{\Lambda}^{i*}=\mb{\Lambda}_R\mb{\Lambda}^{i}.
\end{split}
\end{align}
Using~\eqref{objectivity_firstbasevector},~\eqref{objectivity_smallestrotation3} and~\eqref{largerotations_smallestrotation}, the following relation between the vectors $\mb{g}_{M_{\varphi}2}(\xi)$ and $\mb{g}_{M_{\varphi}3}(\xi)$ of the intermediate triad $\mb{\Lambda}_{M_{\varphi}}(\xi)$ (see~\eqref{triadsr_triadfield}) and their counterparts $\mb{g}_{M_{\varphi}2}^*(\xi)$ and $\mb{g}_{M_{\varphi}3}^*(\xi)$ resulting from the rigid body motion can be derived:
\begin{align}
\begin{split}
\label{objectivity_smallestrotation}
\mb{g}_{M_{\varphi}i}^*(\xi)\!&=\! \mb{g}_{ri}^* \!-\! \frac{\mb{g}_{ri}^{*T}\mb{g}_{1}^*(\xi)}{1 \!+\! \mb{g}_{r1}^{*T} \mb{g}_{1}^*(\xi)}\left(
\mb{g}_1^*(\xi) \!+\! \mb{g}_{r1}^* \right)\\
&=\mb{\Lambda}_R\mb{g}_{ri} \!-\! \frac{\mb{g}_{ri}^{T}\mb{\Lambda}_R^T\mb{\Lambda}_R \mb{g}_{1}(\xi)}{1 \!+\! \mb{g}_{r1}^{T}\mb{\Lambda}_R^T\mb{\Lambda}_R \mb{g}_{1}(\xi)}\left(
\mb{\Lambda}_R\mb{g}_1(\xi) \!+\! \mb{\Lambda}_R\mb{g}_{r1} \right)\!=\!\mb{\Lambda}_R\mb{g}_{M_{\varphi}i}(\xi), \,\,\,\, i=2,3.
\end{split}
\end{align}
From~\eqref{objectivity_firstbasevector} and~\eqref{objectivity_smallestrotation}, it can be concluded that the intermediate triad field is rigidly rotated:
\begin{align}
\begin{split}
\label{objectivity_smallestrotation2}
\mb{\Lambda}_{M_{\varphi}}^*(\xi)=\mb{\Lambda}_R\mb{\Lambda}_{M_{\varphi}}(\xi).
\end{split}
\end{align}
If the transformation property $\mb{\Lambda} \! \exp{\!(\mb{S}[\mb{a}])}\mb{\Lambda}^T\!\!=\!\!\exp{\!(\mb{\Lambda}\mb{S}[\mb{a}])} \, \forall \, \mb{\Lambda} \!\in\! S\!O(3), \mb{a} \!\in\! \Re^3$ together with~\eqref{objectivity_firstbasevector},~\eqref{objectivity_smallestrotation3} and~\eqref{objectivity_smallestrotation2} is considered, the following result can be derived from the fourth equation of the triad interpolation scheme~\eqref{triadsr_triadfield}:
\begin{align}
\begin{split}
\label{objectivity_nodalrelativeangles}
\exp{\!(\mb{S}[\varphi^{i*} \mb{g}_1^{i*}])} \!&=\! \exp{\!(\mb{S}[\varphi^{i*} \mb{\Lambda}_R \mb{g}_1^{i}])} \dot{=}\mb{\Lambda}^{i*} \mb{\Lambda}_{M_{\varphi}}^{*T}(\xi^i) \!=\! \\ 
\mb{\Lambda}_R\mb{\Lambda}^{i} \mb{\Lambda}_{M_{\varphi}}^{T}(\xi^i)\mb{\Lambda}_R^T
\!&=\!\mb{\Lambda}_R\exp{\!(\mb{S}[\varphi^{i} \mb{g}_1^{i}])}\mb{\Lambda}_R^T
\!=\!\exp{\!(\mb{S}[\varphi^{i} \mb{\Lambda}_R \mb{g}_1^{i}])} 
\,\,\,\, \rightarrow \,\,\,\, \varphi^{i*}\!=\!\varphi^{i}.
\end{split}
\end{align}
Thus, the interpolation $\varphi(\xi)$ in~\eqref{triadsr_triadfield} is unchanged by the rigid body motion, i.e. $\varphi^*(\xi)\!=\!\varphi(\xi)$. Together with equations~\eqref{objectivity_firstbasevector} and~\eqref{objectivity_smallestrotation2}, the desired result for $\mb{\Lambda}^{*}(\xi)$ as already stated in~\eqref{requirements} can be derived:
\begin{align}
\label{objectivity_srtriadfield}
\begin{split}
\!\!\!\!\!\!
&\mb{\Lambda}^*(\xi) 
\!=\! \exp{\!(\mb{S}[\varphi^*(\xi) \mb{g}_1^*(\xi)])} \, \mb{\Lambda}_{M_{\varphi}}^*(\xi)
\!=\! \exp{\!(\mb{S}[\varphi(\xi) \mb{\Lambda}_R \mb{g}_1(\xi)])} \mb{\Lambda}_R \mb{\Lambda}_{M_{\varphi}}(\xi)  \!\!\!\!\!\! \\ \!\!\!\!\!\!
\!=&\mb{\Lambda}_R \exp{\!(\mb{S}[\varphi(\xi) \mb{g}_1(\xi)])} \mb{\Lambda}_R^T \mb{\Lambda}_R \mb{\Lambda}_{M_{\varphi}}(\xi)
\!=\! \mb{\Lambda}_R \exp{\!(\mb{S}[\varphi(\xi) \mb{g}_1(\xi)])} \mb{\Lambda}_{M_{\varphi}}(\xi)
\!=\! \mb{\Lambda}_R \mb{\Lambda}(\xi).\,\, \square \!\!\!\!\!\!
\end{split}
\end{align}
In the reformulations made in~\eqref{objectivity_srtriadfield}, again use has been made of the transformation property from above and~\eqref{triadsr_triadfield}. According to~\eqref{objectivity_srtriadfield}, the triad interpolation proposed in Section~\ref{sec:spatialdiscretization_triadsr} fulfills the requirement of objectivity. 
As mentioned above and derived in the third line of~\eqref{consequencesofrequirements2}, the fulfillment of~\eqref{requirements} guarantees for objective deformation measures, provided that these are consistently derived from the triad interpolation. In order to verify this latter restriction, the two individual contributions $K_{M1}$ and $\varphi^{\prime}$ appearing in the third line of~\eqref{consequencesofrequirements2} shall be subject to a closer investigation. Besides the relation $\varphi^*(\xi)\!=\!\varphi(\xi) \rightarrow \varphi^{*\prime}(\xi)\!=\!\varphi^{\prime}(\xi)$, which has already been deduced above, also the torsion $K_{M1}^*$ of the intermediate system can be calculated for the configuration resulting from the rigid body motion:
\begin{align}
\label{objectivity_triadsr_totaltorsion}
\begin{split}
K_{M_{\varphi}1}^* \!=\! -\frac{\boldsymbol{\kappa}^{*T} \! \mb{g}_{1}^{I*}}{1\!+\!\mb{g}_{1}^{*T}\mb{g}_{1}^{I*}}\!=\! -\frac{\boldsymbol{\kappa}^{T} \mb{\Lambda}_R^T \mb{\Lambda}_R \mb{g}_{1}^{I}}{1\!+\!\mb{g}_{1}^{T}\mb{\Lambda}_R^T\mb{\Lambda}_R\mb{g}_{1}^{I}}
\!=\! -\frac{\boldsymbol{\kappa}^{T} \! \mb{g}_{1}^{I}}{1\!+\!\mb{g}_{1}^{T}\mb{g}_{1}^{I}}\!=\!K_{M_{\varphi}1}.\,\,\,\, \square
\end{split}
\end{align}
As expected, $K_{M_{\varphi}1}$ is not affected by the rigid body motion and the torsion $K_{1}^*\!=\!K_{M1}^*\!+\!\varphi^{* \prime}\!=\!K_{M1}\!+\!\varphi^{\prime}\!=\!K_{1}$ remains unchanged. This underlines the objectivity of the interpolation~\eqref{triadsr_triadfield} and the consistency of the torsion measure.

%
\subsubsection{Avoidance of locking effects}
\label{sec:spatialdiscretization_locking}
%

It is well-known that purely displacement-based finite elements are prone to locking. Locking effects particularly relevant for geometrically exact beam formulations are shear locking as well as membrane locking. While shear locking can - by definition - only appear in beam formulations of Simo-Reissner type, membrane locking has already been observed for geometrically linear Kirchhoff beams (see e.g. \cite{armero2012a} or \cite{armero2012b}). In general, membrane locking refers to the inability of elements to exactly reproduce inextensibility, viz. a vanishing axial strain $\epsilon \! \equiv \! 0$, for curved structures such as shells or beams. This behavior can be traced back to a coupling between the kinematic quantities describing the axial tension mode and the curved geometry. While the focus of the subsequent investigations lies on membrane locking, at least some remarks concerning shear locking will be made at the end of this section.\\

\paragraph{Characterization of locking} One possible definition of locking is the deterioration of the spatial convergence rate in dependence of a certain key parameter. Subsequent investigations will reveal that the element slenderness ratio $\zeta_{ele}\!:=\!l_{ele}/R$ plays the role of such a key parameter associated with the membrane locking effect of slender beams. One question of interest is how the liability of a finite element formulation to locking can be assessed in a quantitative manner. From a mathematical point of view, this question can be answered by investigating the stability of the finite element formulation. For example, for mixed finite element formulations, a stability criterion is given by the well-known Ladyshenskaya-Babuska-Brezzi (LBB) condition, also denoted as inf-sup condition (see \cite{brezzi1974,brezzi1991}). Since a direct and general analysis of such conditions can often be intricate, also numerical inf-sup tests have been suggested in the literature (see e.g. \cite{iosilevich1997,bathe2000}). From a mechanical point of view, locking is typically explained by the occurrence of parasitic stresses, viz. the occurrence of modes in the discrete solution that are not part of the analytical solution. Consequently, the question if a formulation is prone to locking or not can also be answered by investigating proper representative test cases for parasitic stresses. Besides these \textit{mathematica}l and \textit{mechanical} interpretations of locking, a third, namely a \textit{numerical} perspective (see \cite{koschnick2004}) can be helpful. From a numerical point of view, locking can be seen as the consequence of an over-constrained system of equations. As introduced in \cite{hughes2000}, the so-called constraint ratio allows for some, at least heuristic, evaluation of the locking behavior of a finite element formulation. The constraint ratio is defined as the ratio of the total number of equilibrium equations $n_{eq}$ to the total number of constraints $n_{eq,c}$:
\begin{align}
r:=\frac{n_{eq}}{n_{eq,c}}.
\end{align}
In order to analyze the locking behavior, the constraint ratio $r$ of the continuous problem and the constraint ratio $r_h$ of the discretized problem evaluated for an infinite number of elements have to be compared. The underlying proposition is that elements with $r_h\!<\!r$ (especially $r_h\!<\!1$, which means that more constraints than degrees of freedom are present) have a tendency to lock, whereas values $r_h\!>\!r$ of the constraint ratio indicate that not enough constraint equations are available in order to reproduce the constraint in an accurate manner. Following this hypothesis, the case $r_h\!=\!r$ has to be regarded as the optimal constraint ratio. Throughout this contribution, the relevant locking phenomena will be analyzed based on a mechanical as well as a numerical perspective. The corresponding concepts are applicable in a straightforward manner. In future work, the stability of the most promising element formulations also has to be investigated in a mathematically rigorous manner, either based on direct analysis or on numerical stability tests.\\   

\paragraph{Membrane locking} In~\cite{meier2015}, the effect of membrane locking in the context of geometrically exact Kirchhoff beam elements based on a Hermite centerline interpolation according to Section~\ref{sec:spatialdiscretization_hermite} has been investigated in detail. Here, only the main results shall be recaptulated. In~\cite{meier2015}, it has been found that the state of (exactly) vanishing axial strains can only be represented for straight configurations but not for abritariy curved configurations. Furthermore, the amount of parasitic axial strain energy occuring in states of pure pending has been shown to increase quadratically with the beam element slenderness ratio leading to a progressively (over-) stiff system answer. The constraint ratios of the space-continuous and discrete 2D Kirchhoff-Love beam problem have been determined to $r=2>r_h=1$, indicating that the considered element formulation is prone to membrane locking. Different solution strategies exist in order to get rid of these locking effects. In~\cite{meier2015}, the approach of Assumed Natural Strains (ANS, see e.g. \cite{hughes1977}), Reduced Integration (RI, see e.g. \cite{noor1981}) and an alternative procedure proposed in~\cite{meier2015} and denoted as Minimally Constrained Strains (MCS), have been compared. The proposed MCS scheme can be characterized as an assumed strain/B-bar approach (see e.g. \cite{simo1986b}), and it can be derived in a variationally consistent manner. Concretely, the contribution $G_{\epsilon}$ of the axial tension to the weak form of the mechanical equilibirum equations is replaced by $G_{\bar{\epsilon}}$:
\begin{align}
\label{MCSdef1}
\begin{split}
G_{\epsilon}\!&=\!\underbrace{\delta \mb{r}^{\prime T} \! \mb{g}_1}_{\delta \epsilon}  EA \underbrace{ \left(||\mb{r}^{\prime}||\!-\!1\right)}_{\epsilon} \quad \rightarrow \quad G_{\bar{\epsilon}}\!=\!\delta \bar{\epsilon} EA \bar{\epsilon}, \\
 \bar{\epsilon}(\xi)\!&=\!\sum_{i=1}^{n_{CP}} L^{i}(\xi)\epsilon(\xi^i),
\quad \delta \bar{\epsilon}(\xi)\!=\!\sum_{k=1}^{n_{CP}} L^{i}(\xi) \delta \epsilon(\xi^i), 
\quad \tilde{\Pi}_{int,\epsilon}\!=\!\frac{1}{2}EA\bar{\epsilon}^2.
\end{split}
\end{align}
In \eqref{MCSdef1}, $\xi^i$ denotes the parameter coordinate and $n_{CP}$ the number of Collocation Points (CPs) where the ''original`` strains are evaluated. Moreover, $L^{i}(\xi)$ are Lagrange shape functions of polynomial order $n_{CP}\!-\!1$. Linearizing~\eqref{MCSdef1} at the undeformed, straight configuration $C_0$, yields the B-bar structure typical for geometrically linear finite elements:
\begin{align}
\label{MCSdef2}
\begin{split}
\!\!\!\!\!\!L(G_{\epsilon})|_{C_0} \!\!:=\! \delta \hat{\mbd{d}}^{ T} \!\!\underbrace{\mbd{H}^{\prime T} \!\mb{t}_0}_{\mbd{B}^T}\! K \!\underbrace{\mb{t}_0^T \mbd{H}^{\prime}}_{\mbd{B}} \! \Delta \hat{\mbd{d}}, \,\, L(G_{\bar{\epsilon}})|_{C_0} \!\!=\! \delta \hat{\mbd{d}}^{ T} \! \bar{\mbd{B}}^T \! \! K \bar{\mbd{B}} \Delta \hat{\mbd{d}}, \,\, \bar{\mbd{B}}(\xi) \!&=\!\!\! \sum_{i=1}^{n_{CP}}\! L^{i}\!(\xi)\mbd{B}(\xi^i).\!\!\!\!\!\!
\end{split}
\end{align}
In the last equation, the additional abbreviation $K\!:=\!EA$ has been introduced in order to shorten the notation.\\

\hspace{0.2 cm}
\begin{minipage}{15.0 cm}
\textbf{Remark:} Actually, the residual vector stated in~\eqref{elestrongkirchhoff_residual} consists of functional expressions that cannot be integrated exactly by Gauss integration. However, in numerical simulations, the deviations in the results for a number of Gauss points $n_G \geqslant 4$ turned out to be very small as compared to the 
discretization error. Therefore, the notion ''exact integration`` will be used whenever four or more Gauss points are applied.\\
\end{minipage}

\hspace{0.2 cm}
\begin{minipage}{15.0 cm}
\textbf{Remark:} In the last section, it has been shown that the strain field $\epsilon(\xi)$ is objective and path-independent, meaning that the strain field does not change as a consequence of a rigid body motion and that its value at a certain configuration is independent from the deformation path leading to that 
configuration. Since the assumed strain field $\bar{\epsilon}(\xi)$ represents a pure re-interpolation of the original strain field $\epsilon(\xi^i)$ evaluated at fixed collocation
 points $\xi^i$, the former will also fulfill objectivity and path-independence.\\
\end{minipage}

In~\cite{meier2015}, the number and location of CPs has been chosen according to $n_{CP}\!=\!1$ and $\xi^1\!=\!-1$, $\xi^2\!=\!1$, $\xi^3\!=\!0$. In order to motivate this choice, the constraint ratio resulting from the MCS method and different sets of CPs as well as for the alternative methods RI and ANS will be presented in the following. In case of a strain re-interpolation such as given by~\eqref{MCSdef1}, it is sufficient that the constraint of vanishing axial strains is fulfilled at the collocation points in order to end up with a vanishing axial strain energy. Since the Hermite interpolation~\eqref{hermite_interpolation} provides a $C^1$-continuous centerline curve, the first derivative $\mb{r}^{\prime}$ and consequently also the axial tension $\epsilon$ is $C^0$-continuous at the element boundaries. Thus, exactly one constraint equation $||\mb{r}^{\prime}(\xi^i)|| \dot{=} 1$ results from each (interior and element boundary) CP.\\

\hspace{0.2 cm}
\begin{minipage}{15.0 cm}
\textbf{Remark:} The chosen CPs are motivated by strain-continuity: If the axial strains were not $C^0$-continuous at the element boundaries, each element boundary node would provide two constraint equations, one for the previous element $(e)$ and one for the subsequent element $(e\!+\!1)$, i.e. $||\mb{r}^{\prime}(\xi^{(e)}\!=\!1)||\dot{=}1$ and $||\mb{r}^{\prime}(\xi^{(e+1)}\!=\!-1)||\dot{=}1$.\\
\end{minipage}

In the following table, the constraint ratios resulting from the three methods ANS, RI and MCS are compared for different choices concerning number and locations of collocation or Gauss points. As shown in~\cite{meier2015}, the ANS approach has to be based on CPs with vanishing axial strain values for a pure bending state, which requires four CPs.

\begin{table}[h!]
\centering
\begin{tabular}{|p{1.0cm}|p{1.5cm}|p{1.5cm}|p{1.5cm}|p{1.5cm}|p{1.5cm}|p{1.5cm}|p{1.5cm}|} \hline
- & MCS-3b & MCS-2b & MCS-3i & MCS-2i & ANS-4i & RI-3i & RI-2i \\ \hline
$n_{eq,c}$ & $2n_{ele}\!+\!1$ & $n_{ele}\!+\!1$ & $3n_{ele}$ & $2n_{ele}$ & $4n_{ele}$ & $3n_{ele}$ & $2n_{ele}$ \\ \hline
$r_h/r$ & $1$ & $2$ & $2/3$ & $1$ & $0.5$ & $2/3$ & $1$ \\ \hline
\end{tabular}
\caption{Quantitative comparison of different "anti-locking" methods.}
\label{tab:comparison_membranelockingmethods}
\end{table}

The index $i$ in Table~\ref{tab:comparison_membranelockingmethods} indicates that all collocation or Gauss points lie in the elements interior while variants that are marked by an index $b$ also employ the element boundary nodes $\xi\!=\!\pm 1$. According to the statements made above, the variants with $r_h/r\!=\!1$ represent the constraints associated with the axial tension in an optimal manner, whereas variants with $r_h/r\!<\!1$ have a tendency to lock. Consequently, at first glance, the variants MCS-3b, MCS-2i and RI-2i seem to be equally suitable. However, as derived in~\cite{meier2015}, the variants MCS-2i and RI-2i will lead to an underconstrained system of equations, allowing for zero-energy modes and yielding a singular system of equations and a rank deficient stiffness matrix in the straight configuration, while the variant MCS-3b exactly provides the minimal number of constraint equations required in order to avoid such zero-energy modes in the straight configuration. For that reason, this method has been denoted as method of Minimally Constrained Strains. Similarly, an optimal constraint ratio of $r_h\!=\!r\!=\!4$ and the avoidance of zero-energy modes can be verified for this choice CPs in the general 3D case.\\ 

\hspace{0.2 cm}
\begin{minipage}{15.0 cm}
\textbf{Remark:} Of course, locking would be avoided if the beam elements could exactly represent the internal energy associated with pure bending. According to~\cite{meier2015}, the MCS method fulfills this requirement only for straight configurations while arbitrary curved configurations will yield a slightly over-constrained system of equations. Thus, the state of constant curvature and vanishing axial tension cannot be displayed exactly. However, it has been shown numerically that membrane locking is avoided. This result is reasonable, since the variant MCS-3i still fulfills the optimal constraint ratio $r/r_h\!=\!1$ for arbitrarily curved configurations.\\
\end{minipage}

\hspace{0.2 cm}
\begin{minipage}{15.0 cm}
\textbf{Remark:} As an alternative to the variant MCS-3i, also a reduced Gauss-Lobatto integration scheme with integration points $\xi^1\!=\!-1$, $\xi^2\!=\!1$ and $\xi^3\!=\!0$ could be applied to the axial tension term of the weak form, yielding the same number of constraint equations $n_{eq,c}\!=\!2n_{ele}\!+\!1$ as the MCS approach. However, within this contribution, the MCS method will be preferred due to its arguably more consistent variational basis and the uniform integration scheme resulting for the individual work contributions of the weak form.\\
\end{minipage}

All in all, it can be concluded that the choice of the element boundary nodes and the element midpoint as CPs of the MCS method leads to the minimal possible number of constraint equations and to an optimal constraint ratio. Consequently, a successful avoidance of locking effects could be confirmed for this approach. A similar effectiveness can be expected for a reduced Gauss-Lobatto integration of the axial tension terms with integration points $\xi^1\!=\!-1$, $\xi^2\!=\!1$ and $\xi^3\!=\!0$. On the contrary, the minimal number of three integration points possible for a reduced Gauss-Legendre integration scheme leads to an increased constraint ratio of $r_h\!=\!3\!>\!r$ and consequently to a suboptimal locking behavior. While the MCS and RI methodologies aim at a reduction of the number of constraint equations, the functional principle of the ANS approach is different: There, parameter coordinates have to be determined where the constraint is already correctly fulfilled by the original element formulation (when applying the latter to a representative test case). These parameter coordinates are typically chosen as CPs for the ANS approach. Such a procedure can avoid locking effects in a manner that is independent of the number of constraint equations. The drawback of the ANS method is that the positions of these points may change for general deformation states in the geometrically nonlinear regime, which might considerably deteriorate the effectiveness of this approach. All these theoretical considerations recommend the proposed MCS approach as method of choice in order to avoid locking effects for the Kirchhoff beam elements considered in this contribution. The numerical results presented in~\cite{meier2015} have confirmed this prediction.\\

\hspace{0.2 cm}
\begin{minipage}{15.0 cm}
\textbf{Review:} In all the three categories isotropic, straight and anisotropic large-deformation geometrically exact Kirchhoff-Love finite elements available in the literature (see requirement 4b) in Table~\ref{tab:comparison_requirements}), the consequences of membrane locking have been observed. However, a rigorous treatment seems to be missing in these works. For example, in~\cite{weiss2002b}, oscillations of the membrane forces have been observed and cured by means of a special force averaging procedure. However, this procedure seems to be rather a post-processing step than an invasion in the actual finite element formulation that would also improve the final displacement solution. In~\cite{zhang2015}, the axial tension term of the weak form is replaced by an averaged element-wise constant approximation of $\epsilon(\xi)$ in order to be able of exactly representing the solution of a straight beam under axial load. However, the influence of this procedure on membrane locking effects in curved configurations, i.e. configurations that are actually relevant for membrane locking, has not been investigated. Finally,~\cite{greco2016} proposes a mixed finite element formulation in combination with a multi patch B-spline approach in order to treat the locking effects. Interestingly, the third-order variant of these B-spline patches could be identified as Hermite interpolation comparable to~\eqref{hermite_interpolation}, which turned out to be favorable for the avoidance of locking as compared to single B-spline patches due to the lower inter-element continuity enforced by the former approach. However, the geometrical compatibility equations of the multi patch approach as well as the considered numerical examples seem only to cover the geometrically linear regime. In contrast, the numerical examples in~\cite{meier2015} and Section~\ref{sec:elenumericalexamples} of this work confirm that this gap could be closed by the proposed MCS method, which successfully cures membrane locking in 3D large-deformation problems.\\
\end{minipage}

\paragraph{Shear locking} The phenomenon of shear locking does not lie in the focus of the current work. Nevertheless, the cause for such locking effects shall at least be briefly compared with the situation already discussed for membrane locking. Shear locking denotes the inability of a finite element to exactly represent the state of vanishing shear strains. This situation can again be illustrated by means of a pure bending example. It is assumed that the beam centerline is discretized either by Lagrange polynomials according to~\eqref{lagrange_interpolation} or by Hermite polynomials according to~\eqref{hermite_interpolation} and that the triad interpolation is given by~\eqref{triadrelrot_interpolation}. It has already been stated that the triad interpolation~\eqref{triadrelrot_interpolation} can exactly represent constant curvatures. On the other hand, it has been shown that the Hermite centerline representation (and also its counterpart based on Lagrange polynomials) cannot exactly display the state of constant curvature. This means, the state $\mb{r}^{\prime T}\!(\xi)\mb{g}_2(\xi) \!\equiv\! \mb{r}^{\prime T}\!(\xi)\mb{g}_3(\xi)\!\equiv\!0$ cannot be displayed exactly in combination with a constant curvature. In other words, the interpolation spaces applied to the translation (i.e. to the beam centerline) and to the rotation field (i.e. to the triad field) do not optimally match in the sense that a state of constant curvature and vanishing shear deformation cannot be represented exactly. Similar to membrane locking, the ratio of the shear stiffness to the bending stiffness increases quadratically with the beam element slenderness ratio. Thus, again the element slenderness ratio represents the key parameter for this locking effect. Further 2D investigations on membrane and shear locking effects in geometrically linear and nonlinear shear-deformable beam element formulations can for example be found in \cite{ibrahimbegovic1993}.\\

\hspace{0.2 cm}
\begin{minipage}{15.0 cm}
\textbf{Review:} At the end of this section, also a view considerations shall be made concerning element patch tests (requirement 4d) in Table~\ref{tab:comparison_requirements}). Such a patch test for geometrically exact beams could for example require the exact representability of a state of constant 3D curvature (bending and torsion) and constant axial tension (as well as constant shear-deformation in case of Simo-Reissner formulations). The beam centerline curve associated with such a configuration is given by a helix with constant slope as investigated in the numerical example of Section~\ref{sec:examples_purebending3d}. In the following, it will be distinguished, if only the strain energy associated with such a state can exactly be represented or if additionally also the beam geometry in terms of centerline and triad field of such a state can exactly be represented. In Sections~\ref{sec:elestrongkirchhoff} and~\ref{sec:examples_purebending3d}, it will be verified theoretically and numerically that the SK element formulations based on strong enforcement of the Kirchhoff constraint and proposed in Section~\ref{sec:elestrongkirchhoff} can in general neither represent the strain energy nor the beam geometry associated with such a state. Due to the identical measure $\kappa$ of centerline curvature and the comparable centerline discretizations  employed by the isotropic, straight and anisotropic Kirchhoff-Love formulations from the literature, it can directly be concluded that also none of these formulations can fulfill this patch test, neither in the energetic nor in the geometrical sense (see again Table~\ref{tab:comparison_requirements}). On the other hand, it will be verified theoretically and numerically that the WK element formulations based on weak enforcement of the Kirchhoff constraint (see Section~\ref{sec:eleweakkirchhoff}) as well as the Simo-Reissner formulation proposed by Crisfield and Jeleni\'c~\cite{crisfield1999,jelenic1999} (denoted as CJ element in Section~\ref{sec:elesimoreissner}) can represent the exact energy state of the patch test due to the employed geodesic triad interpolation (and an appropriate collocation scheme such as MCS or reduced integration for coupling the centerline and the triad field), but not the exact geometry.\\

There exist a few very recent 3D Simo-Reissner beam element formulations that can exactly represent such a patch case in the energetic but also in the geometrical sense. Unfortunately, an explicite analytic representation of the discretization underlying these formulations in form of a helicoidal interpolation applied to the triad and to the centerline field is only possible for first-order interpolations~\cite{cesarek2012,sonneville2014,sonneville2014b}. As emphasized in~\cite{linn2016}, such a helicoidal scheme can be identified as one geodesic interpolations on the \textit{semidirect} product manifold $S \! E(3) :=\Re^3 \rtimes S \! O(3)$, denoted as Special Euclidean group, while e.g. the formulation of Crisfield and Jeleni\'c~\cite{crisfield1999,jelenic1999} consists of two individual geodesic interpolations, a linear one with constant slope on the Euclidean vector space $\Re^3$ as well as a spherical interpolation with constant curvature on the Special Orthogonal group $S \! O(3)$, together composing a interpolation scheme on the \textit{direct} product manifold $\Re^3 \times S \! O(3)$. An extension of these helicoidal interpolations to higher-order schemes is e.g. given by~\cite{zupan2003,zupan2006}. There, a closed-form representation of the interpolation has not been possible anymore and the triad and centerline field have been generated out of the strain field in an implicit manner via numerical integration. In contrast to Simo-Reissner formulations, Kirchhoff-Love element formulations should typically be of higher-order in order to guarantee for $C^1$-continuity and reasonable convergence orders (see also Section~\ref{sec:spatialdiscretization_convergence}). It is not clear if the increased numerical effort required for implicitly defining higher-order helicoidal fields via numerical integration could overcompensate possible advantages in terms of higher approximation quality when developing comparable Kirchhoff-Love element formulations that exactly fulfill the patch test in the energetic and geometrical sense. At least, it seems to be a promising direction of future research to compare formulations of this kind with the Kirchhoff-Love elements proposed here.
\end{minipage}

%
\subsubsection{Optimal convergence orders}
\label{sec:spatialdiscretization_convergence}
%

In order to compare the convergence behavior of different finite element formulations, a well-defined error measure is required. Thereto, the following relative $L^2$-error $||e||^2_{rel}$ will be considered in the numerical examples of Section~\ref{sec:elenumericalexamples}:
\begin{align}
\label{rele2error}
||e||^2_{rel} = \frac{1}{u_{max}} \sqrt{ \frac{1}{l}\int_0^l ||\mb{r}_h-\mb{r}_{ref}||^2 ds}.
\end{align}
In~\eqref{rele2error}, $\mb{r}_h$ denotes the numerical solution of the beam centerline for a certain discretization. For all examples without analytic solution, the standard choice for the reference solution $\mb{r}_{ref}$ is a numerical solution via the WK-TAN element (see Section~\ref{sec:eleweakkirchhoff_variant2}) employing a spatial discretization that is by a factor of four finer than the finest discretization shown in the convergence plot. The normalization with the element length $l$ makes the error independent of the length of the considered beam. The second normalization leads to a more convenient relative error measure, which relates the $L^2$-error to the maximal displacement $u_{max}$ of the load case. For some examples, also the relative energy error
\begin{align}
\label{energy_error}
||e||_{e,rel} = \frac{\Pi_{int,h}-\Pi_{int,ref}}{\Pi_{int,ref}},
\end{align}
will be considered. Here, $\Pi_{int,h}$ and $\Pi_{int,ref}$ represent the stored energy functions (see e.g.~\ref{storedenergyfunctionkirchhoff}) associated with a certain discretization and with the reference solution. Before the convergence plots of selected numerical examples will be considered, the optimal convergence rates in the norms~\eqref{rele2error} and~\eqref{energy_error} expected for the different beam element formulations shall be briefly discussed. The convergence in the energy error (which is minimized by the finite element method) is dominated by the highest derivative $m$ of the primary variable fields $\mb{r}(\xi)$ and $\boldsymbol{\Lambda}(\xi)$ occurring in the energy (see e.g.~\ref{storedenergyfunctionkirchhoff}) and consequently in the weak form. Since the employed hyperelastic stored energy functions represent quadratic forms in the derivatives of these primary variable fields, the convergence rate in the energy error yields:
\begin{align}
\label{energy_convergenceorder}
||e||_e = \mathcal{O}(h^{2(k-m+1)}).
\end{align}
Here, $h\!:=\!l_{ele}$ is the element length and $k$ the polynomial degree completely represented by the trial functions. In \cite{strang2008}, the convergence rate of the $L^2$-error for the Ritz solution to a variational problem of order $m$ is shown to be:
\begin{align}
\label{convergenceorder}
||e||^2 = \mathcal{O}(h^{k+1} + h^{2(k-m+1)}).
\end{align}
The second term in~\eqref{convergenceorder} represents the dependence of the $L^2$-error convergence rate on the energy error convergence, reflecting the variational basis of the finite element method. The first term represents the pure polynomial approximation of the trial functions with respect to the considered primary variable field, e.g. $\mb{r}(\xi)$. In most cases, e.g. when displacement-based solid elements are considered, the first exponent is smaller than the second one and dominates the overall discretization error. For this reason, only the first term is considered by many authors. However, in the following, it will be shown that especially for Kirchhoff type beam elements also the second term of~\eqref{convergenceorder} is important. Thereto, the expected convergence rates for Reissner and Kirchhoff type beam elements shall briefly be discussed.\\

\paragraph{Element formulations of Simo-Reissner type} The highest derivative of primary variable fields occurring in the weak form associated with the Simo-Reissner beam problem is $m\!=\!1$. Thus, for the third-order ($k\!=\!3$) Reissner beam elements considered in this work, a convergence rate of six is expected in the energy error~\eqref{energy_error}, while the $L^2$-error is dominated by the first term of~\eqref{convergenceorder} leading to a corresponding optimal convergence rate of four.\\

\paragraph{Element formulations of Kirchhoff-Love type} The subsequently proposed Kirchhoff beam elements will lead to the values $k\!=\!3$ for the polynomial degree of the triad functions and $m\!=\!2$ for the highest arc-length derivative in the weak form. Consequently, the convergence of the energy error is of order four. Furthermore, the exponents of both terms in~\eqref{convergenceorder} take on a value of four, also leading to an expected convergence rate of four for the $L^2$-error. Thus, also the second, energy-related term has to be considered for Kirchhoff problems of this kind. For that reason, at least polynomials of order three should be chosen as trial functions for Kirchhoff beam elements: Reducing the polynomial degree from $k\!=\!3$ to $k\!=\!2$ would lead to an undesirable decline in the $L^2$-convergence rate from four to two. Thus, the third-order Kirchhoff beam elements proposed in this work can be regarded as approximations of the lowest order that is reasonable from a numerical point of view. However, the fact that both exponents of~\eqref{convergenceorder} take on the same value for $k\!=\!3$ also means that in some cases, usually for examples involving complex deformation states and strain distributions, the second, energy-related term might determine the overall error-level. Thus, in order to fully exploit the approximation power of the employed discretization, it can often be sensible to apply trial functions of increased polynomial degree $k\!>\!3$. In this case, the first term of~\eqref{convergenceorder} will determine the overall error level for sufficiently fine discretizations (since the second term converges faster) and consequently the discretization error is exclusively limited by the approximation power of the applied polynomial order. The extension of the proposed Kirchhoff beam elements to Hermite interpolations of order $k\!>\!3$ is possible in a straightforward manner and will be treated in future research work. In the numerical example of Section~\ref{sec:examples_purebending2d}, a first proof of concept will be given for such an extension.\\ 

\hspace{0.2 cm}
\begin{minipage}{15.0 cm}
\textbf{Remark:} Based on these considerations the question arises, which trial function orders $k_{SR}$ and $k_{KL}$ of formulations based on the Simo-Reissner and the Kirchhoff-Love theory have to be chosen in numerical examples in order to perform a reasonable comparison of their convergence behavior. The answer to this question depends on the primary interest, which might either lie in the $L^2$- or in the energy convergence. Here, the third-order Kirchhoff elements are compared to third-order Reissner elements, leading to equal $L^2$- but to different energy error rates. Alternatively, one could compare the third-order Kirchhoff elements with second-order Reissner elements, leading to equal energy error rates but to different $L^2$-error rates.\\
\end{minipage}

\hspace{0.2 cm}
\begin{minipage}{15.0 cm}
\textbf{Review:} Later in Section~\ref{sec:elestrongkirchhoff}, a quadratic interpolation will be applied to the relative angle field $\varphi_h(\xi)$ of the interpolation scheme~\eqref{triadsr_triadfield}. Since the orientation of the material triad field is determined by the relative angle $\varphi_h(\xi)$ as well as the tangent vector field $\mb{r}_h^{\prime}(\xi)$, with the latter being a polynomial of order two, this second-order interpolation is sufficient for triad field discretization. Even a higher polynomial degree for $\varphi_h(\xi)$ could not further improve the exact polynomial representation of rotational strains, which is of first order. On the other hand, if $\varphi_h(\xi)$ was only interpolated linearly, only the constant term of rotational strains could be exactly represented and the second term in~\eqref{convergenceorder} would dominate the $L^2$-convergence leading to a decline in the expected optimal order to two. In Meier et al.~\cite{meier2014}, these considerations have been confirmed numerically. The Kirchhoff-Love formulations of straight type (see requirement 4a) in Table~\ref{tab:comparison_requirements}) also apply only a first-order interpolation for the relative angle $\varphi_h(\xi)$ and a comparable intermediate triad definition. Thus, also for these formulations a decline in the optimal convergence order from for to two is expected.
\end{minipage}

%
\subsubsection{Conservation properties}
\label{sec:spatialdiscretization_conservation}
%

Since the finite element solution converges towards the corresponding analytic solution in the limit of fine spatial discretizations, elementary properties of the analytic solution such as conservation of linear momentum (or rather equilibrium of forces in statics), conservation of angular momentum (or rather equilibrium of moments in statics) as well as conservation of energy (or rather balance of external and internal work for non-conservative problems) will also be fulfilled by the numerical solution for $l_{ele} \rightarrow 0$. However, often it is desirable to provide such properties already for arbitrarily rough spatial discretizations. The question, if these properties of the space-continuous problem are inherited by the spatially discretized problem, will later be investigated for the different beam element formulations proposed in subsequent sections. Thereto, use will be made of the fact that the discretized weak form of the balance equations is fulfilled for arbitrary values of the nodal primary variable variations. Choosing the nodal primary variable variations such that the associated virtual motion represents a rigid body translation given by
\begin{align}
\label{rigid_translation}
(\delta \mb{r}_{h}(\xi),\delta \boldsymbol{\theta}_{h}(\xi)) \!\equiv\!  (\mb{u}_0,\mb{0}) \in \mathcal{V}_h \quad \text{with} \quad \mb{u}_0^{\prime} = \mb{0},
\end{align}
allows to investigate the conservation of linear momentum. The special choice~\eqref{rigid_translation} leads to $\delta \mb{r}_{h}^{\prime}(\xi) \!\equiv\! \delta \boldsymbol{\theta}_{h}^{\prime}(\xi)\!\equiv\! \mb{0}$ and to vanishing contributions of the internal forces and moments in the discretized weak forms~\eqref{weakformspatial} and~\eqref{weakform3} associated  with the Simo-Reissner and the Kirchhoff-Love beam theory. Inserting~\eqref{rigid_translation} into these weak forms yields
\begin{align}
\label{conservation_linmomentum_ele}
\dot{ \mb{l} }=\mb{f}_{ext}
\quad \text{with} \quad \mb{l}:= \int \limits_0^{l} \tilde{\mb{l}} ds, \, \, \mb{f}_{ext}=\int \limits_0^{l} \mb{\tilde{f}} ds + \Big[\mb{f}_{\sigma}\Big]_{\Gamma_{\sigma}} \hspace{-0.1cm},
\end{align} 
and to exact conservation of linear momentum $\mb{l}\!=$ const. for the unloaded system, viz. if $\mb{f}_{ext}\!=\!\mb{0}$. Since reaction forces at Dirichlet supports are also included in $\mb{f}_{ext}$,~\eqref{conservation_linmomentum_ele} is equivalent to the equilibrium of forces in the static case, i.e. if $ \mb{l} \!=\!\mb{0}$. Similarly, a choice of the nodal primary variable variations representing a rigid body rotation given by
\begin{align}
\label{rigid_rotation}
(\delta \mb{r}_{h}(\xi),\delta \boldsymbol{\theta}_{h}(\xi)) \!\equiv\!  (\mb{w}_0 \times \mb{r}_{h}(\xi),\mb{w}_0) \in \mathcal{V}_h \quad \text{with} \quad \mb{w}_0^{\prime} = \mb{0},
\end{align}
allows to investigate the angular momentum. Relation~\eqref{rigid_rotation} leads to $\delta \mb{r}_{h}^{\prime}(\xi) \!\equiv\! \mb{w}_0 \!\times\! \mb{r}_{h}^{\prime}(\xi)$ as well as $\delta \boldsymbol{\theta}_{h}^{\prime}(\xi) \!\equiv\! \mb{0}$ and again to vanishing contributions of the internal forces and moments in the discrete versions of the weak forms~\eqref{weakformspatial} and~\eqref{weakform3} associated  with the Simo-Reissner and the Kirchhoff-Love beam theory. Inserting~\eqref{rigid_rotation} into these weak forms yields:
\begin{align}
\label{conservation_angmomentum_ele}
\!\!\!\!\!\!\dot{ \mb{h} } \!=\!\mb{m}_{ext},
\,\,\,\, \mb{h} \!:=\! \!\int \limits_0^l \!(\tilde{\mb{h}} \!+\! \mb{r} \!\times\! \tilde{\mb{l}}) ds, \, \, 
\mb{m}_{ext}\!=\!\!\int \limits_0^l \left(\mb{r} \times \mb{\tilde{f}} \!+\! \mb{\tilde{m}}\right) ds 
\!+\! \Big[\mb{r} \times \mb{f}_{\sigma} \!+\! \mb{m}_{\sigma}\Big]_{\Gamma_{\sigma}} \hspace{-0.1cm},\!\!\!\!\!\!
\end{align} 
and to exact conservation of angular momentum $ \mb{h} \!=$ const. for the unloaded system, viz. if $\mb{m}_{ext}\!=\!\mb{0}$. 
Since possible reaction moments at Dirichlet supports are also included in $\mb{m}_{ext}$, relation~\eqref{conservation_angmomentum_ele} is equivalent to the equilibrium of moments in the static case, i.e. if $\mb{h} \!=\!\mb{0}$. Finally, a choice of the nodal primary variable variations according to
\begin{align}
\label{conservation_velocity}
(\delta \mb{r}_{h}(\xi),\delta \boldsymbol{\theta}_{h}(\xi)) \!\equiv\!  (\dot{\mb{r}}_{h}(\xi),\mb{w}_h(\xi)) \in \mathcal{V}_h,
\end{align}
allows to investigate the mechanical power balance. Inserting~\eqref{conservation_velocity} into the discrete versions of the weak forms~\eqref{weakformspatial} and~\eqref{weakform3} associated  with the Simo-Reissner and the Kirchhoff-Love theory, and making use of the relations~$\dot{\tilde{\Pi}}_{int}\!=\!\mb{w}^{\prime T} \mb{m}\!+\!(\dot{\mb{r}}^{\prime} \!-\! \mb{w} \!\times\! \mb{r}^{\prime})^T \mb{f}$ and~$\dot{\tilde{\Pi}}_{kin}\!=\!\mb{w}^T \! \mb{c}_{\rho} \mb{a}$ (see Section~\ref{sec:reissnerhamilton}) as well as $\mb{w}^T\mb{S}(\mb{w}) \mb{c}_{\rho} \mb{w}\!=\!0$, yields the following relation
\begin{align}
\label{conservation_energy_ele}
\dot{\Pi}_{kin}+\dot{\Pi}_{int} = P_{ext} 
\quad \text{with} \quad 
P_{ext} = 
\int \limits_0^l
  \Bigg[ 
 \dot{\mb{r}}^T \mb{\tilde{f}}
 + \mb{w}^T \! \mb{\tilde{m}}
  \Bigg] ds
  +\Bigg[\dot{\mb{r}}^T \! \mb{f}_{\sigma} + \mb{w}^T \! \mb{m}_{\sigma} \Bigg]_{\Gamma_{\sigma}},
\end{align}
and consequently to exact energy conservation $\Pi_{kin}\!+\!\Pi_{int}\!=$ const. for the unloaded system, viz. if $P_{ext}\!=\!0$. So far, it has been shown that exact conservation of linear momentum, angular momentum and energy (see equations~\eqref{conservation_linmomentum_ele},~\eqref{conservation_angmomentum_ele} and~\eqref{conservation_energy_ele}) can be guaranteed for the spatially discretized (and time-continuous) problem, provided the special choices~\eqref{rigid_translation}, \eqref{rigid_rotation} and \eqref{conservation_velocity} for the translational and rotational variation fields are contained in the discrete weighting space $\mathcal{V}_h$ of the considered finite element formulation. In the following sections, the question if~\eqref{rigid_translation}, \eqref{rigid_rotation} and \eqref{conservation_velocity} can indeed be represented by the discrete weighting functions, will be investigated for the proposed beam element formulations. Of course, also the time integration scheme influences the conservation properties of the fully discrete system considerably. However, the investigation of this factor does not lie within the scope of this work. In general, it can be stated that the three conservation properties considered above will be fulfilled by a finite element formulation if i) the formulation is objective and path-independent and ii) the test functions are consistently derived in a Bubnov-Galerkin manner. The first pre-requisite ensures that a unique potential exists in form of a discrete stored-energy function that is path-independent and invariant under rigid body motions. The second pre-requisite ensures that the corresponding contribution to the weak form represents exact increments of this potential, which fulfills the third requirement~\eqref{conservation_energy_ele} per definition, but also the first and second requirement~\eqref{conservation_linmomentum_ele} and~\eqref{conservation_angmomentum_ele}, since infinitesimal rigid body translations and rotations according to~\eqref{rigid_translation} and~\eqref{rigid_rotation} will lead to vanishing increments of this objective potential. However, the opposite conclusion does obviously not hold true: Also Petrov-Galerkin formulations can at least fulfill the first and second requirement~\eqref{conservation_linmomentum_ele} and~\eqref{conservation_angmomentum_ele} if the chosen test functions can represent rigid body translations and rotations~\eqref{rigid_translation} and~\eqref{rigid_rotation} (see e.g. CJ element of Section~\ref{sec:elesimoreissner} or WK element of Section~\ref{sec:eleweakkirchhoff}).\\

\hspace{0.2 cm}
\begin{minipage}{15.0 cm}
\textbf{Review:} Following the argumentation from above, it can be expected that the objective and path-independent Bubnov-Galerkin FEM discretizations employed in the formulations of the straight type (see requirement 4) of Table~\ref{tab:comparison_requirements}) will in general fulfill the above conservation properties as long as the variation and discretization process required to derive the test functions is conducted in a consistent manner and no additional approximations (e.g. small tension assumptions according to $||\mb{r}^{}|| \approx 1$ as discussed in~\cite{zhao2012}) are applied that could spoil this consistency. The situation is slightly different for the anisotropic formulations: The non-objective and path-dependent character of the employed discrete deformation measures might in general lead to non-vanishing energy increments for infinitesimal rigid body translations and rotations according to~\eqref{rigid_translation} and~\eqref{rigid_rotation}. The exact representation of these non-vanishing energy increments resulting from consistently derived Bubnov-Galerkin test functions may in this case exactly be the reason that the conservation properties~\eqref{conservation_linmomentum_ele} and~\eqref{conservation_angmomentum_ele} will be spoiled. Eventually, also the existing Kirchhoff-Love formulations of the straight type in Table~\ref{tab:comparison_requirements} shall be discussed. There, a standard polynomial interpolation based on Lagrange shape functions is applied to the tangential spin vector component $\delta \Theta_1$ similar to the Petrov-Galerkin variant~\eqref{kirchhoffkinematics_spinvector_variant1}. Consequently, as shown in equation~\eqref{elestrongkirchhoff_rigid_rotation3}, also these formulations cannot represent constant spin vector distributions and exact conservation of angular momentum~\eqref{conservation_angmomentum_ele}.
\end{minipage}

\hspace{0.2 cm}
\begin{minipage}{15.0 cm}
It can be shown that these straight formulations indeed represent Petrov-Galerkin formulations, which in turn can also not guarantee for the conservation of energy~\eqref{conservation_energy_ele}. To verify this statement, the following considerations are made: If the discrete spin vector field employed in~\cite{boyer2004} is consistent with the underlying triad interpolation in a Bubnov-Galerkin sense is not straight-forward to be answered, since no explicit triad interpolation scheme is given (and required) by the considered isotropic formulations. However, at least the arc-length derivative of the discrete triad field is defined by the employed torsion interpolation, which is based on Lagrange polynomials in~\cite{boyer2004} or~\cite{weiss2002b}. The consistent Bubnov-Galerkin interpolation can then be identified based on the second compatibility condition in~\eqref{torsionfree_K1andW1_3} an reads:
\begin{align}
\label{conservation_energy_ele}
\delta \Theta_{1h}^{\prime}(\xi) \!=\! \delta {K}_{1h}(\xi) \!-\! \delta {\mb{g}}_{1h}(\xi)^T\!(\mb{g}_{1h}(\xi) \!\times\! \mb{g}_{1h}^\prime(\xi)), \,\,\,
\end{align}
Since ${K}_{1h}(\xi)$ has been interpolated via Lagrange polynomials, $\mb{r}_{h}^\prime(\xi)$ has been interpolated via Hermite polynomials and the relation $\mb{g}_{1h}^\prime(\xi)\!=\!\mb{r}_{h}^\prime(\xi)/||\mb{r}_{h}^\prime(\xi)||$ holds, a consistent Bubnov-Galerkin interpolation of the tangential spin vector component $\delta \Theta_{1h}(\xi)$ fulfilling~\eqref{conservation_energy_ele} cannot be based on a pure polynomial interpolation as done in the mentioned references. Consequently, these interpolations are of Petrov-Galerkin type.
\end{minipage}

\section{Simo-Reissner Beam Element}
\label{sec:elesimoreissner}

Throughout this work, the Reissner type beam element formulation proposed by Crisfield and Jeleni\'{c}~\cite{crisfield1999,jelenic1999}, in the following referred to as CJ element, will serve as reference formulation for numerical comparisons. In the next section, the main constituents required to derive the element residual vector will be presented. In the subsequent Sections~\ref{sec:elesimoreissner_spatialdiscretization_locking} and~\ref{sec:elesimoreissner_spatialdiscretization_conservation}, this element formulation will be investigated with respect to possible locking effects and the fulfillment of mechanical conservation properties as introduced in Sections~\ref{sec:spatialdiscretization_locking} and~\ref{sec:spatialdiscretization_conservation}.

%
\subsection{Element residual vector}
\label{sec:elesimoreissner_spatialdiscretization_residual}
%

In this section, the element residual vector of the CJ element will be derived from the general, space-continuous Simo-Reissner beam problem. First, all trial and weighting functions are replaced by their discrete counterparts taken from the finite-dimensional trial subspace $(\mb{r}_h,\boldsymbol{\Lambda}_h) \in \mathcal{U}_h \subset \mathcal{U}$ and the weighting subspace $(\delta \mb{r}_h, \delta \boldsymbol{\theta}_h) \in \mathcal{V}_h \subset \mathcal{V}$. In the following, $n_n$-noded finite elements with the vectors of nodal primary variables $\hat{\mbd{x}}_{CJ}\!:=\!(\hat{\mb{d}}^{1T}\!,\hat{\boldsymbol{\psi}}^{1T}\!,...,\hat{\mb{d}}^{n_n T}\!,\hat{\boldsymbol{\psi}}^{n_n T})^T$ and $\delta \hat{\mbd{x}}_{CJ}\!:=\!(\delta \hat{\mb{d}}^{1T}\!,\delta \hat{\boldsymbol{\theta}}^{1T}\!,...,\delta \hat{\mb{d}}^{n_n T}\!,\delta \hat{\boldsymbol{\theta}}^{n_n T})^T$ are considered. The centerline interpolation is based on Lagrange polynomials of order $n_n\!-\!1$ according to Section~\eqref{sec:spatialdiscretization_lagrange}, i.e. $n_r\!=\!n_n$. Furthermore, the rotation field interpolation follows equation~\eqref{triadrelrot_interpolation} of Section~\ref{sec:spatialdiscretization_triadrelrot} based on $n_{\Lambda}\!=\!n_n$ nodes, in combination with a Petrov-Galerkin approach for the spin vector discretization given by~\eqref{petrovspininterpolation}. In contrast to the original works~\cite{crisfield1999,jelenic1999}, here the modified generalized-$\alpha$ scheme of Section~\ref{sec:temporaldiscretization} is employed for time integration, defining the velocities and accelerations required for the inertia forces $\mb{f}_{\rho}$ and moments $\mb{m}_{\rho}$ as stated in~\eqref{simoreissner_inertia}. Inserting these interpolation schemes as presented in the previous sections into the weak form of the balance equations~\eqref{weakformspatial} yields the element residual vector contributions $\mbd{r}_{CJ,\hat{\mb{d}}}$ and $\mbd{r}_{CJ,\hat{\boldsymbol{\theta}}}$ according to
\begin{align}
\label{reissner_residual}
\begin{split}
 & \mbd{r}_{CJ,\hat{\mb{d}}} = \int \limits_{-1}^{1} \left( \mbd{L}^{\! \prime T}\mb{f} - \mbd{L}^{\!T} \tilde{\mb{f}}_{\rho} 
 \right) J(\xi) d \xi -\Big[\mbd{L}^{\!T} \mb{f}_{{\sigma}} \Big]_{\varGamma_{\sigma}},\\
& \mbd{r}_{CJ,\hat{\boldsymbol{\theta}}} = \int \limits_{-1}^{1} \left( \mbd{L}^{ \! \prime T} \mb{m} -\mbd{L}^{\! T}\mb{S(\mb{t})} \mb{f}
-\mbd{L}^{T} \tilde{\mb{m}}_{\rho} \right) J(\xi) d \xi - \Big[ \mbd{L}^{ \! T} \mb{m}_{{\sigma}} \Big]_{\varGamma_{\sigma}}\!\!\!\!\!\!=\mb{0}.
\end{split}
\end{align}
The subscripts $(.)_{\hat{\mb{d}}}$ and $(.)_{\hat{\boldsymbol{\theta}}}$ distinguish the residual vector contributions associated with the variations $\delta \hat{\mb{d}}^{i}$ and $\delta \hat{\boldsymbol{\theta}}^{i}$. Within this work, a linearization $\Delta \mbd{r}_{CJ}\!=\!\mbd{k}_{CJ} \Delta \hat{\mbd{x}}_{CJ}$ based on multiplicative rotation increments $\Delta \hat{\boldsymbol{\theta}}^{i}$ according to $\Delta \hat{\mbd{x}}_{CJ}\!:=\!(\Delta \hat{\mb{d}}^{1T}\!,\Delta \hat{\boldsymbol{\theta}}^{1T}\!,...,\Delta \hat{\mb{d}}^{n_n T}\!,\Delta \hat{\boldsymbol{\theta}}^{n_n T})^T$ as given in~\cite{jelenic1999} is employed. In dynamics, the element residual vector $\mbd{r}_{CJ}$ and stiffness matrix $\mbd{k}_{CJ}$ slightly differ from the original work~\cite{jelenic1999} due to the applied time integration scheme.\\

%
\subsection{Avoidance of locking effects}
\label{sec:elesimoreissner_spatialdiscretization_locking}
%

In~\cite{crisfield1999} and~\cite{jelenic1999}, the authors proposed a reduced Gauss integration scheme in order to avoid shear locking and membrane locking in the range of high beam slenderness ratios. Thereto, $n_n\!-\!1$ integration points have been employed for the integration of the internal force contribution of a $n_n$-noded element. The effectiveness of this procedure will be verified in subsequent numerical examples and shall be briefly motivated by the following considerations: The 3D Simo-Reissner beam problem is based on $n_{eq}\!=\!6$ differential equations~\eqref{simoreissner_equilibriumspatial} describing the beam problem and pointwise $n_{eq,c}\!=\!3$ constraint equations in order to represent the state of vanishing axial strains $\mb{r}^{\prime T}\mb{g}_1 \!\equiv\! 1$ and vanishing shear strains $\mb{r}^{\prime T}\mb{g}_2 \!\equiv\! \mb{r}^{\prime T}\mb{g}_3 \!\equiv\! 1$ prevalent in a pure bending problem. Consequently, the constraint ratio of the space-continuous problem yields $r\!=\!n_{eq}/n_{eq,c}\!=\!2$. As consequence of the reduced integration, the discrete number of constraint equations takes on a value of $n_{eq,c}\!=\!3\cdot (n_n\!-\!1) \cdot n_{ele}$. Given the total number of equations $n_{eq}\!=\!6(n_n\!-\!1) \cdot n_{ele}$ after application of proper Dirichlet conditions, the constraint ratio of the discrete problem results in:
\begin{align}
\label{elesimoreissner_constraintratio}
r_h=\lim_{n_{ele} \rightarrow \infty}  \frac{3\cdot (n_n\!-\!1) \cdot n_{ele}}{6(n_n\!-\!1) \cdot n_{ele}} = 2 = r.
\end{align}
Relation~\eqref{elesimoreissner_constraintratio} yields the optimal constraint ratio for this element formulation. Consequently, no locking effects are expected. The investigations made so far can be further refined by realizing that the CJ element can exactly represent the internal energy associated with a 3D pure bending state. In order to understand this statement, the internal energy is split into contributions stemming from torsion and bending and into contributions stemming from axial tension and shear deformation, i.e. $\Pi_{int,h}\!=\!\Pi_{int,\Omega,h}\!+\!\Pi_{int,\Gamma,h}$. For a pure bending state, the energy contribution $\Pi_{int,\Gamma,h}$ has to vanish, and thus the total internal energy of a pure bending state is given by $\Pi_{int,\Omega,h}$, which in turn is uniquely defined by the curvature vector field $\mb{K}_h(\xi)\!=$ const. In order to represent the desired (constant) distribution of the curvature vector field $\mb{K}_h(\xi)$, which is possible for the employed triad interpolation~\eqref{triadrelrot_interpolation}, only $(n_n\!-\!1) \cdot n_{ele}$ of the $(n_n\!-\!1) \cdot n_{ele}\!+\!1$ nodal rotation vectors have to arise properly, while the one remaining nodal rotation vector describes rotational rigid body modes of the beam. Since the curvature vector field is defined via the arc-length derivative of the rotation field, this one remaining nodal rotation vector can also be interpreted as integration constant resulting from an integration of the curvature field. Additionally, the nodal position vectors have to arise in a way such that $\Pi_{int,\Gamma,h}\!=\!0$. While the requirement $\boldsymbol{\Gamma}_h(\xi)\!\equiv\!\mb{0}$ cannot be fulfilled exactly for the employed triad trial function spaces, the reduced Gauss integration scheme applied for the CJ element yields a finite number of $3 \!\cdot\! (n_n\!-\!1) \!\cdot\! n_{ele}$ constraint equations in order to satisfy $\Pi_{int,\Gamma,h} \!=\! 0$. Thus, similar to the rotation field, only $(n_n\!-\!1) \cdot n_{ele}$ of the $(n_n\!-\!1) \cdot n_{ele}\!+\!1$ nodal position vectors have to arise properly in order to fulfill these constraint equations, while the one remaining nodal position vector describes translational rigid body modes of the beam. All in all, the $(n_n\!-\!1) \cdot n_{ele}\!+\!1$ nodal position and rotation vectors can always arise in a way such that a 3D pure bending case can be represented, which consists of a constant curvature vector field $\mb{K}_h(\xi)=$const., a vanishing (reduced integrated) energy contribution $\Pi_{int,\Gamma,h}$ of axial and shear strains as well as six superposed rigid body modes. Consequently, the torsion and bending modes represented by $\boldsymbol{\Omega}_h(\xi)$ as well as the axial tension and shear values at the Gauss points represented by $\boldsymbol{\Gamma}_h(\xi_{GP})$ are non-competing and no locking effects are expected. These considerations can easily be extended to arbitrary curvature fields $\mb{K}_h(\xi)$ that are representable by the employed triad interpolation and arbitrary fields $\boldsymbol{\Gamma}_h(\xi)$ for which the term $\boldsymbol{\Gamma}_h^T \boldsymbol{\Gamma}_h$ as occurring in the energy integral can be integrated exactly by the reduced Gauss integration scheme. In Sections~\ref{sec:examples_purebending2d} and~\ref{sec:examples_purebending3d}, the expected result that the discrete hyperelastic energies associated with pure bending states in 2D and in 3D can be represented exactly by this beam element formulation will be verified by means of corresponding numerical test cases.

%
\subsection{Conservation properties}
\label{sec:elesimoreissner_spatialdiscretization_conservation}
%

In the following, it will be investigated if the Simo-Reissner beam element formulation proposed by Crisfield and Jeleni\'{c} and repeated in the section above can represent the variational fields~\eqref{rigid_translation}, \eqref{rigid_rotation} and \eqref{conservation_velocity} required in order to guarantee conservation of linear momentum, conservation of angular momentum as well as conservation of energy. The representation of a rigid body translation~\eqref{rigid_translation} is trivial and given by the nodal primary variable variations
\begin{align}
\label{relesimoreissner_igid_translation}
\delta \hat{\mb{d}}^i = \mb{u}_0, \quad \delta \hat{\boldsymbol{\theta}}^i=\mb{0} \,\,\, \text{for} \,\,\, i=1,...,n_n.
\end{align}
Similarly, a rigid body rotation~\eqref{rigid_rotation} can be displayed by the nodal primary variable variations
\begin{align}
\label{relesimoreissner_igid_rotation}
\delta \hat{\mb{d}}^i = \mb{w}_0 \times \hat{\mb{d}}^i , \quad \delta \hat{\boldsymbol{\theta}}^i=\mb{w}_0 \,\,\, \text{for} \,\,\, i=1,...,n_n.
\end{align}
It follows from~\eqref{relesimoreissner_igid_translation} and~\eqref{relesimoreissner_igid_rotation}, that conservation of linear and angular momentum can be guaranteed. This statement holds for both spin-vector discretizations~\eqref{crisfieldspininterpolation} and~\eqref{petrovspininterpolation} and for the discretized centerline variation~\eqref{lagrange_testfunctions} since all of these variants fulfill proper completeness conditions and can exactly represent a constant vector field $\delta \boldsymbol{\theta}\!=\!\mb{w}_0$. If the nodal velocities and angular velocities of the time-continuous problem are chosen as primary variable variations
\begin{align}
\label{relesimoreissner_energy}
\delta \hat{\mb{d}}^i =  \dot{\mb{d}}^i , \quad \delta \hat{\boldsymbol{\theta}}^i=\mb{w}^i \,\,\, \text{for} \,\,\, i=1,...,n_n,
\end{align}
only the Bubnov-Galerkin variant~\eqref{crisfieldspininterpolation} leads (per definition) to an exact representation of the rates of the spatially discretized hyperelastic and kinetic energy and consequently to exact energy conservation for the spatially discretized, time-continuous problem. On the contrary, the Petrov-Galerkin variant~\eqref{petrovspininterpolation}, which has been employed in~\eqref{reissner_residual}, is not variationally consistent with the triad interpolation~\eqref{triadrelrot_interpolation} occurring in the discrete internal and kinetic energies. Consequently, the weak form~\eqref{reissner_residual} does not represent exact energy rates of the spatially discretized problem.

%

\section{Motivation for "shear-free" beam theories}
\label{sec:reissnermotivationshearfree}
%

Geometrically exact Simo-Reissner beam elements unify high computational efficiency and accuracy. In fields of application where thick beams are involved and the effect of shear deformation is important, they are favorable as compared to the Kirchhoff type counterparts. However, with increasing beam slenderness ratio $\zeta\!=\!l/R$, the shear contribution to the overall beam deformation decreases. Furthermore, it is exactly the avoidance of the high stiffness contributions of the shear modes which makes the Kirchhoff-Love theory of thin beams not only applicable, but also favorable in the range of high beam slenderness ratios. In this brief section, possible benefits of applying Kirchhoff type beam elements in the range of high slenderness ratios will be illustrated and, at least approximately, quantified. In Section~\ref{sec:elenumericalexamples}, most of these effects will also be investigated and verified by means of numerical examples.\\

\textbf{1: Improved stability properties of time integration scheme}: The dynamic equations of motion of highly slender beams typically result in very stiff Partial Differential Equations (PDEs). With increasing beam slenderness ratio, the ratio between high eigenfrequencies (associated with shear modes), intermediate eigenfrequencies (associated with axial tension and twisting modes) and low eigenfrequencies (associated with bending modes) increases considerably. As a consequence, the stability requirement of explicit time integration schemes leads to very small critical  time step sizes as compared to the large oscillation periods of the bending modes. On the contrary, implicit time integration schemes can provide unconditional stability in the linear regime of small deformations. However, in the large deformation regime, also their performance is considerably deteriorated by such high-frequency contributions. Despite the stability aspect, high-frequent modes are strongly affected by the time discretization error and should be avoided as long as no high-frequency analysis is required by a specific application. In order to illustrate the relevant frequency spectrum, in the following, the proportionalities of the eigenfrequencies resulting from pure bending $\omega_b$, pure torsion  $\omega_t$, pure axial tension  $\omega_a$ and pure shearing  $\omega_s$ are given for the linearized beam problem:
\begin{align}
\label{simoreissner_modefrequencies}
\!\!\!\!\!\!
\omega_b \!\sim\! \sqrt{\frac{EI}{\rho A l^4}} \!\sim\!  \frac{1}{\zeta} \sqrt{\frac{E}{\rho l^2}}, \,\,\,
\omega_t \!\sim\! \sqrt{\frac{GI_T}{\rho I_P l^2}} \!\sim\! \sqrt{\frac{E}{\rho l^2}}, \,\,\,
\omega_a \!\sim\! \sqrt{\frac{E}{\rho l^2}}, \,\,\,
\omega_s \!\sim\! \sqrt{\frac{GA}{\rho I}} \!\sim\! \zeta \sqrt{\frac{E}{\rho l^2}}.\!\!\!\!\!\!
\end{align}
According to the relations~\eqref{simoreissner_modefrequencies}, the ratio of axial and torsional eigenfrequencies to bending eigenfrequencies increases linearly with increasing slenderness ratio. The ratio of shear eigenfrequencies to bending eigenfrequencies increases quadratically with increasing slenderness ratio. Thus, from a theoretical point of view, the avoidance of shear modes could already improve the numerical behavior considerably. Since the numerical examples of Section~\ref{sec:elenumericalexamples} mainly focus on static analysis and only present a brief outlook on possible dynamic investigations, a numerical verification of these theoretical considerations does not lie within the scope of this work. However, numerical investigations of this question available in the literature confirm the predicted trend: Lang and Arnold~\cite{lang2012} investigated the geometrically nonlinear oscillations of a slender beam, which has been modeled by means of the geometrically exact Simo-Reissner theory and discretized via finite differences (see also~\cite{lang2011}). In order to measure the influence of high-frequency modes on time integration stability, the maximal possible time step sizes of the applied explicit time integration scheme have been determined for three different mechanical beam models: The full shear-deformable and extensible Simo-Reissner beam formulation, a beam formulation subject to the Kirchhoff constraint of vanishing shear deformation, and a beam formulation subject to the Kirchhoff constraint and an additional inextensibility constraint enforcing vanishing axial tension. For the roughest discretization considered in this numerical experiment and an investigated slenderness ratio of $\zeta\!=100$, a rather moderate slenderness ratio as compared to many applications as mentioned in Section~\ref{sec:intro}, the time step size could be increased by a factor of $\approx 100$ when abstaining from the shear mode and by a further factor of $\approx 5$ when additionally abstaining from the axial tension mode. These results indicate the potential of the Kirchhoff type beam formulations. Furthermore, they suggest that the first step towards an extensible Kirchhoff beam formulation might already represent the essential one with respect to numerical savings.\\

\textbf{2: Improved performance of (iterative) linear solvers}: According to the previous argumentation, a high ratio of the highest to the lowest dynamical eigenfrequencies, measured by the dynamic spectral radius, deteriorates the performance of time integration schemes. In a similar manner, the performance of iterative linear solvers (see e.g. \cite{quarteroni2000}) decreases with increasing ratio of the highest to the lowest eigenvalue of the tangent stiffness matrix, a measure for the condition number of this matrix. Furthermore, even if direct linear solvers are applied, very high condition numbers might considerably limit the achievable numerical accuracy. Especially in dynamics, where such round-off errors tend to accumulate, these effects are undesirable. In the following, the influence of the different deformation modes on the condition number is investigated. For simplicity, the physical units of the considered beam problem are chosen such that the element length lies in the range of $l_{ele} \!\approx\! 1$. Since the element length, or better the element Jacobian, typically enters the element formulation with different exponents occurring in different stiffness matrix entries, an element length of $l_{ele} \!\approx\! 1$ seems to be a reasonable choice with respect to conditioning. In this case, the resulting contributions to the element stiffness matrix of a Simo-Reissner beam element formulation that has been linearized with respect to the straight configuration, typically obeys the following proportionalities:
\begin{align}
\label{simoreissner_modestiffnesses}
\!\!\!\!\!\!
k_b \!\sim\! EI \!\sim\!ER^4, \,\,\,
k_t \!\sim\! GI_T \!\sim\!ER^4, \,\,\,
k_a \!\sim\! EA \!\sim\!ER^2, \,\,\,
k_s \!\sim\! GA \!\sim\!ER^2.\!\!\!\!\!\!
\end{align}
Again, $k_b$, $k_t$, $k_a$ and $k_s$ denote stiffness contributions from bending, torsion, axial tension and shear modes. As long as $l_{ele} \!\approx\! 1$ holds and the discretization is kept fixed, the cross-section radius $R$ decreases linearly with increasing slenderness ratio $\zeta$. According to~\eqref{simoreissner_modestiffnesses}, it is expected that the ratio of high stiffness contributions (from shear and axial tension modes) to low stiffness contributions (from torsional and bending modes), and also the condition number, increases quadratically with the beam slenderness ratio $\zeta$. Furthermore, it is expected that a pure neglect of shear modes is not sufficient in order to improve conditioning. Thus, the supplementation of the proposed Kirchhoff-Love formulations by an additional inextensibility constraint seems to be beneficial in order to get also rid of the axial stiffness contributions and to improve also the performance of linear solvers. However, the formulation of an additional inextensibility constraint does not lie in the focus of this contribution (see also the remark at the end of this section).\\

\textbf{3: Improved performance of nonlinear solvers}: The relation between the performance of a nonlinear solver, e.g. of a Newton-Raphson scheme, and the conditioning of the considered problem, e.g. measured via the condition number of the tangent stiffness matrix, is not that clear as in the case of linear solvers. Nevertheless, typically, it is at least expected that the performance of tangent-based nonlinear solvers also deteriorates for ill-conditioned problems showing slope differences of the target function by several orders of magnitude when stepping in different directions (e.g. in directions that activate shear and axial tension modes or in directions that activate bending and torsional modes). Interestingly, all numerical examples investigated in this work will confirm the trend that the nonlinear solver performance of the considered Reissner type beam elements deteriorates drastically with increasing slenderness ratio while the total number of Newton iterations required by the Kirchhoff type formulations remains almost unchanged.\\

\textbf{4: Reduced system size}: Kirchhoff type beam element formulations do not require any degrees of freedom for representing the mode of shear deformation. It can be expected that, at least as long as no convergence deteriorating phenomena such as locking occur, the same polynomial approximation and the same discretization error level can be guaranteed with fewer degrees of freedom. This prediction will be confirmed by the numerical examples in Section~\ref{sec:elenumericalexamples}.\\

\textbf{5: Smooth geometry representation: } The proposed Kirchhoff-Love beam elements will be based on $C^1$-continuous centerline interpolations. These interpolations will eventually result in smooth beam-to-beam contact kinematics, a property that is highly desirable in order to yield efficient and robust contact algorithms (see e.g.~\cite{meier2015b,meier2015c}).\\

\textbf{6: Abstaining from algorithmic treatment of large rotations}: It has already been mentioned earlier in this work that the proposed Kirchhoff-Love beam formulations provide an ideal basis for the derivation of reduced beam models which are valid under certain restrictions concerning external loads and initial geometry. For example in~\cite{meier2015}, a torsion-free beam element formulation could be derived (and extended to dynamic problems in~\cite{meier2015b}) that is based on a pure centerline representation and can consequently avoid the treatment of large rotations and associated degrees of freedom. Thus, many steps within a nonlinear finite element algorithm that are typically complicated by the presence of large rotations (e.g. spatial interpolation, time discretization, non-symmetric tangent stiffness matrix, non-constant and non-symmetric mass matrix, incremental and iterative configuration updates) are comparable to those of standard solid finite elements for this torsion-free formulation. For further details, the reader is referred to~\cite{meier2015,meier2015b}.\\

These sources of potential benefits were the motivation for the development of shear-free element formulations based on the Kirchhoff-Love theory of thin beams. Different realizations of such element formulations, e.g. based on a weak or on a strong enforcement of the Kirchhoff constraint, will be presented in the next sections. The influence of the aforementioned aspects on the resulting numerical behavior will be verified in Section~\ref{sec:elenumericalexamples} via proper test cases.\\

\hspace{0.2 cm}
\begin{minipage}{15.0 cm}
\textbf{Remark:} As discussed above, the performance of iterative linear solvers could be improved by supplementing the proposed Kirchhoff type element formulations by an additional inextensibility constraint. Unfortunately, in contrast to the Kirchhoff constraint, there is no straightforward way to enforce the inextensibility constraint directly through a special choice of the primary variables or by a collocation approach that would allow for Lagrange multiplier elimination on element level, as long as the interpolation property $\mb{r}(x^i)=\hat{\mb{d}^i}$ with $i=1,2$ has to be fulfilled for the position vector field $\mb{r}(s)$ at the element boundary nodes. This statement can easily be illustrated by considering a straight beam element of arbitrary order. In order to avoid zero-energy modes, inextensibility means in such a case that the solution for the nodal position vectors at the two boundary nodes cannot arise independently, but have to fulfill a constraint (e.g. $||\mb{\hat{\mb{d}^1}}\!-\!\mb{\hat{\mb{d}^2}}||\dot{=}l_{ele}$).
\end{minipage}

\section{Kirchhoff-Love Beam Element Based on Strong Constraint Enforcement}
\label{sec:elestrongkirchhoff}

In this section, a finite element formulation based on a strong enforcement of the Kirchhoff constraint is presented. In Section~\ref{sec:elestrongkirchhoff_variant2}, a variant based on nodal triads parametrized via tangent vectors according to Section~\ref{sec:spatialdiscretization_nodaltriadsrotation_smallestrotation} is investigated. In Section~\ref{sec:elestrongkirchhoff_variant1}, the transition to a rotation vector-based parametrization as in Section~\ref{sec:spatialdiscretization_nodaltriads_rotationvector} is conducted. Similar to the CJ element presented in the previous section, also for this element formulation, the avoidance of possible membrane locking effects as well as the fulfillment of mechanical conservation properties will be verified in Sections~\ref{sec:elestrongkirchhoff_spatialdiscretization_locking} and~\ref{sec:elestrongkirchhoff_spatialdiscretization_conservation}. In~\ref{anhang:elestrongkirchhoff_dirichletconditions} the suitability of the tangent-based and rotation vector-based nodal triad parametrizations for the modeling of practically relevant Dirichlet boundary conditions and joints is evaluated.

\subsection{Residual vector of tangent-based parametrization}
\label{sec:elestrongkirchhoff_variant2}

Similar to the Simo-Reissner case, the trial and weighting functions are replaced by their discrete counterparts taken from the finite-dimensional trial subspace $(\mb{r}_h,\varphi_h) \in \mathcal{U}_h \subset \mathcal{U}$ and the weighting subspace $(\delta \mb{r}_h, \delta \Theta_{1h}) \in \mathcal{V}_h \subset \mathcal{V}$. In the following, $3$-noded elements with the nodal primary variables $\hat{\mbd{x}}_{TAN}\!:=\!(\hat{\mb{d}}^{1T}\!,\hat{\mb{t}}^{1T}\!,\hat{\varphi}^1, \hat{\mb{d}}^{2T}\!,\hat{\mb{t}}^{2T}\!,\hat{\varphi}^2, \hat{\varphi}^3)^T$ as well as the set of nodal primary variable variations $\delta \hat{\mbd{x}}_{TAN}\!:=\!(\delta \hat{\mb{d}}^{1T}\!, \delta \hat{\mb{t}}^{1T}\!, \delta \hat{\Theta}_{1}^1, \delta \hat{\mb{d}}^{2T}\!,\delta \hat{\mb{t}}^{2T}\!,\delta \hat{\Theta}_{1}^2, \delta \hat{\Theta}_{1}^3)^T$ are considered (see also Figure~\ref{fig:nodenum2}). The centerline interpolation is based on Hermite polynomials according to Section~\ref{sec:spatialdiscretization_hermite} and completely defined by the two element boundary nodes. The rotation field interpolation follows equation~\eqref{triadsr_triadfield}. Concretely, a quadratic rotation interpolation based on three nodes, thus also involving the element center node, is considered. Since the orientation of the material triad field is determined by the relative angle $\varphi_h(\xi)$ as well as the tangent vector field $\mb{r}_h^{\prime}(\xi)$, with the latter being a
polynomial of order two, this second-order interpolation is sufficient for triad field discretization. In Meier et al.~\cite{meier2014}, it has been confirmed that a higher interpolation order $n_{\Lambda}\!>\!3$ will not further improve the approximation quality while a lower interpolation order $n_{\Lambda}\!<\!3$ will lead to a decline in the convergence rate. The time integration of Section~\ref{sec:temporaldiscretization} is employed, thus leading to the inertia forces $\mb{f}_{\rho}$ and moments $\mb{m}_{\rho}$ given in~\eqref{simoreissner_inertia}. Inserting these discretizations into equation~\eqref{weakform3} and taking advantage of the spin vector interpolation given by~\eqref{kirchhoffkinematics_spinvector_variant2} eventually yields the element residual vector of the Bubnov-Galerkin variant of this element formulation:
\begin{align}
\label{elestrongkirchhoff_residual}
\begin{split}
 \!\!\!\!\!\! & \mbd{r}_{SK-TAN+CS,\hat{\mb{d}}} \!=\!\! \int \limits_{-1}^{1} \!\! \left( \left[ \mbd{v}_{\theta_{\perp}}^{\prime} \!\!+\! \mbd{v}_{\theta_{\parallel d}}^{\prime} \right] \! \mb{m} \!+\! \mbd{v}_{\epsilon} F_1 \!-\! \mbd{H}^T \tilde{\mb{f}}_{\rho} \!-\! \left[ \mbd{v}_{\theta_{\perp}} \!\!+\! \mbd{v}_{\theta_{\parallel d}} \right] \!\! \tilde{\mb{m}}_{\rho} \right) \! J(\xi) d \xi \\
  & \hspace{2.7cm}-\! \Bigg[\mbd{H}^T \mb{f}_{\sigma} \!+\! \left[ \mbd{v}_{\theta_{\perp}} \!\!+\! \mbd{v}_{\theta_{\parallel d}} \right] \mb{m}_{\sigma} \Bigg]_{\varGamma_{\sigma}}\!\!, \!\!\!\!\! \\ \!\!\!\!\!\!
& \mbd{r}_{SK-TAN+CS,\mb{\hat{\Theta}}_1} \!=\!\!\int \limits_{-1}^{1} \!\! \left( \mbd{v}_{\theta_{\parallel \Theta}}^{\prime} \mb{m}
-\mbd{v}_{\theta_{\parallel \Theta}} \tilde{\mb{m}}_{\rho} \right) J(\xi) d \xi - \Bigg[ \mbd{v}_{\theta_{\parallel \Theta}} \mb{m}_{\sigma} \Bigg]_{\varGamma_{\sigma}}\!\!\!\!\!\! \quad \text{with} \quad  
\mbd{v}_{\epsilon}\!=\! \frac{\mbd{H}^{\prime T} \mb{t}}{||\mb{t}||}.\!\!\!\!\!\!
\end{split}
\end{align}
The Bubnov-Galerkin formulation~\eqref{elestrongkirchhoff_residual} can be transformed into a Petrov-Galerkin variant based on the spin interpolation scheme~\eqref{kirchhoffkinematics_spinvector_variant1} by simply omitting the terms $\mb{v}_{\theta_{\parallel d}}$, which yields:
\begin{align}
\label{elestrongkirchhoff_residual_petrov}
\begin{split}
 \!\!\!\!\!\! & \mbd{r}_{SK-TAN,\hat{\mb{d}}} \!=\!\! \int \limits_{-1}^{1} \!\! \left( \mbd{v}_{\theta_{\perp}}^{\prime} \! \mb{m} \!+\! \mbd{v}_{\epsilon} F_1 \!-\! \mbd{H}^T \tilde{\mb{f}}_{\rho} \!-\! \mbd{v}_{\theta_{\perp}} \! \tilde{\mb{m}}_{\rho} \right) \! J(\xi) d \xi
 \!-\! \Bigg[\mbd{H}^T \mb{f}_{\sigma} \!+\! \mbd{v}_{\theta_{\perp}} \mb{m}_{\sigma} \Bigg]_{\varGamma_{\sigma}}\!\!, \!\!\!\!\! \\ \!\!\!\!\!\!
& \mbd{r}_{SK-TAN,\mb{\hat{\Theta}}_1} \!=\!\!\int \limits_{-1}^{1} \!\! \left( \mbd{v}_{\theta_{\parallel \Theta}}^{\prime} \mb{m}
-\mbd{v}_{\theta_{\parallel \Theta}} \tilde{\mb{m}}_{\rho} \right) J(\xi) d \xi - \Bigg[ \mbd{v}_{\theta_{\parallel \Theta}} \mb{m}_{\sigma} \Bigg]_{\varGamma_{\sigma}}\!\!\!\!\!\! \quad \text{with} \quad  
\mbd{v}_{\epsilon}\!=\! \frac{\mbd{H}^{\prime T} \mb{t}}{||\mb{t}||}.\!\!\!\!\!\!
\end{split}
\end{align}
In order to avoid membrane locking in the range of high element slenderness ratios, the following re-interpolation of the axial tension $\epsilon$ and its variation $\delta \epsilon$ based on the MCS procedure~\eqref{MCSdef1} (see also~\cite{meier2015}) is applied: 
\begin{align}
\label{MCS}
\begin{split}
\bar{F}_{1}\!&=\!EA\bar{\epsilon}, 
\,\, \bar{\epsilon}(\xi)\!=\!\!\sum_{i=1}^{3} L^i(\xi)\epsilon(\xi^i),
\,\, \delta \bar{\epsilon}(\xi)\!=\!\!\sum_{i=1}^{3} L^i(\xi) \delta \epsilon(\xi^i),\\
\epsilon(\xi^i)&=\left(||\mb{r}^{\prime}||-1\right)_{(\xi^i)}, \,\,
\delta \epsilon(\xi^i)= \left(\! \delta \mb{r}^{\prime T} \frac{\mb{r}^{\prime}}{||\mb{r}^{\prime}||}\right)_{\!\!(\xi^i)}\!\!, \,\,
\bar{\mbd{v}}_{\epsilon}\!=\!\!\sum_{i=1}^{3} L^i(\xi) \mbd{v}_{\epsilon}(\xi^i).
\end{split}
\end{align}
Thus, a locking-free finite element formulation can be obtained by simply replacing the axial force $F_1$ by $\bar{F}_{1}$ and the discrete axial tension variation operator $\mbd{v}_{\epsilon}$ by $\bar{\mbd{v}}_{\epsilon}$ in the discrete weak form~\eqref{elestrongkirchhoff_residual}. The discrete expression for the internal energy associated with the modified axial tension $\bar{\epsilon}$ eventually reads:
\begin{align}
\label{MCS_axialstrainenergy}
\tilde{\Pi}_{int,\bar{\epsilon}}=\frac{1}{2}EA \bar{\epsilon}^2.
\end{align}
The element formulation based on the degrees of freedom $\hat{\mbd{x}}_{TAN}$ and $\delta \hat{\mbd{x}}_{TAN}$, the residual~\eqref{elestrongkirchhoff_residual_petrov} and~\eqref{MCS} will be denoted as SK-TAN element, which stands for "Strong Kirchhoff constraint enforcement combined with nodal triad parametrization via nodal TANgents". Correspondingly,  the combination of the degrees of freedom $\hat{\mbd{x}}_{TAN}$ and $\delta \hat{\mbd{x}}_{TAN}$, the residual~\eqref{elestrongkirchhoff_residual} and~\eqref{MCS} will be denoted as SK-TAN+CS element, referring to the Consistent Spin vector interpolation underlying the Bubnov-Galerkin variant. It has to be emphasized that the replacement of the original axial tension terms with the corresponding MCS terms according to~\eqref{MCS} is standard for \textit{all} Kirchhoff type beam element formulations considered in this work. Only in examples where for comparison reasons also variants without MCS method are considered, an additional abbreviation ...-MCS, e.g. SK-TAN-MCS, is employed. For the SK-TAN element, a linearization $\Delta \mbd{r}_{SK-TAN}\!=\!\mbd{k}_{SK-TAN} \Delta \hat{\mbd{x}}_{TAN}$ based on the increment vector $\Delta \hat{\mbd{x}}_{TAN}\!:=\!(\Delta \hat{\mb{d}}^{1T}\!, \Delta \hat{\mb{t}}^{1T}\!, \Delta \hat{\varphi}^1, \Delta \hat{\mb{d}}^{2T}\!,\Delta \hat{\mb{t}}^{2T}\!,\Delta \hat{\varphi}^2, \Delta \hat{\varphi}^3)^T$ will be employed (see~\ref{anhang:linsktan}). In contrary to the multiplicative rotation variations $\delta \hat{\Theta}_{1}^i$ occurring in $\delta \mbd{r}_{SK-TAN}$, the quantities $\Delta \hat{\varphi}^i$ represent additive rotation increments of the nodal relative angles $\hat{\varphi}^i$.

\subsection{Residual vector of rotation vector-based parametrization}
\label{sec:elestrongkirchhoff_variant1}

In some scenarios, e.g. applications where complex rotational Dirichlet or coupling conditions should be prescribed at the element boundary, it can be beneficial to employ the alternative parametrization of the triads $\boldsymbol{\Lambda}^1$ and $\boldsymbol{\Lambda}^2$ at the element boundary nodes via rotation vectors according to Section~\ref{sec:spatialdiscretization_nodaltriads_rotationvector}. In such a case, an alternative set of nodal primary variables given by $\hat{\mbd{x}}_{ROT}\!:=\!(\hat{\mb{d}}^{1T}\!,\hat{\boldsymbol{\psi}}^{1T}\!,\hat{t}^1, \hat{\mb{d}}^{2T}\!,\hat{\boldsymbol{\psi}}^{2T}\!,\hat{t}^2, \hat{\varphi}^3)^T$ and $\delta \hat{\mbd{x}}_{ROT}\!:=\!(\delta \hat{\mb{d}}^{1T}\!, \delta \hat{\boldsymbol{\theta}}^{1T}\!, \delta \hat{t}^1, \delta \hat{\mb{d}}^{2T}\!,\delta \hat{\boldsymbol{\theta}}^{2T}\!,\delta \hat{t}^{2}, \delta \hat{\Theta}_{1}^3)^T$ can be employed. Here, $\hat{\boldsymbol{\psi}}^1$ and $\hat{\boldsymbol{\psi}}^2$ represent the rotation vectors associated with these boundary triads, $\delta \hat{\boldsymbol{\theta}}^1$ and $\delta \hat{\boldsymbol{\theta}}^2$ are the corresponding spin vectors and $\hat{t}^1$ and $\hat{t}^2$ represent the magnitudes of the nodal tangents, i.e. $\hat{t}^1\!:=\!||\mb{t}^1||$ and $\hat{t}^2\!:=\!||\mb{t}^2||$, as introduced in Section~\ref{sec:largerotations_sr}. In this case, the nodal tangents of the Hermite interpolation~\eqref{hermite_interpolation} are no primary variables anymore, but have to be expressed by $\hat{\boldsymbol{\psi}}^1$ and $\hat{t}^1$ as well as $\hat{\boldsymbol{\psi}}^2$ and $\hat{t}^2$ (see also the transformation rule~\eqref{largerotations_transformation_tandpsi}):
 \begin{align}
\label{elestrongkirchhoff_variant1_tandpsi}
 \hat{\mb{t}}^i \rightarrow \mb{t}^i = \hat{t}^i \exp{ \! ( \mb{S}( \hat{\boldsymbol{\psi}}^i) )} \mb{E}_1 \quad \text{for} \quad i=1,2.
\end{align}
The transformation between the variations $(\delta \hat{\boldsymbol{\theta}}^i, \delta \hat{t}^i)$ as well as $(\delta \hat{\mb{t}}^i, \delta \hat{\Theta}_{1}^i)$ for $i=1,2$ is given by the transformation matrices $\tilde{\mb{T}}$ and $\tilde{\mb{T}}^{-1}$ according to~\eqref{largerotations_transformation_tandpsi} and leads to the following relation:
\begin{align}
\label{elestrongkirchhoff_trafovariations}
\delta \hat{\mbd{x}}_{TAN}\!=\!\tilde{\mbd{T}}_{\hat{\mbd{x}}} \delta \hat{\mbd{x}}_{ROT} \quad \text{with} \quad
\tilde{\mbd{T}}_{\hat{\mb{x}}} \!=\!
   \left(
   \begin{array}{ccccc}
   \mb{I}_3 &&&&\\
   &\tilde{\mb{T}}^1&&&\\
   &&\mb{I}_3&&\\
   &&&\tilde{\mb{T}}^2&\\
   &&&&1
   \end{array}
   \right),
\end{align}
where all blank entries are zero. The required transformation matrices $\tilde{\mb{T}}^1$ as well as $\tilde{\mb{T}}^2$ follow from equation~\eqref{largerotations_transformation_tandpsi} as:
\begin{align}
\label{elestrongkirchhoff_trafovariations2}
   \tilde{\mb{T}}^1\!=\!
   \left(
   \begin{array}{cc}
   \!\!\!-\hat{t}^1 \mb{S}(\mb{g}_1^1) & \mb{g}_{1}^1\!\!\\
   \!\mb{g}_{1}^{1T} & 0\!\!
   \end{array}
   \right)\!, \,\,\,\,
   \tilde{\mb{T}}^2\!=\!
   \left(
   \begin{array}{cc}
   \!\!\!-\hat{t}^2 \mb{S}(\mb{g}_1^2) & \mb{g}_{1}^2\!\!\\
   \!\mb{g}_{1}^{2T} & 0\!\!
   \end{array}
   \right) \quad \text{with} \quad
   \mb{g}_{1}^i\!=\! \exp{ \! ( \mb{S}( \hat{\boldsymbol{\psi}}^i) )} \mb{E}_1.
\end{align}
In order to simplify the transformation between the sets of degrees of freedom considered in Sections~\ref{sec:elestrongkirchhoff_variant2} and~\ref{sec:elestrongkirchhoff_variant1}, the residual $\mbd{r}_{TAN,\hat{\mb{d}}}$ and $\mbd{r}_{TAN,\hat{\mb{\Theta}}_1}$ according to~\eqref{elestrongkirchhoff_residual} are slightly reordered and the sought-after residual $\mbd{r}_{ROT}$ is introduced:
 \begin{align}
 \begin{split}
\label{elestrongkirchhoff_residualreordering}
\mbd{r}_{TAN,\hat{\mb{d}}}&:=(\mb{r}_{TAN,\hat{\mb{d}}^{1}}^T,\mb{r}_{TAN,\hat{\mb{t}}^{1}}^T,\mb{r}_{TAN,\hat{\mb{d}}^{2}}^T,\mb{r}_{TAN,\hat{\mb{t}}^{2}}^T)^T,\\ 
\mbd{r}_{TAN,\mb{\hat{\Theta}}_1}&:=(r_{TAN,\hat{\Theta}_1^{1}},r_{TAN,\hat{\Theta}_1^{2}},r_{TAN,\hat{\Theta}_1^{3}})^T, \\
\mbd{r}_{TAN}&:=(\mb{r}_{TAN,\hat{\mb{d}}^{1}}^T,\mb{r}_{TAN,\hat{\mb{t}}^{1}}^T,r_{TAN,\hat{\Theta}_1^{1}}, \mb{r}_{TAN,\hat{\mb{d}}^{2}}^T,\mb{r}_{TAN,\hat{\mb{t}}^{2}}^T,r_{TAN,\hat{\Theta}_1^{2}},r_{TAN,\hat{\Theta}_1^{3}})^T, \\
\mbd{r}_{ROT}&:=(\mb{r}_{ROT,\hat{\mb{d}}^{1}}^T,\mb{r}_{ROT,\hat{\boldsymbol{\theta}}^{1}}^T,r_{ROT,\hat{t}^{1}}, \mb{r}_{ROT,\hat{\mb{d}}^{2}}^T,\mb{r}_{ROT,\hat{\boldsymbol{\theta}^{2}}}^T,r_{ROT,\hat{t}^{2}},r_{ROT,\hat{\Theta}_1^{3}})^T.
\end{split}
\end{align}
Inserting relation~\eqref{elestrongkirchhoff_trafovariations} into the virtual work contribution resulting from one beam element yields:
\begin{align}
\label{elestrongkirchhoff_trafovariations3}
\delta \hat{\mbd{x}}_{TAN}^T \mbd{r}_{TAN} = \delta \hat{\mbd{x}}_{ROT}^T \tilde{\mbd{T}}_{\hat{\mb{x}}}^T \mbd{r}_{TAN} \,\dot{=}\, \delta \hat{\mbd{x}}_{ROT}^T \mbd{r}_{ROT} \quad \rightarrow \quad \mbd{r}_{ROT} = \tilde{\mbd{T}}_{\hat{\mb{x}}}^T \mbd{r}_{TAN}.
\end{align}
According to~\eqref{elestrongkirchhoff_trafovariations2} and~\eqref{largerotations_tildetinvmatrixsr}, the matrix $\tilde{\mb{T}}$ and its inverse $\tilde{\mb{T}}^{-1}$ are well-defined as long as $\hat{t}^1 \neq 0$ and $\hat{t}^2 \neq 0$. The physical interpretation of $\hat{t}^i = 0$ is that an arc-segment on the beam centerline at the position of the node $i$ with initial length $ds$ would be compressed to a length of zero. Since such a scenario is impossible from a physical point of view, these requirements are assumed to be fulfilled. Consequently, the transformation from the residual vector $\mbd{r}_{TAN}$ to the residual vector $\mbd{r}_{ROT}$ is based on a non-singular matrix $\tilde{\mbd{T}}_{\hat{\mb{x}}}$. The same statement holds for the transformation of the global residual vectors $\mbd{R}_{TAN}$ and $\mbd{R}_{ROT}$ via the matrix $\tilde{\mbd{T}}_{\mbd{X}}$ which represents an assembly of the element matrices $\tilde{\mbd{T}}_{\hat{\mbd{x}}}$. Based on these considerations, the following relation can be established:
\begin{align}
\label{elestrongkirchhoff_trafovariations4}
\mbd{R}_{ROT} \!=\! \tilde{\mbd{T}}_{\mbd{X}}^T \mbd{R}_{TAN}, \,\,
\tilde{\mbd{T}}_{\mbd{X}} \!\in\! \Re^{n_{X}} \!\!\times \! \Re^{n_{X}}, \,\, \text{rank}(\tilde{\mbd{T}}_{\mb{X}})\!=\!n_{X}\quad \rightarrow \quad 
\mbd{R}_{TAN} \!=\! \mb{0} \Leftrightarrow \mbd{R}_{ROT} \!=\! \mb{0}.
\end{align}
As long as a unique solution of $\mbd{R}_{TAN}\!=\!\mb{0}$ exists, the solution of $\mbd{R}_{ROT} \!=\! \mb{0}$ will lead to the same mechanical equilibrium configuration. In other words, the pure re-parametrization performed from Section~\ref{sec:elestrongkirchhoff_variant2} to Section~\ref{sec:elestrongkirchhoff_variant1} will not change the results of the discretized beam problem. Nevertheless, the transformation matrix $\tilde{\mbd{T}}_{\mb{X}}$ depends on the primary degrees of freedom in a nonlinear manner and has to be considered within a consistent linearization procedure.\\

Throughout this work, the element formulation based on the degrees of freedom $\hat{\mbd{x}}_{ROT}$ and $\delta \hat{\mbd{x}}_{ROT}$, the residual~\eqref{elestrongkirchhoff_residual_petrov} transformed via~\eqref{elestrongkirchhoff_trafovariations3} together with the MCS approach~\eqref{MCS} will be denoted as SK-ROT element, which stands for "Strong Kirchhoff constraint enforcement combined with nodal triad parametrization via nodal ROTation vectors". Correspondingly,  the combination of the degrees of freedom $\hat{\mbd{x}}_{ROT}$ and $\delta \hat{\mbd{x}}_{ROT}$, the residual~\eqref{elestrongkirchhoff_residual_petrov} transformed via~\eqref{elestrongkirchhoff_trafovariations3} together with the MCS approach~\eqref{MCS} will be denoted as SK-ROT+CS element. Since the SK-ROT and the SK-TAN element yield the same finite element solution (expressed via different nodal primary variables), the following theoretical investigations concerning locking behavior and conservation properties will only be performed for the SK-TAN element. For the SK-ROT element, a linearization $\Delta \mbd{r}_{SK-ROT}\!=\!\mbd{k}_{SK-ROT} \Delta \hat{\mbd{x}}_{ROT}$ based on the increment vector $\Delta \hat{\mbd{x}}_{ROT}\!:=\!(\Delta \hat{\mb{d}}^{1T}\!, \Delta \hat{\boldsymbol{\theta}}^{1T}\!, \Delta \hat{t}^1, \Delta \hat{\mb{d}}^{2T}\!,\Delta \hat{\boldsymbol{\theta}}^{2T}\!,\Delta \hat{t}^{2}, \Delta \hat{\varphi}^3)^T$ will be employed (see~\ref{anhang:linskwkrot}). Here, $\Delta \hat{\boldsymbol{\theta}}^{1}$ and $\Delta \hat{\boldsymbol{\theta}}^{2}$ represent multiplicative increments while $\Delta \hat{\varphi}^3$ can be identified as an additive increment.

%
\subsection{Avoidance of locking effects}
\label{sec:elestrongkirchhoff_spatialdiscretization_locking}
%

In order to investigate the locking behavior of the SK-TAN element, the investigations already made in Section~\ref{sec:spatialdiscretization_locking} only have to be extended from 2D to the general 3D case. In 3D, the Kirchhoff beam problem is described by $n_{eq}\!=\!4$ differential equations~\eqref{condensedequilibrium} and constrained by $n_{eq,c}\!=\!1$ constraint equation in case a pure bending state shall be represented. Thus, the constraint ratio of the space-continuous problem yields $r\!=\!n_{eq}/n_{eq,c}\!=\!4$. Due to the employed MCS method, the discrete number of constraint equations takes on a value of $n_{eq,c}\!=\!2n_{ele}\!+\!1$. Given the total number of equations $n_{eq}\!=\!8n_{ele}\!+\!1$ after application of proper Dirichlet boundary conditions, the discrete constraint ratio yields:
\begin{align}
\label{elestrongkirchhoff_constraintratio}
r_h=\lim_{n_{ele} \rightarrow \infty}  \frac{8n_{ele}\!+\!1}{2n_{ele}\!+\!1} = 4 = r.
\end{align}
Relation~\eqref{elestrongkirchhoff_constraintratio} yields the optimal constraint ratio for this element formulation. Consequently, no locking effects are expected. Furthermore, it has been shown in Section~\ref{sec:spatialdiscretization_locking} that the requirement of representing a straight beam configuration with arbitrary distribution of $\bar{\epsilon}_h(\xi)$ yields a number of independent equations that equals the number of degrees of freedom. Consequently, such a state can be represented exactly and no zero-energy modes associated with this state have to be expected. The extension of this statement to 3D is straightforward and will not be further investigated here. Moreover, it has also been stated in Section~\ref{sec:spatialdiscretization_locking} that the discrete hyperelastic energy $\Pi_{int,h}$ associated with a pure 2D bending case cannot be displayed exactly by the SK-TAN element. Of course, this statement still holds in 3D. In Sections~\ref{sec:examples_purebending2d} and~\ref{sec:examples_purebending3d}, the expected result that the discrete hyperelastic energies associated with pure bending states in 2D and in 3D cannot be represented in an exact manner by this beam element formulation will be verified by means of corresponding numerical test cases. There, it will also be shown that this property leads to a slightly increased discretization error level as compared to the subsequently derived WK-TAN element. However, this observation is independent from the element slenderness ratio and cannot be attributed to membrane locking.

%
\subsection{Conservation properties}
\label{sec:elestrongkirchhoff_spatialdiscretization_conservation}
%

Also for the proposed SK-TAN beam element with strong enforcement of the Kirchhoff constraint, it shall be investigated if the variational fields~\eqref{rigid_translation}, \eqref{rigid_rotation} and \eqref{conservation_velocity} required for conservation of linear momentum, conservation of angular momentum and conservation of energy can be represented by the corresponding discrete weighting subspace $\mathcal{V}_h$. The representation of a rigid body translation~\eqref{rigid_translation} is given by the nodal primary variable variations
\begin{align}
\label{elestrongkirchhoff_rigid_translation}
\delta \hat{\mb{d}}^j = \mb{u}_0, \quad \delta \hat{\mb{t}}^j = \mb{0}, \quad \delta \hat{\Theta}_1^i=0 \,\,\, \text{for} \,\,\, j=1,2; \,\,\, \text{and} \,\,\, i=1,2,3.
\end{align}
This result can be verified by inserting the choices for $\delta \hat{\mb{d}}^j$ and $\delta \hat{\mb{t}}^j$ made in~\eqref{elestrongkirchhoff_rigid_translation} into the Hermite interpolation~\eqref{testfunctions} and making use of the completeness conditions underlying the Hermite polynomials (see~\cite{meier2014}), which yields $\delta \mb{r}_h(\xi)\!=\!\mb{u}_0\!=$ const. as well as $\delta \mb{r}_h^{\prime}(\xi)\!=\!\delta \mb{g}_{1h}^{\prime}(\xi)\!=\!\mb{0}$. Inserting these relations together with $\delta \hat{\Theta}_1^i\!=\!0$ into either~\eqref{kirchhoffkinematics_spinvector_variant1} or~\eqref{kirchhoffkinematics_spinvector_variant2} results in the required vanishing of the discrete spin vector field. Thus, both the Petrov-Galerkin as well as the Bubnov-Galerkin variant for the spin vector interpolation lead to an exact conservation of the linear momentum. Next, a rigid body rotation~\eqref{rigid_rotation} has to be displayed by the following choice of nodal primary variable variations
\begin{align}
\label{elestrongkirchhoff_rigid_rotation2}
\delta \hat{\mb{d}}^j = \mb{w}_0 \times \hat{\mb{d}}^j, \quad 
\delta \hat{\mb{t}}^j = \mb{w}_0 \times \hat{\mb{t}}^j, \quad
\delta \hat{\Theta}_1^i=\mb{g}_1^T(\xi^i) \mb{w}_0 \,\,\, \text{for} \,\,\, j=1,2; \,\,\, \text{and} \,\,\, i=1,2,3.
\end{align}
Inserting $\delta \hat{\mb{d}}^j$ and $\delta \hat{\mb{t}}^j$ according to~\eqref{relesimoreissner_igid_rotation} into~\eqref{testfunctions} gives the desired result $\delta \mb{r}_h(\xi)\!=\!\mb{w}_0\!\times\!\mb{r}_h(\xi)$. In Section~\ref{sec:spatialdiscretization_triadsr} (see~\eqref{elestrongkirchhoff_rigid_rotation}-\eqref{elestrongkirchhoff_deltatheta1diff}), it has been shown that based on the nodal values~\eqref{elestrongkirchhoff_rigid_rotation2} the Bubnov-Galerkin interpolation~\eqref{kirchhoffkinematics_spinvector_variant2} can exactly represent such a "virtual" rigid body rotation, while the Petrov-Galerkin variant~\eqref{kirchhoffkinematics_spinvector_variant1} cannot. Thus, only the Bubnov-Galerkin interpolation can guarantee for exact conservation of angular momentum. This result will be confirmed by subsequent numerical examples. Finally, the conservation of energy has to be investigated. If the nodal velocities and angular velocities of the time-continuous problem are chosen as nodal primary variable variations,
\begin{align}
\label{elestrongkirchhoff_igid_rotation}
\delta \hat{\mb{d}}^j =  \dot{\mb{d}}^j, \quad
\delta \hat{\mb{t}}^j =  \dot{\mb{t}}^j, \quad
\delta \hat{\Theta}_1^i = \mb{g}_1^T(\xi^i) \mb{w}^i \,\,\, \text{for} \,\,\, j=1,2; \,\,\, \text{and} \,\,\, i=1,2,3,
\end{align}
again, only the Bubnov-Galerkin variant~\eqref{kirchhoffkinematics_spinvector_variant2} leads (per definition) to an exact representation of the rates of the discrete internal and kinetic energy and to exact energy conservation for the spatially discretized, time-continuous problem. In contrast, the Petrov-Galerkin variant~\eqref{kirchhoffkinematics_spinvector_variant1} is not variationally consistent with the triad interpolation~\eqref{triadsr_triadfield} underlying the discrete energies and cannot guarantee for energy conservation of the time-continuous problem.

%
\section{Kirchhoff-Love Beam Element Based on Weak Constraint Enforcement}
\label{sec:eleweakkirchhoff}
%

As alternative to the formulation presented in the last section, a beam element will now be presented that is based on the weak fulfillment of the Kirchhoff constraint. Thus, the basis for the intended element formulation is provided by the Simo-Reissner beam theory. In a first step, a finite element formulation of Simo-Reissner type with a $C^1$-continuous centerline representation will be derived in Section~\ref{sec:eleweakkirchhoff_reissner}. Afterwards, the Kirchhoff constraint of vanishing shear strains will be enforced in order to end up with a finite element formulation of Kirchhoff type. Following the derivations in Section~\ref{sec:weakkirchhoff}, the weak statement of the Kirchhoff constraint~\eqref{weakkirchhoff_weakconstraintpotential} can be realized by introducing spatial interpolations for the Lagrange multipliers and their variations, i.e. by choosing a proper discrete trial space $(\lambda_{\Gamma,2,h},\lambda_{\Gamma,3,h}) \in \mathcal{U}_{\lambda_{\Gamma{23}},h} \in \mathcal{U}_{\lambda_{\Gamma{23}}}$ and a proper discrete weighting space $(\delta \lambda_{\Gamma,2,h},\delta \lambda_{\Gamma,3,h}) \in \mathcal{V}_{\lambda_{\Gamma{23}},h} \in \mathcal{V}_{\lambda_{\Gamma{23}}}$. The resulting nonlinear system of discrete equilibrium equations will contain discrete Lagrange multipliers as additional nodal primary variables and exhibit a saddle point type structure. In order to avoid the additional effort of solving a large system of equations with saddle point structure, a slightly different approach is chosen here. In the next section, a modified Reissner type beam element formulation will be presented, which is based on a smooth Hermite centerline interpolation and a MCS type strain re-interpolation not only of the axial strain $\epsilon_h(\xi)$ but also of the shear strains $\Gamma_{2,h}(\xi)$ and $\Gamma_{3,h}(\xi)$. Applying the constraint of vanishing shear strains in a consistent manner directly on the re-interpolated strain fields $\bar{\Gamma}_{2,h}(\xi)$ and $\bar{\Gamma}_{3,h}(\xi)$ yields a collocation point type approach of constraint enforcement which does not require additional Lagrange multipliers (see Section~\ref{sec:eleweakkirchhoff_variant2}). Throughout this contribution, this variant will be preferred since it does neither yield additional Lagrange multiplier degrees of freedom nor a saddle point type system of equations. Also for this element formulation, the two variants concerning nodal rotation parametrization according to Sections~\ref{sec:spatialdiscretization_nodaltriads_rotationvector} and~\ref{sec:spatialdiscretization_nodaltriadsrotation_smallestrotation} will be presented in the following Sections~\ref{sec:eleweakkirchhoff_variant2} and~\ref{sec:eleweakkirchhoff_variant1}.

\subsection{Basic formulation: Hermitian Simo-Reissner element}
\label{sec:eleweakkirchhoff_reissner}

The Reissner type beam element formulated in this section represents an intermediate step in the derivation of a corresponding Kirchhoff type beam element formulation in the next section. The discrete beam centerline representation is given by the Hermite interpolation~\eqref{hermite_interpolation} based on the position and tangent vectors $\hat{\mb{d}}^i$ and $\hat{\mb{t}}^i$ at the two element boundary nodes. Furthermore, the rotation interpolation is given by a three-noded representation of~\eqref{triadrelrot_interpolation} with nodal triads $\boldsymbol{\Lambda}^1$, $\boldsymbol{\Lambda}^2$ and $\boldsymbol{\Lambda}^3$. Again, a finite element formulation will be considered on the basis of a strain re-interpolation similar to the MCS method~\eqref{MCS}. While in the Kirchhoff case~\eqref{MCS}, only the axial strain has been treated, now the entire deformation measure $\boldsymbol{\Gamma}$ will be re-interpolated in order to avoid membrane as well as shear locking:
\begin{align}
\label{reinterpolation_Gamma}
\bar{\mb{\Gamma}}(\xi)=\sum_{i=1}^{3} L^i(\xi)\mb{\Gamma}(\xi^i),
\quad \delta \bar{\mb{\Gamma}}(\xi)=\sum_{i=1}^{3} L^i(\xi) \delta \mb{\Gamma}(\xi^i) \quad \xi^1=-1, \xi^2=1, \xi^3=0.
\end{align}
On the basis of~\eqref{reinterpolation_Gamma}, also the hyper-elastic stored energy function $\tilde{\Pi}_{int}(\mb{\Omega},\mb{\Gamma})$ given in~\eqref{storedenergyfunction} has to be replaced by $\tilde{\Pi}_{int}(\mb{\Omega},\bar{\mb{\Gamma}})$. Now, one can introduce the following set of degrees of freedom $\hat{\mbd{x}}_{HSR}\!:=\!(\hat{\mb{d}}^{1T}\!,\hat{\mb{t}}^{1T}\!,\hat{\boldsymbol{\psi}}^{1T}\!,\hat{\mb{d}}^{2T}\!,\hat{\mb{t}}^{2T}\!,\hat{\boldsymbol{\psi}}^{2T},\hat{\boldsymbol{\psi}}^{3T})^T$ as well as the associated variation vector $\delta \hat{\mbd{x}}_{HSR}\!:=\!(\delta \hat{\mb{d}}^{1T}\!,\delta \hat{\mb{t}}^{1T}\!,\delta \hat{\boldsymbol{\theta}}^{1T}\!,\delta \hat{\mb{d}}^{2T}\!,\delta \hat{\mb{t}}^{2T}\!,\delta \hat{\boldsymbol{\theta}}^{2T}\!,\delta \hat{\boldsymbol{\theta}}^{3T})^T$. Based on the weak form~\eqref{weakformspatial}, \eqref{reinterpolation_Gamma}, the definitions~\eqref{variationspatialandmaterialdeformationmeasures} and~\eqref{materialdeformationmeasures},~\eqref{hermite_interpolation} and~\eqref{triadrelrot_interpolation}, the element residual vector can be derived as follows:
\begin{align}
\label{eleweakkirchhoff_reissner_residual}
\begin{split}
 \mbd{r}_{HSR,\hat{\mb{d}}} &= \int \limits_{-1}^{1} \left( \mbd{v}_{\boldsymbol{\Gamma}_1} \bar{\mb{F}} - \mbd{L}^{\!T} \tilde{\mb{f}}_{\rho} \right) J(\xi) d \xi -\Big[\mbd{L}^{\!T} \mb{f}_{{\sigma}} \Big]_{\varGamma_{\sigma}}\\
\mbd{r}_{HSR,\hat{\boldsymbol{\theta}}} &= \int \limits_{-1}^{1} \left( \mbd{L}^{\! \prime T} \mb{m} + \mbd{v}_{\boldsymbol{\Gamma}_2}\bar{\mb{F}}
-\mbd{L}^{\!T} \tilde{\mb{m}}_{\rho} \right) J(\xi) d \xi - \Big[ \mbd{L}^{\!T} \mb{m}_{{\sigma}} \Big]_{\varGamma_{\sigma}}\!\!\!\!\!\!=\mb{0},\\
\bar{\mb{F}}& =\mb{C}_F\bar{\mb{\Gamma}}, \,\,\,\, \mbd{v}_{\boldsymbol{\Gamma}_1}\!=\!\! \sum_{i=1}^{3} L^i(\xi) \left(\mbd{H}^{\prime T} \! \boldsymbol{\Lambda}\right)_{\! (\xi^i)}, \,\,\,\,
\mbd{v}_{\boldsymbol{\Gamma}_2}\!=\!\! \sum_{i=1}^{3} L^i(\xi) \left(\mbd{L}^{T} \! \mb{S}(\mb{t}) \boldsymbol{\Lambda} \right)_{\! (\xi^i)}.
\end{split}
\end{align}
This element formulation could e.g. be applied to problems of thick beams with higher continuity requirements (e.g. beam contact). However, here, the formulation~\eqref{eleweakkirchhoff_reissner_residual} solely represents an intermediate step in the derivation of Kirchhoff beam elements with weak enforcement of the Kirchhoff constraint as performed in the next two sections.

\subsection{Residual vector of tangent-based parametrization}
\label{sec:eleweakkirchhoff_variant2}

Due to~\eqref{reinterpolation_Gamma}, the general weak constraint enforcement of Section~\eqref{sec:weakkirchhoff} can be simplified:
\begin{align}
\label{vanishing_shearstrains}
\bar{\Gamma}_2(\xi) \, \equiv \, \bar{\Gamma}_3(\xi) \, \equiv \, 0 \,\, \rightarrow \,\, \mb{r}^{\prime T}\!(\xi^i)\mb{g_2}(\xi^i) \, \dot{=} \, \mb{r}^{\prime T} \! (\xi^i) \mb{g_3} (\xi^i) \, \dot{=} \, 0 \quad \text{for} \quad i=1,2,3.
\end{align}
According to \eqref{vanishing_shearstrains}, the Kirchhoff constraint is exactly fulfilled at the three collocation points. In the following, a parametrization is chosen that directly fulfills these constraints without the need for additional Lagrange multipliers. Thereto, the same set of nodal degrees of freedom $\hat{\mbd{x}}_{TAN}\!=\!(\hat{\mb{d}}^{1T}\!,\hat{\mb{t}}^{1T}\!,\hat{\varphi}^1, \hat{\mb{d}}^{2T}\!,\hat{\mb{t}}^{2T}\!,\hat{\varphi}^2, \hat{\varphi}^3)^T$ as well as the set of nodal primary variable variations $\delta \hat{\mbd{x}}_{TAN}\!=\!(\delta \hat{\mb{d}}^{1T}\!, \delta \hat{\mb{t}}^{1T}\!, \delta \hat{\Theta}_{1}^1, \delta \hat{\mb{d}}^{2T}\!,\delta \hat{\mb{t}}^{2T}\!,\delta \hat{\Theta}_{1}^2, \delta \hat{\Theta}_{1}^3)^T$ as in Section~\ref{sec:elestrongkirchhoff_variant2} are chosen. In case of a Bubnov-Galerkin approach, the following discrete spin vector field results from the triad interpolation~\eqref{triadrelrot_interpolation} in combination with the Kirchhoff constraint according to~\eqref{vanishing_shearstrains} (see also~\eqref{crisfieldspininterpolation}):
\begin{align}
\label{eleweakkirchhoff_spinvector1}
\delta \boldsymbol{\theta}_h(\xi) \!=\!\! \sum_{i=1}^{3} \tilde{\mb{I}}^i(\xi) \delta \boldsymbol{\theta}^i \quad 
\text{with} \quad
\delta \boldsymbol{\theta}^i= \left(\! \delta \hat{\Theta}_1^i \mb{g}_1(\xi^i) + \frac{\mb{S}[\mb{r}^{\prime}(\xi^i)] \delta \mb{r}^{\prime}(\xi^i)}{||\mb{r}^{\prime}(\xi^i)||^2}\right).
\end{align}
Since the Kirchhoff constraint is exactly fulfilled at the three element nodes, the constrained variant~\eqref{kirchhoffkinematics_spinvector} of the nodal spin vectors $\delta \boldsymbol{\theta}^i$ has been combined with~\eqref{crisfieldspininterpolation}. Based on the alternative Petrov-Galerkin approach~\eqref{petrovspininterpolation}, the Kirchhoff constraint given by the relations~\eqref{vanishing_shearstrains} yields the following expression for the spin vector field:
\begin{align}
\label{eleweakkirchhoff_spinvector2}
\delta \boldsymbol{\theta}_h(\xi) \!=\!\! \sum_{i=1}^{3} L^i(\xi) \delta \boldsymbol{\theta}^i \quad 
\text{with} \quad
\delta \boldsymbol{\theta}^i= \left(\! \delta \hat{\Theta}_1^i \mb{g}_1(\xi^i) + \frac{\mb{S}[\mb{r}^{\prime}(\xi^i)] \delta \mb{r}^{\prime}(\xi^i)}{||\mb{r}^{\prime}(\xi^i)||^2}\right).
\end{align}
Similar to the CJ element, only the latter version~\eqref{eleweakkirchhoff_spinvector2} will be employed throughout this contribution. The final finite element residual vector resulting from the discretized counterparts of these fields reads:
\begin{align}
\label{eleweakkirchhoff_residual}
\begin{split}
  \mbd{r}_{WK-TAN,\hat{\mb{d}}}\!&=\!\!\int \limits_{-1}^{1} \! \left( \bar{\mbd{v}}_{\theta_{\perp}}^{\prime} \mb{m} + \bar{\mbd{v}}_{\epsilon} \bar{F}_{1} - \mbd{H}^T \tilde{\mb{f}}_{\rho} 
-\bar{\mbd{v}}_{\theta_{\perp}} \tilde{\mb{m}}_{\rho} \right) \! J(\xi) d \xi -\Bigg[\mbd{H}^T \mb{f}_{{\sigma}} \Bigg]_{\varGamma_{\sigma}} \!\!\!\!\!\! - \Bigg[ \bar{\mbd{v}}_{\theta_{\perp}} \mb{m}_{{\sigma}} \Bigg]_{\varGamma_{\sigma}}\!\!\!\!\!\!, \\
 \mbd{r}_{WK-TAN,\hat{\mb{\Theta}}_1}\!&=\!\!\int \limits_{-1}^{1} \! \left( \bar{\mbd{v}}_{\theta_{\parallel \Theta}}^{\prime} \mb{m}
-\bar{\mbd{v}}_{\theta_{\parallel \Theta}} \tilde{\mb{m}}_{\rho} \right) \! J(\xi) d \xi - \Bigg[ \bar{\mbd{v}}_{\theta_{\parallel \Theta}} \mb{m}_{{\sigma}} \Bigg]_{\varGamma_{\sigma}}\!\!\!\!\!\!, \\ 
\bar{\mbd{v}}_{\theta_{\perp}}\!&=\!\! -\sum_{i=1}^{3} L^i(\xi) \mbd{v}_{\theta_{\perp}}(\xi^i), \quad
\bar{\mbd{v}}_{\epsilon}\!=\!\!\sum_{i=1}^{3} L^i(\xi) \mbd{v}_{\epsilon}(\xi^i), \\
\bar{\mbd{v}}_{\theta_{\parallel \Theta}}\!&=\!\! \sum_{i=1}^{3} L^i(\xi) \mbd{v}_{\theta_{\parallel \Theta} }(\xi^i) \quad \text{with} \quad
\bar{\mbd{v}}^{\prime}_{...}\!\!=\!\!\sum_{i=1}^{3} \frac{L^i_{,\xi}(\xi)}{J(\xi)} \mbd{v}_{...} (\xi^i).
\end{split}
\end{align}
In the following, the formulation based on the degrees of freedom $\hat{\mbd{x}}_{TAN}$ and $\delta \hat{\mbd{x}}_{TAN}$ and on the residual~\eqref{eleweakkirchhoff_residual} will be denoted as WK-TAN element, which stands for "Weak Kirchhoff constraint enforcement combined with nodal triad parametrization via nodal TANgents". For the WK-TAN element, a linearization $\Delta \mbd{r}_{WK-TAN}\!=\!\mbd{k}_{WK-TAN} \Delta \hat{\mbd{x}}_{TAN}$ based on the increment vector $\Delta \hat{\mbd{x}}_{TAN}\!:=\!(\Delta \hat{\mb{d}}^{1T}\!, \Delta \hat{\mb{t}}^{1T}\!, \Delta \hat{\varphi}^1, \Delta \hat{\mb{d}}^{2T}\!,\Delta \hat{\mb{t}}^{2T}\!,\Delta \hat{\varphi}^2, \Delta \hat{\varphi}^3)^T$ is employed (see~\ref{anhang:linwktan}).\\

\hspace{0.2 cm}
\begin{minipage}{15.0 cm}
\textbf{Remark: } Actually, a collocation type approach has been applied in order to enforce the Kirchhoff constraint. Nevertheless, the notion "weak constraint enforcement" is kept throughout this work since such a procedure still represents the basis of the space-continuous formulation. Moreover, the difference to the formulation based on "strong constraint enforcement" of Section~\ref{sec:elestrongkirchhoff} shall be emphasized by this naming. 
\end{minipage} 

\subsection{Residual vector of rotation vector-based parametrization}
\label{sec:eleweakkirchhoff_variant1}

Also for the element formulation of Section~\ref{sec:eleweakkirchhoff_variant2} based on weak Kirchhoff constraint enforcement, a coordinate transformation from $\hat{\mbd{x}}_{TAN}$ and $\delta \hat{\mbd{x}}_{TAN}$ to the alternative primary variables $\hat{\mbd{x}}_{ROT}\!:=\!(\hat{\mb{d}}^{1T}\!,\hat{\boldsymbol{\psi}}^{1T}\!,\hat{t}^1, \hat{\mb{d}}^{2T}\!,\hat{\boldsymbol{\psi}}^{2T}\!,\hat{t}^2, \hat{\varphi}^3)^T$ and $\delta \hat{\mbd{x}}_{ROT}\!:=\!(\delta \hat{\mb{d}}^{1T}\!, \delta \hat{\boldsymbol{\theta}}^{1T}\!, \delta \hat{t}^1, \delta \hat{\mb{d}}^{2T}\!,\delta \hat{\boldsymbol{\theta}}^{2T}\!,\delta \hat{t}_{1}, \delta \hat{\Theta}_{1}^3)^T$ can be performed. The transformation rule for the element residual vector $\mbd{r}_{ROT} \!=\! \tilde{\mbd{T}}_{\hat{\mbd{x}}}^T \mbd{r}_{TAN}$ is identical to~\eqref{elestrongkirchhoff_trafovariations3} in Section~\ref{sec:elestrongkirchhoff_variant1}. Throughout this work, the element formulation based on the degrees of freedom $\hat{\mbd{x}}_{ROT}$ and $\delta \hat{\mbd{x}}_{ROT}$ and on the element residual vector~\eqref{eleweakkirchhoff_residual} transformed via~\eqref{elestrongkirchhoff_trafovariations3} will be denoted as WK-ROT element, which stands for "Weak Kirchhoff constraint enforcement combined with nodal triad parametrization via nodal ROTation vectors". For the WK-ROT element, a linearization $\Delta \mbd{r}_{WK-ROT}\!=\!\mbd{k}_{WK-ROT} \Delta \hat{\mbd{x}}_{ROT}$ based on the increment vector $\Delta \hat{\mbd{x}}_{ROT}\!:=\!(\Delta \hat{\mb{d}}^{1T}\!, \Delta \hat{\boldsymbol{\theta}}^{1T}\!, \Delta \hat{t}^1, \Delta \hat{\mb{d}}^{2T}\!,\Delta \hat{\boldsymbol{\theta}}^{2T}\!,\Delta \hat{t}_{1}, \Delta \hat{\varphi}^3)^T$ will be employed (see~\ref{anhang:linskwkrot}).

%
\subsection{Avoidance of locking effects}
\label{sec:eleweakkirchhoff_spatialdiscretization_locking}
%

In the investigation of the locking behavior of the proposed WK-TAN element, many results already derived in Section~\ref{sec:elesimoreissner_spatialdiscretization_locking} and~\ref{sec:elestrongkirchhoff_spatialdiscretization_locking} can be re-used. Since the numbers $n_{eq}$ and $n_{eq,c}$ for the space-continuous as well as for the discrete problem are identical to the SK-TAN element, it can readily be concluded that also the WK-TAN element formulation shows an optimal constraint ratio of $r\!=\!r_h\!=\!4$ and that no membrane locking effects are expected for this element.\\

Similar to Section~\ref{sec:elesimoreissner_spatialdiscretization_locking}, it shall be shown that also the WK-TAN/-ROT elements can exactly represent the internal energy associated with a 3D pure bending state. This time, the internal energy is split into contributions stemming from torsion and bending and into contributions stemming from axial tension, i.e. $\Pi_{int,h}\!=\!\Pi_{int,\Omega,h}\!+\!\Pi_{int,\epsilon,h}$. For a pure bending state, the energy contribution $\Pi_{int,\epsilon,h}$ has to vanish, and thus the total internal energy of a pure bending state is given by $\Pi_{int,\Omega,h}$, which is uniquely defined by the curvature vector field $\mb{K}_h(\xi)\!=$ const. In order to represent the desired (constant) distribution of the curvature vector field $\mb{K}_h(\xi)$, which is possible for the employed triad interpolation~\eqref{triadrelrot_interpolation}, only $2n_{ele}$ of the $2n_{ele}\!+\!1$ nodal triads have to arise properly, while the one remaining nodal triad describes rotational rigid body modes of the beam. Although the nodal triads are not necessarily parametrized by nodal rotation vectors, still three conditions result from each of these $2n_{ele}$ nodal triads, thus resulting in a total of $n_{eq,\Omega}\!=\!6n_{ele}$ conditions. Additionally, the axial strains at the collocation points have to vanish in order to yield a vanishing contribution \mbox{$\Pi_{int,\epsilon,h}\!=\!0$}. This requirement results in $n_{eq,\epsilon}\!=\!2n_{ele}\!+\!1$ additional conditions that have to be fulfilled at the collocation points. If again six further conditions are considered in order to superpose arbitrary rigid body modes (representing the minimally required number of Dirichlet boundary conditions in static problems), the total number of $n_{eq}\!=\!8n_{ele}\!+\!7$ equations equals the total number of $n_{uk}\!=\!7(n_{ele}\!+\!1)\!+\!n_{ele}$ unknowns contained in the global vector $\mbd{X}$ for the considered WK-TAN/ROT elements. Thus, in case a unique FEM solution is existent, a 3D pure bending case can be represented exactly. Similar to the CJ element, the torsion and bending modes represented by $\boldsymbol{\Omega}_h(\xi)$ as well as the axial tension values at the collocation points represented by $\epsilon(\xi_{CP})$ are non-competing and no locking effects are expected. Again, these considerations can easily be extended to arbitrary curvature fields $\mb{K}_h(\xi)$ that are representable by the employed triad interpolation and arbitrary second-order polynomials $\bar{\epsilon}(\xi)$ according to~\eqref{MCS}. In Sections~\ref{sec:examples_purebending2d} and~\ref{sec:examples_purebending3d}, the expected result that the discrete hyperelastic energies associated with pure bending states in 2D and in 3D can exactly be represented by the WK-TAN/ROT elements will be verified by means of corresponding numerical test cases.

%
\subsection{Conservation properties}
\label{sec:eleweakkirchhoff_spatialdiscretization_conservation}
%
Since the WK-TAN beam element proposed above basically combines the triad interpolation and the spin vector interpolation $\delta \boldsymbol{\theta}_h(\xi)$ of the CJ element (see Section~\ref{sec:elesimoreissner}) with the centerline interpolation and its variation $\delta \mb{r}_h^{\prime}(\xi)$ applied to the SK-TAN element of Section~\ref{sec:elestrongkirchhoff}, the corresponding conservation properties can directly be concluded from the investigations of Sections~\ref{sec:elesimoreissner_spatialdiscretization_conservation} and~\ref{sec:elestrongkirchhoff_spatialdiscretization_conservation}. Consequently, this element will exactly fulfill conservation of linear and angular momentum. Conservation of energy can only be guaranteed for the spatially discretized, time-continuous problem in case the Petrov-Galerkin spin vector interpolation~\eqref{eleweakkirchhoff_spinvector2} is replaced by its Bubnov-Galerkin counterpart~\eqref{eleweakkirchhoff_spinvector1}.

\section{Numerical Examples}
\label{sec:elenumericalexamples}

In this section, the previously proposed beam element formulations will be investigated numerically by means of proper test cases. All simulations results presented in the following rely on a software implementation of the proposed finite element formulations and numerical algorithms within the in-house finite element research code BACI (cf. Wall and Gee~\cite{wall2012}), developed jointly at the Institute for Computational Mechanics at the Technical University of Munich. While most of the numerical examples (see Sections~\ref{sec:examples_objectivity}-\ref{sec:examples_helix}) are considered as quasi-static problems, eventually, in Section~\ref{sec:examples_freeoscillationselbow}, also a dynamic test case is investigated. In a first step, these numerical examples aim to verify the principle applicability and accuracy of the proposed general and reduced Kirchhoff-Love beam elements in the range of different beam slenderness ratios. This verification crucially relies on detailed comparisons with analytic reference solutions, benchmark tests known form the literature as well as numerical reference solutions generated by means of well-established geometrically exact beam element formulations of Simo-Reissner type. More specifically, also the essential requirements formulated in Section~\ref{sec:spatialdiscretization_requirements} such as objectivity and path-independence, avoidance of locking effects, consistent spatial convergence behavior as well as the fulfillment of conservation properties will be verified for the different beam element formulations presented in Sections~\ref{sec:elesimoreissner}-\ref{sec:eleweakkirchhoff}. Finally, based on the arguments given in Section~\ref{sec:reissnermotivationshearfree}, the focus will also lie on detailed comparisons of Reissner and Kirchhoff type beam element formulations for example with respect to the resulting discretization error level or the performance of the Newton-Raphson scheme.\\

Since the Kirchhoff type beam element formulations based on a tangent-based triad parametrization and the formulations based on a rotation vector-based triad parametrization (compare e.g. Sections~\ref{sec:elestrongkirchhoff_variant2} and~\ref{sec:elestrongkirchhoff_variant1} or Sections~\ref{sec:eleweakkirchhoff_variant2} and~\ref{sec:eleweakkirchhoff_variant1}) have been shown to yield identical FEM solutions, only the former category will be investigated with respect to spatial discretization errors. Furthermore, for all examples without analytic solution, the standard choice for the reference solution $\mb{r}_{ref}$ (see also~\eqref{rele2error} of Section~\ref{sec:spatialdiscretization_convergence}) is a numerical solution via the WK-TAN element (see Section~\ref{sec:eleweakkirchhoff_variant2}) employing a spatial discretization that is by a factor of four finer than the finest discretization shown in the corresponding convergence plot. In order to achieve a good comparability among the different geometries and load cases, a standard set of geometrical and constitutive parameters has been applied in all simulations unless stated otherwise. This standard set consists of a beam with initial length $l\!=\!1000$ and square
cross-section with side length $R$. These parameters lead to a cross-section area of $A\!=\!R^2$ and to moments of inertia of area of $I_{2}\!=\!I_{3}\!=:\!I\!=\!R^4 / 12$ and $I_T\!=\!R^4 / 6$. Different beam slenderness ratios $\zeta\!:=\! l / R$ are generated by varying the value $R\!=\!0.1,1.0,10,100$ of the cross-section side length. The standard choice for the constitutive parameters is $E\!=\!1.0$ and $G\!=\!0.5$, thus leading to $EI\!=\!GI_P\!=\!R^4 / 12$. For all numerical examples considered in the following sections, a Newton-Raphson scheme based on consistent linearization has been applied in order to solve the set of nonlinear equations resulting from the temporally and spatially discretized weak form of the balance equations. As convergence criteria, the Euclidean norms of the displacement increment vector $\Delta {\mbd{X}}_{n+1}^{k+1}$ and of the residual vector $\mbd{R}({\mbd{X}}_{n+1}^{k+1})$ are checked. For convergence, these norms have to fall below prescribed tolerances $\delta_{\mbd{R}}$ and $\delta_{\mbd{X}}$, i.e. $||\mbd{R}({\mbd{X}}_{n+1}^{k+1})||<\delta_{\mbd{R}}$ and $||\Delta {\mbd{X}}_{n+1}^{k+1}||<\delta_{\mbd{X}}$. Typical convergence tolerances chosen for the subsequent examples are in the range of $\delta_{\mbd{X}}\!=\!10^{-8}$ as well as $\delta_{\mbd{R}}\!=\!10^{-7}, 10^{-9}, 10^{-11}, 10^{-13}$ for the slenderness ratios $\zeta\!=\!10,100,1000,10000$. For the (quasi-)static problems presented in the following sections, the external loads are applied on the basis of an incremental procedure, where $N$ shall denote the number of load steps and $\Delta t$ the load step size. As long as nothing is stated to the contrary, the following simple procedure is applied in order to adapt the load step size during the static simulation in an efficient manner: Initially, a comparatively small load step size $\Delta t_0$ is chosen, e.g. according to $N_0\!=\!1$. If the Newton-Raphson scheme has not converged within a prescribed number of $n_{iter,max}$ iterations, the step size is halved and the load step is repeated. This procedure is repeated until convergence can be achieved. Then, after \textit{four} converging load steps on the low step size level, the step size is doubled again. Also this procedure of successively doubling the step size after four converging load steps at the current step size level is repeated until the original step size $\Delta t_0$ is reached again. This procedure will not only drastically increase the overall computational efficiency, it also allows for comparatively objective and fair comparisons of the performance of the Newton-Raphson scheme for different element formulations. In subsequent numerical examples, such comparisons will be made on the basis of the accumulated number of Newton iterations
\begin{align}
\label{loadstepadaption2}
\begin{split}
n_{iter,tot}\!:=\!\sum \limits_{n=1}^N n_{iter,n},
\end{split}
\end{align}
required to solve the entire problem. Here, $n_{iter,n}$ is the number of iterations required for load step $t_n$. In "load step adaption scheme" above, non-converging steps are considered in the total number of iterations with $n_{iter,n}\!=\!n_{iter,max}$.

\subsection{Example 1: Verification of objectivity}
\label{sec:examples_objectivity}

\begin{figure}[h!]
 \centering
  \subfigure[Problem setup.]
   {
    \includegraphics[height=0.35\textwidth]{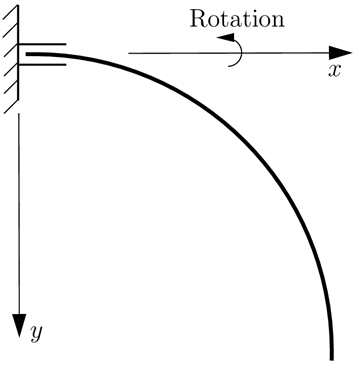}
    \label{fig:objectivity_a}
   }
   \hspace{0.05\textwidth}
   \subfigure[Internal energy due to imposed rigid body rotation.]
   {
    \includegraphics[width=0.58\textwidth]{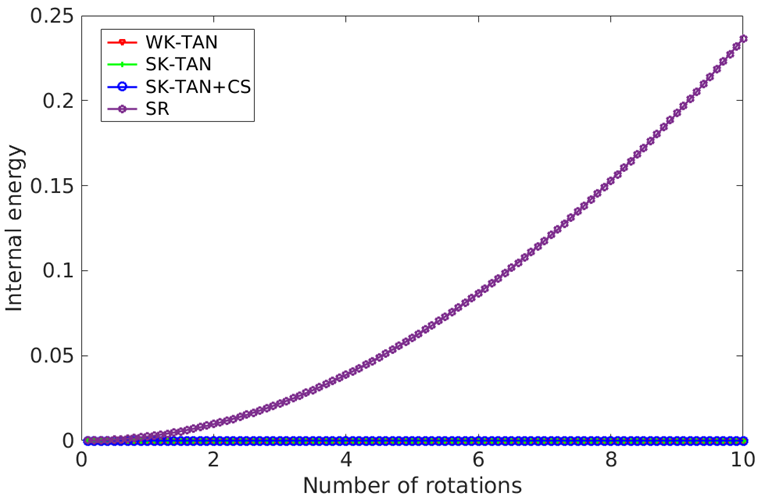}
    \label{fig:objectivity_b}
   }
  \caption{Objectivity test: Rigid body rotation of an initially stress-free quarter circle.}
  \label{fig:objectivity}
\end{figure}

The objectivity of the Kirchhoff beam element formulations proposed in Sections~\ref{sec:elestrongkirchhoff} and~\ref{sec:eleweakkirchhoff} has already been proven theoretically. In order to verify these results numerically, the following test case will be investigated (see Figure \ref{fig:objectivity_a}): At the clamped end of an initially curved beam with slenderness ratio $\zeta=10$, whose stress-free centerline configuration equals a quarter circle, a Dirichlet rotation with respect to the global $x$-axis is imposed. For the presented quasi-static example, a total rotation angle of $20 \pi$, increasing linearly over $100$ load steps, is prescribed. In order to investigate objectivity, the normalized internal (hyperelastic) energy is plotted over the total number of rotations (see Figure \ref{fig:objectivity_b}) for the WK-TAN and SK-TAN element as well as for the Bubnov-Galerkin variant of the SK-TAN element formulation with Consistent Spin (SK-TAN+CS) vector interpolation according to~\eqref{elestrongkirchhoff_residual}. For comparison reasons, also the non-objective SR Kirchhoff beam element formulation investigated in Meier et al.~\cite{meier2014} will be considered (see also the last remark at the end of Section~\ref{sec:spatialdiscretization_triadsr}). For clearness, the internal energy~$\Pi_{int}$ is normalized by the factor $\Pi_{int,r}\!=\!0.5 EI \pi^2 / (4l)$, which is equal to the amount of mechanical work that is required to bend the initially stress-free quarter circle into a straight beam by means of a discrete, external end-moment. Of course, the internal energy should vanish for a beam that is merely rotated out of its stress-free initial configuration.\\

From Figure \ref{fig:objectivity_b}, it becomes obvious, however, that the internal energy of the SR
formulation increases over the number of rotations, which is a clear indication for the already theoretically predicted non-objectivity. Within $10$ rotations, the normalized energy reaches a value of almost $\Pi_{int,r} / 4$, which results in a clearly visible deformation of the initial quarter circle. On the contrary, the internal energy of all the other investigated element formulations results in a value that is zero up to machine precision. Finite element formulations based on such interpolation schemes as the investigated SR element might show reasonable results for static test cases (see e.g.~\cite{meier2014}). However, especially in dynamic problems involving considerable rigid body motions, non-physical results as well as a drastic deterioration of the conservation properties investigated in Section~\ref{sec:spatialdiscretization_conservation} can follow from the application of such element formulations.

\subsection{Example 2: Pure bending in 2D}
\label{sec:examples_purebending2d}

The examples shown in this section exclusively focus on 2D geometries and load cases. The section is sub-divided into two subsections: In Section~\ref{sec:examples_purebending2d_locking}, two load cases, a pure pending case as well as a combined moment-and-force load case, yielding geometrically nonlinear, but still moderate centerline deformations, are considered. This section aims at the investigation of membrane locking effects and at the comparison of different anti-locking tools, especially of the MCS method proposed in Section~\ref{sec:spatialdiscretization_locking}. In Section~\ref{sec:examples_purebending2d_comparison}, again a pure bending and a combined moment-and-force load case will be considered. However, due to higher load factors, the resulting degree of deformation is further increased as compared to the examples of Section~\ref{sec:examples_purebending2d_locking}. This higher degree of deformation reveals clear differences in the approximation quality of the WK and SK Kirchhoff beam element variants. Besides the comparison of these two variants, also a first proof of concept for the development of higher-order Hermitian Kirchhoff elements is given.

\subsubsection{Comparison of different anti-locking methods}
\label{sec:examples_purebending2d_locking}

\begin{figure}[ht]
 \centering
  \subfigure[Straight beam bent by end-moment.]
   {
    \includegraphics[width=0.43\textwidth]{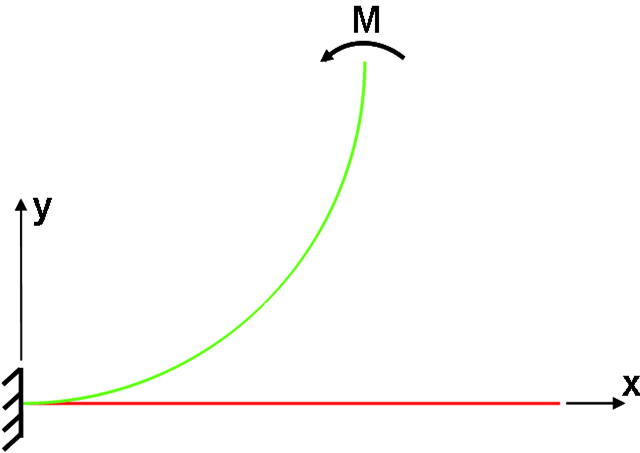}
    \label{fig:end-moment_complete}
   }
   \hspace{0.02\textwidth}
   \subfigure[Straight beam bent by end-moment and -force.]
   {
    \includegraphics[width=0.43\textwidth]{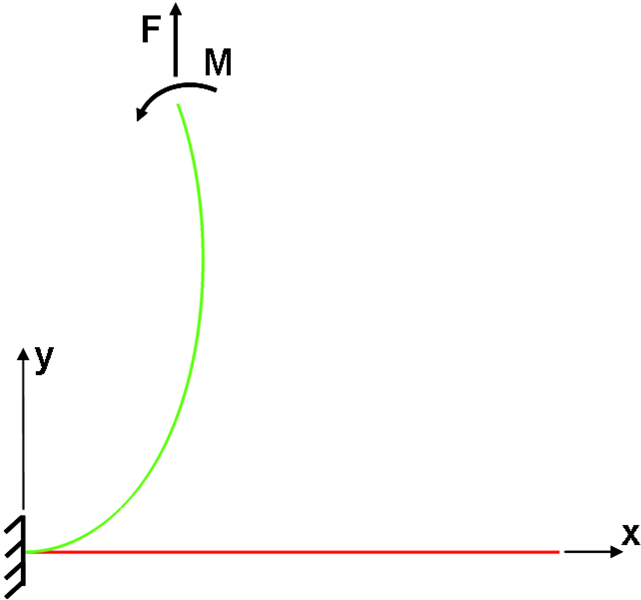}
    \label{fig:end-momentandforce_complete}
   }
   \caption{Initial and deformed configuration of an initially straight beam for two load cases.}
  \label{fig:planearcs1}
\end{figure}

An initially straight beam is clamped at one end. Two different load cases will be analyzed: The first load case M is identical to the example analyzed in Section~\ref{sec:spatialdiscretization_locking} and solely consists of a discrete end-moment $\mb{M}=(0,0,M)^T$ applied in one load step. The moment $M=EI \pi / (2l)$ exactly bends the beam into a quarter-circle shaped arc. In the second load case M+F, the end-moment and an additional tip force $\mb{F}=(0,F,0)^T$
in global $y$-direction are applied in one load step. The initial and deformed geometries for these load cases are illustrated in Figure \ref{fig:planearcs1}.\\

\begin{figure}[t!!!]
 \centering
  \subfigure[Reference: Analytic, different slenderness ratios.]
   {
    \includegraphics[width=0.48\textwidth]{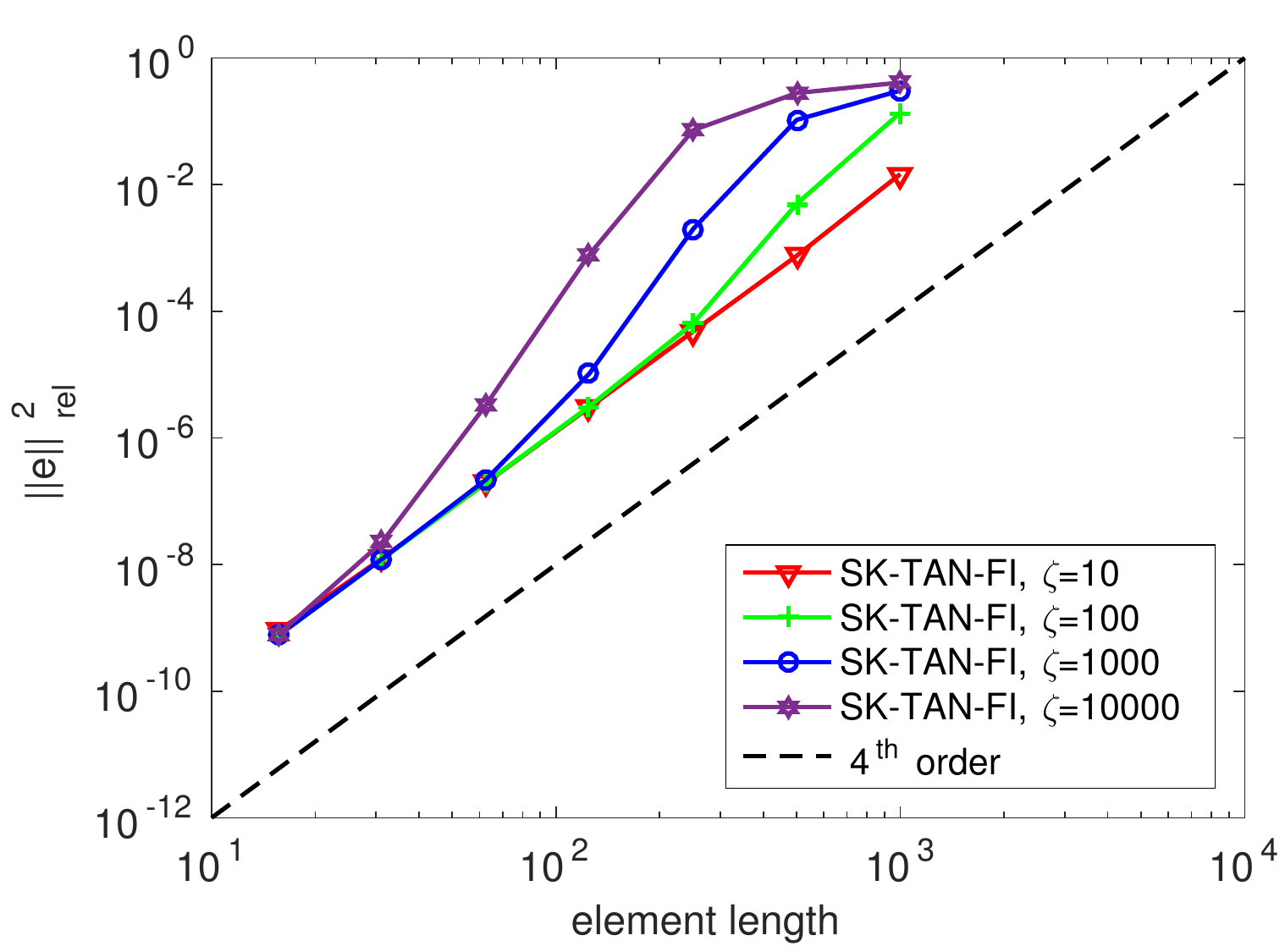}
    \label{fig:quartercircle_convergence_a}
   }
   \subfigure[Reference: Analytic, different slenderness ratios.]
   {
    \includegraphics[width=0.48\textwidth]{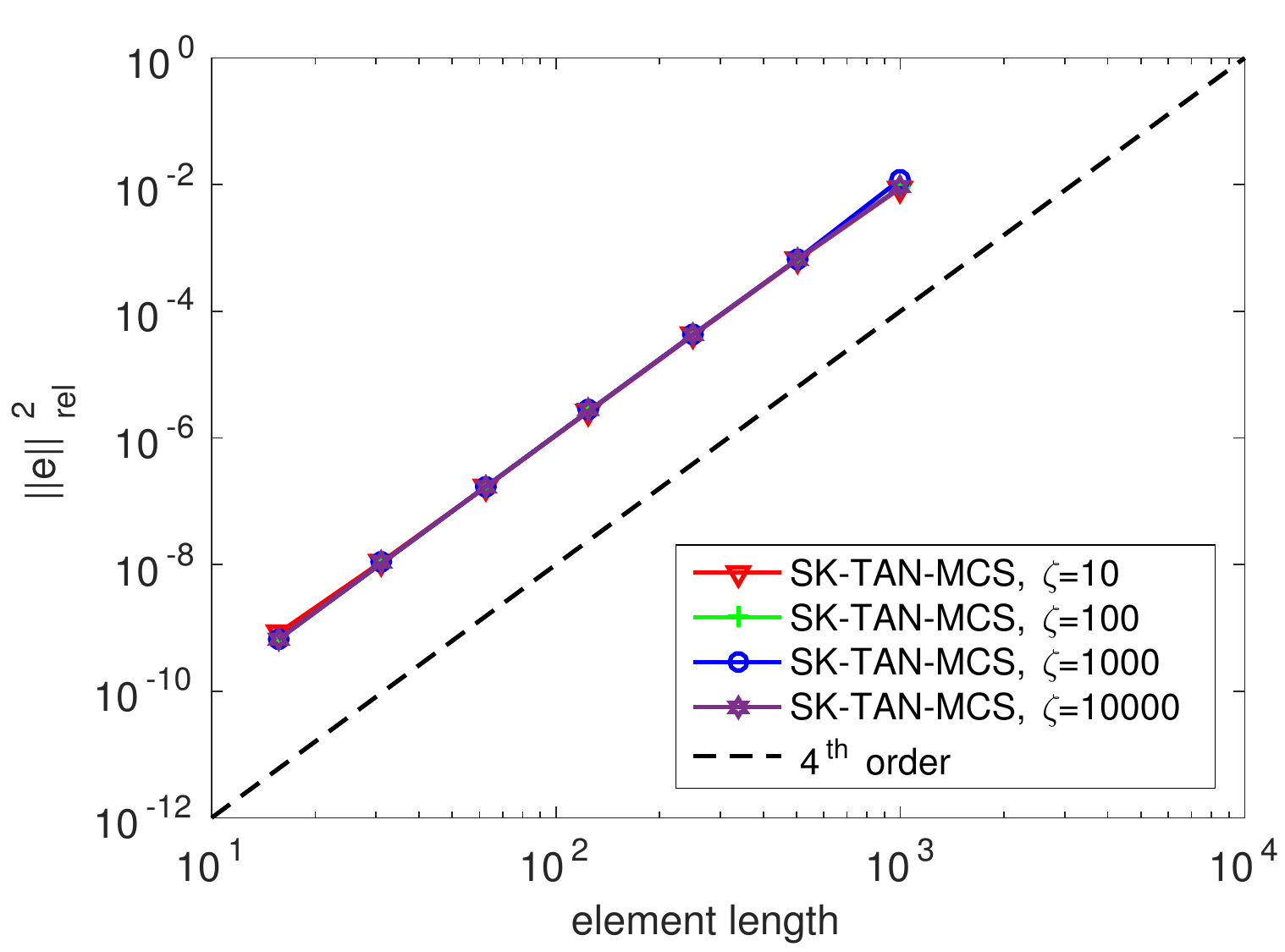}
    \label{fig:quartercircle_convergence_b}
   }
   \subfigure[Reference: Analytic, $\zeta\!=\!10000$.]
   {
    \includegraphics[width=0.48\textwidth]{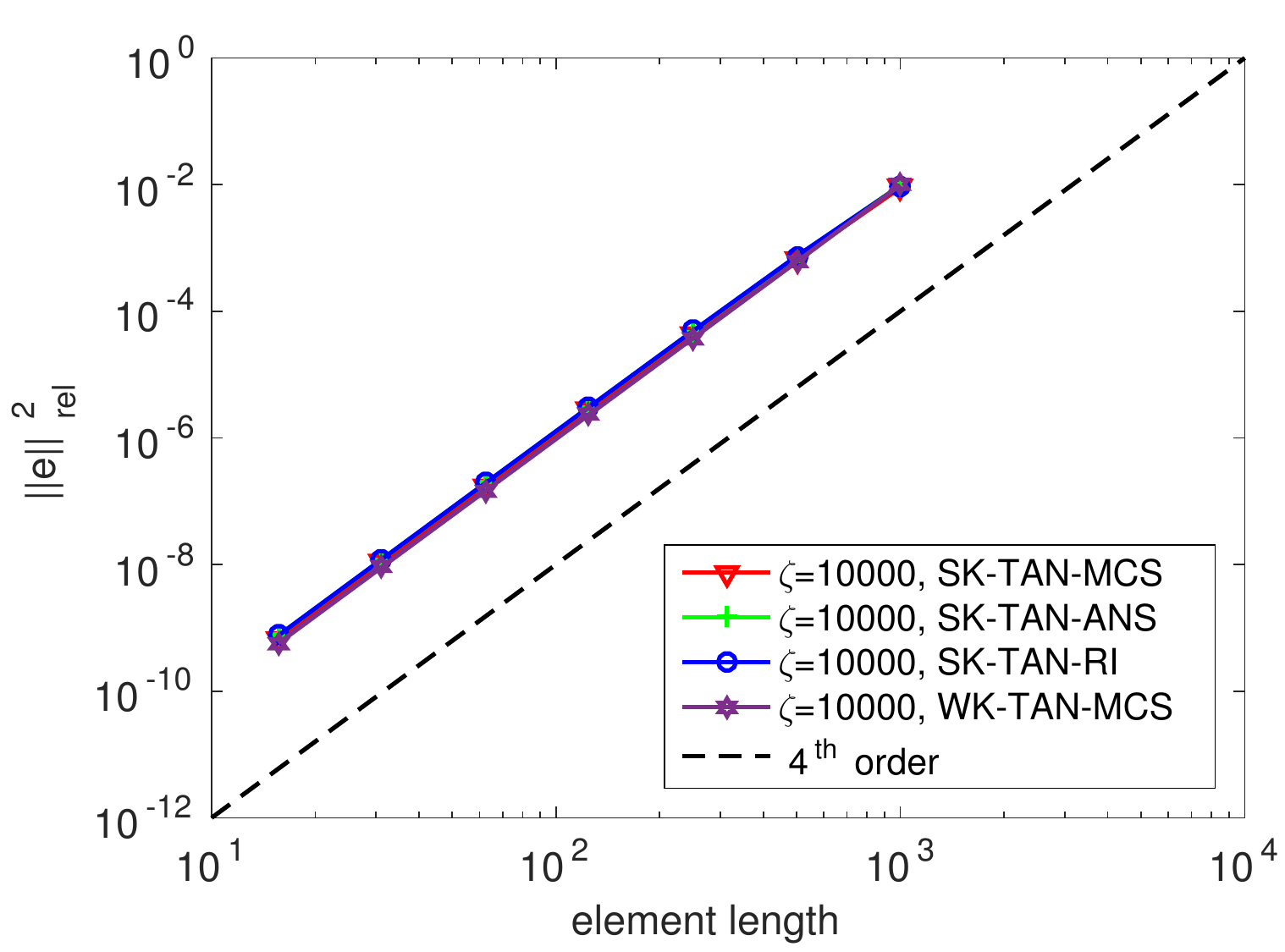}
    \label{fig:quartercircle_convergence_c}
   }
   \subfigure[Reference: WK-TAN-MCS, $\zeta\!=\!10000$.]
   {
    \includegraphics[width=0.48\textwidth]{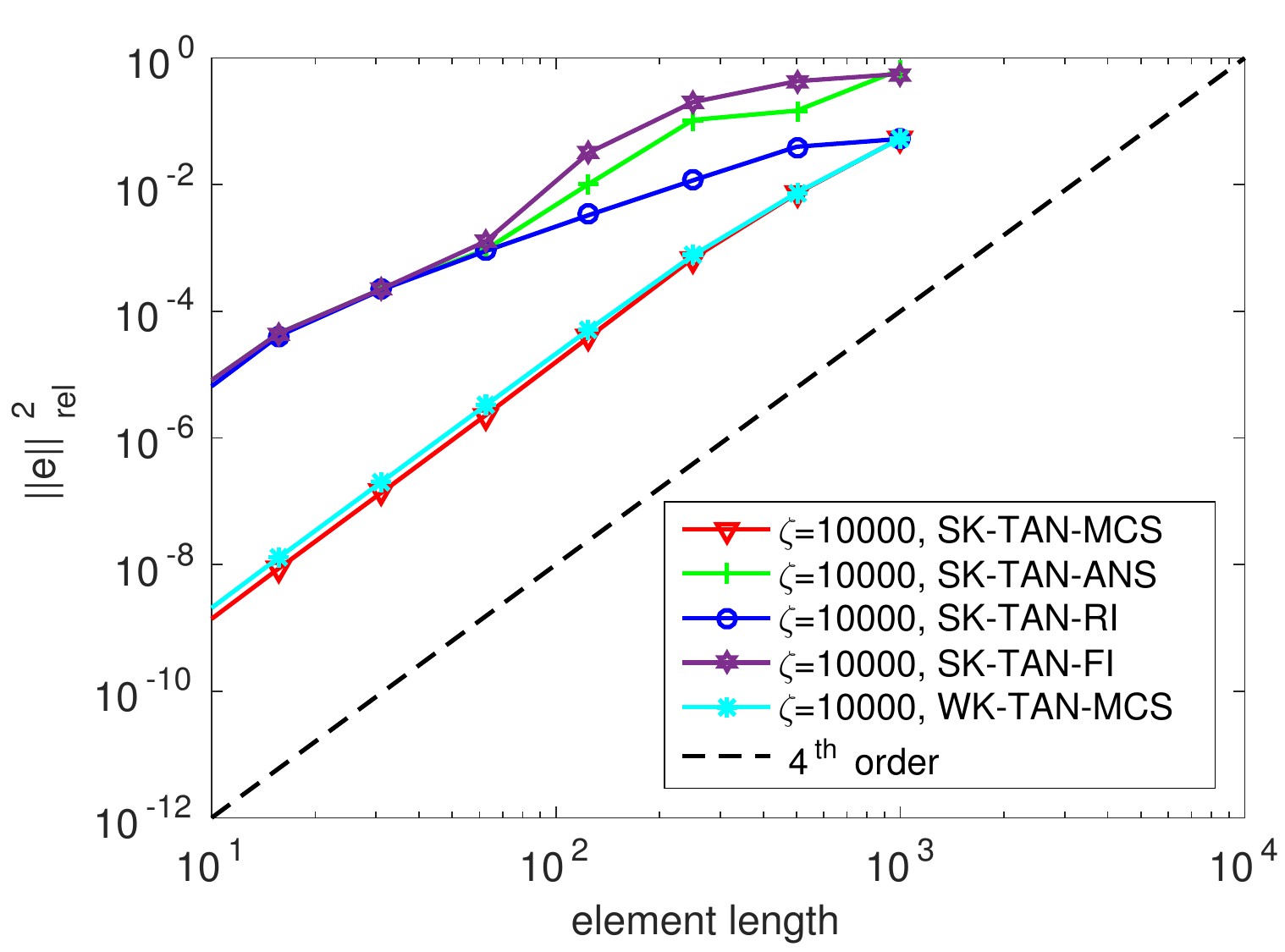}
    \label{fig:quartercircle_convergence_d}
   }
  \caption{Straight beam subject to the load cases M and M+F: Relative $L^2$-error.}
  \label{fig:quartercircle_convergence}
\end{figure}

While the standard parameters and the slenderness ratios $\zeta\!=\!10,100,1000,10000$ have been chosen for the load case M, only the highest and therefore most critical slenderness ratio $\zeta\!=\!10000$ in combination with an external force of $F\!=\!1.0 \cdot 10^{-10} \!\approx\! 10 M / l$
has been investigated for the load case M+F. In a first step, the SK-TAN element formulation according to~\eqref{elestrongkirchhoff_residual_petrov} is applied in combination with a full Gauss-Legendre integration (''SK-TAN-FI``) with $n_G\!=\!4$, a reduced Gauss-Legendre integration (''SK-TAN-RI``) with $n_G\!=\!3$, a classical assumed natural strain approach (''SK-TAN-ANS``) as well as the MCS method according to~\eqref{MCS} based on $n_G\!=\!4$ integration points (''SK-TAN-MCS``) (see Section~\ref{sec:spatialdiscretization_locking} and~\cite{meier2015} for further details on these variants). In Figure~\ref{fig:quartercircle_convergence_a}, the relative $L^2$-error of the load case M and different slenderness ratios is plotted with respect to an analytic reference solution. For spatial discretization, the variant SK-TAN-FI based on meshes with $1,2,4,8,16,32$ and $64$ elements has been applied. Accordingly, the convergence is slowed down dramatically with increasing slenderness ratio. If the beam is e.g. discretized by one finite element ($l_{ele}\!=\!1000$), the relative error increases almost by two orders of magnitude when enhancing the slenderness ratio from $\zeta\!=\!10$ to $\zeta\!=\!10000$. However, Figure~\ref{fig:quartercircle_convergence_a} also reveals that this effect decreases with decreasing element sizes and almost completely disappears for discretizations with more than $32$ elements. The reason for this behavior lies in the fact that the element slenderness ratio $\zeta_{ele}\!=\!l_{ele} / R$ is the key-parameter for the observed locking effect and that the latter also decreases with decreasing element sizes. However, for typical engineering applications with relative error bounds in the range of $1 \%$, the effect is by no means negligible. For sufficiently fine discretizations the expected convergence order of four is reached. In Figure~\ref{fig:quartercircle_convergence_b}, the relative $L^2$-error is plotted for the same slenderness ratios as before, but for the variant SK-TAN-MCS, which is supplemented by the MCS method according to~\eqref{MCS}. As expected, the locking effect completely disappears for all investigated slenderness ratios. However, as shown in Figure \ref{fig:quartercircle_convergence_c}, for the load case~M and the highest investigated slenderness ratio of $\zeta\!=\!10000$, the same effect can alternatively be achieved by applying a simple reduced integration procedure (variant SK-TAN-RI) or a classical ANS approach (variant SK-TAN-ANS). On the contrary to load case M, Figure~\ref{fig:quartercircle_convergence_d} reveals that no distinctive improvement of the locking behavior can be obtained by these alternative methods for the load case M+F: Both the ANS approach as well as the reduced integration scheme can only slightly alleviate the locking effect in the range of rather coarse discretizations as compared to the variant SK-TAN-FI. The MCS approach, however, completely eliminates the error offset due to membrane locking also for this load case. The explanation for this observation is obvious and in agreement with the statements of Section~\ref{sec:spatialdiscretization_locking}: Similar to the working principle of the MCS method, the reduced integration scheme can alleviate locking by reducing the number of constraint equations.
Yet, as shown in Section~\ref{sec:spatialdiscretization_locking}, the MCS method leads to a lower number of constraint equations as compared to the simple reduced integration scheme, which makes the latter method less effective. While the reduced integration scheme seems to be sufficient for the load case M, a special case yielding symmetric curvature distributions within the elements, the more general deformed configurations resulting from the load case F+M already demonstrate the limits of this simple method. On the other hand, the working principle of the standard ANS method does not aim at a reduction of the number of constraint equations, but rather at an evaluation of the critical axial tension term at selected collocation points with vanishing parasitic strains. In the geometrically nonlinear regime of large deformations, the parameter space positions of these optimal collocation points are deformation-dependent. Obviously, the load case~M+F already leads to a change in these positions up to an extent that almost completely destroys the working principle and impact of the ANS method. Summing up, it is to say that based on these two examples the proposed MCS method seems to be superior to standard methods such as reduced integration or ANS in terms of locking avoidance when combined with the considered geometrically exact Kirchhoff beam elements. In example 8 of Section~\ref{sec:examples_helix}, a further comparison of these different anti-locking methodologies on the basis of a very general problem setting involving 3D deformation states and an initially curved geometry will be presented, which will confirm this result. For completeness, in Figures~\ref{fig:quartercircle_convergence_c} and~\ref{fig:quartercircle_convergence_d}, also the $L^2$-error of the WK-TAN-MCS element based on a weak enforcement of the Kirchhoff constraint according to~\eqref{eleweakkirchhoff_residual} has been plotted. For the load cases M and M+F, this formulation yields a comparable convergence behavior and discretization error level as the SK-TAN-MCS element. Next, it will be shown that this behavior will change with increasing deformation. In the following, the MCS method will be employed per default and the abbreviation ...-...-MCS omitted.

\subsubsection{Comparison of different element formulations}
\label{sec:examples_purebending2d_comparison}

\begin{figure}[ht]
 \centering
  \subfigure[Straight beam bent by end-moment.]
   {
    \includegraphics[width=0.47\textwidth]{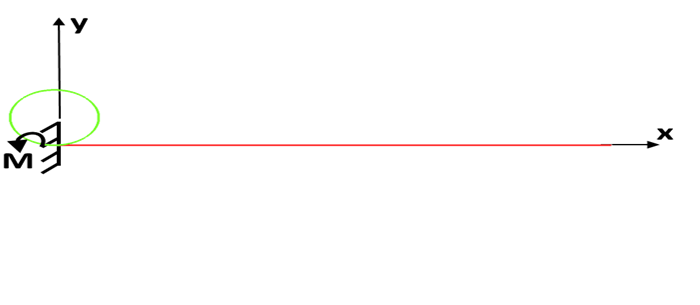}
    \label{fig:end-moment_doublecircle}
   }
   \hspace{0.01\textwidth}
   \subfigure[Straight beam bent by end-moment and -force.]
   {
    \includegraphics[width=0.47\textwidth]{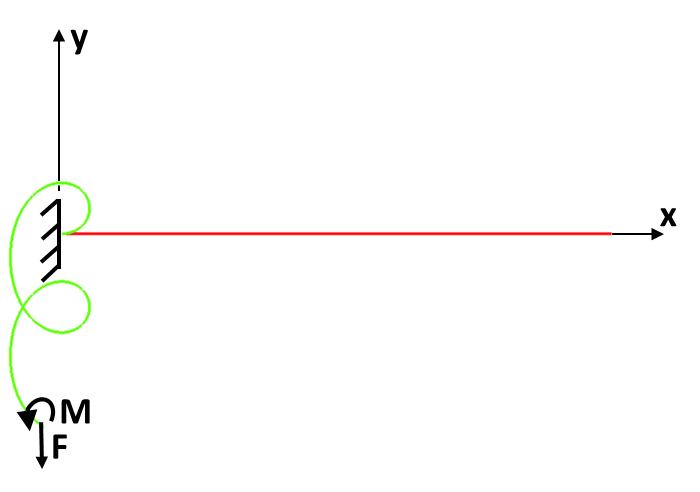}
    \label{fig:end-momentandforce_doublecircle}
   }
   \caption{Initial and deformed configuration of an initially straight beam for two load cases.}
  \label{fig:planearcs2}
\end{figure}

In the convergence plots investigated for the two load cases M and M+F, no noteworthy differences between the SK and WK element could be observed. In order to investigate the difference in the two general approaches of enforcing the Kirchhoff constraint in a strong or in a weak manner further in detail and to perform first comparisons with geometrically exact beam element formulations of Simo-Reissner type, two additional load cases will be considered (see Figure~\ref{fig:planearcs2}): The first load case considered in this section, in the following denoted as $\tilde{M}$, simply increases the magnitude of the external moment by a factor of eight as compared to the previous load case M, i.e. $\tilde{M}\!=\!8M$, thus leading to a deformed geometry that is represented by a double circle (see Figure~\ref{fig:end-moment_doublecircle}). Since the contribution of Simo and Vu-Quoc~\cite{simo1986}, this load case has been established as a standard test case for geometrically exact beam element formulations. Finally, in a fourth load case, denoted as $\tilde{M}$+$\tilde{F}$, the end-moment $\tilde{M}\!=\!8M$ is supplemented by a tip force $\mb{F}=(0,\tilde{F},0)^T$ in global $y-$direction, whose magnitude is this time exactly chosen as $\tilde{F}\!=\!10\tilde{M}/l\!=\!0.08M$. The initial geometry as well as the final configuration of this last load case is illustrated are Figure~\ref{fig:end-momentandforce_doublecircle}.\\

\begin{figure}[ht]
 \centering
  \subfigure[Reference: Analytic, $L^2$-error.]
   {
    \includegraphics[width=0.48\textwidth]{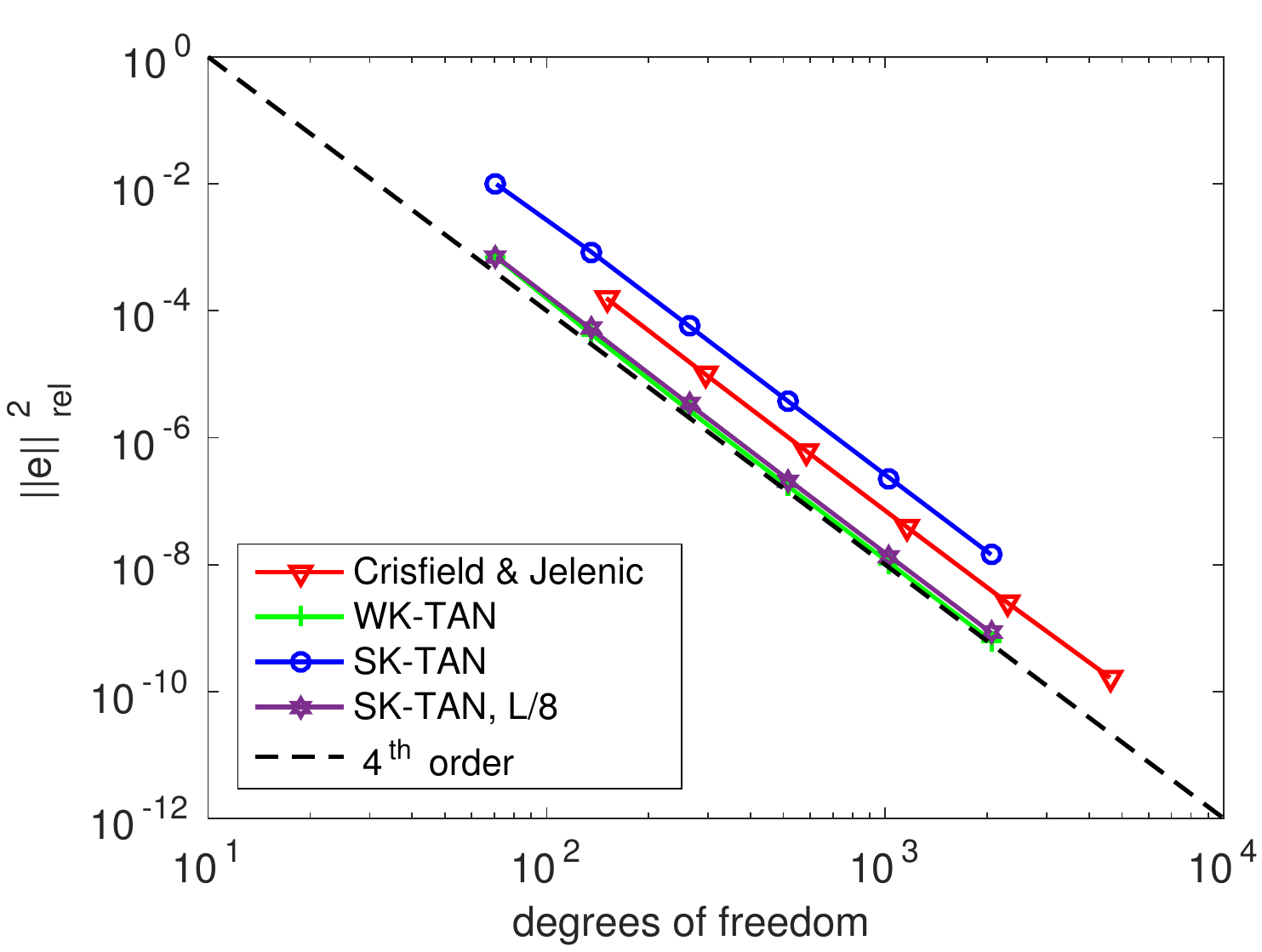}
    \label{fig:doublecirce_l2_1}
   }
   \subfigure[Reference: Analytic, energy error.]
   {
    \includegraphics[width=0.48\textwidth]{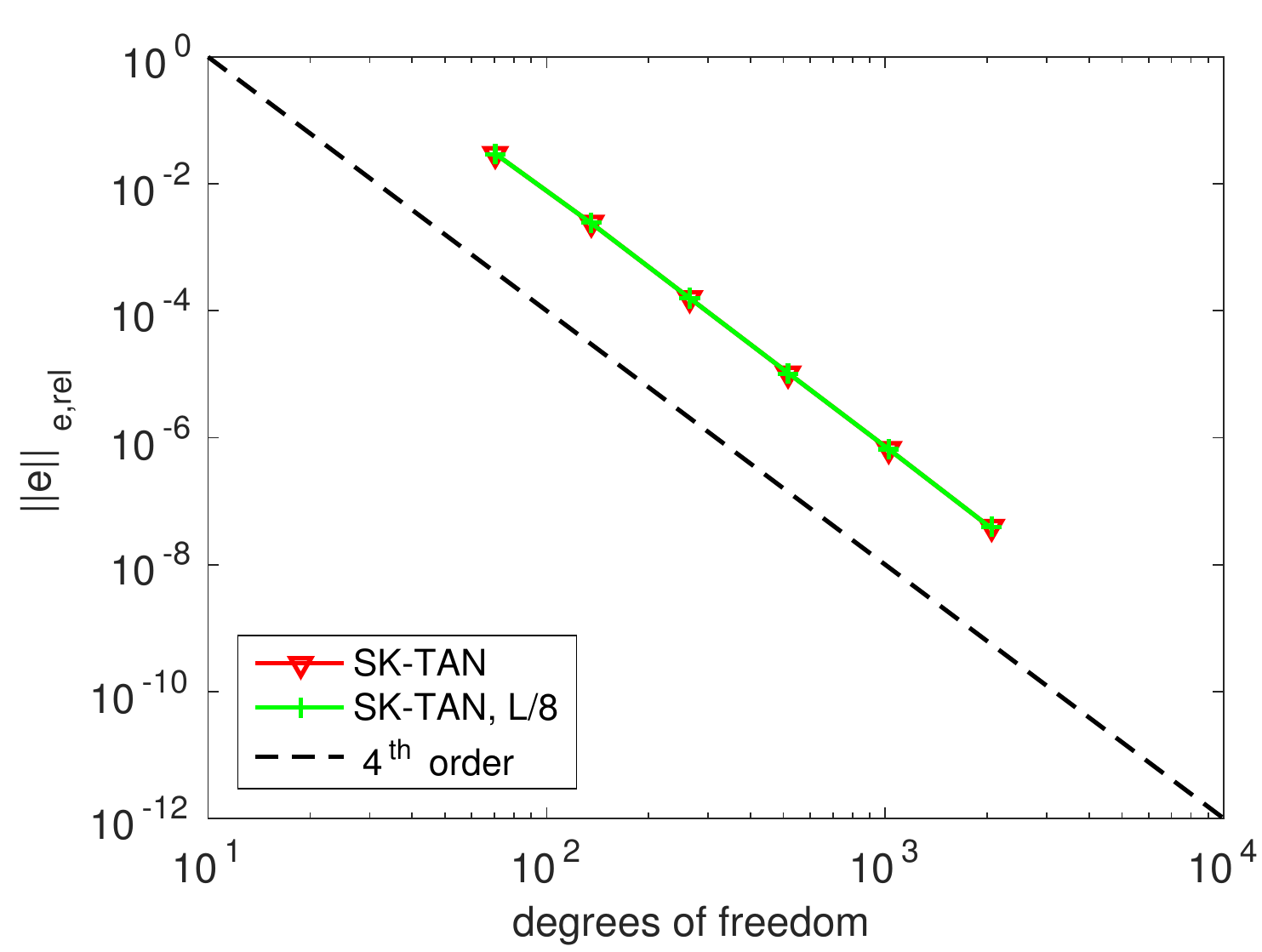}
    \label{fig:doublecirce_l2_2}
   }
  \caption{Straight beam and load case $\tilde{M}$: convergence of $L^2$- and energy error.}
  \label{fig:doublecircle_convergence1}
\end{figure}

\begin{figure}[t!!!]
 \centering
  \subfigure[Reference: Crisfield \& Jeleni\'{c}, $\zeta\!=\!100$.]
   {
    \includegraphics[width=0.48\textwidth]{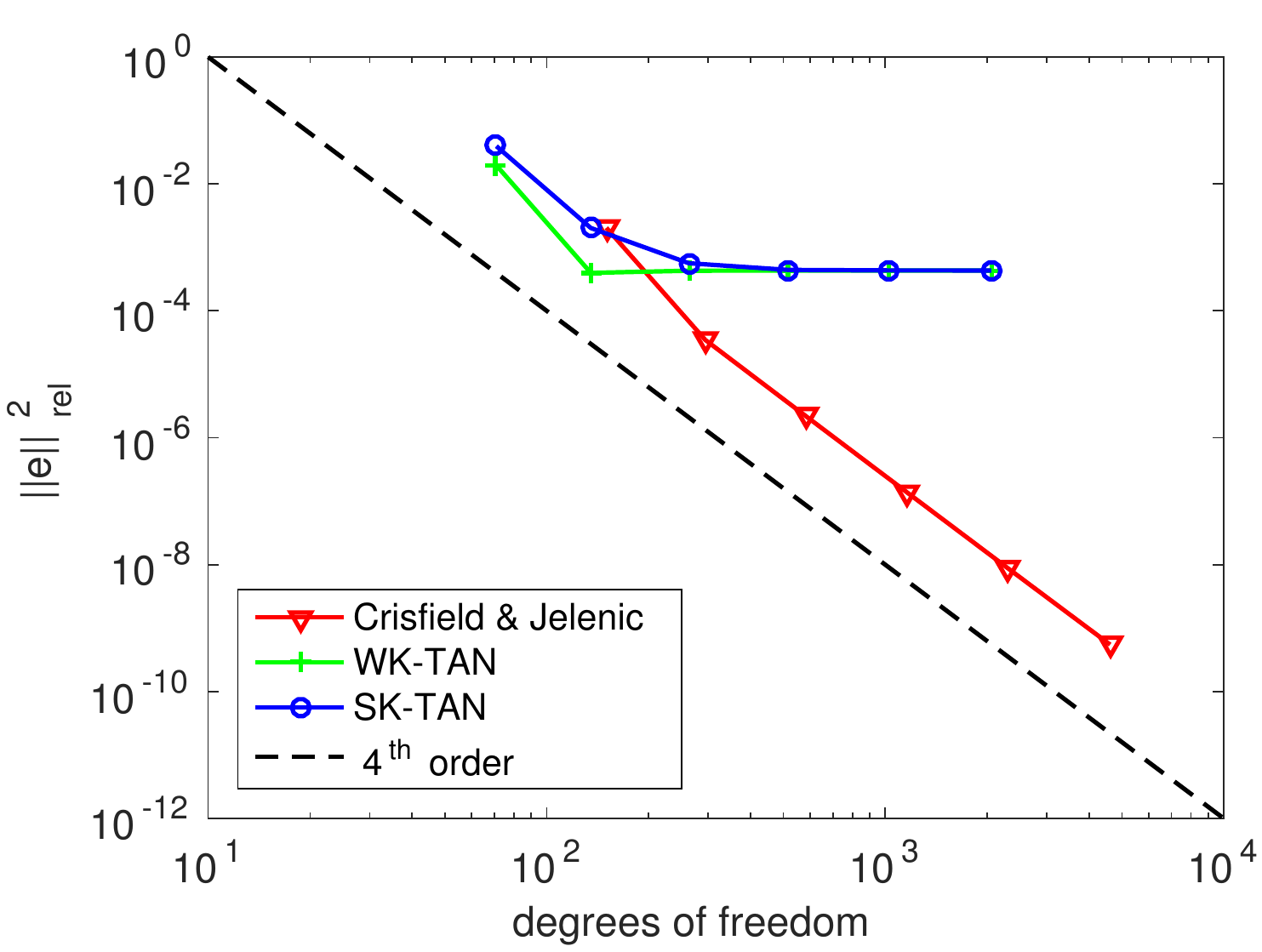}
    \label{fig:doublecircewithforce_l2_1}
   }
   \subfigure[Reference: Crisfield \& Jeleni\'{c}, $\zeta\!=\!10000$.]
   {
    \includegraphics[width=0.48\textwidth]{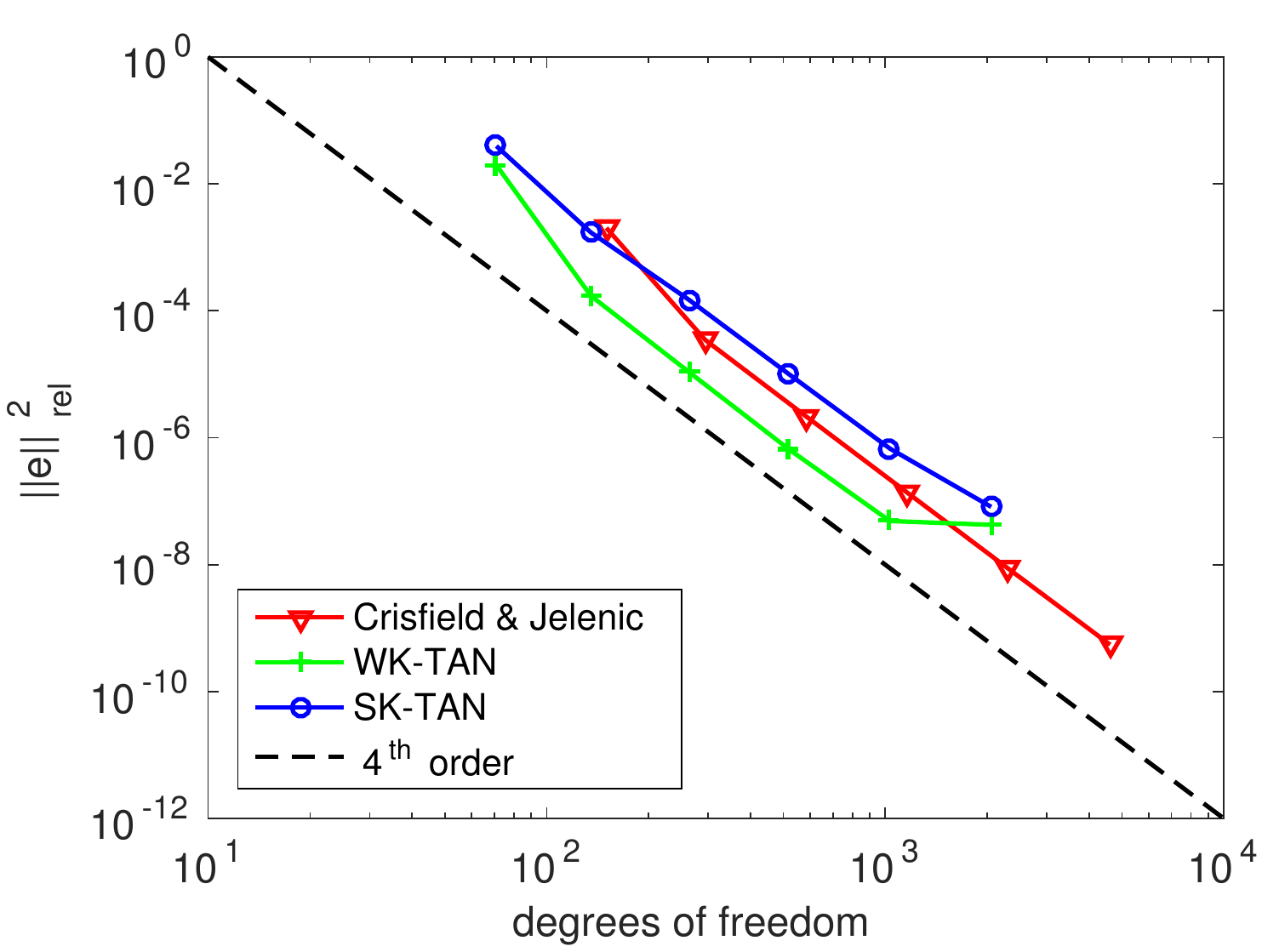}
    \label{fig:doublecircewithforce_l2_2}
   }
  \caption{Load case $\tilde{M}$+$\tilde{F}$: $L^2$-error for different element formulations.}
  \label{fig:doublecircewithforce_convergence1}
\end{figure}

\begin{figure}[t!!!]
 \centering
  \subfigure[Moderate slenderness ratio: $\zeta\!=\!100$.]
   {
    \includegraphics[width=0.48\textwidth]{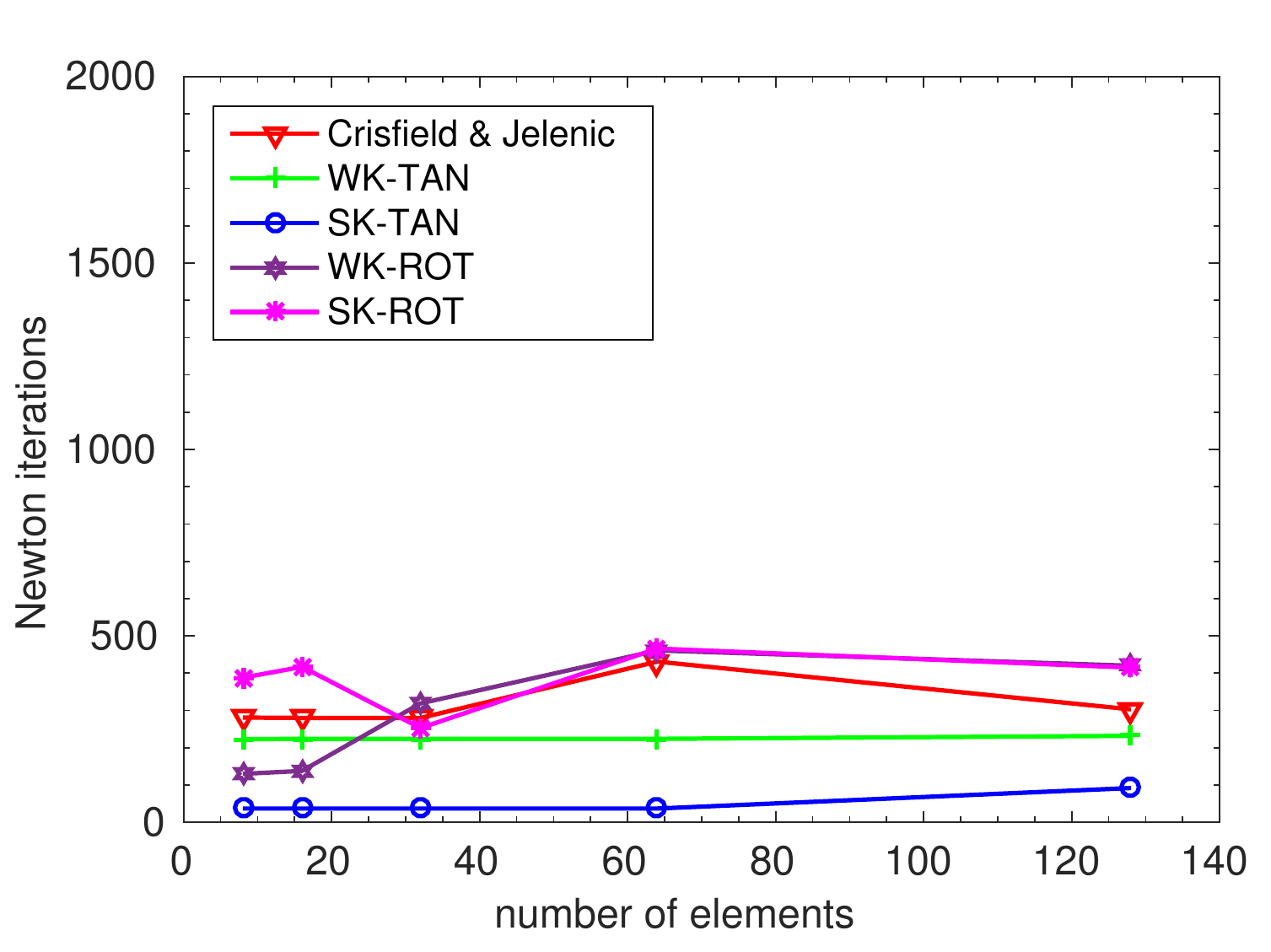}
    \label{fig:doublecircewithforce_newton1}
   }
   \subfigure[High slenderness ratio: $\zeta\!=\!10000$.]
   {
    \includegraphics[width=0.48\textwidth]{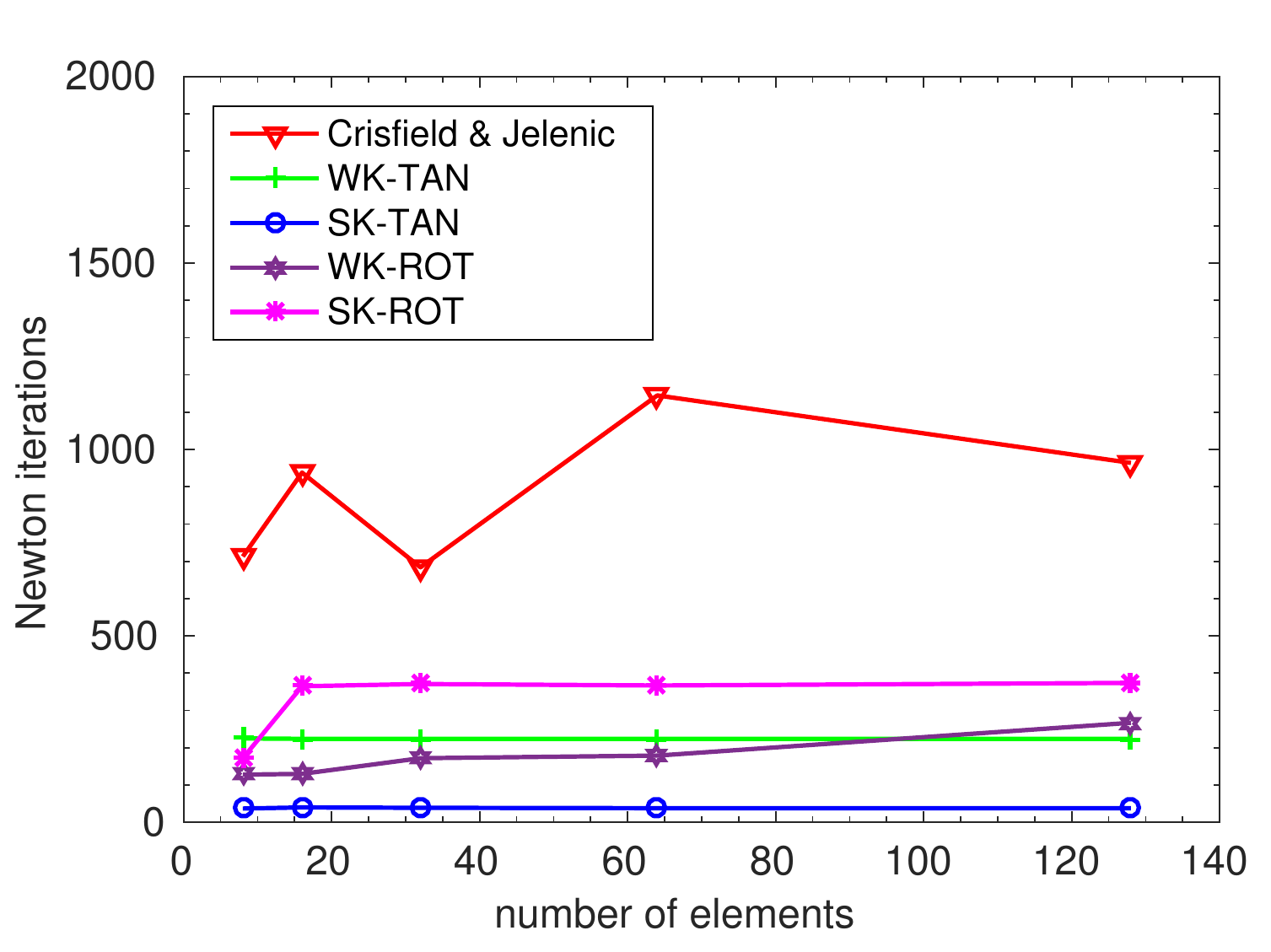}
    \label{fig:doublecircewithforce_newton2}
   }
  \caption{Load case $\tilde{M}$+$\tilde{F}$: Total number of Newton iterations.}
  \label{fig:doublecircewithforce_newton}
\end{figure}

In Figure~\ref{fig:doublecirce_l2_1}, the $L^2$-error resulting from the load case $\tilde{M}$ has been plotted for the WK-TAN and the SK-TAN element as well as for the Reissner type CJ beam element formulation proposed by Crisfield and Jelinic~\cite{crisfield1999,jelenic1999} and presented in Section~\ref{sec:elesimoreissner}. Here, discretizations with $8,16,32,64,128$ and $256$ elements have been employed. These discretizations are comparable to the load case M since similarly to that example, also here, the roughest discretization is based on \textit{one} finite element per $90^{\circ}$-arc segment of the analytic solution. In order to enable a reasonable comparison of different element formulations, here and in the following, the discretization error will be plotted over the total number of degrees of freedom resulting from the respective finite element discretization. Since no shear deformation is present for this example, the Reissner and  Kirchhoff type elements converge towards the same analytic solution. All element formulations exhibit the expected optimal convergence order of four, indicated by the black dashed line. The WK-TAN element shows the expected result that Kirchhoff element formulations can represent the same discretization error level with less degrees of freedom as compared to the Reissner type element formulation (see Section~\ref{sec:reissnermotivationshearfree}). Furthermore, for this example, it can even be shown that the lines representing the discretization error of the WK-TAN element and of the CJ element formulation would be almost identical if the discretization error was plotted solely over the degrees of freedom associated with the centerline interpolation. Thus, the observable difference in Figure~\ref{fig:doublecirce_l2_1} is a pure result of the additional rotational degrees of freedom required for Reissner type element formulations in order to represent shear deformation. Such a behavior is expected for this pure bending example since the two considered element formulations can exactly represent the internal energy associated with a pure bending state (see Sections~\ref{sec:elesimoreissner_spatialdiscretization_locking} and~\ref{sec:eleweakkirchhoff_spatialdiscretization_locking}). Consequently, the discretization error contribution stemming from the second term in~\eqref{convergenceorder} vanishes, the finite element problem degenerates to a pure problem of polynomial curve approximation represented by the first term in~\eqref{convergenceorder} and, thus, the discretization error plotted over the number of centerline DoFs yields similar results for the Lagrange centerline interpolation of the Reissner type element and the Hermite centerline interpolation of the Kirchhoff type element. The situation is completely different for the SK-TAN element, which cannot exactly represent the internal energy associated with a pure bending state. A closer investigation would confirm the expected result that the SK-TAN element exhibits a remaining error in the length-specific hyperelastic stored energy~\eqref{storedenergyfunctionkirchhoff}, which is more or less constant along the beam length. Based on this finding, it can easily be answered why the discretization error level of the SK-TAN element applied to the load case $\tilde{M}$ is considerably increased as compared to the first load case M (while an identical level of the length-specific $L^2$-error has been observed for the WK-TAN and the CJ element): By the FEM, solely the distribution of the second centerline derivative $\mb{r}^{\prime \prime}$ is optimized in order to yield a minimal energy error within the beam domain $\Omega_l$, while the centerline field $\mb{r}$ itself is only constrained at the clamped end of the beam. Thus, with increasing distance from the clamped end, the discretization error in the centerline field $\mb{r}$, resulting from a two-fold integration of the (more or less) constant error in the second derivative $\mb{r}^{\prime \prime}$ along an increasing arc-length segment, also increases. Consequently, by assuming comparable errors in the length-specific energy for comparable discretizations (i.e. the same number of finite elements representing the same angle segment of the analytic solution), a higher length-specific discretization error is expected for the load case $\tilde{M}$ as compared to the load case M. Figure~\ref{fig:doublecirce_l2_2} confirms the expected result that the energy error of the SK-TAN element does not vanish for this example and exhibits a convergence order of four. Furthermore, it is shown that the length-specific energy error "averaged" along the entire beam length is identical to the length-specific energy error "averaged" only along the first eighth  of the beam (representing a quarter circle). As consequence of the error accumulation described above, the length-specific $L^2$-error is lower (and similar to  load case M) if it is only "averaged" along the first eighth of the beam (see Figure~\ref{fig:doublecirce_l2_1}).
\begin{figure}[t!!!]
 \centering
  \subfigure[Reference: Analytic, third-order elements.]
   {
    \includegraphics[width=0.48\textwidth]{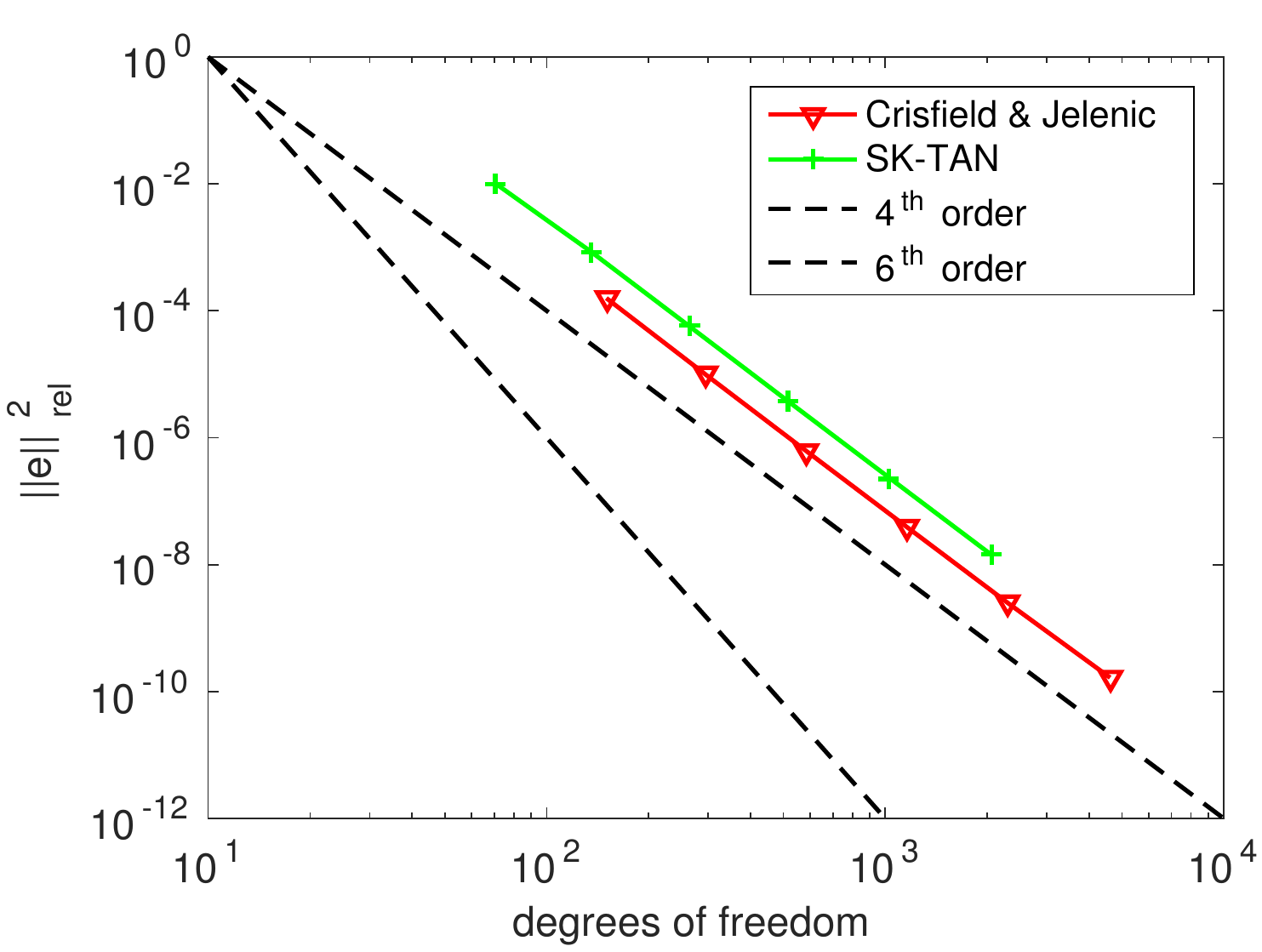}
    \label{fig:doublecirce_l2_3}
   }
   \subfigure[Reference: Analytic, fifth-order elements.]
   {
    \includegraphics[width=0.48\textwidth]{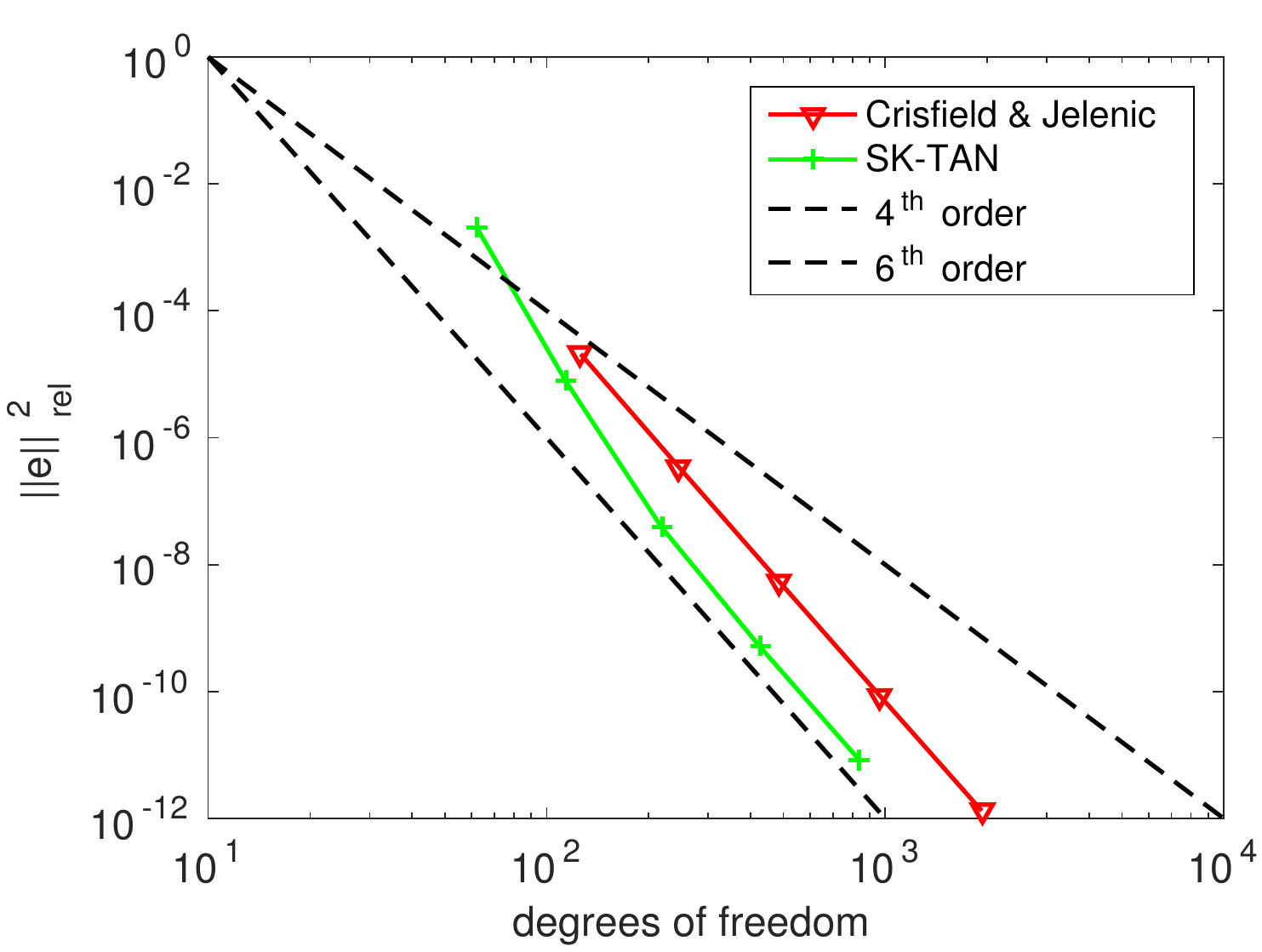}
    \label{fig:doublecirce_l2_4}
   }
  \caption{Straight beam and load case $\tilde{M}$: convergence of third- and fifth-order elements.}
  \label{fig:doublecircle_convergence2}
\end{figure}
From a rather mathematical point of view, the increased discretization error level of the SK-TAN element can be explained by a high level of the second, energy-related term in~\eqref{convergenceorder} that dominates the overall discretization error. This behavior, in turn, is a pure consequence of the fact that the two exponents $k\!+\!1$ and $2(k\!-\!m\!+\!1)$ in~\eqref{convergenceorder} are identical for trial functions of polynomial degree $k\!=\!3$. However, for polynomial degrees $k\!>\!3$, the second term in~\eqref{convergenceorder} is expected to converge with a higher rate and consequently, for sufficiently fine discretizations, the first term reflecting the pure polynomial approximation power will determine the overall discretization error level. In this range, a lower discretization error per DoF can be expected for the Kirchhoff type beam element formulations independently of the beam length, the complexity of the deformation state or the type of boundary conditions. For a first proof of principle, in Figure~\ref{fig:doublecirce_l2_4}, the $L^2$-error resulting from the CJ element with fifth-order Lagrange interpolation as well as from the SK-TAN element based on a fifth-order Hermite interpolation are depicted: While for very rough discretizations, the energy-related error contribution with higher convergence rate still seems to dominated the overall discretization error of the SK-TAN element, the expected optimal gap between the Reissner discretization error (higher level) and the Kirchhoff discretization error (lower level) can be observed. For comparison reasons, in Figure~\ref{fig:doublecirce_l2_3}, the results of the corresponding third-order variants of Figure~\ref{fig:doublecirce_l2_1} are repeated. Since this work focuses on the development of third-order Kirchhoff beam elements, no further details on the construction of higher-order Hermite polynomials (either by introducing additional nodes or by considering higher-order derivatives) will be given at this point. However, it is expected that a comparable behavior as illustrated in Figure~\ref{fig:doublecirce_l2_4}, can also be achieved for the other test cases considered in the following, if fifth-order SK and WK elements are employed. A detailed investigation of general geometrically exact beam element formulations of Kirchhoff-Love type with polynomial degree $k\!>\!3$ will be considered in future research work. Eventually, also the convergence behavior of the fourth load case shall be investigated. In Figures~\ref{fig:doublecircewithforce_l2_1} and~\ref{fig:doublecircewithforce_l2_2} the $L^2$-error of the CJ, the WK-TAN and the SK-TAN element is plotted for the two beam slenderness ratios $\zeta\!=\!100$ and $\zeta\!=\!10000$. Since no closed-form analytic solution has been available for this example, a numerical reference solution based on the element formulation of Crisfield and Jeleni\'{c} has been employed. As a consequence of shear deformation induced by the tip force $\tilde{F}$, the result derived from the Reissner and Kirchhoff type beam element formulations will differ in the limit of very fine discretizations $h \rightarrow 0$. This "model error" of the shear-free Kirchhoff elements becomes visible in form of a kink in the convergence diagram and a certain cutoff error level that remains constant even for arbitrarily fine discretizations. As expected, the model difference between the Simo-Reissner and the Kirchhoff-Love beam theory decreases with increasing beam slenderness ratio, a property that is reflected by a lower cutoff error level for the higher slenderness ratio $\zeta\!=\!10000$. For the lower slenderness ratio $\zeta\!=\!100$, the relative error distinguishing the Kirchhoff from the Reissner solution lies below $10^{-3}$, which can be assumed as reasonable approximation for many engineering applications. For the high slenderness ratio $\zeta\!=\!10000$, the relative error between these two models is smaller than $10^{-7}$. For the investigated cases of $\zeta\!=\!100$ and $\zeta\!=\!10000$, the cutoff error scales almost quadratically with the slenderness ratio, which would be the expected result for the solution of the geometrically linear theory. This result is remarkable for this highly nonlinear example.
From Figures~\ref{fig:doublecircewithforce_l2_1} and~\ref{fig:doublecircewithforce_l2_2}, it can again be observed that all element formulations exhibit the expected convergence rate of four, that the error level of the SK-TAN element lies slightly above and that the error level of the WK-TAN element lies below the error level of the CJ element. Despite the fact that the Simo-Reissner formulations yield the more general solutions, which also contain the effects of shear deformation, the Kirchhoff type WK-TAN element formulation will be considered as numerical reference solution for all remaining examples throughout this work. Such a procedure seems to be sensible, since within this contribution, the convergence behavior of the Kirchhoff type elements and not of the Reissner type elements shall be studied. Nevertheless, the model error of the Kirchhoff type beam elements is still observable, but this time in form of a kink and a remaining cutoff error level in the convergence plots of the Reissner type formulation.
\begin{figure}[t!!!]
 \centering
   \subfigure[Reference: WK-TAN, $\zeta\!=\!100$.]
   {
    \includegraphics[width=0.48\textwidth]{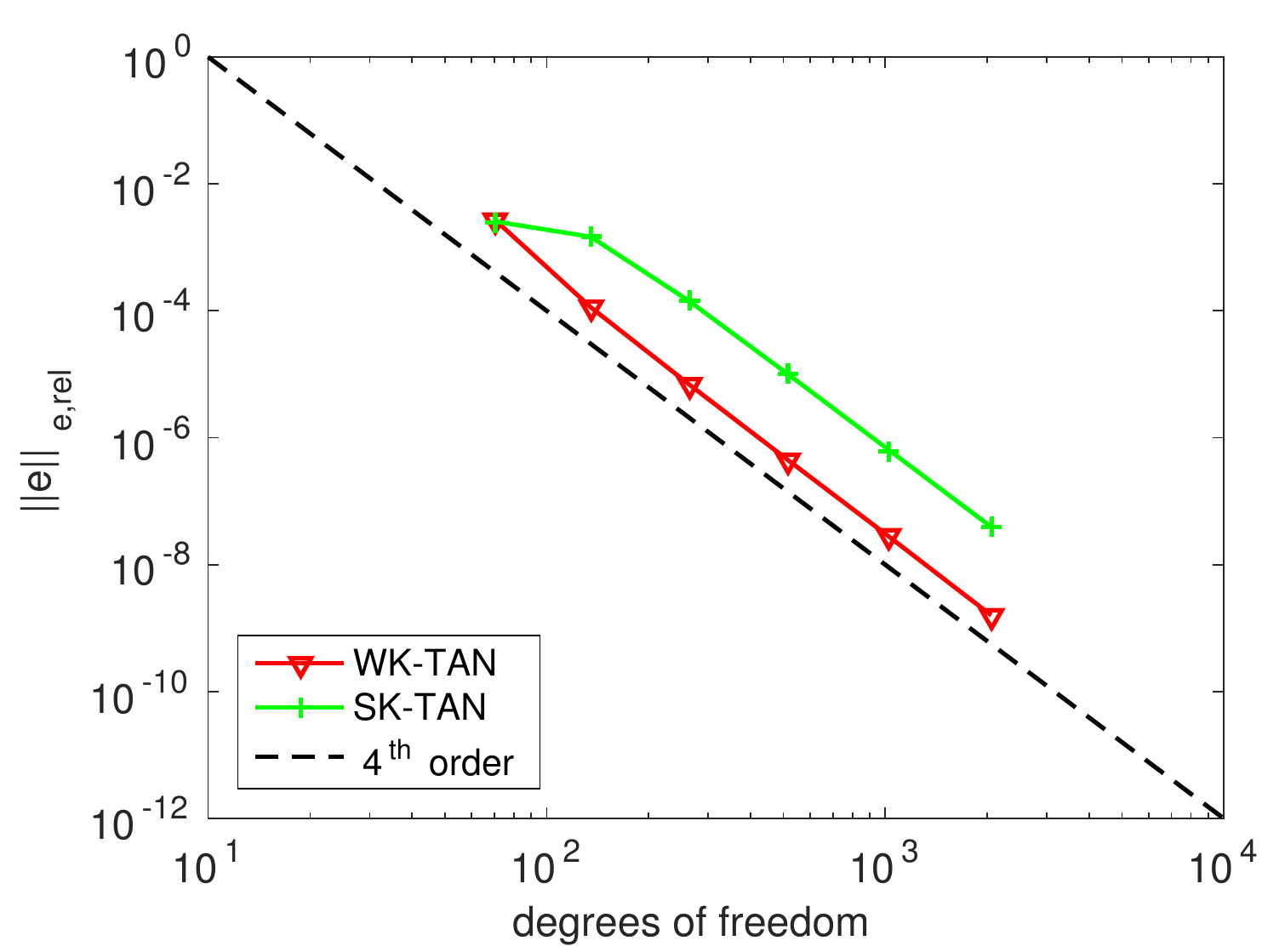}
    \label{fig:doublecircewithforce_e2_1}
   }
   \subfigure[Reference: WK-TAN, $\zeta\!=\!10000$.]
   {
    \includegraphics[width=0.48\textwidth]{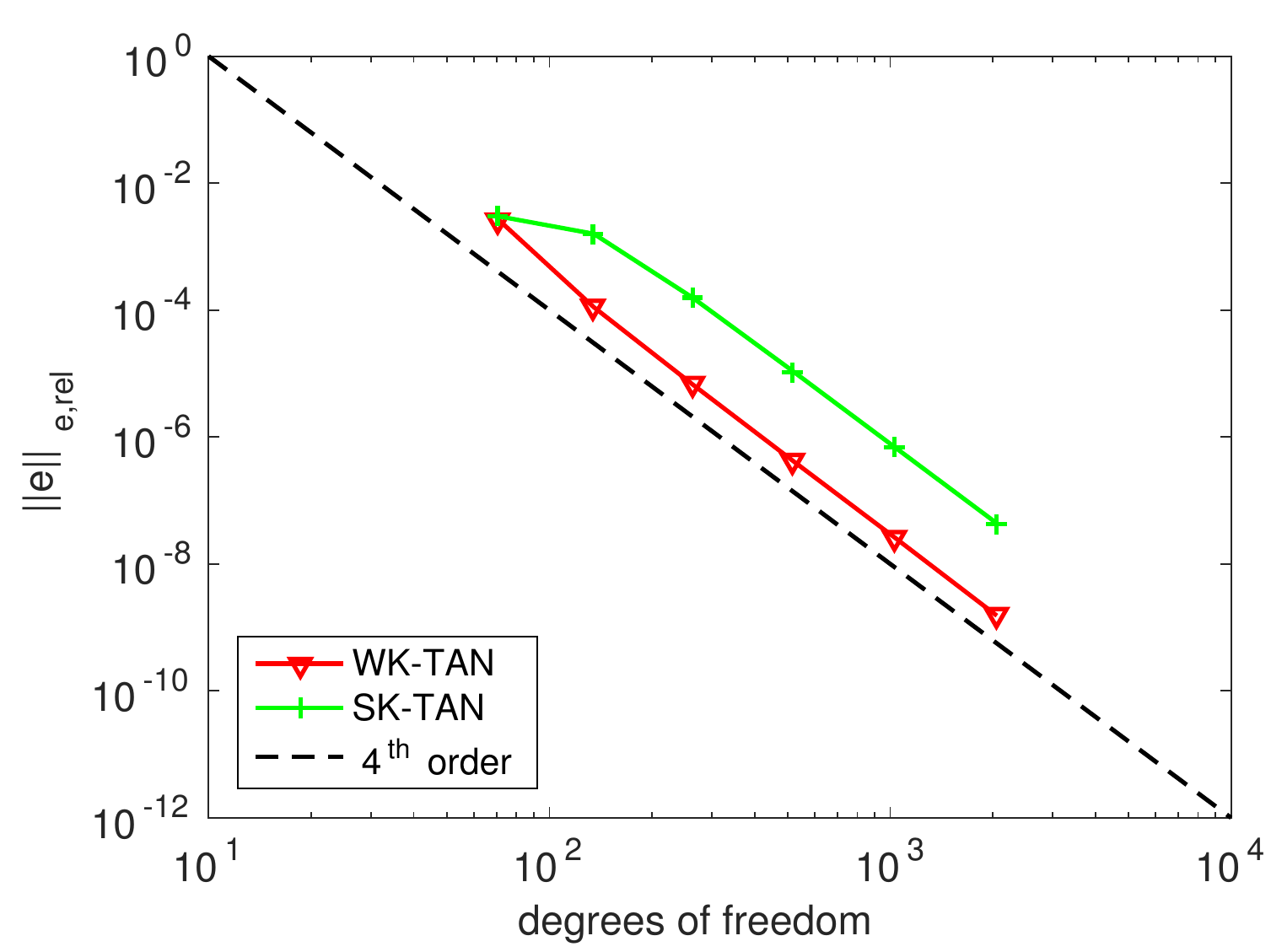}
    \label{fig:doublecircewithforce_e2_2}
   }
  \caption{Load case $\tilde{M}$+$\tilde{F}$: Energy-error for different element formulations.}
  \label{fig:doublecircewithforce_convergence2}
\end{figure}
While the WK-TAN element was able to exactly represent the internal energy of the load case $\tilde{M}$, now, for the load case $\tilde{M}$+$\tilde{F}$, the energy convergence of the formulations WK-TAN and SK-TAN can be compared (see Figures~\ref{fig:doublecircewithforce_e2_1} and~\ref{fig:doublecircewithforce_e2_2} for the two slenderness ratios $\zeta\!=\!100$ and $\zeta\!=\!10000$). Both element formulations exhibit the expected convergence order of four, and similar to the load case $\tilde{M}$, the WK-TAN element yields a better approximation of the internal energy than the SK-TAN element. This is observable in form of a lower energy error level in Figures~\ref{fig:doublecircewithforce_e2_1} and~\ref{fig:doublecircewithforce_e2_2} and the reason for the lower $L^2$-error level visible in Figures~\ref{fig:doublecircewithforce_l2_1} and~\ref{fig:doublecircewithforce_l2_2}. A possible explanation for the better performance of the WK-TAN element may be found by considering the interaction of the employed translational and rotational interpolation schemes: As shown in Section~\ref{sec:eleweakkirchhoff_spatialdiscretization_locking}, the number of unknowns equals the number of equations required for the WK-TAN element to (energetically) represent a pure bending state. This means that an exact representation of the internal energy associated with states of constant axial tension, bending curvature and torsion is possible. This property does not hold for the SK-TAN element (see e.g. Section~\ref{sec:elestrongkirchhoff_spatialdiscretization_locking}), where the corresponding system of equations that has to be fulfilled for representing a pure bending state is slightly over-constrained. While the difference between the WK and the SK elements in the $L^2$-error level is expected to vanish with higher-order trial functions, such an effect can in general not be predicted for the energy error. Eventually, it has to be mentioned that for both the $L^2$-error as well as the energy error plots of the load case $\tilde{M}$+$\tilde{F}$, no difference between the slenderness ratios $\zeta\!=\!100$ and $\zeta\!=\!10000$ is evident, which again underlines the successful avoidance of membrane locking and the effectiveness of the MCS method. Finally, also the performance of the Newton-Raphson scheme shall be investigated and compared between Kirchhoff and Reissner type element formulations (see Figure~\ref{fig:doublecircewithforce_newton}). Since the computationally expensive steps of solving a nonlinear system of equations and evaluating the tangent stiffness matrix have to be conducted in every Newton iteration, a reduction in the total number of Newton iterations $n_{iter,tot}$ as defined in~\eqref{loadstepadaption2} would considerably increase the overall efficiency of the numerical algorithm. In Figures~\ref{fig:doublecircewithforce_newton1} and~\ref{fig:doublecircewithforce_newton2}, the total number of Newton iterations $n_{iter,tot}$  of the load case $\tilde{M}$+$\tilde{F}$ in combination with slenderness ratios of $\zeta\!=\!100$ and $\zeta\!=\!10000$ has been plotted for the element formulations CJ, WK-TAN, SK-TAN, WK-ROT and SK-ROT and different spatial discretizations. While the final FEM solutions have been shown to be independent from the choice of nodal rotation parametrization, the number of Newton iterations required for the SK/WK-ROT and SK/WK-TAN variants might differ considerably. Therefore, also the Newton performance of these variants has been investigated. For solving the highly nonlinear beam problem, the load step adaption scheme mentioned above based on an initial number of $N_0\!=\!2$ load steps has been employed.
By comparing Figures~\ref{fig:doublecircewithforce_newton1} and~\ref{fig:doublecircewithforce_newton2}, one realizes that the Newton performance of the Kirchhoff type element formulations is rarely influenced by the considered slenderness ratio, while the number of Newton iterations required by the Reissner element increases drastically with increasing slenderness ratio. Furthermore, it seems that the SK/WK-TAN variants require fewer Newton iterations than the SK/WK-ROT variants. These trends will be confirmed, and even more pronounced, in more general 3D examples presented in subsequent sections. In summary, the following conclusions can be drawn from the 2D examples considered in this section: 1) The proposed Kirchhoff elements yield accurate results with acceptable model errors for slenderness ratios of $\zeta\!\geq\!100$ and a model error that decreases quadratically with increasing beam slenderness ratio. 2) The expected convergence orders of four in the $L^2$- as well as in the energy error could be confirmed for all investigated Kirchhoff elements. 3) In combination with the MCS method, none of the considered element formulations exhibited an influence of the element slenderness ratio on the resulting discretization error. This result confirms the effectiveness of the MCS method in the avoidance of membrane locking. 4) The discretization error level of the WK-TAN element lies below the error level of the Reissner type element and also below the error level of the SK-TAN element. The increased error level of the SK-TAN element has been shown to vanish with higher polynomial degree $k\!>\!3$ of the trial functions. 5) While the total number of Newton iterations required by the Reissner type element formulations considerably increases with increasing beam slenderness ratio, the number of iterations remains more or less constant for the Kirchhoff type formulations. These conclusions drawn from these 2D tests will be confirmed by the 3Dl examples investigated in the following sections.

\subsection{Example 3: Pure bending in 3D}
\label{sec:examples_purebending3d}

\begin{figure}[ht]
 \centering
  \subfigure[Initial and deformed geometry]
   {
     \includegraphics[height=0.35\textwidth]{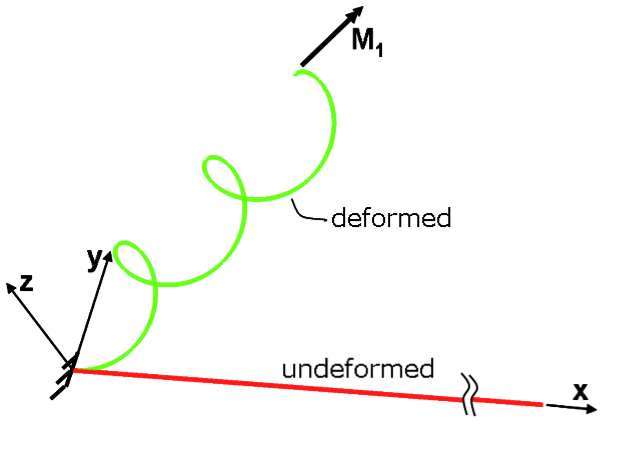}
    \label{fig:straighttohelix_1}
   }
   \subfigure[reative $L^2$-error, reference: analytic.]
   {
    \includegraphics[width=0.48\textwidth]{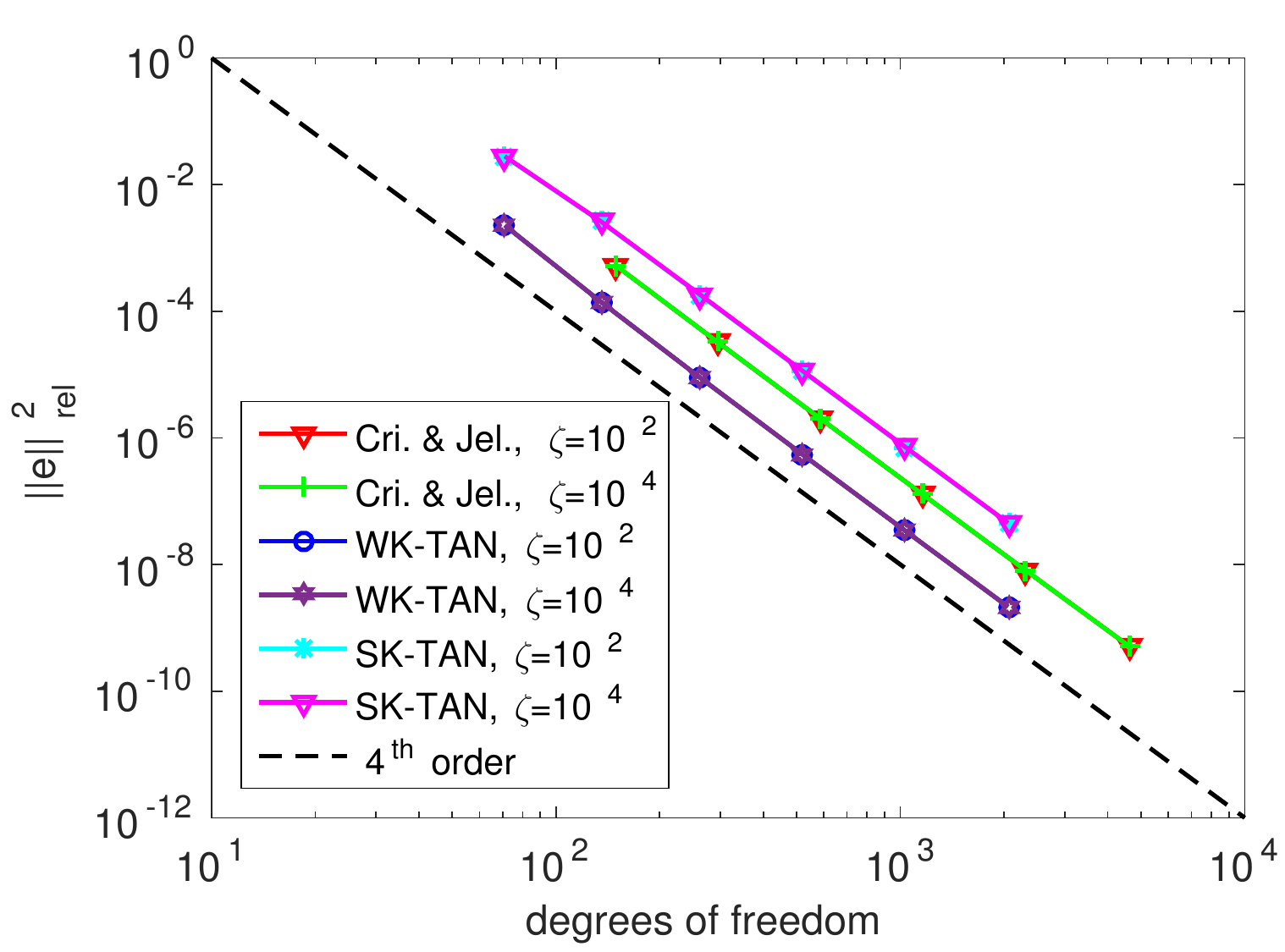}
    \label{fig:convergence_straighttohelix_1}
   }
   \caption{Straight beam loaded with a 3-dimensional discrete end-moment.}
  \label{fig:convergence_straighttohelix}
\end{figure}

In this section, the 3D extension of the pure bending examples (load cases M and $\tilde{M}$) of the last section will be considered. Again, the focus lies on an initially straight, clamped beam of standard length $l\!=\!1000$ investigated for the two slenderness ratios $\zeta\!=\!100$ and $\zeta\!=\!10000$. However, this time, the beam is loaded by a 3D end-moment $\mb{M}_1\!:=\!(M,0,M)^T$, with $M \!=\! 10$ for $\zeta\!=\!100$ and $M \!=\! 10^{-7}$ for $\zeta\!=\!10000$, which contains an additional moment component in beam length direction inducing torsion. The initial and deformed configuration are illustrated in Figure~\ref{fig:straighttohelix_1}.\\

\begin{figure}[t!!!]
 \centering
  \subfigure[Moderate slenderness ratio: $\zeta=100$.]
   {
    \includegraphics[width=0.48\textwidth]{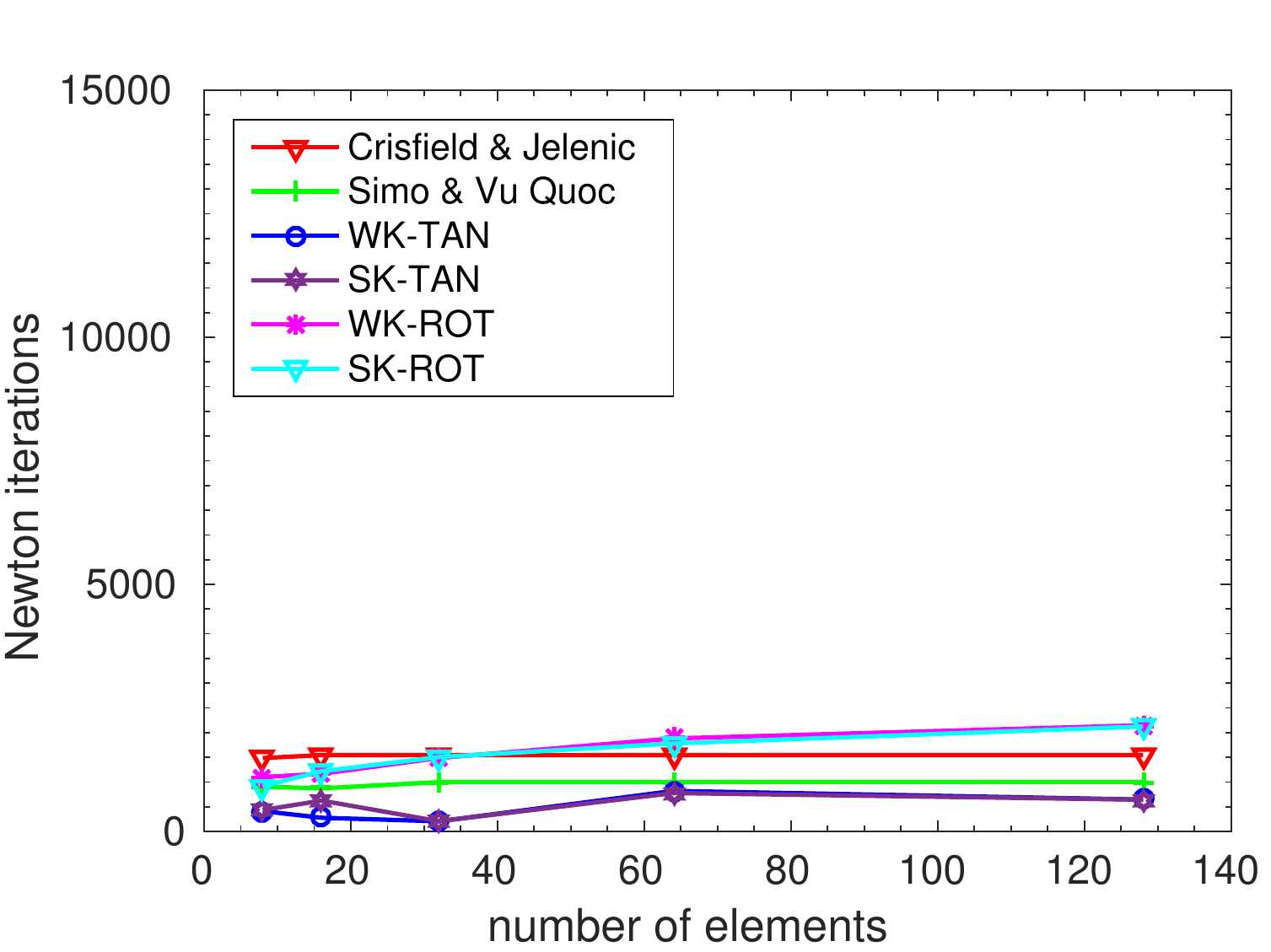}
    \label{fig:straighttohelix_newton1}
   }
   \subfigure[High slenderness ratio: $\zeta=10000$.]
   {
    \includegraphics[width=0.48\textwidth]{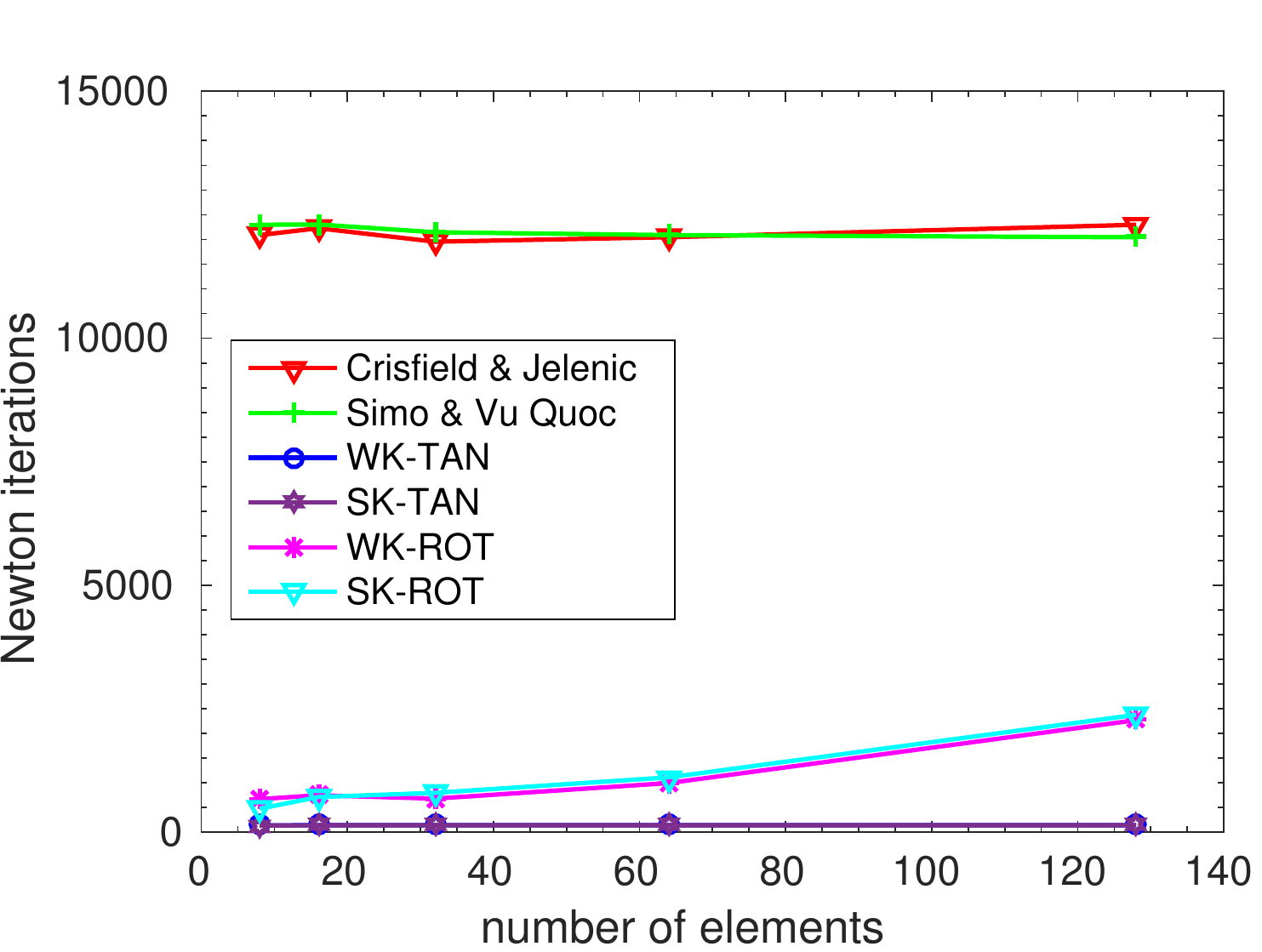}
    \label{fig:straighttohelix_newton2}
   }
  \caption{Load case ''straight to helix``: Total number of Newton iterations.}
  \label{fig:straighttohelix_newton}
\end{figure}

As argued in~\cite{meier2015}, the analytic solution of this example is given by the following space curve representation:
\begin{align}
\label{r_analyt_sth}
  \mb{r}(s) = R_0
  \left(
   \begin{array}{c}
   \frac{1}{\sqrt{2}} \left(\sin{\beta} + \beta \right) \\
    \left(1-\cos{\beta}\right)\\ 
    \frac{1}{\sqrt{2}} \left( \beta -\sin{\beta} \right)
   \end{array}
   \right)
   \quad \text{with} \quad
   R_0 = \frac{EI}{2M}
   \quad \text{and} \quad
   \beta = \frac{s}{\sqrt{2} R_0}.
\end{align}
This solution represents a helix whose mid-axis points into the $(1,0,1)^T$-direction, viz. in the direction of the applied external moment. The special parameter choice of this example leads to a radius $R_0$ of the enveloping cylinder that is identical to the slope of the helix. In Figure \ref{fig:convergence_straighttohelix_1}, the relative $L^2$-error resulting from the two investigated slenderness ratios is plotted for the element formulations CJ, WK-TAN and SK-TAN as well as spatial discretizations based on $8,16,32,64,128$ and  $256$ elements. Again, all element formulations exhibit the expected convergence order of four and the discretization error level of the WK-TAN element is lower, whereas the discretization error of the SK-TAN element is slightly higher than for the CJ element. Furthermore, no visible difference can be observed between the discretization error levels associated with the two different slenderness ratios. Due to the choice $GI_T\!=\!EI_2\!=\!EI_3$, it can easily be verified that this example results in an analytic solution exhibiting vanishing axial tension and shear deformation as well as a constant spatial and material curvature vector along the entire beam pointing into the direction of the external moment vector, i.e. $\mb{k}\!=\!\mb{K}\!=\!M/(EI)(1,0,1)^T\!=$ const. Thus, already the roughest discretizations of the CJ and WK-TAN elements can exactly represent the hyperelastic stored energy function for this pure bending case, which can be interpreted as a simple 3D patch test for geometrically exact beams. Finally, also the number of Newton iterations shall be investigated (see Figure~\ref{fig:straighttohelix_newton}). In order to enable more general conclusions, this time, a second Reissner type beam element formulation, which is based on a completely different triad interpolation scheme, has additionally been included in the comparison. Concretely, this element represents an "interpretation" formulated by Crisfield~\cite{crisfield1997a} (see Chapter~17.2) of the original variant proposed by Simo and Vu-Quoc~\cite{simo1986}, in the following denoted as SV element. This time, the load step adaption scheme presented above based on an initial number of $N_0\!=\!10$ load steps has been employed.
Similarly to the 2D case considered in the last section, the Newton performance of the Reissner type element formulations drastically deteriorates with increasing slenderness ratio whereas the performance of the Kirchhoff type elements remains unchanged (or is even slightly improved in case of the WK/SK-TAN variants).  Concretely, for the slenderness ratio $\zeta\!=\!10000$, all investigated discretizations of the WK/SK-TAN elements exhibit a remarkably constant number of $n_{iter,tot}\!=\!140\!\pm\!4$ iterations, while the total number of iterations required by the WK/SK-ROT elements increases from $n_{iter,tot}\!\approx\!500$ to $n_{iter,tot}\!\approx\!2000$ with increasing number of elements and the total number of iterations required by the Reissner type formulations is almost by two orders of magnitude higher than for the WK/SK-TAN elements and lies constantly above a value of $n_{iter,tot}\!>\!12000$. Seemingly, the considerable difference between the WK/SK-TAN elements and the Reissner type element formulations can be attributed to two different effects: Firstly, the parametrization of nodal triads via tangent vectors seems to be more good-natured than the parametrization based on nodal rotation vectors. This effect already gets visible as difference between the WK/SK-TAN and the WK/SK-ROT variants and seems to be more or less independent from the beam slenderness ratio. Secondly, the high stiffness contributions resulting from the shear mode seem to considerably deteriorate the Newton convergence in the range of high slenderness ratios: This effect becomes obvious as difference between the WK/SK-ROT elements and the Reissner type elements. The linearizations of all of these four elements are based on multiplicative updates of nodal rotation vectors. This observation is emphasized by the two elements types WK-ROT and CJ, which additionally exhibit the same triad interpolation. These two elements only differ in the centerline interpolation (based on Lagrange or Hermite polynomials), which is not expected to influence the Newton convergence in such a drastic manner, and the fact that the WK-ROT element additionally enforces the constraint of vanishing shear strains. Consequently, the avoidance of shear modes seems to be the main reason for the considerably improved performance of the Kirchhoff type element formulations. Finally, the observation that the total number of Newton iterations required by the WK/SK-ROT variants increases with increasing number of elements is only of secondary practical interest since the discretizations relevant for practical applications are located in the range of small element numbers (on the left of Figure~\ref{fig:straighttohelix_newton2}). The observations made so far, will be confirmed by the subsequent 3D examples.\\

\hspace{0.2 cm}
\begin{minipage}{15.0 cm}
\textbf{Remark:} For some of the discretizations investigated in Figure~\ref{fig:straighttohelix_newton}, solutions of the SK-TAN element could already be found in one load step. However, since in these cases no convergence could be achieved for simulations based on two or three load steps, the solution of the problem by means of one load step can rather be regarded as a "lucky shot" than as a representative convergence behavior. In order to avoid a biased comparison resulting from such effects, the initial number of load steps has been increased to $N_0\!=\!10$. By this means and the mentioned load step adaption scheme, an evaluation and comparison process is intended that is as fair and objective as possible. Nevertheless, this example shows that an absolute statement concerning the robustness of the nonlinear solution scheme based on a single example / discretization and a deliberately chosen "good-natured" step size, as sometimes done in the literature, is questionable. Here, the degree of arbitrariness is intended to be minimized by employing an automated scheme for determining the optimal load step size, by comparing the results of different test cases, different discretizations, different element types (here Reissner and Kirchhoff type beam elements) as well as different representatives for each element type. In order to avoid biased results as consequence of incorrect linearizations, the results of the Reissner type elements derived on the basis of an analytic representation of the consistent tangent stiffness matrix have been verified by simulations on the basis of a consistent tangent stiffness matrix derived via an automatic differentiation tool.\\
\end{minipage}

\subsection{Example 4: Verification of path-independence}
\label{sec:example_pathindepence}

\begin{figure}[ht]
    \centering
    \includegraphics[width=0.7\textwidth]{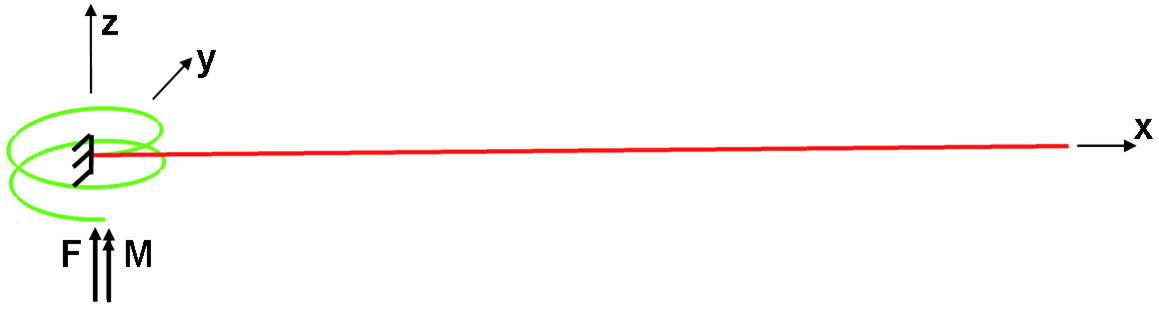}
    \caption{Problem setup: Initially straight beam bent by a discrete end-moment and -force.}
    \label{fig:path1}
\end{figure}

In Section~\ref{sec:spatialdiscretization_objectivity}, the fundamental property of objectivity has already been verified for the proposed Kirchhoff beam elements. In this section, it will be shown that these element formulations are also path-independent, i.e. for beam problems whose analytic solution is independent from a specific loading path, these beam element formulations also yield a discrete solution that is independent from a specific loading path. As numerical test case for path independence, an initially straight clamped beam with initial length $l\!=\!1000$ and slenderness ratio $\zeta\!=\!100$ (thus $R\!=\!10$) is considered that is loaded by an end-moment $\mb{M}\!=\!(0,0,M)^T$, with the moment $M\!=\!4 EI \pi / l \approx 10.47$ being defined such that it exactly bends the beam into a ''double-circle``, and an additional end-force $\mb{F}\!=\!(0,0,F)^T$, with $F\!=\!0.01 \!\approx\! M / l$. Again, for comparison reasons, also the case of an increased slenderness ratio $\zeta\!=\!10000$ with correspondingly adapted loads $M\!=\!4 EI \pi / l \approx 1.047 \cdot 10^{-7}$ and $F\!=\! 10^{-10} \!\approx\! M / l$ will be investigated. The problem setup as well as the deformed
configuration for this example are shown in Figure \ref{fig:path1}. In the following, two different possibilities how to apply these tip loads are investigated: In a first load case, the moment and
the force are applied simultaneously (load case ''sim``), while, in a second load case, the moment and the force are applied successively (load case ''suc``). In the latter case, the external moment is
increased linearly from zero to $M$ within the pseudo-time interval $t\!\in\![0;0.5]$, whereas the external force is increased linearly from zero to $F$ within the pseudo-time interval $t \in [0.5;1.0]$.\\

\begin{figure}[t!!!]
 \centering
  \subfigure[Step t=0.25 (sim).]
   {
    \includegraphics[width=0.23\textwidth]{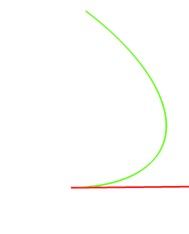}
    \label{fig:differentpaths_1}
   }
     \subfigure[Step t=0.50 (sim).]
   {
    \includegraphics[width=0.23\textwidth]{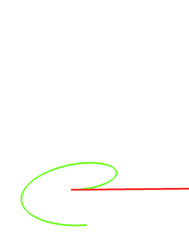}
    \label{fig:differentpaths_2}
   }
   \subfigure[Step t=0.75 (sim).]
   {
    \includegraphics[width=0.23\textwidth]{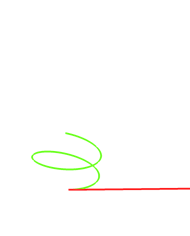}
    \label{fig:differentpaths_3}
   }
     \subfigure[Step t=1.0 (sim).]
   {
    \includegraphics[width=0.23\textwidth]{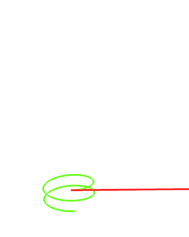}
    \label{fig:differentpaths_4}
   }
     \subfigure[Step t=0.25 (suc).]
   {
    \includegraphics[width=0.23\textwidth]{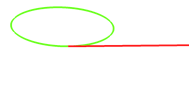}
    \label{fig:differentpaths_5}
   }
     \subfigure[Step t=0.50 (suc).]
   {
    \includegraphics[width=0.23\textwidth]{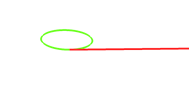}
    \label{fig:differentpaths_6}
   }
   \subfigure[Step t=0.75 (suc).]
   {
    \includegraphics[width=0.23\textwidth]{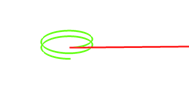}
    \label{fig:differentpaths_7}
   }
     \subfigure[Step t=1.0 (suc).]
   {
    \includegraphics[width=0.23\textwidth]{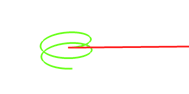}
    \label{fig:differentpaths_8}
   }
  \caption{Deformed configurations for simultaneous (sim) and successive (suc) loading of an straight beam with a moment $M$ and a force $F$.}
  \label{fig:differentpaths}
\end{figure}

\begin{figure}[b!!!]
 \centering
  \subfigure[Moderate slenderness ratio: $\zeta=100$.]
   {
    \includegraphics[width=0.48\textwidth]{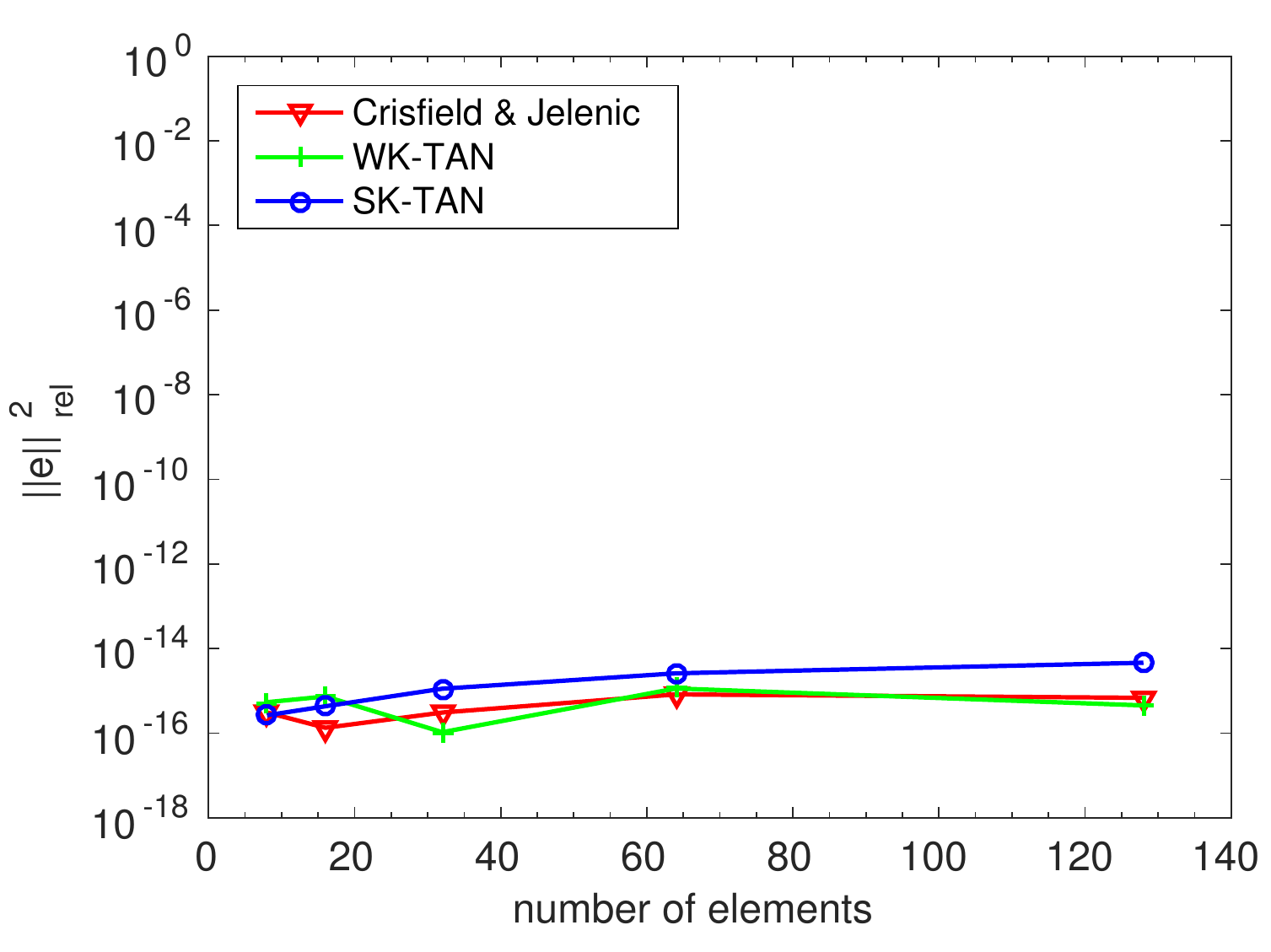}
    \label{fig:path_differenceerror_1}
   }
   \subfigure[High slenderness ratio: $\zeta=10000$.]
   {
    \includegraphics[width=0.48\textwidth]{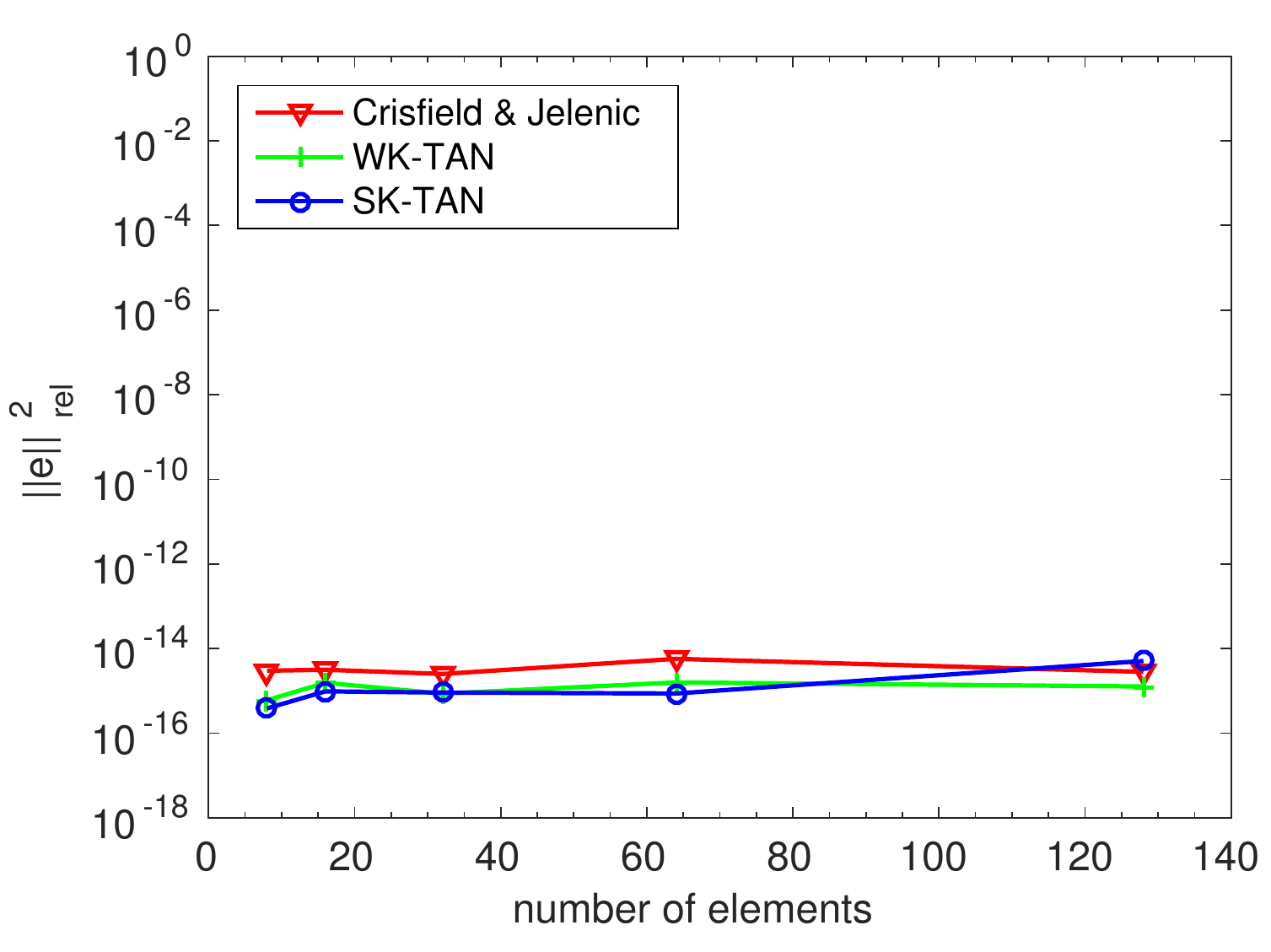}
    \label{fig:path_differenceerror_2}
   }
  \caption{$L^2$-error between the load cases "Simultaneously" and "Successively".}
  \label{fig:path_differenceerror}
\end{figure}

\begin{figure}[t!!!]
 \centering
   \subfigure[Reference: WK-TAN, $\zeta=100$.]
   {
    \includegraphics[width=0.48\textwidth]{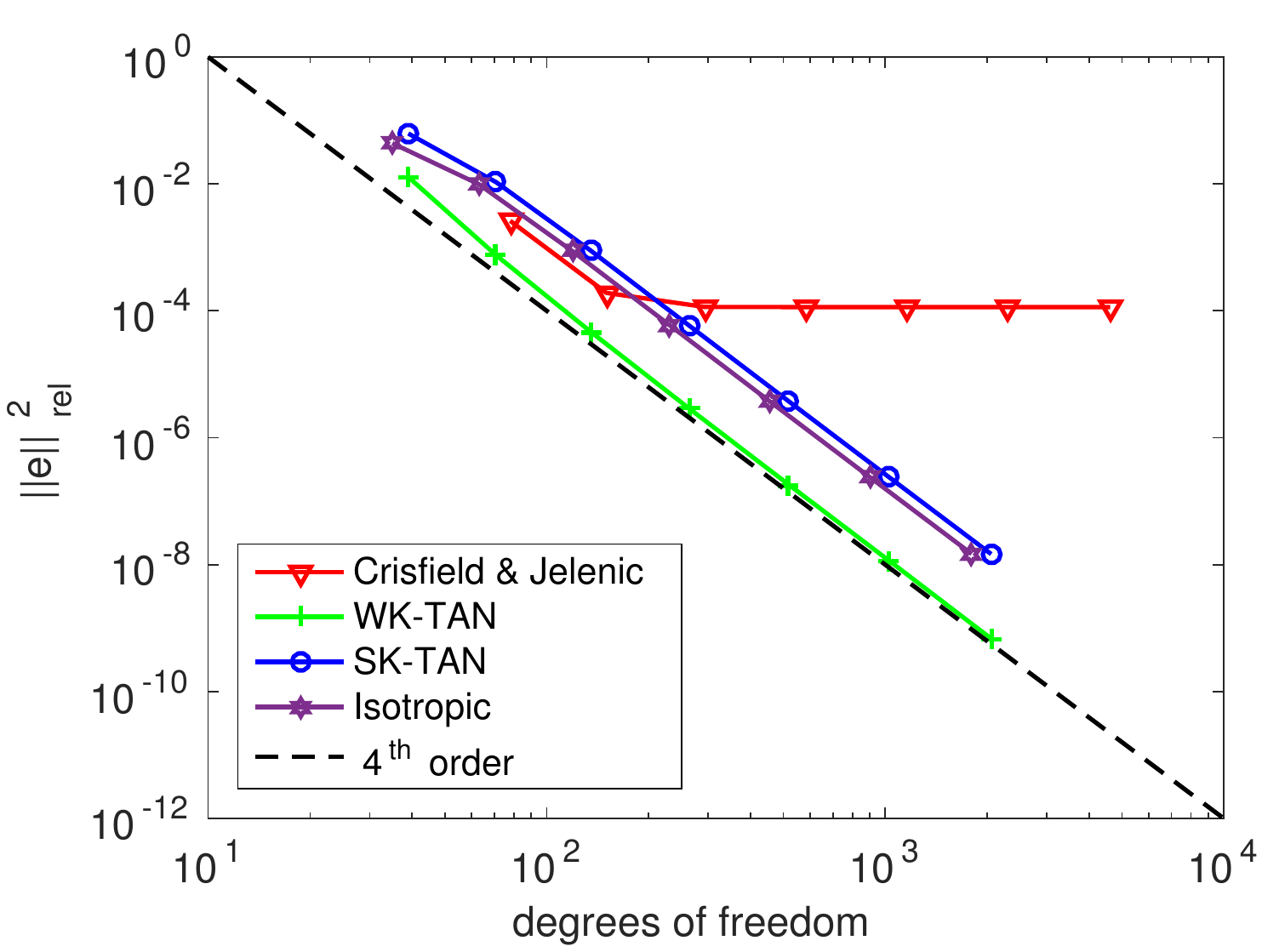}
    \label{fig:path_convergence_1}
   }
   \subfigure[Reference: WK-TAN, $\zeta=10000$.]
   {
    \includegraphics[width=0.48\textwidth]{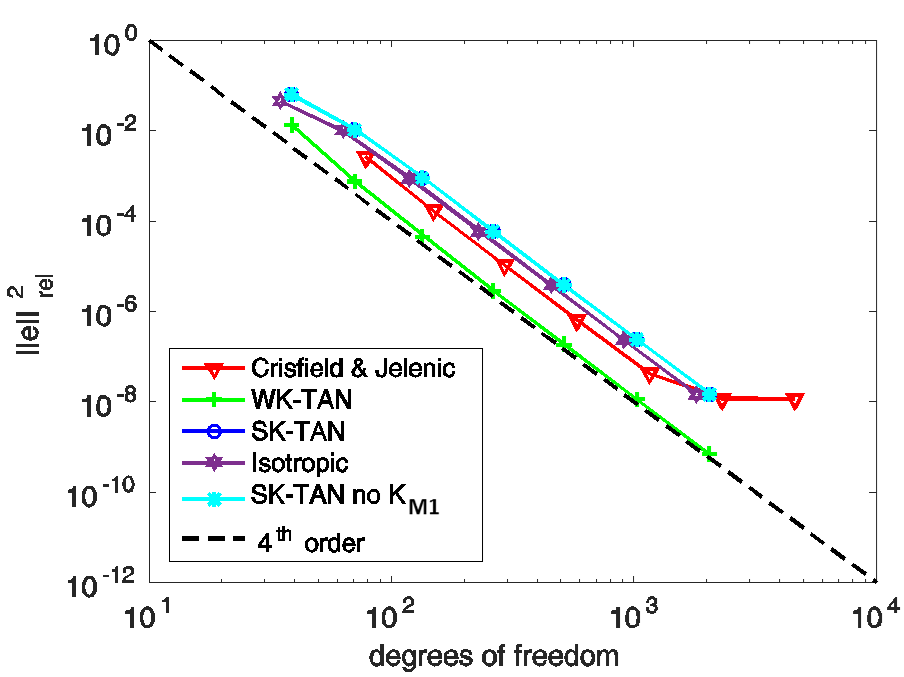}
    \label{fig:path_convergence_2}
   }
  \caption{Path independence: $L^2$-error for different element formulations.}
  \label{fig:path_convergence}
\end{figure}

\begin{figure}[h!]
 \centering
  \subfigure[Moderate slenderness ratio: $\zeta=100$.]
   {
    \includegraphics[width=0.48\textwidth]{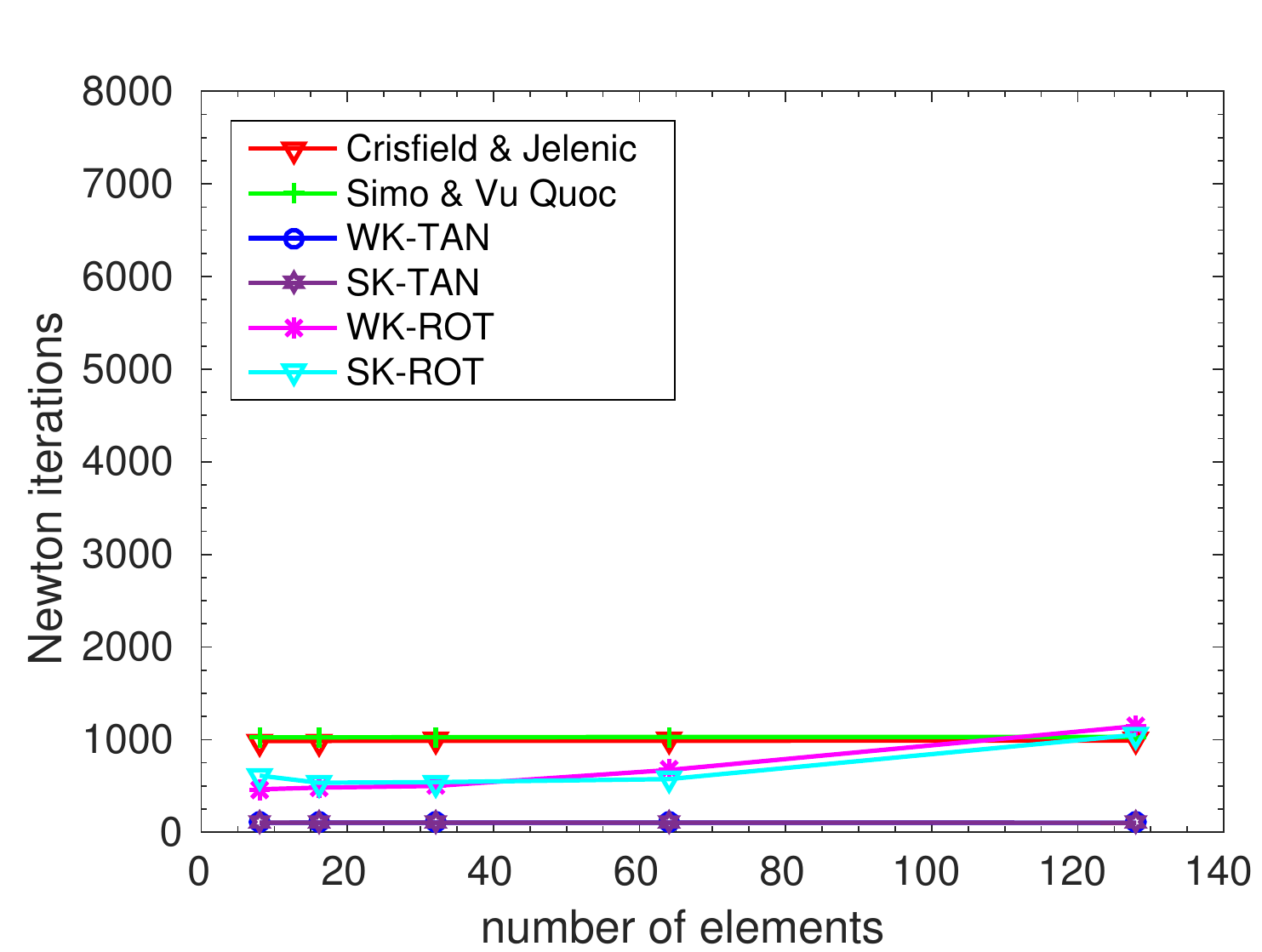}
    \label{fig:path_newton1}
   }
   \subfigure[High slenderness ratio: $\zeta=10000$.]
   {
    \includegraphics[width=0.48\textwidth]{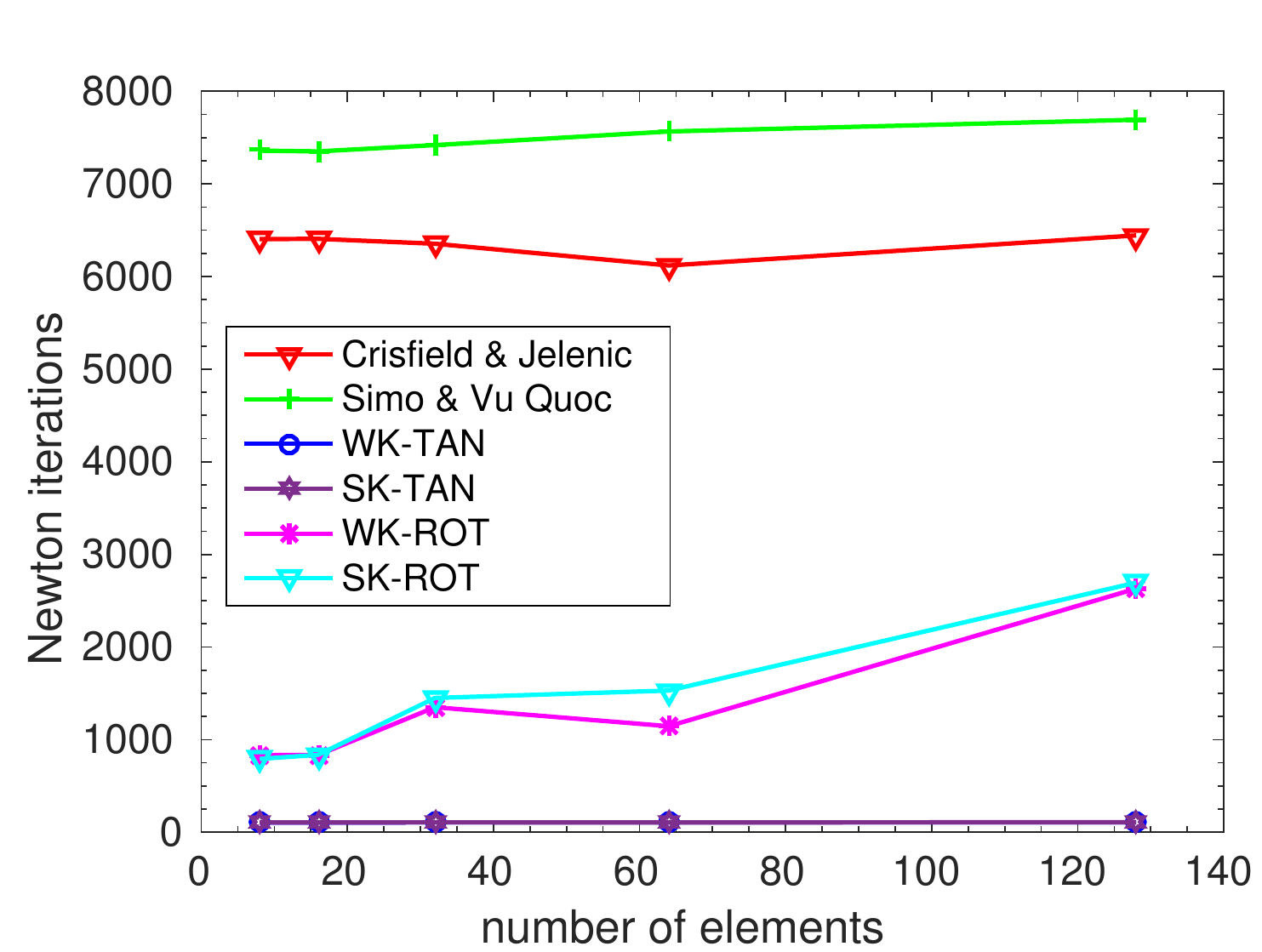}
    \label{fig:path_newton2}
   }
  \caption{Load case ''Simultaneously``: Total number of Newton iterations.}
  \label{fig:path_newton}
\end{figure}

The deformed shapes of both load cases have been plotted for the pseudo-time steps $t=0.25$, $t=0.5$, $t=0.75$ and $t=1.0$ in Figure \ref{fig:differentpaths}. Apparently, the two
load cases lead to different deformation paths, but to an identical final configuration. It contradicts intuition that this final deformed configuration
lies completely in the half space with $z\!\leq\!0$, although the tip force points into the positive $z$-direction. For the case of small forces $F$, this observation has been verified by deriving an analytical solution based on a linearization of the equilibrium equations with respect to the double-circle configuration resulting from the end-moment $M$ (see also~\cite{meier2015}). Furthermore, this observation is in agreement with the results obtained in~\cite{ibrahimbegovic1997} and~\cite{battini2002}, where a similar example based on a slightly modified parameter choice has been analyzed.
In order to investigate possible path dependence effects also in a quantitative manner, the relative $L^2$-error has been calculated between the solution
$\mb{r}_{h,suc}$ of the load case ''suc`` for a certain discretization and the solution $\mb{r}_{h,sim}$ of the load case ''sim`` for the same centerline
discretization. Thus, basically the relative $L^2$-error definition of equation \eqref{rele2error} has been applied, with $\mb{r}_h\!=\!\mb{r}_{h,suc}$ 
and $\mb{r}_{ref}\!=\!\mb{r}_{h,sim}$. The results obtained for the two different slenderness ratios and the investigated element formulations CJ, SK-TAN 
and WK-TAN are illustrated in Figure~\ref{fig:path_differenceerror}. Accordingly, for all investigated element types, discretizations and slenderness ratios, this error 
vanishes up to machine precision, which verifies the path independence of these formulations.
For completeness, Figure~\ref{fig:path_convergence} represents the discretization error resulting from the CJ, SK-TAN and WK-TAN elements for the considered slenderness ratios of $\zeta\!=\!100$ and $\zeta\!=\!10000$. There, the observations already made in earlier examples with respect to convergence rate, discretization error level and cutoff error between Kirchhoff and Reissner type element formulations are confirmed. For comparison reasons, also the $L^2$-error resulting from the reduced isotropic beam element formulation proposed in~\cite{meier2015} has been plotted. The result is very similar to the SK-TAN element. The only reason why the discretization error level is slightly lower for the isotropic than for the SK-TAN element lies in the twist interpolation, which only requires two DoFs for the isotropic element but three DoFs for the SK-TAN element. In Section~\ref{sec:spatialdiscretization_triadsr}, the importance of a consistent torsion~\eqref{triadsr_totaltorsion} of the intermediate triad field has been emphasized. Now, in Figure~\ref{fig:path_convergence_2}, the discretization error for a variant (SK-TAN no $K_{M1}$) has been plotted, where exactly this torsion term has been neglected. Surprisingly, the resulting discretization error level is identical to the "correct" SK-TAN element formulation. How can this contradiction be explained? In order to answer this question, it has to be realized that the actual triad orientation is not important for isotropic examples in order to yield a consistent centerline convergence. It can easily be verified that only the mechanical torsion has to be represented correctly, which is also the functional principle of the isotropic beam element. If the torsion of the intermediate triad field is neglected, the total torsion is solely represented by the derivative of the relative angle field $\varphi(\xi)$. Consequently, the relative angle arises in a way such that the total torsion is represented exactly, which in turn results in an inconsistent triad orientation. However, since for isotropic beams, only the torsion, but not the triad orientation, enters the weak form, the final result for the beam centerline is correct. Later in Section~\ref{sec:examples_arcsegment}, it will be shown that the situation changes for anisotropic beams, i.e. beams with initial curvature or with anisotropic cross-section shapes. There, the neglect of the intermediate triad torsion will indeed lead to an inconsistent centerline solution resulting in a decreased spatial convergence rate. Furthermore, this investigation explains why certain Kirchhoff element formulations available in the literature, which accidentally neglect this torsion term, nevertheless produce correct results and consistent convergence rates for the centerline solution as long as isotropic beam problems are considered.
Finally, in Figure~\ref{fig:path_newton}, the total number of Newton iterations required by the different finite element formulations is plotted for the load case "sim" and the two investigated slenderness ratios. Here, the load step adaption scheme based on $N_0\!=\!10$ has been employed. The obtained results are very similar to the last section: The Newton performance of the Reissner type element formulations drastically deteriorates with increasing slenderness ratio whereas the performance of the Kirchhoff type elements remains unchanged.  For the slenderness ratio $\zeta\!=\!10000$, all investigated discretizations of the WK/SK-TAN elements exhibit a remarkably constant number of $n_{iter,tot}\!=\!107\!\pm\!1$ iterations, while the total number of iterations required by the WK/SK-ROT elements increases from $n_{iter,tot}\!\approx\!800$ to $n_{iter,tot}\!\approx\!2500$ with increasing number of elements. Again, the total number of iterations required by the Reissner type beam element formulations is almost by two orders of magnitude higher than for the WK/SK-TAN elements and lies constantly above a value of $n_{iter,tot}\!>\!6000$ for the CJ element formulation and $n_{iter,tot}\!>\!7000$ for the SV element formulation. 

\subsection{Example 5: Arc-segment with out-of-plane load}
\label{sec:examples_arcsegment}

Besides the objectivity test in Section~\ref{sec:examples_objectivity}, all examples investigated in the previous sections were based on isotropic geometries, i.e. straight beams with quasi-circular cross-sections $EI_2\!=\!EI_3$. Now, an initially curved beam will be considered. The initial geometry is represented by a $45^{\circ}$-degree circular arc-segment with curvature radius $r_0\!=\!100$ that lies completely in the global $x$-$y$-plane and that is clamped at one end. The section constitutive parameters of the beam result from a quadratic cross-section shape with side length $R\!=\!1$ and a Young's modulus of $E\!=\!10^7$ as well as a shear modulus of $G\!=\!0.5 \cdot 10^7$. This initial geometry is loaded by an out-of-plane force $\mb{f}\!=\!(0,0,f_z)^T$ in global $z$-direction with magnitude $f_z\!=\!600$. This example has initially been proposed by Bathe and Bolourchi~\cite{bathe1979} and can meanwhile be considered as standard benchmark test for geometrically exact beam element formulations that has been investigated by many authors (see e.g.~\cite{simo1986,cardona1988,ibrahimbegovic1995a,dvorkin1988,crisfield1990,jelenic1999,crivelli1993,schulz2001,eugster2013,romero2004,romero2008,avello1991}). While the original definition of the slenderness ratio yields a value of $\zeta\!=\!l/R\!=\!100\pi/4$ for this example, a slightly modified definition of the slenderness ratio according to $\tilde{\zeta}\!=\!r_0/R\!=\!100$ is employed in the following.
\begin{figure}[b!!!]
 \centering
  \includegraphics[width=0.5\textwidth]{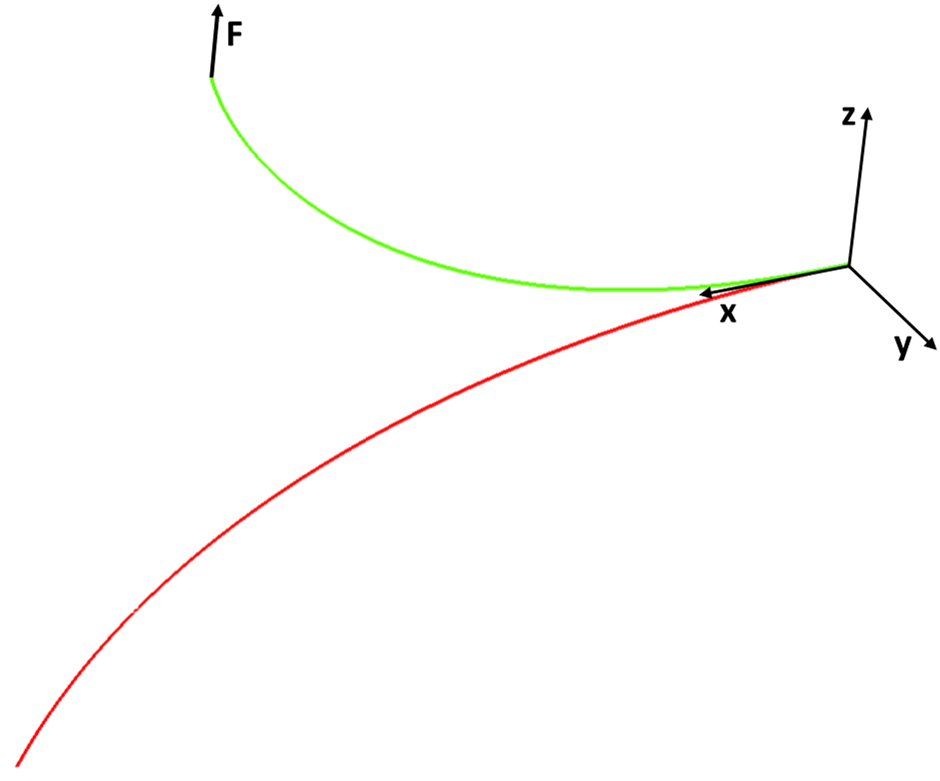}
  \caption{Arc-segment with out-of-plane force: Initial (red) and final (green) configuration.}
  \label{fig:arcsemgent_geometry}
\end{figure}
For comparison reasons, also a second variant of this example with increased slenderness ratio $\tilde{\zeta}\!=\!r_0/R\!=\!10000$, i.e. $R\!=\!0.01$, and adapted force $f_z\!=\!6\cdot 10^{-6}$ will be investigated. The initial and deformed geometry are illustrated in Figure~\ref{fig:arcsemgent_geometry}. In Tables~\ref{tab:arcsegment_displacements2} and~\ref{tab:arcsegment_displacements3}, the tip displacements resulting from the two slenderness ratios and different discretizations with WK-TAN and SK-TAN Kirchhoff type elements as well as with Reissner type elements of Crisfield and Jeleni\'{c} and Simo and Vu-Quoc are plotted. Due to rough spatial discretizations, and in some cases also due to additional model simplifications, the corresponding values derived in the literature for the case $\tilde{\zeta}\!=\!100$ show a comparatively large variation. On the contrary, the deviation in the results displayed in Table~\ref{tab:arcsegment_displacements2} is smaller than $0.1\%$ for all investigated formulations. The fact that these results have been derived by representatives of different beam theories, i.e. of the Simo-Reissner and of the Kirchhoff-Love theory, indicates their correctness. While the Reissner and Kirchhoff values resulting from a discretization with 32 elements coincide up to the fourth significant digit for the case $\tilde{\zeta}\!=\!100$, the corresponding values are identical in all seven significant digits displayed for the case of the high slenderness ratio $\tilde{\zeta}\!=\!10000$.
\begin{table}[t!]
\centering
\begin{tabular}{|p{4.5cm}|p{2.0cm}||p{2.0cm}|p{2.0cm}|p{2.0cm}|} \hline
Formulation & $\#$ Elements &$u_x(l)$&$u_y(l)$&$u_z(l)$ \\ \hline
Crisfield \& Jeleni\'{c} & 32 & 47.15044 & 15.68480 & 53.47486 \\ \hline
Crisfield \& Jeleni\'{c} & 8 & 47.15044 & 15.68480 & 53.47486 \\ \hline
Simo \& Vu-Quoc & 8 & 47.14634 & 15.69146 & 53.47362 \\ \hline
WK-TAN & 32 & 47.15215 & 15.68535 & 53.47176 \\ \hline
WK-TAN & 8 & 47.15178 & 15.68510 & 53.47225 \\ \hline
SK-TAN & 8 & 47.15201 & 15.68557 & 53.47216 \\ \hline
\end{tabular}
\caption{Case $\tilde{\zeta}=100$: tip displacement and relative error for different formulations.}
\label{tab:arcsegment_displacements2}
\end{table}
\begin{table}[t!]
\centering
\begin{tabular}{|p{4.5cm}|p{2.0cm}||p{2.0cm}|p{2.0cm}|p{2.0cm}|} \hline
Formulation & $\#$ Elements &$u_x(l)$&$u_y(l)$&$u_z(l)$ \\ \hline
Crisfield \& Jeleni\'{c} & 32 & 47.15129 & 15.68508 & 53.46860 \\ \hline
Crisfield \& Jeleni\'{c} & 8 & 47.15129 & 15.68508 & 53.46860 \\ \hline
Simo \& Vu-Quoc. & 8 & 47.14719 & 15.69174 & 53.46736 \\ \hline
WK-TAN & 32 & 47.15129 & 15.68508 & 53.46860  \\ \hline
WK-TAN & 8 & 47.15093 & 15.68482 &  53.46908 \\ \hline
SK-TAN & 8 & 47.15115 & 15.68530 & 53.46900 \\ \hline
\end{tabular}
\caption{Case $\tilde{\zeta}=10000$: tip displacement and relative error for different formulations.}
\label{tab:arcsegment_displacements3}
\end{table}
\begin{figure}[b!!!]
 \centering
   \subfigure[Reference: WK-TAN, $\tilde{\zeta}=100$.]
   {
    \includegraphics[width=0.48\textwidth]{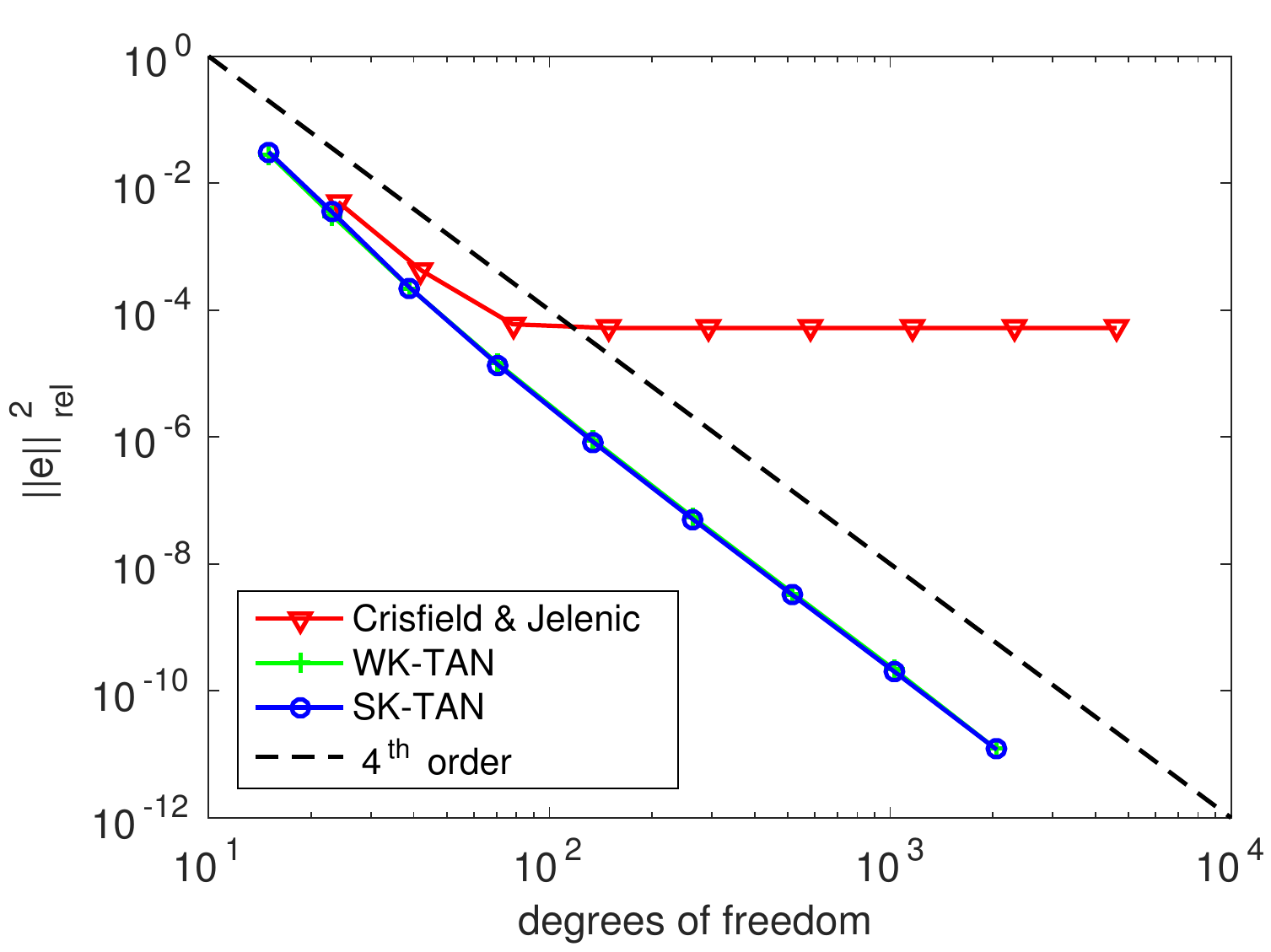}
    \label{fig:arcsegment_convergence_1}
   }
   \subfigure[Reference: WK-TAN, $\tilde{\zeta}=10000$.]
   {
    \includegraphics[width=0.48\textwidth]{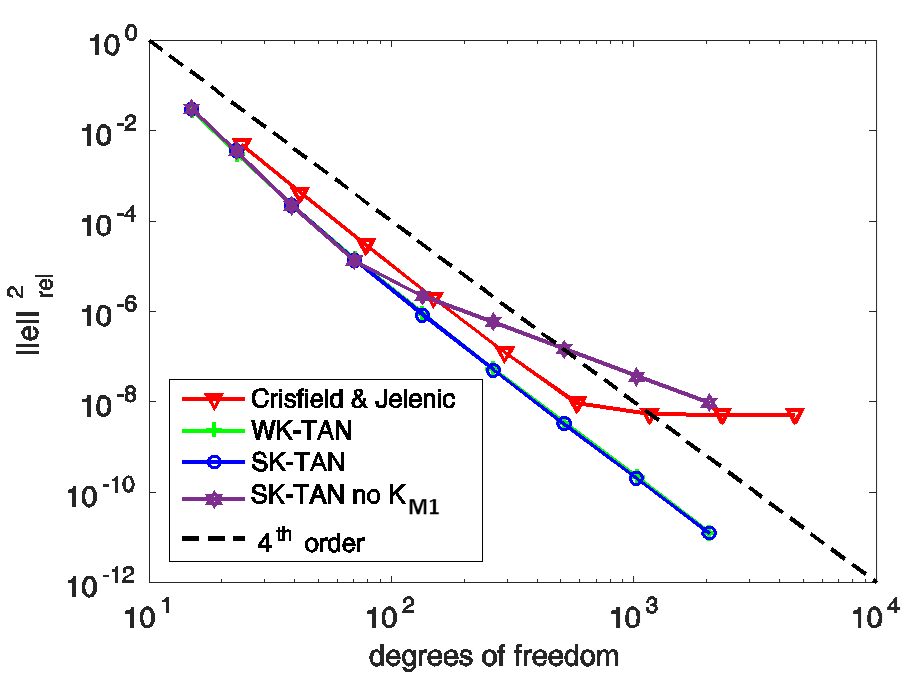}
    \label{fig:arcsegment_convergence_2}
   }
  \caption{Arc-segment with out-of-plane force: $L^2$-error for different formulations.}
  \label{fig:arcsegment_convergence}
\end{figure}
The observations described above are also confirmed by the convergence plots in Figure~\ref{fig:arcsegment_convergence}. All formulations yield the expected convergence orders, and, similar to the last example, the SK-TAN element exhibits an identical discretization error level as the WK-TAN element, since again no multiple centerline loops are involved for this example. Furthermore, similar to the example of Section~\ref{sec:examples_purebending3d}, also a variant (SK-TAN no $K_{M1}$) has been investigated where the torsion of the intermediate triad field has been omitted. While the omission of this term did not influence the convergence order observed in Section~\ref{sec:examples_purebending3d}, this inconsistency yields a decline in the convergence rate from four to two for the anisotropic example considered here. This underlines the importance of consistently considering this term (see also Section~\ref{sec:examples_purebending3d} for further explanation).
\begin{figure}[t!!!]
 \centering
  \subfigure[Number of Newton iterations for $\tilde{\zeta}=100$.]
   {
    \includegraphics[width=0.48\textwidth]{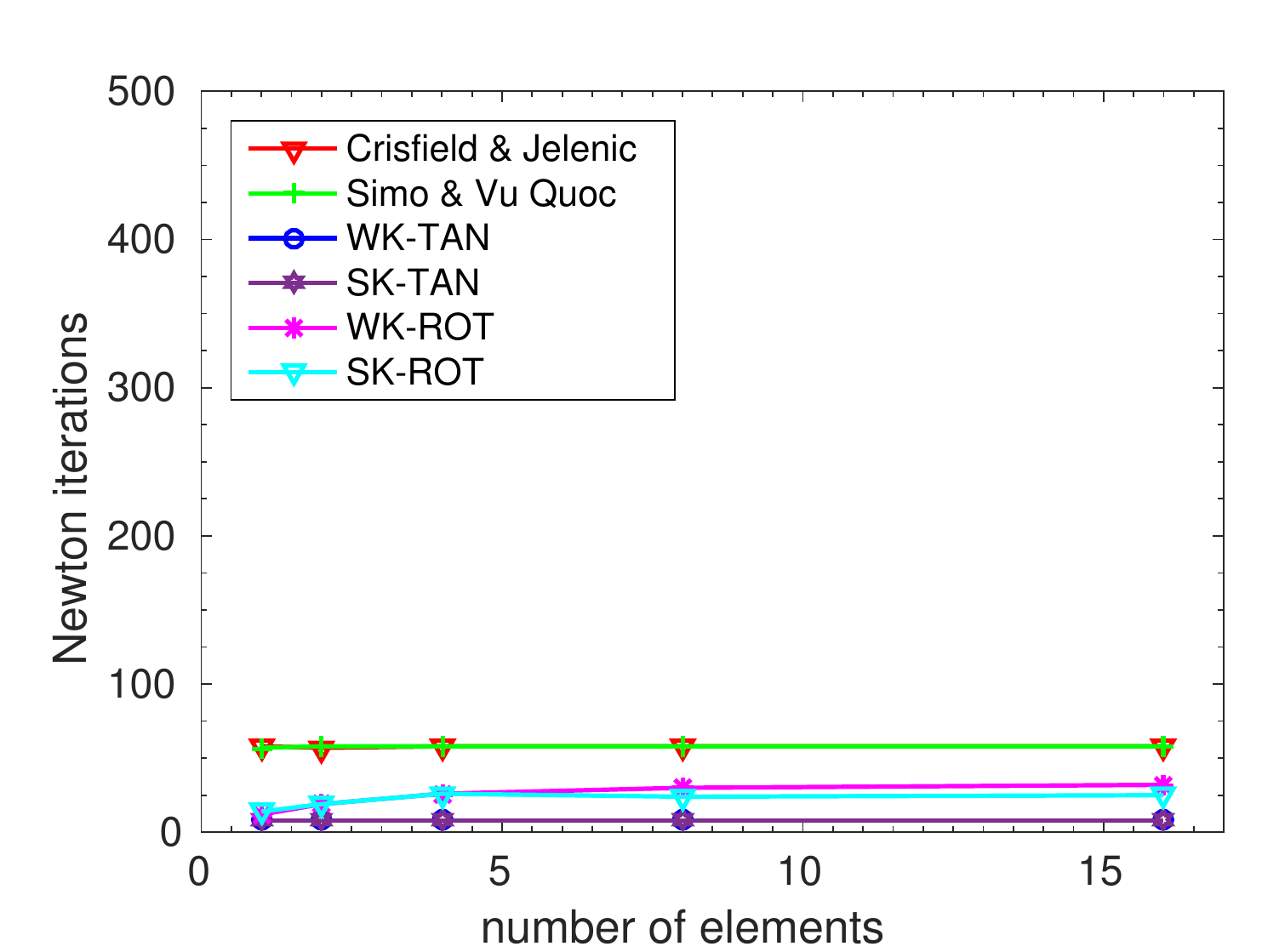}
    \label{fig:arcsegment_newton_1}
   }
   \subfigure[Number of Newton iterations for $\tilde{\zeta}=10000$.]
   {
    \includegraphics[width=0.48\textwidth]{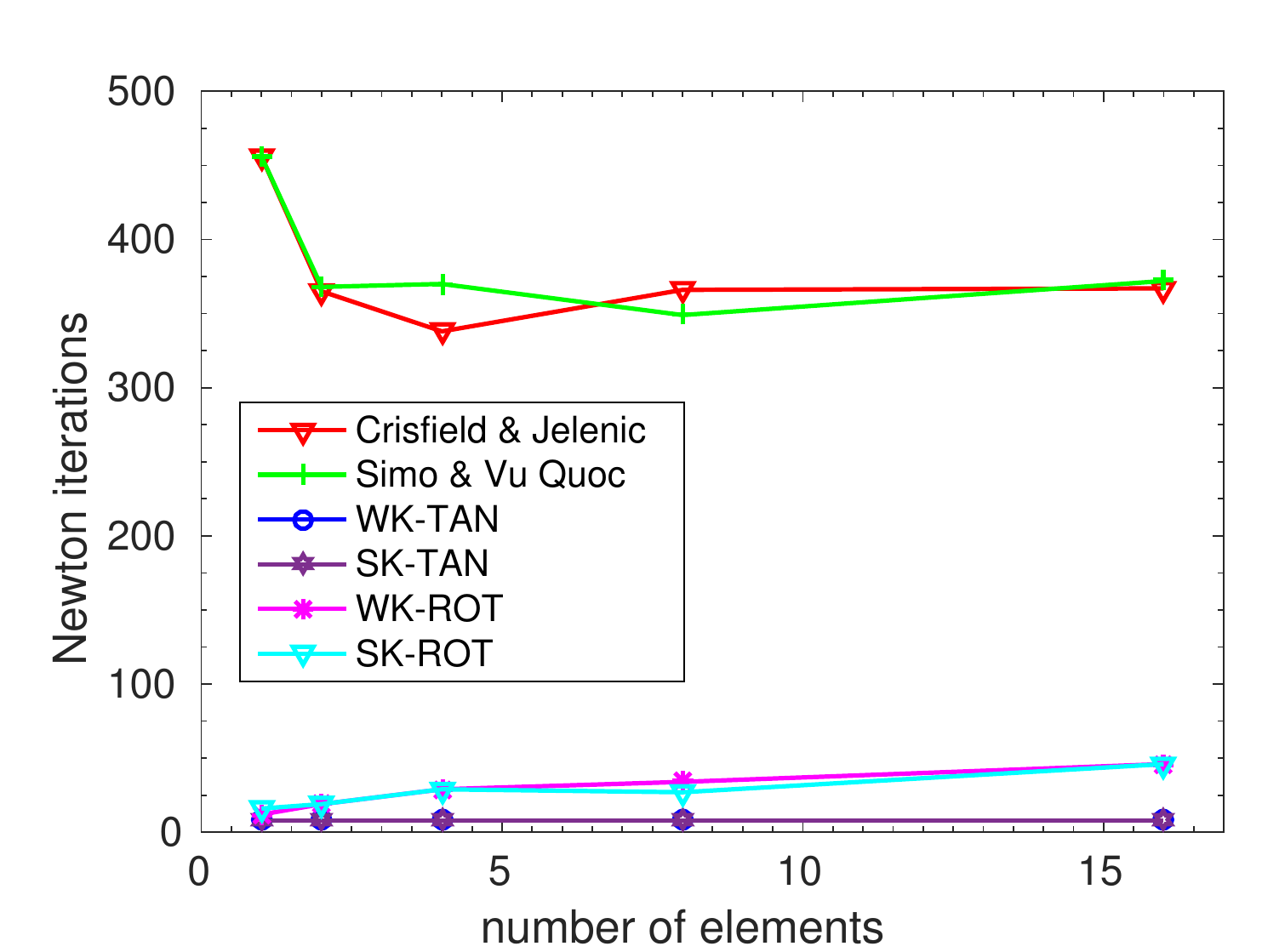}
    \label{fig:arcsegment_newton_2}
   }
   \subfigure[Number of load steps for $\tilde{\zeta}=100$.]
   {
    \includegraphics[width=0.48\textwidth]{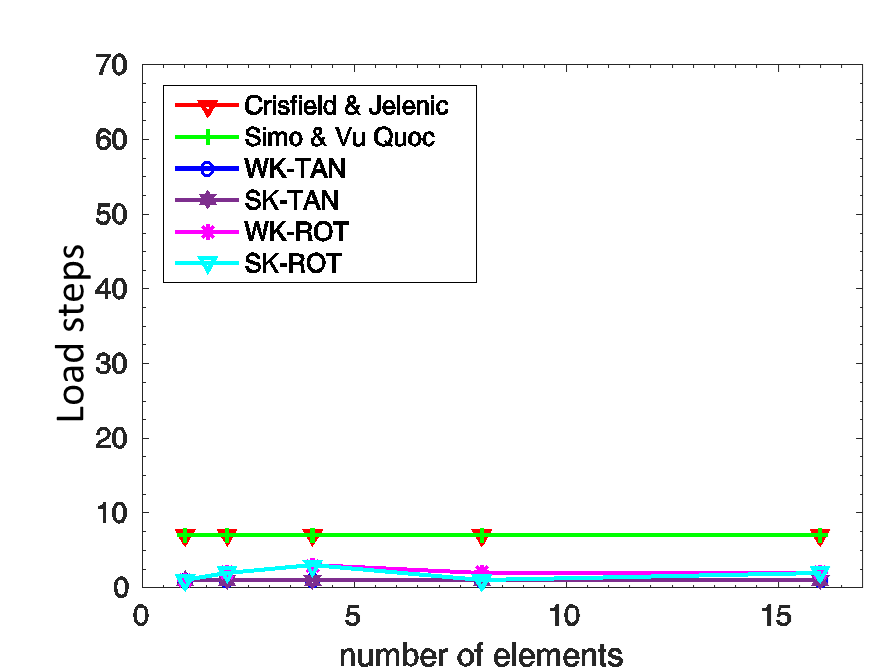}
    \label{fig:arcsegment_newton_3}
   }
   \subfigure[Number of load steps for $\tilde{\zeta}=10000$.]
   {
    \includegraphics[width=0.48\textwidth]{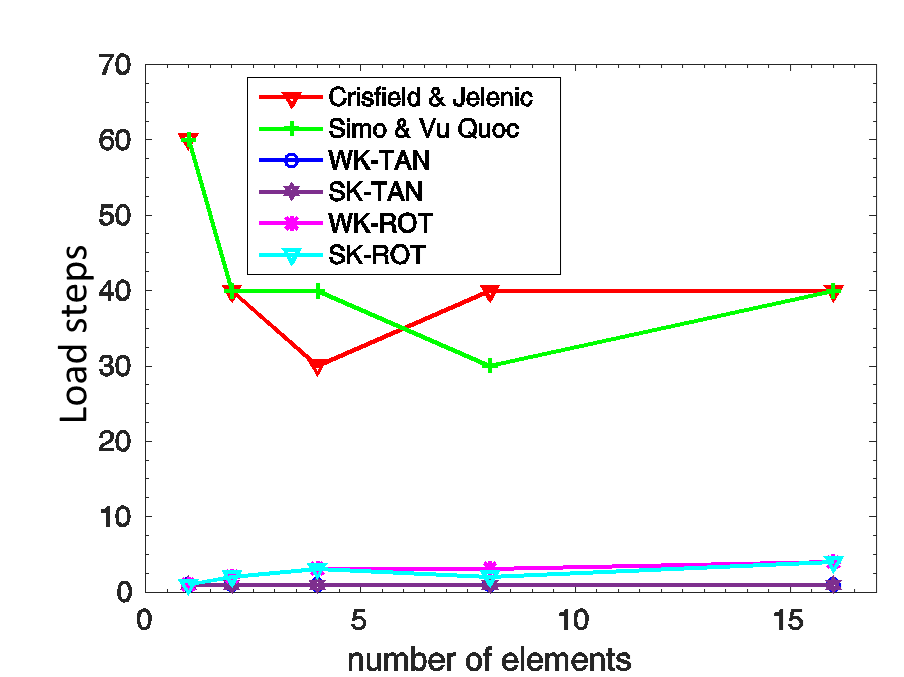}
    \label{fig:arcsegment_newton_4}
   }
  \caption{Arc-segment with out-of-plane force: Number of Newton iterations and load steps.}
  \label{fig:arcsegment_newton}
\end{figure}
Also for this example, the performance of the Newton-Raphson scheme will be evaluated. However, in order to enable a comparison with the values available in the literature, this time, no load step adaption scheme is employed, but the following alternative procedure in order to determine the maximal constant load step size $\Delta t\!=\!$ const.: Starting with a scheme based on one load step $N_0\!=\!1$, the number of load steps is increased by one, i.e. $N_{new}\!=\!N_{old}\!+\!1$, in the range $N = 1,...,10$ and increased by increments of 10, i.e. $N_{new}\!=\!N_{old}\!+\!10$, in the range $N>10$, until Newton convergence is achieved for all load steps. In order to avoid "lucky shots" (see the remark at the end of this section and also the remark in Section~\ref{sec:examples_purebending3d}), a maximal load step size and an associated minimal number of load steps $N_{min}$ is accepted only if also the next incrementation step of the load step size according to the procedure described above leads to Newton convergence for all load steps. In Figure~\ref{fig:arcsegment_newton}, the total number of Newton iterations as well as the minimal number of load steps $N_{min}$ resulting from the maximal constant load step size are plotted for the two different slenderness ratios. The results are similar to the observations made in previous sections, however, with a smaller difference between the SK/WK-TAN elements and the SK/WK-ROT elements. Concretely, the beam problem with slenderness ratio $\tilde{\zeta}\!=\!100$ is solved in 1 load step and a total of 8 iterations for the SK/WK-TAN discretizations, in 1-3 load steps and a total of 10-40 iterations for the SK/WK-ROT discretizations and in 7 load steps and a total of 57-58 iterations for the discretizations based on Reissner type elements. For the slenderness ratio $\tilde{\zeta}\!=\!10000$, the problem is again solved in 1 load step and a total of 8 iterations for the SK/WK-TAN discretizations, in 1-4 load steps an a total of 10-50 iterations for the SK/WK-ROT discretizations and in 30-60 load steps and a total of 
350-450 iterations for the Reissner discretizations. In Table~\ref{tab:arcsegment_newton_literature}, the corresponding values reported in the literature for the slenderness ratio $\tilde{\zeta}\!=\!100$ are summarized. As already mentioned earlier, a direct comparison of these results is difficult since it is not clear which procedure has been applied by the different authors in order to determine the minimal number of Newton iterations (e.g. if it was required that also "subsequent refinement steps" have to be convergent or if, on the contrary, also singular occurrences of convergence for special, good-natured loading paths were accepted).\\

Nevertheless, the numbers summarized in Table~\ref{tab:arcsegment_newton_literature} should at least give a first impression on the behavior of the Newton-Raphon scheme resulting from different finite element formulations. Accordingly, only a few formulations can solve the problem in less than 20 iterations. Furthermore, for the case $\tilde{\zeta}\!=\!100$, only the Reissner type formulation proposed in~\cite{jelenic1995} and investigated in~\cite{jelenic1999} yields a lower number of Newton iterations than the SK/WK-TAN elements. However, as shown in~\cite{jelenic1999}, this beam element formulation is non-objective and path-dependent. Moreover, for all examples investigated so far, the real advantage of the Kirchhoff type formulations occurred especially for the high slenderness ratio $\tilde{\zeta}\!=\!10000$, which has not been investigated in the literature. For the range of moderate and high slenderness ratios, it can be concluded that the proposed Kirchhoff beam elements can be considered as very robust and efficient formulations as compared to many (Reissner type) alternatives from the literature.\\

\begin{table}[t!]
\centering
\begin{tabular}{|p{1.7cm}|p{2.15cm}|p{0.9cm}|p{1.2cm}|p{7.8cm}|} \hline
Reference & Elements & $N_{min} $ &$n_{iter,tot}$& Remark \\ \hline
\cite{crivelli1993}  & 8 first-order & 6 & - & Number of iterations has not been reported. \\ \hline
\cite{simo1986}  & 8 first-order & 3 & 27 &  - \\ \hline
\cite{cardona1988}  & 8 first-order & 6 & 47 &  - \\ \hline
\cite{schulz2001}  & 8 first-order & 3 & 30 & Application of a standard Newton scheme. \\ \hline
\cite{schulz2001}  & 8 first-order & 2 & 11 & Application of an accelerated Newton scheme. \\ \hline
\cite{bathe1979}  & 8 first-order & 60 & - & Number of iterations has not been reported. \\ \hline
\cite{crisfield1990}   & 8 first-order & 3 & 16&  - \\ \hline
\cite{jelenic1999} & 8 first-order & 1 & 4 & Non-objective variant proposed in~\cite{jelenic1995}. \\ \hline \hline
CJ  & 8 first-order & 7 & 59 &  Objective variant proposed in~\cite{jelenic1999}. \\ \hline
CJ  & 8 third-order & 7 & 58 & Objective variant proposed in~\cite{jelenic1999}. \\ \hline
SV  & 8 first-order & 7 & 52 & Interpretation of the formulation of \cite{simo1986}. \\ \hline
SV  & 8 third-order & 7 & 58 & Interpretation of the formulation of \cite{simo1986}. \\ \hline
SK-TAN  & 8 third-order & 1 & 8 & Kirchhoff type beam element formulation.\\ \hline
WK-TAN  & 8 third-order & 1 & 8 & Kirchhoff type beam element formulation.\\ \hline
SK-ROT  & 8 third-order & 1 & 24 & Kirchhoff type beam element formulation.\\ \hline
WK-ROT  & 8 third-order & 2 & 30 & Kirchhoff type beam element formulation.\\ \hline
\end{tabular}
\caption{Case $\tilde{\zeta}=100$: Number of load steps and Newton iterations from the literature (top) and from this work (bottom).}
\label{tab:arcsegment_newton_literature}
\end{table}

\hspace{0.2 cm}
\begin{minipage}{15.0 cm}
\textbf{Remark:} Maybe the reader is wondering why the SV element formulation, which is an interpretation of the element formulation proposed in~\cite{simo1986}, based on an identical discretization with eight first-order elements as investigated in~\cite{simo1986}, required more Newton iterations than reported in that reference. Actually, also in the numerical tests performed here, the nonlinear problem resulting from a discretization with eight first-order SV elements could be solved in three load steps. However, since a subsequent simulation based on four load steps was not convergent, the procedure for the avoidance of  "lucky shots" as explained above has been applied, thus leading to a total of 7 load steps and 52 Newton iterations.
\end{minipage}

\subsection{Example 6: Helix loaded with axial force}
\label{sec:examples_helix}

\begin{figure}[ht]
 \centering
 \includegraphics[width=0.75\textwidth]{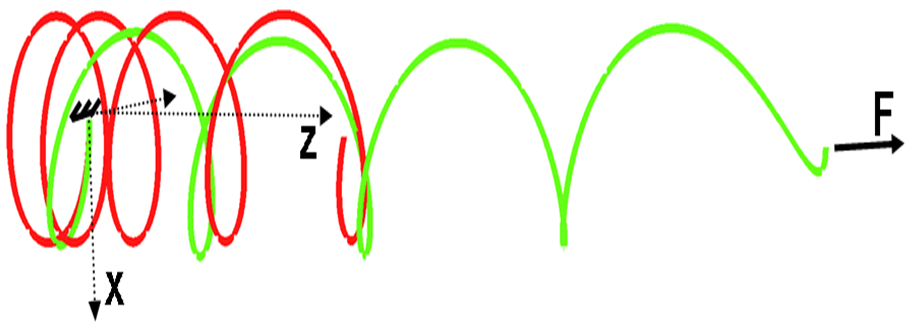}
 \caption{Helix with varying slope loaded with discrete force: Problem setup.}
 \label{fig:3Dhelix_problemsetup}
\end{figure}

In this example, the generality of the initial geometry shall be further increased: A helix with linearly increasing slope, clamped at one of its ends, 
is loaded with a end-force $\mb{F}=(0,0,F)^T$ (see Figure~\ref{fig:3Dhelix_problemsetup} for illustration).\\

\begin{figure}[b!!!]
 \centering
   \subfigure[Reference: WK-TAN, $\zeta=100$.]
   {
    \includegraphics[width=0.48\textwidth]{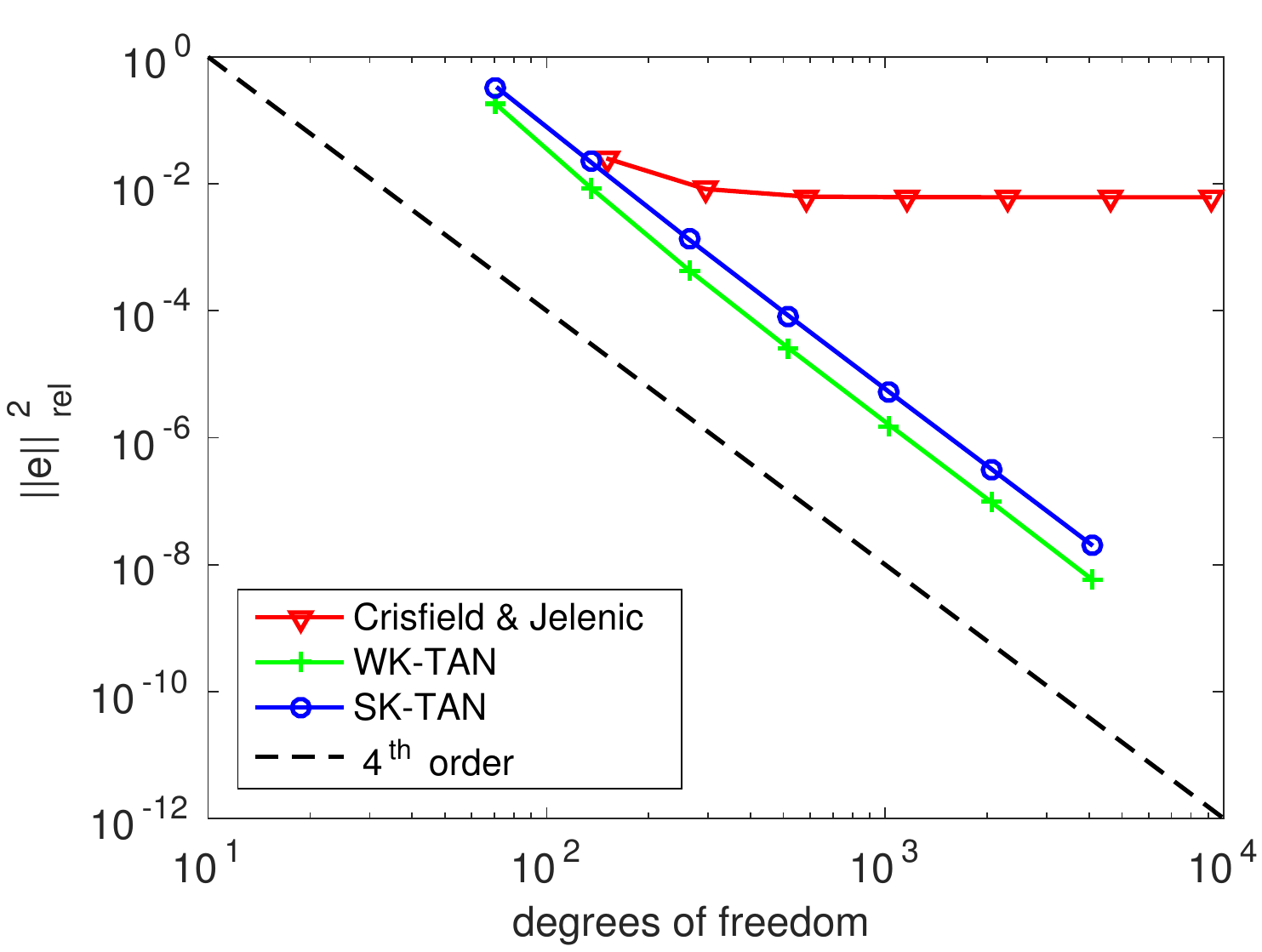}
    \label{fig:helix_convergence_1}
   }
   \subfigure[Reference: WK-TAN, $\zeta=10000$.]
   {
    \includegraphics[width=0.48\textwidth]{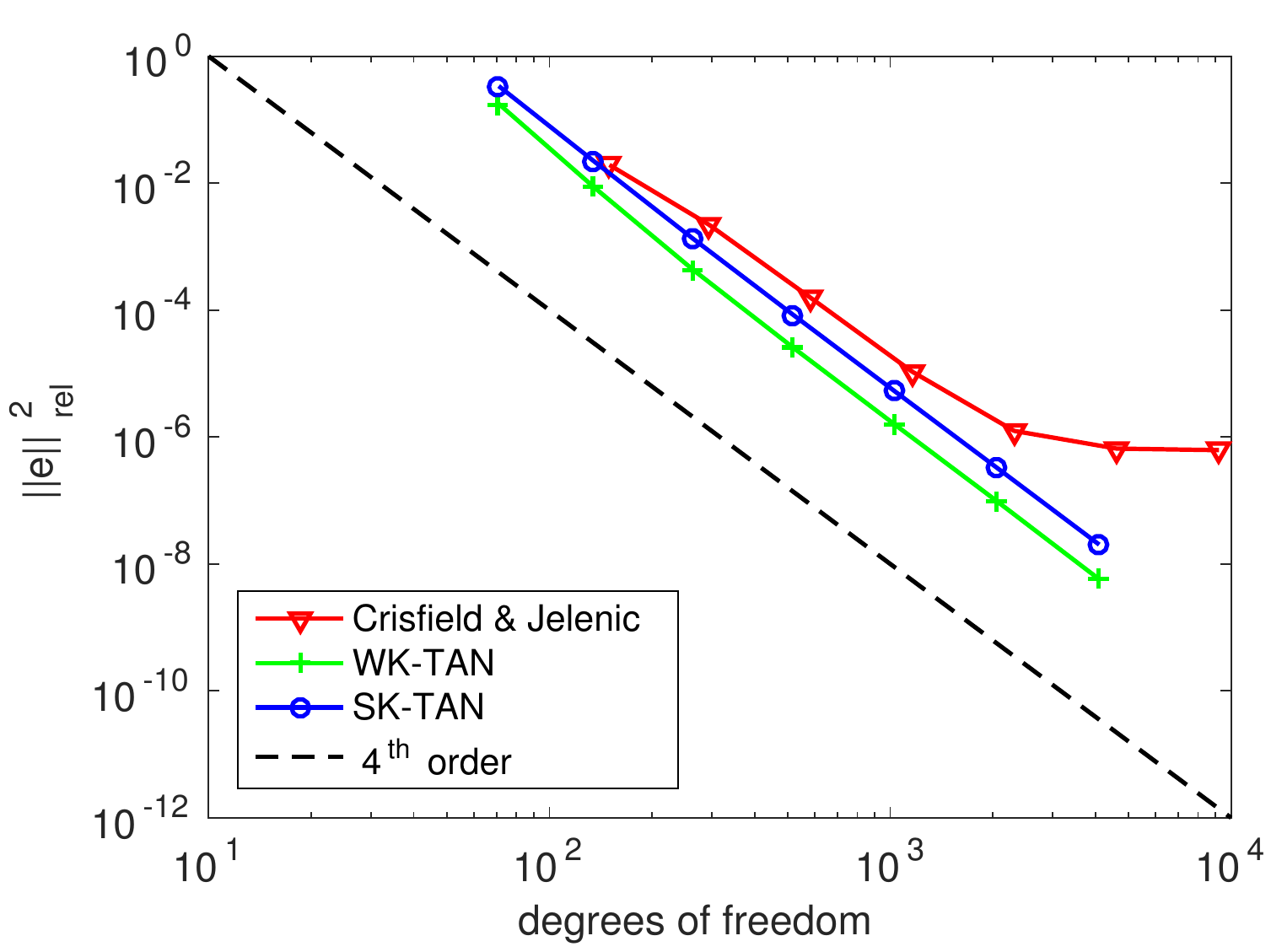}
    \label{fig:helix_convergence_2}
   }
  \caption{Helix loaded with axial force: $L^2$-error for different formulations.}
  \label{fig:helix_convergence}
\end{figure}

\begin{figure}[t!!!]
 \centering
   \subfigure[Reference: WK-TAN, $\zeta=100$.]
   {
    \includegraphics[width=0.48\textwidth]{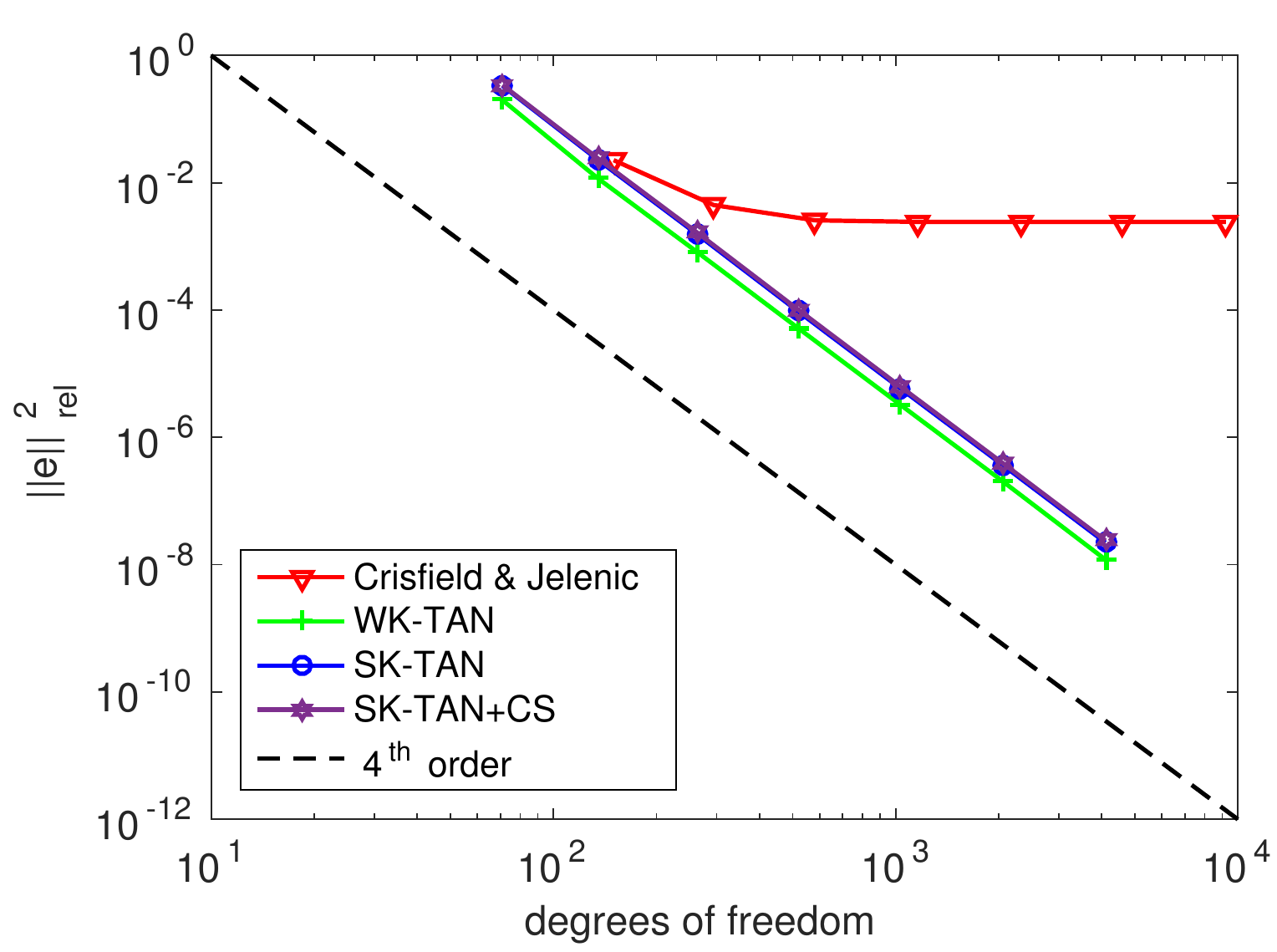}
    \label{fig:3Dhelixwithtwist_convergence_1}
   }
   \subfigure[Reference: WK-TAN, $\zeta=10000$.]
   {
    \includegraphics[width=0.48\textwidth]{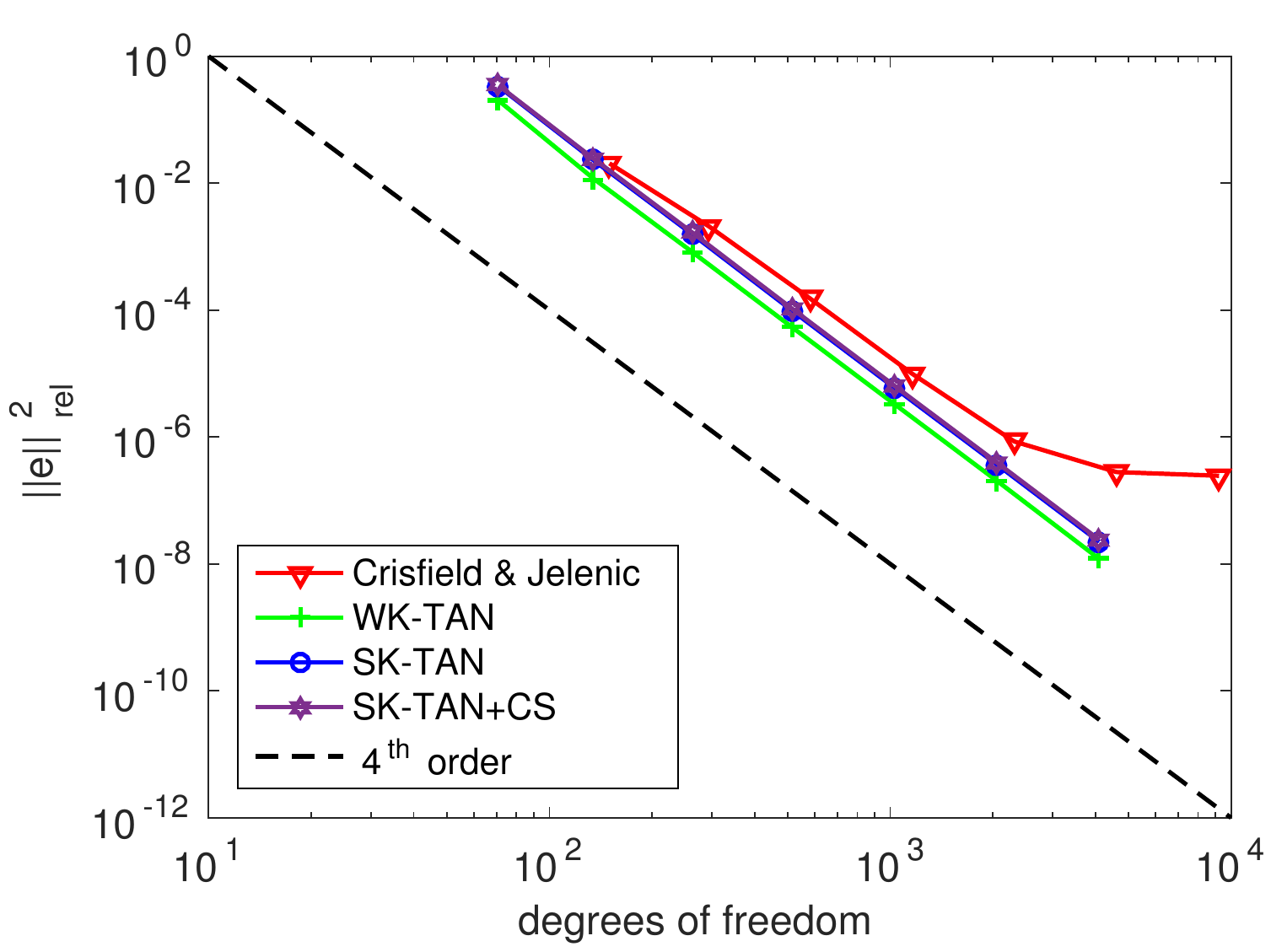}
    \label{fig:3Dhelixwithtwist_convergence_2}
   }
  \caption{Twisted helix loaded with axial force: $L^2$-error for different formulations.}
  \label{fig:3Dhelixwithtwist_convergence}
\end{figure}

The space curve representing initial geometry of the helix can be described via the following analytic representation:
\begin{align}
\label{r0_analyt_3Dhelix}
  \!\!\!\!\!\! \mb{r}_0(\beta) \!=\! R_0 \!
  \left(
   \begin{array}{c}
   \sin{\beta} \\
   \cos{\beta} \!-\!1 \\ 
    \frac{6}{81 \pi^2} \beta^2
   \end{array}
   \right)\!,
   \,\,\,
   R_0 \!=\! \frac{l}{6\sqrt{\left(\frac{3 \pi}{4}\right)^2 \!+\! 1} \!+\! \frac{27 \pi^2}{8} \ln{\left(\frac{4}{3\pi}\!+\!\sqrt{1\!+\!\left(\frac{4}{3\pi}\right)^2}\right)}}
   \approx 34.36. \!\!\!\!\!\!
\end{align}
The radius $R_0$ of the enveloping cylinder of the helix is chosen such that the helix exactly consists of $4.5$ loops, i.e. $\beta \in[0;9 \pi]$, along the standard
length of $l\!=\!1000$. Also this example is investigated for two different slenderness ratios $\zeta\!=\!100$ and $\zeta\!=\!10000$ with associated axial forces $F\!=\!2 \!\cdot\! 10^{-1}$ as well as $F\!=\!2 \!\cdot\! 10^{-9}$. The ratio of these forces is chosen identical to the ratio of the bending stiffnesses of the cases $\zeta_1\!=\!100$ and $\zeta_2\!=\!10000$ leading to the comparable values $u_{z,max,1} \!\approx\! 267$ and $u_{z,max,2} \!\approx\! 266$ for the maximal tip-displacement in $z$-direction. In Figure \ref{fig:helix_convergence}, the resulting relative $L^2$-error of the CJ, SK-TAN and WK-TAN element is plotted for discretizations with $16,32,64,128,256$ and $512$ elements. All element formulations show the expected convergence rate of four, the discretization error level of the WK-TAN element is slightly lower than the discretization error level of the SK-TAN element. However, both discretization error levels lie below the error level of the Reissner type CJ element. In the authors' former contribution, also the different anti-locking approaches already investigated for the 2D case in Section~\ref{sec:examples_purebending2d_locking} have been compared for this 3D helix example. The derived results and the drawn conclusion are similar as for this 2D example. For further details on this comparison, the interested reader is referred to this reference.
\begin{figure}[t!!!]
 \centering
 \includegraphics[width=0.99\textwidth]{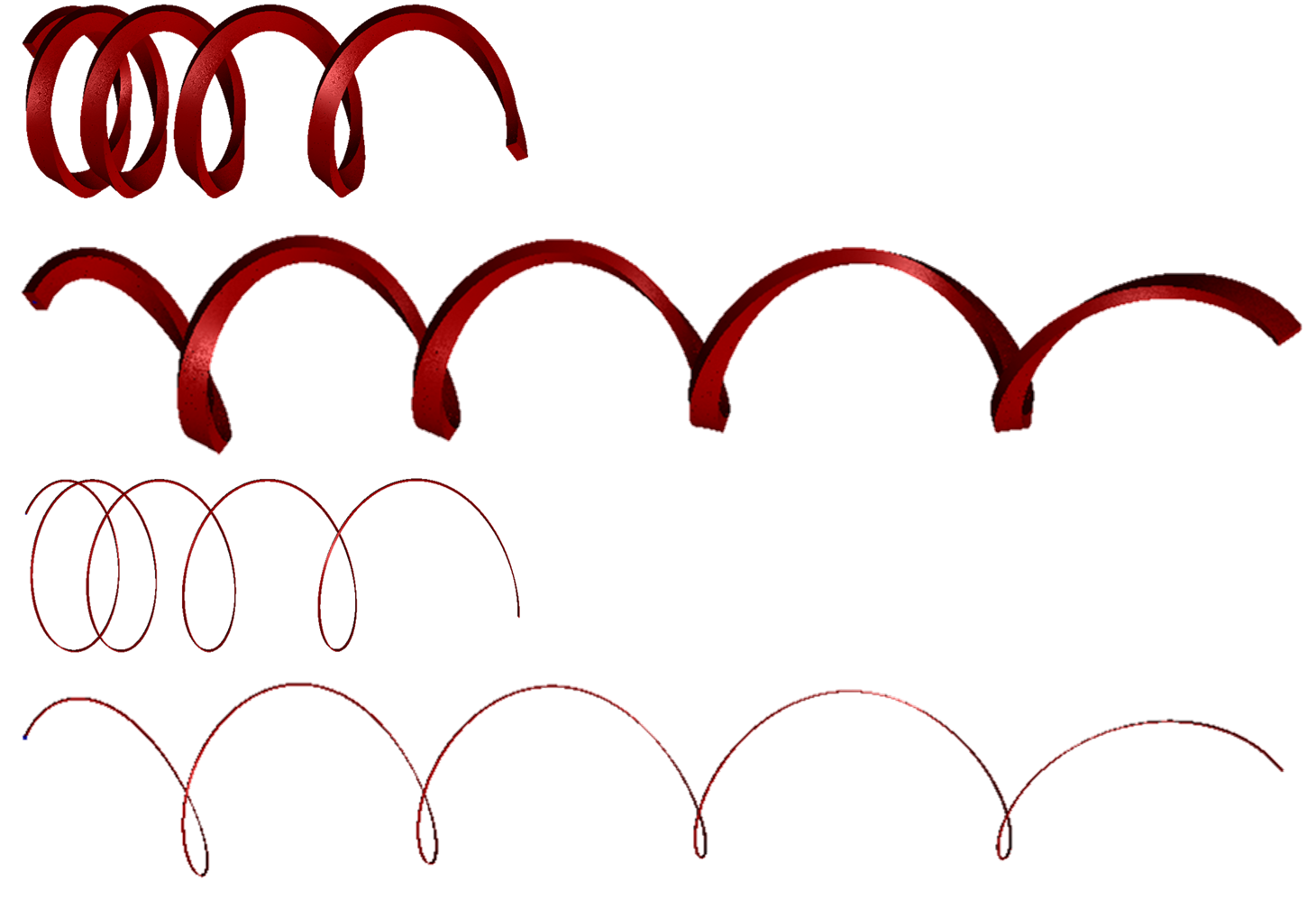}
 \caption{Twisted helix with axial force: Initial and final shape for $\zeta \!=\!100$ and $\zeta\!=\!1000$ (visualization based on $\zeta\!=\!1000$).}
 \label{fig:3Dhelixwithtwist_problemsetup}
\end{figure}
Finally, the helix problem shall be even more generalized by not only accounting for 3D initial curvatures but also for anisotropic cross-section shapes with initial twist as illustrated in Figure~\ref{fig:3Dhelixwithtwist_problemsetup}. Again, the two slenderness ratios $\zeta\!=\!100$ and $\zeta\!=\!10000$ have been investigated. The square cross-section of the last example is extended to a rectangular cross-section shape with dimensions $b\!=\!10$ and $h\!=\!5$ and an assumed torsional moment of inertia $I_T\!\approx\!3.2875\!\cdot\!10^{-2}$ for the case $\zeta\!=\!100$. The case $\zeta\!=\!10000$ is defined by $b\!=\!0.1$ and $h\!=\!0.05$ and an assumed torsional moment of inertia $I_T\!\approx\!3.2875\!\cdot\!10^{-6}$. The external forces have been chosen as $F\!=\!5\!\cdot\!10^{-2}$ for $\zeta\!=\!100$ and $F\!=\!5 \cdot 10^{-10}$ for $\zeta\!=\!10000$ and the initial twist as one twist rotation per helix loop. The resulting $L^2$-error plotted in Figure~\ref{fig:3Dhelixwithtwist_convergence} again shows a consistent convergence behavior similar to Figure~\ref{fig:helix_convergence}. Additionally, in Figure~\ref{fig:3Dhelixwithtwist_problemsetup}, also the Bubnov-Galerkin variant (SK-TAN+CS) of the SK-TAN element with consistent spin vector interpolation has been plotted. Accordingly, no visible difference compared to the Petrov-Galerkin (SK-TAN) variant can be observed. In a last step, also the balances of forces and moments are investigated for this most general example of Figure~\ref{fig:3Dhelixwithtwist_problemsetup}. In Table~\ref{tab:3Dhelixwithtwist_reactionforces}, the reaction forces and moments at the clamped end of the helix at $s\!=\!0$ and the force and moment contributions (with respect to the point $s\!=\!0$) resulting from the external load applied at $s\!=\!l$ are plotted for discretizations with eight elements. It can easily be verified that the balance of forces and moments is exactly fulfilled by the variants CJ, WK-TAN and SK-TAN+CS, while the Petrov-Galerkin variant SK-TAN only fulfills the balance of forces but not the balances of moments. This confirms the prediction made in Section~\ref{sec:elestrongkirchhoff_spatialdiscretization_conservation}.
\begin{table}[h!]
\centering
\begin{tabular}{|p{1.2cm}|p{3cm}|p{3cm}|p{3cm}|p{3cm}|} \hline
 & Cri. \& Jel. & WK-TAN & SK-TAN & SK-TAN+CS \\ \hline
$F_1(0)$ & 0.0 & 0.0 & 0.0 & 0.0  \\ \hline
$F_1(l)$ & 0.0 & 0.0 & 0.0 & 0.0 \\ \hline
$F_2(0)$ & 0.0 & 0.0 & 0.0 & 0.0 \\ \hline
$F_2(l)$ & 0.0 & 0.0 & 0.0 & 0.0 \\ \hline
$F_3(0)$ & 5.00000000E-6 & 5.00000000E-6 & 5.00000000E-6 & 5.00000000E-6 \\ \hline
$F_3(l)$ & 5.00000000E-6 & 5.00000000E-6 & 5.00000000E-6 & 5.00000000E-6 \\ \hline
$M_1(0)$ & -1.64350142E-4 & -1.54617971E-4 & -1.51929992E-4 & -1.65509346E-4  \\ \hline
$M_1(l)$ & -1.64350142E-4 & -1.54617971E-4 & -1.52158798E-4 & -1.65509346E-4 \\ \hline
$M_2(0)$ & -4.41005618E-5 & -8.55776851E-6 & -1.13519642E-5 & -5.43051999E-6 \\ \hline
$M_2(l)$ & -4.41005617E-5 & -8.55776851E-6 & -2.30787384E-5 & -5.43051999E-6 \\ \hline
$M_3(0)$ & 0.0 & 0.0 & 0.0 & 0.0 \\ \hline
$M_3(l)$ & 0.0 & 0.0 & 0.0 & 0.0 \\ \hline
\end{tabular}
\caption{Slenderness $\zeta=10000$: Reaction forces and moments for different formulations.}
\label{tab:3Dhelixwithtwist_reactionforces}
\end{table}

\newpage

\subsection{Example 7: Free oscillations of an elbow cantilever}
\label{sec:examples_freeoscillationselbow}

The final example represents a dynamic test case. The example has initially been investigated in~\cite{simo1988} and subsequently been considered in several contributions in the field of geometrically nonlinear beam element formulations (see e.g.~\cite{jelenic1999} and~\cite{bruels2012}). A right-angle elbow cantilever beam consists of two straight beam segments of length $l\!=\!10$ being rigidly connected at one of their ends. In the initial configuration, the first segment points into global $y$-direction and the second segment into global $x$-direction.
\begin{figure}[ht]
 \centering
   \subfigure[Step 0.]
   {
    \includegraphics[width=0.18\textwidth]{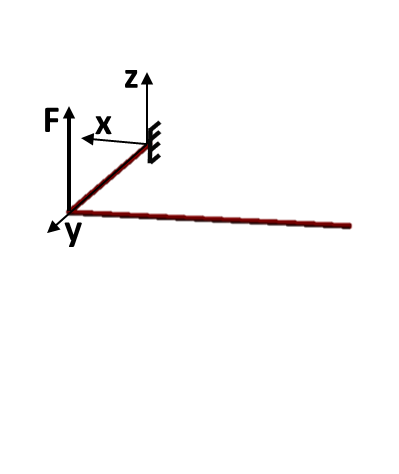}
    \label{ig:ellbow_configs1}
   }
      \subfigure[Step 5.]
   {
    \includegraphics[width=0.18\textwidth]{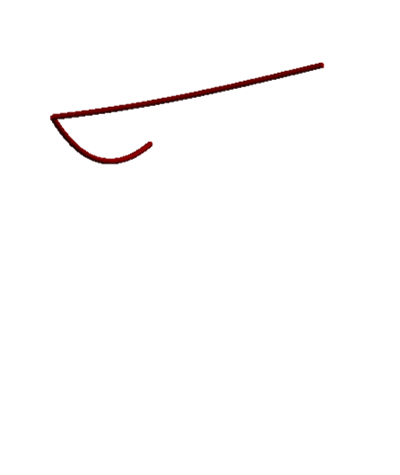}
    \label{ig:ellbow_configs2}
   }
   \subfigure[Step 10.]
   {
    \includegraphics[width=0.18\textwidth]{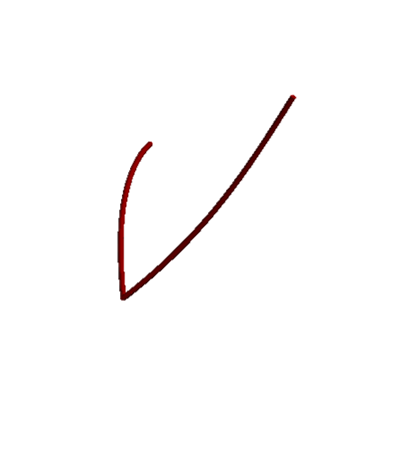}
    \label{ig:ellbow_configs3}
   }
   \subfigure[Step 15.]
   {
    \includegraphics[width=0.18\textwidth]{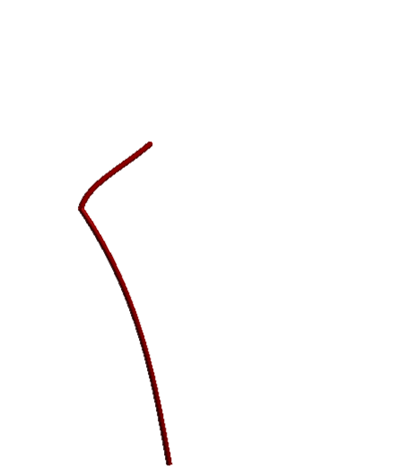}
    \label{ig:ellbow_configs4}
   }
   \subfigure[Step 20.]
   {
    \includegraphics[width=0.18\textwidth]{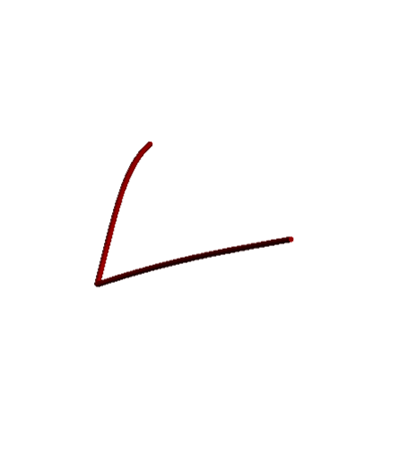}
    \label{ig:ellbow_configs5}
   }
   \subfigure[Step 25.]
   {
    \includegraphics[width=0.18\textwidth]{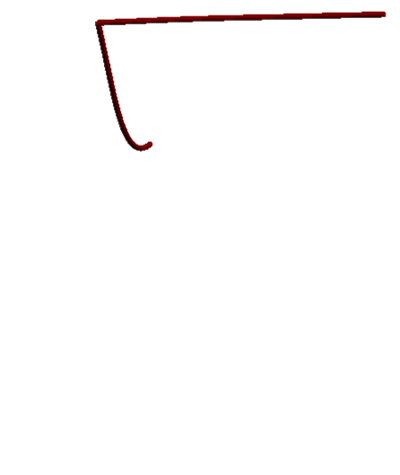}
    \label{ig:ellbow_configs6}
   }
   \subfigure[Step 30.]
   {
    \includegraphics[width=0.18\textwidth]{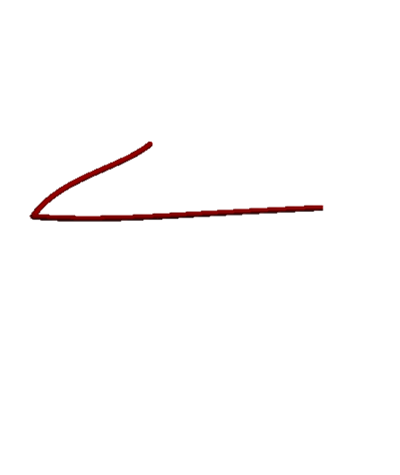}
    \label{ig:ellbow_configs7}
   }
   \subfigure[Step 35.]
   {
    \includegraphics[width=0.18\textwidth]{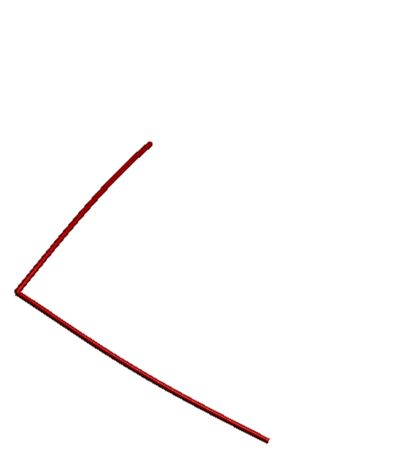}
    \label{ig:ellbow_configs8}
   }
   \subfigure[Step 40.]
   {
    \includegraphics[width=0.18\textwidth]{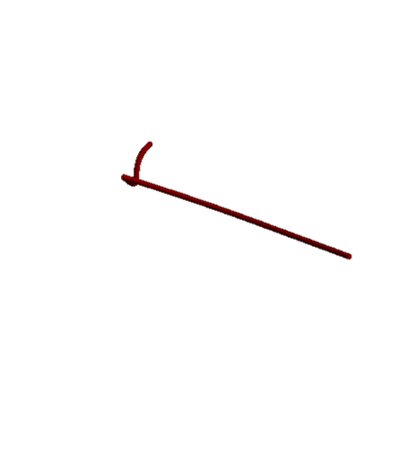}
    \label{ig:ellbow_configs9}
   }
   \subfigure[Step 45.]
   {
    \includegraphics[width=0.18\textwidth]{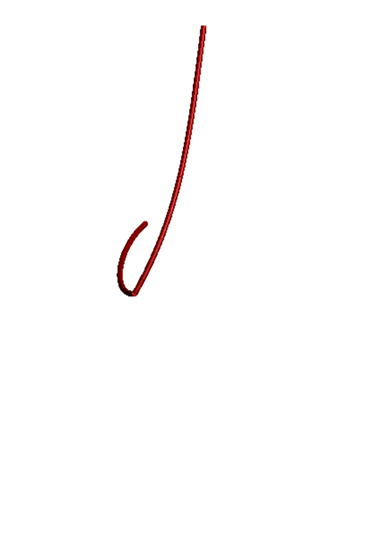}
    \label{ig:ellbow_configs10}
   }
  \caption{Free oscillations of an elbow cantilever: Initial and deformed configurations.}
  \label{fig:ellbow_configs}
\end{figure}
\begin{figure}[h!]
 \centering
   \subfigure[WK-ROT, 16 elements, $\Delta t=0.01$.]
   {
    \includegraphics[width=0.48\textwidth]{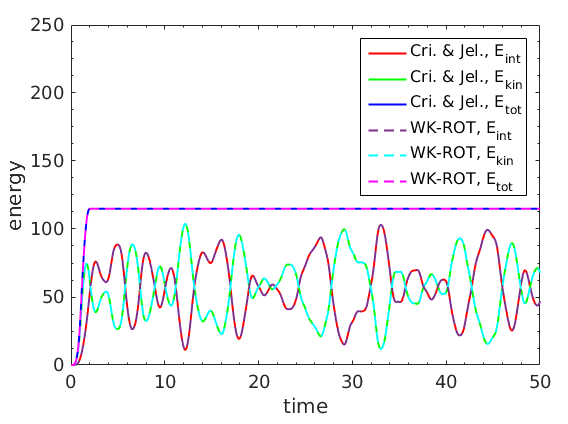}
    \label{fig:ellbow_energy1}
   }
   \subfigure[SK-ROT, 16 elements, $\Delta t=0.01$.]
   {
    \includegraphics[width=0.48\textwidth]{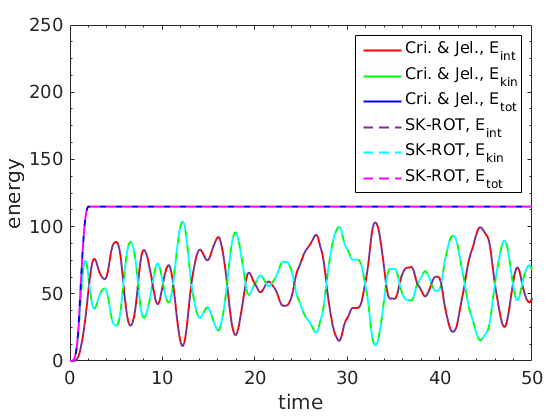}
    \label{fig:ellbow_energy2}
   }
   \subfigure[Cri. \& Jel., 2 elements, $\Delta t=0.25$.]
   {
    \includegraphics[width=0.48\textwidth]{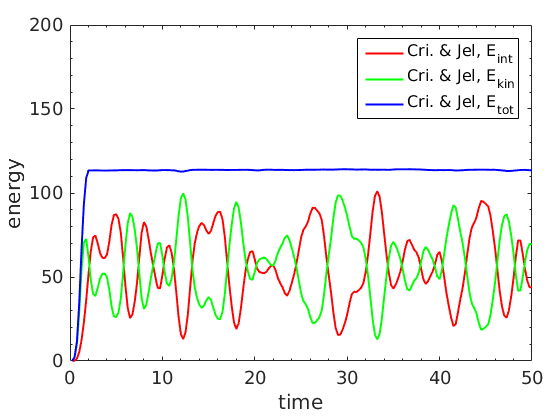}
    \label{fig:ellbow_energy3}
   }
   \subfigure[WK-ROT, 2 elements, $\Delta t=0.25$.]
   {
    \includegraphics[width=0.48\textwidth]{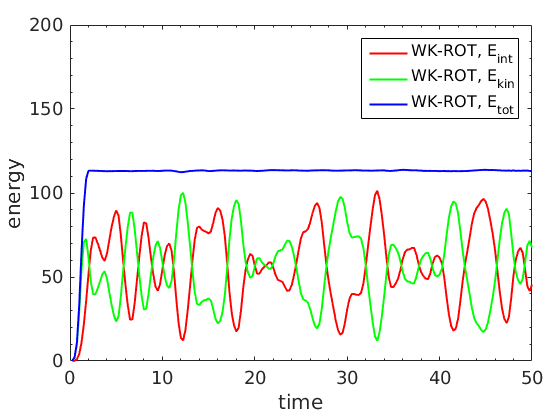}
    \label{fig:ellbow_energy4}
   }
   \subfigure[SK-ROT, 2 elements, $\Delta t=0.25$.]
   {
    \includegraphics[width=0.48\textwidth]{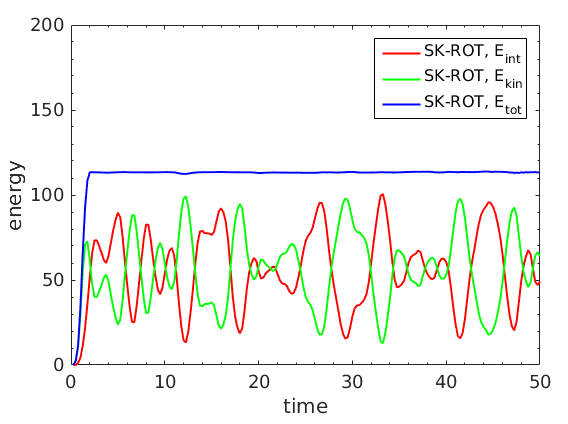}
    \label{fig:ellbow_energy5}
   }
   \subfigure[SK-ROT+CS, 2 elements, $\Delta t=0.25$.]
   {
    \includegraphics[width=0.48\textwidth]{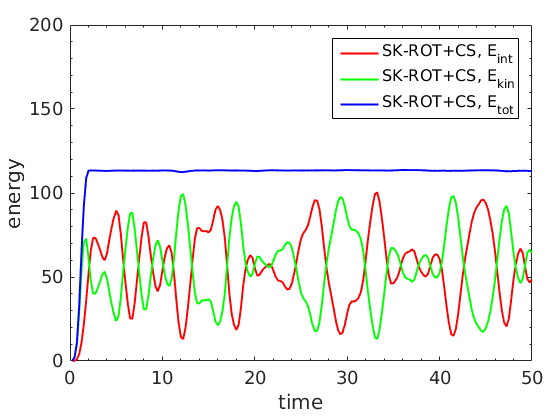}
    \label{fig:ellbow_energy6}
   }
  \caption{Free oscillations of an elbow cantilever: Energy conservation.}
  \label{fig:ellbow_energy}
\end{figure}
The first segment of the cantilever is clamped at the position $s\!=\!0$. In the original work, the cross-section of the beam is described by the section parameters $EA\!=\!GA_2\!=\!GA_3\!=\!10^6$, $GI_T\!=\!EI_2\!=\!EI_3\!=\!10^3$ as well as $\rho A\!=\!1$ and $\rho I_P/2\!=\!\rho I_2\!=\!\rho I_3\!=\!10$. It can easily be verified that these section parameters can for example be represented by a quadratic cross-section with radius $R\!\approx\!0.1$, thus resulting in a slenderness ratio of $\zeta\!\approx\!100$ for each of the two segments, for which the rotational inertia contributions are additionally scaled by a factor of $10^4$. This artificial scaling has been applied in order to emphasize (and properly verify) the rotational inertia contributions, which would otherwise be comparatively small for the chosen slenderness ratio. This cantilever beam is loaded by a discrete force $\mb{F}\!=\!(0,0,F_z)^T$ pointing in global $z$-direction and acting at the rigid corner of the elbow. The magnitude of the force is linearly increased from $F_z\!=\!0$ at $t\!=\!0$ to $F_z\!=\!50$ at $t\!=\!1$ and then linearly decreased to $F_z\!=\!0$ at $t\!=\!2$. In the remaining simulation time until $T\!=\!50$, the cantilever executes geometrically nonlinear free oscillations in 3D space with amplitudes being in the range of the segment lengths. The initial and deformed configurations at different time steps are illustrated in Figure~\ref{fig:ellbow_configs}. For time integration, the modified generalized-$\alpha$ scheme of Section~\ref{sec:temporaldiscretization} with $\rho_{\infty}\!=\!0.95$ has been employed. As spatial discretizations, the CJ element, the WK-ROT element as well as the Petrov-Galerkin variant SK-ROT and the Bubnov-Galerkin variant SK-ROT+CS with consistent spin vector interpolation according to~\eqref{elestrongkirchhoff_residual} have been applied. This example yields a further complexity that has not been present in the previous test cases: The modeling of a rigid beam connection occurring at the corner of the elbow. As already mentioned in earlier sections, the modeling of such kinks in the beam centerline is easier to be realized by the SK/WK-ROT variants. Nevertheless, the resulting solutions are identical as for the SK/WK-TAN variants. In Figure~\ref{fig:ellbow_energy}, the internal, kinetic and total system energy is plotted for different time step sizes and spatial discretizations. For discretizations with 8 WK-ROT and SK-ROT elements per elbow segment and a time step size of $\Delta t=0.01$ as illustrated in Figures~\ref{fig:ellbow_energy1} and~\ref{fig:ellbow_energy2}, no visible oscillations in the total system energy and no visible differences in the energy contributions of the Reissner and Kirchhoff type elements can be observed. In Figures~\ref{fig:ellbow_energy3}-~\ref{fig:ellbow_energy6}, the energy contributions resulting from a larger time step size of $\Delta t=0.25$ and a rougher spatial discretization based on one beam element per elbow segment has been plotted. Accordingly, also for these rough discretizations, the overall system energy is preserved very well. The fact that the total system energy approximation resulting from the (energetically consistent) Bubnov-Galerkin discretization SK-ROT+CS is rarely improved as compared to the Petrov-Galerkin variants CJ, WK-ROT and SK-ROT indicates that for the chosen spatial and temporal discretizations, the influence of the temporal discretization error might dominate the error in the total system energy. As compared to reference~\cite{jelenic1999}, where an identical time step size of $\Delta t\!=\!0.25$ and a comparable spatial discretization consisting of one second-order CJ element per elbow segment has been applied, the oscillations of the total system energy could be considerably decreased and the instability observed there at the end of the considered time interval could be completely avoided. This improvement can be attributed to the applied Lie group extension of the generalized-$\alpha$ scheme  of Section~\ref{sec:temporaldiscretization}, whereas in reference~\cite{jelenic1999} the Lie group extension of a Newmark time integration scheme as proposed in~\cite{simo1986} has been considered. Based on (simplified) Reissner type beam element formulation (see~\cite{lens2008}), similar observations have already been made in reference~\cite{bruels2012}, where the  Lie group extension of the generalized-$\alpha$ scheme has been proposed.

\newpage
  \hspace{5.0cm}
\pagebreak

%
\section{Conclusion}
\label{sec:conclusion}
%

The current work focused on geometrically exact finite elements for highly slender beams. It aimed at the proposal of novel formulations of Kirchhoff-Love type, a detailed review of existing formulations of Kirchhoff-Love and Simo-Reissner type and a careful evaluation and comparison of the proposed and existing formulations. In the authors' recent contribution~\cite{meier2014}, the first 3D large-deformation geometrically exact Kirchhoff-Love beam element formulation that fulfills the essential properties of objectivity and path-independence and that is capable of representing arbitrary initial curvatures and anisotropic cross-section shapes has been proposed. The developed beam element formulation incorporates the modes of axial tension, torsion and anisotropic bending. In the subsequent work~\cite{meier2015}, also the important question of membrane locking has successfully been addressed. The current contribution has extended these methodologies by providing considerable improvements in terms of accuracy and practical applicability as well as a generalization to dynamic problems. Thereto, two alternative interpolation schemes have been proposed: i) The first scheme is based on a strong enforcement of the Kirchhoff constraint (SK) enabled by a novel orthonormal rotation interpolation scheme. ii) The space-continuous theory of the second scheme is based on a weak enforcement of the Kirchhoff constraint (WK). The discrete realization of the Kirchhoff constraint relies on a properly chosen collocation strategy which can entirely abstain from the use of additional Lagrange multipliers. While this second formulation allows for arbitrary rotation interpolations, the investigated numerical realization has employed the well-known orthonormal, geodesic triad interpolation scheme proposed in~\cite{crisfield1999}. Furthermore, for each of these two interpolation schemes, two different sets of nodal rotation parametrizations are proposed, one based on nodal rotation vectors (ROT) and one on nodal tangent vectors (TAN). While these different choices have been shown to yield identical FEM solutions, they differ in the resulting performance of nonlinear solvers and in the effort required for prescribing essential boundary and transition conditions such as rigid joints. The four finite element formulations resulting from a combination of the two interpolation schemes and the two choices of nodal primary variables have been denoted as SK-TAN, SK-ROT, WK-TAN and WK-ROT elements, respectively. Taking advantage of a third-order Hermite interpolation, both element formulations provide a $C^1$-continuous centerline representation. In order to avoid membrane locking effects in the regime of high beam slenderness ratios, the concept of Minimally Constrained Strains (MCS) recently proposed in~\cite{meier2015} has been employed for re-interpolation of the axial tension field.  Eventually, the proposed beam elements are supplemented by an implicit, second-order accurate time integration scheme recently proposed in the literature for time discretization of large rotations. This integration scheme can be identified as Lie group extension of the well-known generalized-$\alpha$ method with comparable properties. The generality and flexibility of this scheme allow for a straightforward combination with the different element formulations considered in this work.\\

The review of existing geometrically exact Kirchhoff-Love beam elements revealed that there are only a few approaches of this kind that are suitable for general 3D, large-deformation problems. These approaches have been categorized in isotropic, straight and anisotropic Kirchhoff-Love formulations. A detailed evaluation of these formulations led to the result that typically only a few of the essential requirements summarized in Table~\ref{tab:comparison_requirements} are fulfillled. On the contrary, for the finite elements proposed in this work, the fulfillment of these essential properties such as objectivity and path-independence, consistent spatial convergence behavior, the avoidance of locking in the high slenderness regime or the conservation of energy and momentum by the spatial discretization scheme have been shown theoretically and verified by means of representative numerical examples. In the context of conservation properties, especially, the influence of applying either Bubnov-Galerkin or Petrov-Galerkin discretizations has been in the focus. Concerning locking behavior, the recently proposed MCS method has been compared with alternative methods known from the literature such as Assumed Natural Strains (ANS) or Reduced Integration (RI) schemes (see also~\cite{meier2015}). In contrast to these alternative methods, the MCS method could effectively avoid any evidence of membrane locking for all investigated load cases and slenderness ratios. In contrast to previously existing Kirchhoff-Love formulations, Simo-Reissner formulations fulfilling the requirements of Table~\ref{tab:comparison_requirements} can be found in the literature very well. However, it has been argued that the shear-free Kirchhoff-Love formulations can provide considerable numerical advantages such as lower spatial discretization error level, improved performance of time integration schemes as well as linear and nonlinear solvers or smooth geometry representation as compared to shear-deformable Simo-Reissner formulations when applied to highly slender beams. On the basis of several numerical examples, detailed and systematic numerical comparisons of the resulting discretization error levels and performance of the nonlinear solver have been performed between the four variants of the proposed geometrically exact Kirchhoff-Love beam elements and two well-established geometrically exact Simo-Reissner beam element formulations known from the literature. Most of the examples have been investigated for the two different slenderness ratios $\zeta\!=\!100$ and $\zeta\!=\!10000$. For the low slenderness ratio $\zeta\!=\!100$, the general model difference between the shear-free Kirchhoff-Love theory and the Simo-Reissner theory of shear-deformable beams, measured in form of the relative $L^2$-error remaining in the limit of arbitrarily fine spatial discretizations, typically lay below $0.1\%$. Also the quadratic decrease of this model difference with increasing slenderness ratio could be confirmed numerically. In all investigated examples, the proposed WK elements have shown a lower discretization error level than the investigated Simo-Reissner beam element formulation. These results confirm the theoretical prediction that Kirchhoff type formulations can achieve the same discretization error level as Reissner type formulations with less degrees of freedom, since no shear deformation has to be represented. Compared to the excellent results of the WK elements, the SK elements showed an increased discretization error level, in some examples even higher as for the Simo-Reissner elements. Based on the underlying convergence theory, this phenomenon could be attributed to the polynomial degree $k\!=\!3$ of the employed trial functions and predicted to vanish for higher-order elements with $k\!>\!3$. This prediction has been confirmed by means of a first numerical test case employing Hermite polynomials of order $k\!=\!5$ which resulted in the expected optimal discretization error level lying below the error level of the Simo-Reissner reference formulation. While most of the investigated examples have been conducted in a quasi-static manner, also one well-known dynamic benchmark test from the literature has been conducted. There, the accuracy of the inertia contributions as well as the energy stability of the employed time integration scheme could be confirmed.\\

Besides the resulting discretization error level, also the total number of Newton-Raphson iterations required to solve the considered test cases by means of the different element formulations and for different slenderness ratios has been analyzed in a systematic manner. For all investigated examples and slenderness ratios, the proposed SK-TAN and WK-TAN elements required less Newton iterations to solve the problem as compared to the two well-established Simo-Reissner formulations chosen as reference. In the small slenderness range $\zeta\!=\!100$, the results of the four proposed Kirchhoff-Love variants and the two investigated Simo-Reissner formulations lay at least in the same order of magnitude. While the behavior of the Kirchhoff-Love formulations remained more or less unchanged, the number of Newton iterations required by the two different Simo-Reissner formulations increased considerably with increasing slenderness ratio. In the investigated examples with slenderness ratio $\zeta\!=\!10000$, this number was up to two orders of magnitude(!) higher for the Simo-Reissner elements as compared to the proposed SK-TAN and WK-TAN elements. Also the number of iterations required by the SK-ROT and WK-ROT elements, which are based on nodal rotation vectors for triad parametrization, has been independent from the considered slenderness ratio, higher as for the SK-TAN and WK-TAN elements but still considerably lower as for the Reissner type elements. Recapitulatory, out of the four proposed Kirchhoff variants, the WK-TAN element, based on a weak enforcement of the Kirchhoff constraint and a triad parametrization via nodal tangent vectors, can be recommended in terms of a low discretization error level and an excellent performance of the Newton-Raphson scheme. Of course, further factors could be considered in a comprehensive comparison. For example, the SK-ROT and WK-ROT elements based on nodal rotation vectors simplify the prescription of Dirichlet conditions. The flexibility of the proposed beam element variants allows to combine the advantages of the two different rotation parametrizations by choosing e.g. the WK-TAN element as basic formulation that provides an excellent Newton Raphson performance and by replacing the nodal tangents by nodal rotation vectors at nodes where complex boundary or coupling conditions have to be prescribed. This can be realized by a simple transformation applied to the residual and stiffness contributions of the relevant node. The abstaining from stiff shear mode contributions underlying the proposed Kirchhoff-Love element formulations may not only yield an improved Newton-Raphson performance. Also the highest eigenfrequency band of slender beams, which is associated with the shear modes, can be avoided by this means. The theoretical considerations made in this work give hope for considerably improved stability properties of numerical time integration schemes when combined with the developed shear-free elements. A future numerical investigation of this topic seems to provide a considerable scientific potential.\\

\appendix

%
\section{Definition of rotational shape function matrices}
\label{anhang:rotshapefunctions}
%

In this appendix, the shape functions $\tilde{\mb{I}}^i(\xi)$ required for the multiplicative rotation increments
\begin{align}
\label{anhang_crisfieldspininterpolation1}
\Delta \boldsymbol{\theta}(\xi) = \sum_{i=1}^{n_{\Lambda}} \tilde{\mb{I}}^i(\xi) \Delta \hat{\boldsymbol{\theta}}^i, \quad
\Delta \boldsymbol{\theta}^{\prime}(\xi) = \sum_{i=1}^{n_{\Lambda}} \tilde{\mb{I}}^{i \prime}(\xi) \Delta \hat{\boldsymbol{\theta}}^i.
\end{align}
associated with the triad interpolation~\ref{triadrelrot_interpolation} and originally derived in~\cite{jelenic1999} shall be presented:
\begin{align}
\label{anhang_crisfieldspininterpolation2}
\begin{split}
\tilde{\mb{I}}^i(\xi)&=L^i(\xi) \boldsymbol{\Lambda}_r \mb{T}^{-1}(\boldsymbol{\Phi}_{lh}(\xi))
\mb{T}(\boldsymbol{\Phi}^i_l)\boldsymbol{\Lambda}_r^T\\
&+\delta^{iI}\boldsymbol{\Lambda}_r\left[
\mb{I_3}-\mb{T}^{-1}(\boldsymbol{\Phi}_{lh}(\xi))
\left\{
\sum \limits_{j=1}^{n_{\Lambda}} L^j(\xi) \mb{T}(\boldsymbol{\Phi}^j_l)
\right\}
\right] \mb{v}^I\boldsymbol{\Lambda}_r^T\\
&+\delta^{iJ}\boldsymbol{\Lambda}_r\left[
\mb{I_3}-\mb{T}^{-1}(\boldsymbol{\Phi}_{lh}(\xi))
\left\{
\sum \limits_{j=1}^{n_{\Lambda}} L^j(\xi) \mb{T}(\boldsymbol{\Phi}^j_l)
\right\}
\right] \mb{v}^J\boldsymbol{\Lambda}_r^T.
\end{split}
\end{align}
In~\eqref{anhang_crisfieldspininterpolation2}, no summation over double indices is applied. The vectors $\mb{v}^I$ and $\mb{v}^J$ are defined as
\begin{align}
\label{anhang_crisfieldspininterpolation3}
\begin{split}
\!\!\!\!\!\!\mb{v}^I\!=\!\frac{1}{2}\left( \mb{I}_3 \!+\!\frac{1}{\Phi^{IJ}}\tan{\left(\frac{\Phi^{IJ}}{4}\right)}\mb{S}(\boldsymbol{\Phi}^{IJ})\right), \,\,\,\,
\mb{v}^J\!=\!\frac{1}{2}\left( \mb{I}_3 \!-\!\frac{1}{\Phi^{IJ}}\tan{\left(\frac{\Phi^{IJ}}{4}\right)}\mb{S}(\boldsymbol{\Phi}^{IJ})\right),
\!\!\!\!\!\!
\end{split}
\end{align}
with the common abbreviation $\Phi^{IJ}\!=\!||\boldsymbol{\Phi}^{IJ}||$. Moreover, the arc-length derivative $\tilde{\mb{I}}^{i \prime}(\xi)$ reads:
\begin{align}
\label{anhang_crisfieldspininterpolation4}
\begin{split}
\!\!\!\!\!\!
&\tilde{\mb{I}}^i(\xi)=L^{i \prime}(\xi) \boldsymbol{\Lambda}_r \mb{T}^{-1}(\boldsymbol{\Phi}_{lh}(\xi)) \mb{T}(\boldsymbol{\Phi}^i_l)\boldsymbol{\Lambda}_r^T
+L^i(\xi) \boldsymbol{\Lambda}_r \mb{T}^{-1}_{,s} (\boldsymbol{\Phi}_{lh}(\xi))
\mb{T}(\boldsymbol{\Phi}^i_l)\boldsymbol{\Lambda}_r^T
\!\!\!\!\!\!\\\!\!\!\!\!\!
&-\delta^{iI} \!
\boldsymbol{\Lambda}_r \!
\left(\!
\mb{T}^{-1}_{,s}(\boldsymbol{\Phi}_{lh}(\xi))\!
\left\{\!
\sum \limits_{j=1}^{n_{\Lambda}} L^j(\xi) \mb{T}(\boldsymbol{\Phi}^j_l)\!
\right\}
\!+\!\mb{T}^{-1}(\boldsymbol{\Phi}_{lh}(\xi)) \!
\left\{ \!
\sum \limits_{j=1}^{n_{\Lambda}} \! L^{j \prime} (\xi) \mb{T}(\boldsymbol{\Phi}^j_l)\!
\right\}\!
\right) \!\mb{v}^I\boldsymbol{\Lambda}_r^T\!\!\!\!\!\!\\\!\!\!\!\!\!
&-\delta^{iJ} \!
\boldsymbol{\Lambda}_r \!
\left(\!
\mb{T}^{-1}_{,s}(\boldsymbol{\Phi}_{lh}(\xi))\!
\left\{\!
\sum \limits_{j=1}^{n_{\Lambda}} L^j(\xi) \mb{T}(\boldsymbol{\Phi}^j_l)\!
\right\}
\!+\!\mb{T}^{-1}(\boldsymbol{\Phi}_{lh}(\xi)) \!
\left\{ \!
\sum \limits_{j=1}^{n_{\Lambda}} \! L^{j \prime} (\xi) \mb{T}(\boldsymbol{\Phi}^j_l)\!
\right\}\!
\right) \!\mb{v}^J\boldsymbol{\Lambda}_r^T.
\!\!\!\!\!\!
\end{split}
\end{align}
Finally, the required arc-length derivative $\mb{T}^{-1}_{,s}(\boldsymbol{\Phi}_{lh}(\xi))$ is given by (see also \cite{jelenic1998,crisfield1997a}):
\begin{align}
\label{anhang_crisfieldspininterpolation5}
\begin{split}
\!\!\!\!\!\!
\mb{T}^{-1}_{,s}(\boldsymbol{\Phi}_{lh}(\xi))&=\boldsymbol{\Phi}_{lh}^T\boldsymbol{\Phi}_{lh}^{\prime}
\frac{{\Phi}_{lh} \sin{{\Phi}_{lh}}-2(1-\cos{{\Phi}_{lh}})}{{\Phi}_{lh}^4}\mb{S}(\boldsymbol{\Phi}_{lh}) +\frac{1-\cos{{\Phi}_{lh}}}{{\Phi}_{lh}^2}\mb{S}(\boldsymbol{\Phi}_{lh}^{\prime})\\
&+\frac{1}{{\Phi}_{lh}^2}\left(1-\frac{\sin{{\Phi}_{lh}}}{{\Phi}_{lh}}\right)
\left(\mb{S}(\boldsymbol{\Phi}_{lh})\mb{S}(\boldsymbol{\Phi}_{lh}^{\prime})+\mb{S}(\boldsymbol{\Phi}_{lh}^{\prime})\mb{S}(\boldsymbol{\Phi}_{lh}) \right)\\
&+\boldsymbol{\Phi}_{lh}^T\boldsymbol{\Phi}_{lh}^{\prime}
\frac{3\sin{{\Phi}_{lh}}-{\Phi}_{lh}(2+\cos{{\Phi}_{lh}})}{{\Phi}_{lh}^5}
\mb{S}(\boldsymbol{\Phi}_{lh})\mb{S}(\boldsymbol{\Phi}_{lh}).
\!\!\!\!\!\!
\end{split}
\end{align}
Here, the abbreviations $\boldsymbol{\Phi}_{lh}\!=\!\boldsymbol{\Phi}_{lh}(\xi)$ as well as ${\Phi}_{lh}\!=\!||\boldsymbol{\Phi}_{lh}(\xi)||$ have been applied. As mentioned in the original work~\cite{jelenic1999}, the limit $\mb{T}^{-1}_{,s}(\boldsymbol{\Phi}_{lh}(\xi)) \rightarrow 0.5 \mb{S}(\boldsymbol{\Phi}_{lh}^{\prime}(\xi))$ can be derived for small angles $\boldsymbol{\Phi}_{lh}(\xi) \rightarrow \mb{0}$.

%
\section{Modeling of Dirichlet boundary conditions and joints}
\label{anhang:elestrongkirchhoff_dirichletconditions}
%

For many applications, the formulation of proper Dirichlet boundary conditions or joints between the nodes of different beam elements are of a high practical relevance. This appendix represents a brief summary, where the possibility of formulating some basic constraint conditions will be investigated for the SK-ROT and the SK-TAN element.

%
\subsection{SK-ROT element}
\label{sec:elestrongkirchhoff_dirichletconditions_skrot}
%

Since the SK-ROT element simplifies the formulation of Dirichlet boundary conditions and kinematic constraints in many practically relevant cases, it will be considered first.\\

\textit{i) Dirichlet boundary conditions:} A simple support at element node $a$ can be realized via
\begin{align}
\label{dirichletconditions_skrot1}
\hat{\mb{d}}^a=\hat{\mb{d}}_u^a=\hat{\mb{d}}_0^a \rightarrow \Delta \hat{\mb{d}}^a=\mb{0}.
\end{align}
If a clamped end should be modeled, also the cross-section orientation has to be fixed, i.e.
\begin{align}
\label{dirichletconditions_skrot2}
\boldsymbol{\Lambda}^a=\boldsymbol{\Lambda}_u^a=\boldsymbol{\Lambda}_0^a, \quad \text{and} \quad
\hat{\boldsymbol{\psi}}^a=\hat{\boldsymbol{\psi}}_0^a \rightarrow
\Delta \hat{\boldsymbol{\theta}} ^a=\mb{0}.
\end{align}
Thus, the modeling of Dirichlet boundary conditions for the employed translational and rotational degrees of freedom is similar to standard finite elements that are purely based on translational degrees of freedom. This procedure can also be extended to inhomogeneous conditions. However, the determination of $\Delta \hat{\boldsymbol{\theta}} ^a$ requires special care in this case:
\begin{align}
\label{dirichletconditions_skrot3}
\begin{split}
\hat{\mb{d}}^a&=\hat{\mb{d}}_u^a(t) \rightarrow \Delta \hat{\mb{d}}_{n+1}^a=\hat{\mb{d}}_{u,n+1}^a-\hat{\mb{d}}_{u,n}^a, \\
\boldsymbol{\Lambda}^a&=\boldsymbol{\Lambda}_u^a(t) \rightarrow
\exp{ \! ( \mb{S}(  \Delta \hat{\boldsymbol{\theta}}_{n+1}^a ) )  } =\boldsymbol{\Lambda}_{u,n+1}^a \boldsymbol{\Lambda}_{u,n}^{a T}.
\end{split}
\end{align}
The multiplicative procedure of the second line simplifies to the additive procedure according to the first line if the prescribed rotation is additive, which only holds for 2D rotations.\\

\textit{ii) Connections:} A simple (moment-free) joint between the two nodes $a$ and $b$ of two connected elements reads:
\begin{align}
\label{dirichletconditions_skrot4}
\hat{\mb{d}}^b=\hat{\mb{d}}^a, \quad \delta \hat{\mb{d}}^b=\delta \hat{\mb{d}}^a, 
\quad \Delta \hat{\mb{d}}^b=\Delta \hat{\mb{d}}^a.
\end{align}
Thus, the degrees of freedom $\hat{\mb{d}}^b$ can be eliminated from the global system of equations in a standard manner by simply assembling the corresponding lines and columns of the global residual vector and stiffness matrix properly. A rigid joint between two elements prescribed at the nodes $a$ and $b$ additionally requires to suppress any relative rotation between the associated nodal triads. It is assumed that these nodal triads differ by some fixed relative rotation $\boldsymbol{\Lambda}_{0}$:
\begin{align}
\label{dirichletconditions_skrot5}
\boldsymbol{\Lambda}^{a}=\boldsymbol{\Lambda}^{b}\boldsymbol{\Lambda}_{0} \quad \text{or} \quad
\exp{ \! ( \mb{S}(  \Delta \hat{\boldsymbol{\theta}}^a ))}=\exp{ \! ( \mb{S}(  \Delta \hat{\boldsymbol{\theta}}^b ))}\boldsymbol{\Lambda}_{0} \quad \rightarrow \quad 
\boldsymbol{\Lambda}_{0}=\boldsymbol{\Lambda}^{bT}\boldsymbol{\Lambda}^{a}.
\end{align}
From~\eqref{dirichletconditions_skrot5}, the following relations between the associated rotation increments can be derived:
\begin{align}
\label{dirichletconditions_skrot6}
\delta \boldsymbol{\Lambda}^{a}=\delta \boldsymbol{\Lambda}^{b}\boldsymbol{\Lambda}_{0} \rightarrow 
\mb{S}(\delta \hat{\boldsymbol{\theta}}^a) \boldsymbol{\Lambda}^{a}= \mb{S}(\delta \hat{\boldsymbol{\theta}}^b) \boldsymbol{\Lambda}^{b}\boldsymbol{\Lambda}_{0} \rightarrow
\delta \hat{\boldsymbol{\theta}}^b = \delta \hat{\boldsymbol{\theta}}^a \rightarrow
\Delta \hat{\boldsymbol{\theta}}^b = \Delta \hat{\boldsymbol{\theta}}^a.
\end{align}
Consequently, also the rotational degrees of freedom $\hat{\boldsymbol{\psi}}^b$ can be eliminated in a standard manner by simply assembling the corresponding lines and columns of the global residual vector and of the global stiffness matrix properly.\\

\hspace{0.2 cm}
\begin{minipage}{15.0 cm}
\textbf{Remark: } It is emphasized that a rigid joint according to~\eqref{dirichletconditions_skrot5} is formulated via right-translation of the rotation tensor $\boldsymbol{\Lambda}_{0}$. This is mandatory since a rigid joint represents a fixed orientation difference between material quantities, i.e. a fixed relative rotation with respect to material axes. A left-translation via
\begin{align}
\label{dirichletconditions_skrot7}
\boldsymbol{\Lambda}^{a}=\boldsymbol{\Lambda}_{0}\boldsymbol{\Lambda}^{b}  \quad \rightarrow \quad 
\boldsymbol{\Lambda}_{0}=\boldsymbol{\Lambda}^{a}\boldsymbol{\Lambda}^{bT} \quad \rightarrow \quad
\delta \hat{\boldsymbol{\theta}}^b = \boldsymbol{\Lambda}_{0}^T \delta \hat{\boldsymbol{\theta}}^a \neq \delta \hat{\boldsymbol{\theta}}^a,
\end{align}
i.e. a fixed relative rotation with respect to spatial axes, has a different physical meaning.\\
\end{minipage}

\hspace{0.2 cm}
\begin{minipage}{15.0 cm}
\textbf{Remark: } If additive increments $\Delta \hat{\boldsymbol{\psi}}^a$ and $\Delta \hat{\boldsymbol{\psi}}^b$ of the rotation vectors $\hat{\boldsymbol{\psi}}^a$ and $\hat{\boldsymbol{\psi}}^b$ instead of the multiplicative increments $\Delta \hat{\boldsymbol{\theta}}^a$ and $\Delta \hat{\boldsymbol{\theta}}^b$ were applied in the linearization process, equation~\eqref{dirichletconditions_skrot6} has to be replaced by:
\begin{align}
\label{dirichletconditions_skrot8}
\Delta \hat{\boldsymbol{\theta}}^b = \Delta \hat{\boldsymbol{\theta}}^a \quad \rightarrow \quad
\Delta \hat{\boldsymbol{\psi}}^b=\mb{T}(\hat{\boldsymbol{\psi}}^b)\mb{T}^{-1}(\hat{\boldsymbol{\psi}}^a)\Delta \hat{\boldsymbol{\psi}}^a \neq \Delta \hat{\boldsymbol{\psi}}^a.
\end{align}
In this case, a direct elimination of the degrees of freedom $\hat{\boldsymbol{\psi}}^b$ via a proper assembly of the global stiffness matrix is not possible. Instead, the corresponding columns have to be scaled with the matrix $\mb{T}(\hat{\boldsymbol{\psi}}^b)\mb{T}^{-1}(\hat{\boldsymbol{\psi}}^a)$.\\
\end{minipage}

\hspace{0.2 cm}
\begin{minipage}{15.0 cm}
\textbf{Remark: } Physically reasonable boundary conditions can be completely defined by the cross-section orientation and centroid position. For all considered types of boundary conditions, the degrees of freedom $\hat{t}^{a}$ and $\hat{t}^{b}$, which are a measure for the nodal axial force, are part of the FEM solution and \textit{must not be prescribed}. 
\end{minipage}

%
\subsection{SK-TAN element}
\label{sec:elestrongkirchhoff_dirichletconditions_sktan}
%

The treatment of the translational degrees of freedom required for the subsequently considered boundary conditions is identical to the last section and will therefore be omitted here.\\

\textit{i) Dirichlet boundary conditions:} In order to model a clamped end with the SK-TAN element, the simplest case of a tangent vector that is parallel to a global base vector, e.g. $\hat{\mb{t}}^a \!\parallel\! \mb{e}_1$, is considered. Then,~\eqref{dirichletconditions_skrot1} has to be supplemented by
\begin{align}
\label{dirichletconditions_sktan1}
\hat{\mb{t}}^{aT} \mb{e}_2 = \hat{\mb{t}}^{aT} \mb{e}_3 = 0 \rightarrow
\Delta \hat{{t}}_2^{a} = \Delta \hat{{t}}_3^{a} = 0 \quad \text{and} \quad
\hat{\varphi}^a = \hat{\varphi}_0^a \rightarrow
\Delta \hat{\varphi}^a = 0.
\end{align}
Here, the representation $\hat{\mb{t}}^{a}\!=\!\hat{{t}}^{a}_i \mb{e}_i$ of the tangent in the global frame $\mb{e}_i$ has been exploited. In order to prescribe boundary conditions with arbitrary triad orientation, the tangent has to be expressed in the basis of the prescribed triad:
\begin{align}
\label{dirichletconditions_sktan2}
\!\!\!\!\!\!
\boldsymbol{\Lambda}^a \!=\! \boldsymbol{\Lambda}_u^a \!=\! \boldsymbol{\Lambda}_0^a, \,\,\,\, \hat{\mb{t}}^{a}\!=\!\hat{T}_i^{a} \mb{g}_i^a \rightarrow
\hat{T}_2^{a}\!=\!\hat{T}_3^{a} \!=\! 0 \rightarrow \Delta \hat{T}_2^{a} \!=\! \Delta \hat{T}_3^{a} \!=\! 0, \,\,\,\,
\hat{\varphi}^a \!=\! \hat{\varphi}_0^a \rightarrow
\Delta \hat{\varphi}^a \!=\! 0.\!\!\!\!\!\!
\end{align}
Consequently, in this case, the equations of the linearized residual vector that are associated with the degrees of freedom $\hat{\mb{t}}^{a}$ have to be transformed by the rotation tensor $\boldsymbol{\Lambda}_0^a$ and the Dirichlet conditions have to be formulated in this rotated coordinate system. Again, the first component~$\hat{T}_1$ of the tangent vector, when expressed in the material frame, represents its magnitude and must not be prescribed. If the Dirichlet conditions are time-dependent, the prescribed evolution of the relative angle has to be adapted, since now the intermediate frame $\mb{\Lambda}_{M_{\hat{\varphi}}}^a$ might change in time:
\begin{align}
\label{dirichletconditions_sktan3}
\!\!\!\!\!\!
\exp{\!(\mb{S}[\hat{\varphi}^a_{n+1} \mb{g}_{1,n+1}^a])} \!=\! \mb{\Lambda}^a_{n+1} \mb{\Lambda}_{M_{\hat{\varphi}},n+1}^{aT} \,\,\,\, \text{with} \,\,\,\,
 \mb{\Lambda}_{M_{\hat{\varphi}},n+1}^a\!=\!\text{sr}(\mb{\Lambda}_{M_{\hat{\varphi}},n}^a,\mb{g}_{1,n+1}^a).\!\!\!\!\!\!
\end{align}
Thus, the required value $\hat{\varphi}^a_{n+1}$ has to be determined based on the prescribed current triad $\mb{\Lambda}^a_{n+1}$ and the intermediate triad $\mb{\Lambda}_{M_{\hat{\varphi}},n}^{a}$ of the last step (see also Section~\ref{sec:spatialdiscretization_nodaltriadsrotation_smallestrotation}). The remaining conditions remain unchanged as compared to~\eqref{dirichletconditions_sktan2}.\\

\textit{ii) Connections:} Based on~\eqref{dirichletconditions_skrot5},~\eqref{dirichletconditions_skrot6} and~\eqref{largerotations_tildetinvmatrixsr}, the following relations between $(\hat{\mb{t}}^a,\hat{\varphi}^a)$ and $(\hat{\mb{t}}^b,\hat{\varphi}^b)$ can be stated:
\begin{align}
\label{dirichletconditions_sktan4}
\begin{split}
\!\!\!\!\!\!
\delta \hat{\mb{t}}^b&=-t^b \mb{S}(\mb{g}_1^b) \underbrace{\delta \boldsymbol{\theta}^b}_{\dot{=}\delta \boldsymbol{\theta}^a} + \mb{g}_{1}^b \delta t^b=
-t^b \mb{S}(\mb{g}_1^b) \left(
\frac{1}{t^a} \mb{S}(\mb{g}_1^a) \delta \hat{\mb{t}}^a + \mb{g}_{1}^a \delta \hat{\Theta}_1^a
\right) + \mb{g}_{1}^b \delta \hat{t}^b, \!\!\!\!\!\! \\ \!\!\!\!\!\!
\delta \hat{\Theta}_1^b&=\mb{g}_{1}^{bT} \underbrace{\delta \boldsymbol{\theta}^b}_{\dot{=}\delta \boldsymbol{\theta}^a}=
\mb{g}_{1}^{bT}\left(
\frac{1}{t^a} \mb{S}(\mb{g}_1^a) \delta \hat{\mb{t}}^a + \mb{g}_{1}^a \delta \hat{\Theta}_1^a
\right)
.\!\!\!\!\!\!
\end{split}
\end{align}
Combining these two relations eventually yields the following total transformation matrix:
\begin{align}
\label{dirichletconditions_sktan5}
\begin{split}
\!\!\!\!\!\!
   \left(
   \begin{array}{c}
   \delta \hat{\mb{t}}^b \\
   \delta \hat{\Theta}_{1}^b
   \end{array}
   \right)=
   \underbrace{
   \left(
   \begin{array}{ccc}
   -t^b \mb{S}(\mb{g}_1^b) \frac{1}{t^a} \mb{S}(\mb{g}_1^a) 
   & -t^b \mb{S}(\mb{g}_1^b) \mb{g}_{1}^a & \mb{g}_{1}^b\\
   \mb{g}_{1}^{bT} \frac{1}{t^a} \mb{S}(\mb{g}_1^a) & \mb{g}_{1}^{bT} \mb{g}_{1}^a & 0
   \end{array}
   \right)
   }_{=:\mb{T}_{RC}}
   \left(
   \begin{array}{c}
   \delta \hat{\mb{t}}^a \\
   \delta \hat{\Theta}_{1}^a\\
   \delta \hat{t}^b
   \end{array}
   \right) 
.\!\!\!\!\!\!
\end{split}
\end{align}
A similar relation can also be formulated for the iterative increments. Since the multiplicative rotation increment components $\Delta \Theta_1 \!=\! \mb{T}_{\Theta_{M1} \! \mb{t}} \Delta \mb{t} \!+\! \Delta \varphi$ (see~\eqref{largerotations_deltaThetaSR1}) have to be expressed by additive increments $\Delta \mb{t}$ and $\Delta \varphi$ for the chosen linearization scheme, an additional transformation is required as compared to~\eqref{dirichletconditions_sktan5}:
\begin{align}
\label{dirichletconditions_sktan6}
\begin{split}
\!\!\!\!\!\!
   \left(
   \begin{array}{c}
   \!\!\!\!\Delta \hat{\mb{t}}^b  \!\!\!\\
   \!\!\!\!\Delta \hat{\varphi}^b \!\!\!
   \end{array}
   \right)\!=\!
   \tilde{\mb{T}}_{RC1}
   \mb{T}_{RC} 
   \tilde{\mb{T}}_{RC2}\!
   \left(
   \begin{array}{c}
   \!\!\!\Delta \hat{\mb{t}}^a \!\!\!\\
   \!\!\!\Delta \hat{\varphi}_{1}^a \!\!\!\\
   \!\!\!\Delta \hat{t}^b \!\!\!
   \end{array}
   \right)\!,
   \,\,\,\,
      \tilde{\mb{T}}_{RC1}\!:=\!
   \left(
   \begin{array}{cc}
   \!\!\!\!\mb{I}_3 & \mb{0}\!\!\!\\
   \!\!\!\!-\mb{T}_{\Theta_{M1} \! \mb{t}} & 1\!\!\!
   \end{array}
   \right)\!
   , \,\,\,\,
   \tilde{\mb{T}}_{RC2}\!:=\!
   \left(
   \begin{array}{ccc}
   \!\!\!\!\mb{I}_3 & \mb{0} & \mb{0}\!\!\!\\
   \!\!\!\!\mb{T}_{\Theta_{M1} \! \mb{t}} & 1 & 0\!\!\!\\
   \!\!\!\!\mb{0}^T & 0 & 1\!\!\!
   \end{array}
   \right)\!.\!\!\!\!\!\!
\end{split}
\end{align}
Equations~\eqref{dirichletconditions_sktan5} and~\eqref{dirichletconditions_sktan6} allow to transform the corresponding lines and columns of the global residual vector and of the global tangent stiffness matrix properly and to eliminate the degrees of freedom $(\hat{\mb{t}}^b,\hat{\varphi}^b)$ from the global system of equations. Again, the magnitude of the tangent vector $\hat{t}^b$ is not influenced by the rigid joint and enters the system of equations as new degree of freedom. While in the last section, the motion of the rigid joint was completely determined by the set $(\hat{\mb{d}}^a, \hat{\boldsymbol{\psi}}^a,\hat{t}^a,\hat{t}^b)$, in this section the alternative set $(\hat{\mb{d}}^a,\hat{\mb{t}}^a,\hat{\varphi}^a,\hat{t}^b)$ is employed.\\

All in all, it can be concluded that the realization of clamped ends with arbitrary orientation or of rigid joints between beams is simpler for the SK-ROT formulation based on nodal rotation vectors. While for these elements such conditions can directly be formulated in the global coordinate system, the tangent vector-based SK-TAN formulation requires an additional transformation of the corresponding lines and columns of the global residual vector and stiffness matrix. In Section~\ref{sec:elenumericalexamples}, some properties of the tangent vector-based variant will become apparent which make this type of formulation favorable for many applications. If certain element nodes require Dirichlet conditions of the type considered here, it is still possible to apply a hybrid approach, and to replace the nodal tangents by nodal rotation vectors as primary variables only at the specific nodes where such conditions are required. All the results derived so far apply in a similar manner to the WK-TAN and WK-ROT elements that will be derived in the next section.

%
\section{Linearization of SK-TAN element}
\label{anhang:linsktan}
%

Before deriving the linearization of the SK-TAN element, some former definitions are repeated:
\begin{align}
\begin{split}
\label{linelestrongkirchhoff_residual_petrov0}
 \mb{t}\!&:=\!\mb{r}^{\prime}, \quad \mb{g}_1\!:=\!\frac{\mb{r}^{\prime}}{||\mb{r}^{\prime}||}, \quad
 \tilde{\mb{t}}\!:=\!\frac{\mb{r}^{\prime}}{||\mb{r}^{\prime}||^2}, \quad
 \mb{g}_1^{\prime}\!=\!\frac{\mb{\mb{r}^{\prime \prime}}}{||\mb{r}^{\prime}||}
 \!-\!
 \frac{(\mb{r}^{\prime T}\mb{r}^{\prime \prime})\mb{r}^{\prime}}{||\mb{r}^{\prime}||^3}, \quad
  \tilde{\mb{t}}^{\prime}\!=\!\frac{\mb{r}^{\prime \prime}}{||\mb{r}^{\prime}||^2}
  \!-\!\frac{2(\mb{r}^{\prime T}\mb{r}^{\prime \prime})\mb{r}^{\prime}}{||\mb{r}^{\prime}||^4}.
\end{split}
\end{align}
These quantities will be required for later derivations. Linearization of~\eqref{linelestrongkirchhoff_residual_petrov0} yields:
\begin{align}
\begin{split}
\label{linelestrongkirchhoff_residual_petrov0b}
  \Delta \mb{g}_1\!&=\!\frac{1}{||\mb{r}^{\prime}||}
 \left(
 \mb{I}_3\!-\!\mb{g}_1 \!\otimes\! \mb{g}_1^T
 \right)\mbd{H}^{\prime} \Delta \hat{\mbd{d}}, \quad \quad \,\,\,
  \Delta \tilde{\mb{t}}\!=\!\frac{1}{||\mb{r}^{\prime}||^2}
 \left(
 \mb{I}_3\!-\!2\mb{g}_1 \!\otimes\! \mb{g}_1^T
 \right)\mbd{H}^{\prime} \Delta \hat{\mbd{d}},\\
 \Delta \mb{g}_1^{\prime}\!&=\!-\frac{(\mb{r}^{\prime T}\mb{r}^{\prime \prime})}{||\mb{r}^{\prime}||^3}
 \left(
 \mb{I}_3\!-\!\mb{g}_1 \!\otimes\! \mb{g}_1^T
 \right)\mbd{H}^{\prime} \Delta \hat{\mbd{d}}\!-\!
 \frac{1}{||\mb{r}^{\prime}||}
 \left(
 \!\mb{g}_1^{\prime} \!\otimes\! \mb{g}_1^T \!+\! \mb{g}_1 \!\otimes\! \mb{g}_1^{\prime T}
 \right)\mbd{H}^{\prime} \Delta \hat{\mbd{d}} \\ & \,\,\,\,\,\, +\!
 \frac{1}{||\mb{r}^{\prime}||}
 \left(
 \mb{I}_3\!-\!\mb{g}_1 \!\otimes\! \mb{g}_1^T
 \right)\mbd{H}^{\prime \prime} \Delta \hat{\mbd{d}},\\
 \Delta \tilde{\mb{t}}^{\prime}\!&=\!-\frac{2(\mb{r}^{\prime T}\mb{r}^{\prime \prime})}{||\mb{r}^{\prime}||^4}
 \left(
 \mb{I}_3\!-\!2\mb{g}_1 \!\otimes\! \mb{g}_1^T
 \right)\mbd{H}^{\prime} \Delta \hat{\mbd{d}}\!-\!
 \frac{2}{||\mb{r}^{\prime}||^2}
 \left(
 \!\mb{g}_1^{\prime} \!\otimes\! \mb{g}_1^T \!+\! \mb{g}_1 \!\otimes\! \mb{g}_1^{\prime T}
 \right)\mbd{H}^{\prime} \Delta \hat{\mbd{d}} \\ & \,\,\,\,\,\, +\!
 \frac{1}{||\mb{r}^{\prime}||^2}
 \left(
 \mb{I}_3\!-\!2\mb{g}_1 \!\otimes\! \mb{g}_1^T
 \right)\mbd{H}^{\prime \prime} \Delta \hat{\mbd{d}}.
\end{split}
\end{align}
In the following, the linearization of the SK-TAN element will be derived. For completeness, the underlying residual vector~\eqref{elestrongkirchhoff_residual_petrov} with inserted strain re-interpolation~\eqref{MCS} is repeated here:
\begin{align}
\label{elestrongkirchhoff_residual_petrov_anhang}
\begin{split}
 \!\!\!\!\!\! & \mbd{r}_{\hat{\mb{d}}} \!=\!\! \int \limits_{-1}^{1} \!\! \left( \mbd{v}_{\theta_{\perp}}^{\prime} \! \mb{m} \!+\! \bar{\mbd{v}}_{\epsilon} \bar{F}_{1} \!-\! \mbd{H}^T \tilde{\mb{f}}_{\rho} \!-\! \mbd{v}_{\theta_{\perp}} \! \tilde{\mb{m}}_{\rho} \right) \! J(\xi) d \xi
 \!-\! \Bigg[\mbd{H}^T \mb{f}_{\sigma} \!+\! \mbd{v}_{\theta_{\perp}} \mb{m}_{\sigma} \Bigg]_{\varGamma_{\sigma}}\!\!\!\!\!\!, \!\!\!\!\! \\ \!\!\!\!\!\!
& \mbd{r}_{\mb{\hat{\Theta}}_1} \!=\!\!\int \limits_{-1}^{1} \!\! \left( \mbd{v}_{\theta_{\parallel \Theta}}^{\prime} \mb{m}
-\mbd{v}_{\theta_{\parallel \Theta}} \tilde{\mb{m}}_{\rho} \right) J(\xi) d \xi - \Bigg[ \mbd{v}_{\theta_{\parallel \Theta}} \mb{m}_{\sigma} \Bigg]_{\varGamma_{\sigma}}\!\!\!\!\!\!.\!\!\!\!\!\!
\end{split}
\end{align}
The linearization of the element residual vector~\eqref{elestrongkirchhoff_residual_petrov_anhang} obeys the following general form:
\begin{align}
\label{linelestrongkirchhoff_residual_petrov1}
\begin{split}
 \!\!\!\!\!\! \Delta\mbd{r}_{\hat{\mb{d}}} \!&=\!\! \int \limits_{-1}^{1} \!\! \left( \Delta \mbd{v}_{\theta_{\perp}}^{\prime} \! \mb{m} \!+\! \mbd{v}_{\theta_{\perp}}^{\prime} \! \Delta \mb{m}  \!+\! \Delta \bar{\mbd{v}}_{\epsilon} \bar{F}_{1}
 \!+\! \bar{\mbd{v}}_{\epsilon} \Delta \bar{F}_{1} \right) \! J(\xi) d \xi \\
  \!\! \!&-\! \int \limits_{-1}^{1} \!\! \left( 
\mbd{H}^T \Delta \mb{f}_{\rho} 
 \!+\! \Delta \mbd{v}_{\theta_{\perp}} \! \tilde{\mb{m}}_{\rho}
 \!+\! \mbd{v}_{\theta_{\perp}} \! \Delta \mb{m}_{\rho} \right) \! J(\xi) d \xi\!-\! \Bigg[\Delta \mbd{v}_{\theta_{\perp}} \! \mb{m}_{\sigma} \Bigg]_{\varGamma_{\sigma}}\!\!\!\!\!\!,\\
 \!\!\!\!\!
 \Delta\mbd{r}_{\mb{\hat{\Theta}}_1} \!&=\!\!\int \limits_{-1}^{1} \!\! \left( \Delta \mbd{v}_{\theta_{\parallel \Theta}}^{\prime} \mb{m} \!+\! \mbd{v}_{\theta_{\parallel \Theta}}^{\prime} \Delta \mb{m} -\Delta \mbd{v}_{\theta_{\parallel \Theta}} \tilde{\mb{m}}_{\rho}
 - \mbd{v}_{\theta_{\parallel \Theta}} \Delta \mb{m}_{\rho} \right) J(\xi) d \xi - \Bigg[ \Delta \mbd{v}_{\theta_{\parallel \Theta}} \mb{m}_{\sigma} \Bigg]_{\varGamma_{\sigma}}\!\!\!\! .\!\!\!\!\!\!
\end{split}
\end{align}
In order to identify the element stiffness matrix $\mbd{k}_{SK-TAN}$,~\eqref{linelestrongkirchhoff_residual_petrov1} has to be brought in the form
\begin{align}
\label{linelestrongkirchhoff_residual_petrov1c}
\begin{split}
 \!\!\!\!\!\! \Delta \mbd{r}_{SK-TAN}=:\mbd{k}_{SK-TAN} \Delta \hat{\mbd{x}}_{TAN}.\!\!\!\!\!\!
\end{split}
\end{align}
The vector $\Delta \hat{\mbd{x}}_{TAN}$ has already been defined in Section~\ref{sec:elestrongkirchhoff_variant2}. The moment-related terms yield:
\begin{align}
\label{linelestrongkirchhoff_residual_petrov2}
\begin{split}
 \Delta \mbd{v}_{\theta_{\perp}}^{\prime} \! \mb{m}
 \!&=\! \mbd{H}^{\prime \prime T} \mb{S}(\mb{m})\Delta \tilde{\mb{t}} \!+\! \mbd{H}^{\prime T} \mb{S}(\mb{m})\Delta \tilde{\mb{t}}^{\prime}, \quad
  \Delta  \mb{m}
 \!=\! -\mb{S}(\mb{m}) \Delta \boldsymbol{\theta} \!+\! \mb{c}_m \Delta \boldsymbol{\theta}^{\prime},\\
 \Delta \mbd{v}_{\theta_{\perp}} \! \tilde{\mb{m}}_{\rho}
 \!&=\! \mbd{H}^{\prime T} \mb{S}(\tilde{\mb{m}}_{\rho})\Delta \tilde{\mb{t}}, \quad \quad \quad \quad \quad \,
 \Delta \mbd{v}_{\theta_{\perp}} \! \mb{m}_{\sigma}
 \!=\! \mbd{H}^{\prime T} \mb{S}(\mb{m}_{\sigma})\Delta \tilde{\mb{t}}, \\ 
 \Delta \mbd{v}_{\theta_{\parallel \Theta}} \tilde{\mb{m}}_{\rho}\!&=\! 
 (\mbd{L}^{\! T}_{\parallel} \!\otimes\! \tilde{\mb{m}}_{\rho}^T) \Delta \mb{g}_1, \quad \quad \quad \quad \,\,
  \Delta \mbd{v}_{\theta_{\parallel \Theta}} \mb{m}_{\sigma}\!=\! 
 (\mbd{L}^{\! T}_{\parallel} \!\otimes\! \mb{m}_{\sigma}^T) \Delta \mb{g}_1, \\
  \Delta \mbd{v}_{\theta_{\parallel \Theta}}^{\prime} \mb{m}\!&=\! 
  (\mbd{L}^{\! \prime T}_{\parallel} \!\otimes\! \mb{m}^T) \Delta \mb{g}_1 \!+\! (\mbd{L}_{\! \parallel}^T \!\otimes\! \mb{m}^T) \Delta \mb{g}_1^{\prime}.
\end{split}
\end{align}
Here, many of the relations already derived in Section~\ref{sec:spatialdiscretization_triadsr} could be re-used. The field of multiplicative rotation vector increments $\Delta \boldsymbol{\theta}$ follows directly from equation~\eqref{kirchhoffkinematics_spinvector_variant2final}:
\begin{align}
\begin{split}
\label{linelestrongkirchhoff_residual_petrov4}
  \Delta \boldsymbol{\theta} \!& =\! \mbd{v}_{\theta_{\parallel \Theta}}^T \Delta \hat{\mbds{\Theta}}_1 \!  \!+\!  (\mbd{v}_{\theta_{\perp}}^T \!+\!  \mbd{v}_{\theta_{\parallel d}}^T) \Delta \hat{\mbd{d}},\\
  \quad \mbd{v}_{\theta_{\parallel \Theta}} \!&=\! \mbd{L}_{\parallel}^{\!T} \!\otimes\! \mb{g}_1^T, \,\,\,
  \mbd{v}_{\theta_{\perp}} \!\!=\! -\mbd{H}^{\prime T} \mb{S}(\tilde{\mb{t}}), \\
  \mbd{v}_{\theta_{\parallel d}} \!&=\! \left(\sum_{i=1}^{n_{\Lambda}} L^i \, \mbd{v}_{1i} \!-\! \mbd{v}_{1} \right) \! \otimes 
  \!\mb{g}_1^T, \,\, 
  \mbd{v}_{1} \!=\! \frac{\mbd{H}^{\prime T}(\xi)(\mb{g}_{1}^I \! \times \! \tilde{\mb{t}}) - 
  \mbd{H}^{\prime T}(\xi_I)(\mb{g}_{1} \! \times \! \tilde{\mb{t}}^I) }{1 \!+\!\mb{g}_{1}^{\, T} \mb{g}_{1}^I}, \\
  \mbd{v}_{1i} \!&=\! \mbd{v}_{1}(\xi_i)\!=\!\frac{ \mbd{H}^{\prime T}(\xi_i)(\mb{g}_{1}^I \! \times \! \tilde{\mb{t}}^i) - 
  \mbd{H}^{\prime T}(\xi_I) (\mb{g}_{1}^i \! \times \! \tilde{\mb{t}}^I) }{1 \!+\!\mb{g}_{1}^{i \, T} \mb{g}_{1}^I}.
\end{split}
\end{align}
In a similar manner, the associated arc-length derivative $\Delta \boldsymbol{\theta}^{\prime}$ follows from equation~\eqref{kirchhoffkinematics_spinvector_variant2final2}:
\begin{align}
\begin{split}
\label{linelestrongkirchhoff_residual_petrov5}
  \!\!\!\!\!
  \Delta \boldsymbol{\theta}^{\prime} \!& =\! \mbd{v}_{\theta_{\parallel \Theta}}^{\prime T} \Delta \hat{\mbds{\Theta}}_1  \!\!+\!   (\mbd{v}_{\theta_{\perp}}^{\prime T} \!\!+\! \mbd{v}_{\theta_{\parallel d}}^{\prime T}) \Delta \hat{\mbd{d}},\\ 
 \mbd{v}_{\theta_{\parallel \Theta}}^{\prime} \!\!& =\! \mbd{L}^{\! \prime T}_{\parallel} \!\otimes\! \mb{g}_1^T \!+\! \mbd{L}_{\! \parallel}^T \!\otimes\! \mb{g}_1^{\prime T}, \,\,
  \mbd{v}_{\theta_{\perp}}^{\prime} \!\!=\! -\mbd{H}^{\prime \prime T} \mb{S}(\tilde{\mb{t}}) \!-\! \mbd{H}^{\prime T} \mb{S}(\tilde{\mb{t}}^{\prime}),\,\,
  \tilde{\mb{t}}^{\prime}\!=\!\frac{\mb{r}^{\prime \prime}}{||\mb{r}^{\prime}||^2}
  \!-\!\frac{2(\mb{r}^{\prime T}\mb{r}^{\prime \prime})\mb{r}^{\prime}}{||\mb{r}^{\prime}||^4},
   \!\!\!\!\! \\ \!\!\!\!\!
  \mbd{v}_{\theta_{\parallel d}}^{\prime} \!&=\! \Big(\sum_{i=1}^{n_{\Lambda}} L^{i \prime} \, \mbd{v}_{1i} \!-\! \mbd{v}_{1}^{\prime} \Big) 
  \!\otimes \!\mb{g}_1^T \!+\! \Big(\sum_{i=1}^{n_{\Lambda}} L^{i} \, \mbd{v}_{1i} \!-\! \mbd{v}_{1} \Big) \! \otimes 
  \!\mb{g}_1^{\prime T}, \!\!\!\!\! \\ \!\!\!\!\!
  \mbd{v}_{1}^{\prime} \!&=\! \frac{
  \mbd{H}^{\prime T}(\xi) (\mb{g}_{1}^I \! \times \! \tilde{\mb{t}}^{\prime}) \!+\! 
  \mbd{H}^{\prime \prime T}(\xi) (\mb{g}_{1}^I \! \times \! \tilde{\mb{t}}) \!-\! 
  \mbd{H}^{\prime T}(\xi_I) (\mb{g}_{1}^{\prime} \! \times \! \tilde{\mb{t}}^I)
  }{1 \!+\!\mb{g}_{1}^{T} \! \mb{g}_{1}^I} \!-\! \frac{(\mb{g}_{1}^{\prime T} \mb{g}_{1}^I)\mb{v}_{1}}{1 \!+\!\mb{g}_{1}^{\, T} \mb{g}_{1}^I}.\!\!\!\!\!
\end{split}
\end{align}
The remaining linearizations required in equation~\eqref{linelestrongkirchhoff_residual_petrov2} have already been derived in~\eqref{linelestrongkirchhoff_residual_petrov0b}. In contrast to the spin vector field $\delta \boldsymbol{\theta}$, the increment field $\Delta \boldsymbol{\theta}$ has to be expressed via additive relative angle increments $\Delta \hat{\varphi}^i$. The required relation is given by~\eqref{DeltaTheta1} and repeated here:
\begin{align}
\begin{split}
\label{linelestrongkirchhoff_residual_petrov6}
  \!\!\!\!\!
 \Delta \hat{\mbds{\Theta}}_1  \!=\! (\hat{\Theta}_1^1,\hat{\Theta}_1^2,\hat{\Theta}_1^3)^T, \quad 
 \Delta \hat{\Theta}_1^i = -\frac{ \bar{\mb{g} }_{1}^{i T} \mb{S}(\mb{g}_{1}^i)}{1+ \mb{g}_{1}^{i T} \bar{\mb{g}}_{1}^i}  \frac{\mbd{H}^{\prime}(\xi^i) \Delta \hat{\mbd{d}}}{|| \mb{t}^i ||} + \Delta \hat{\varphi}^i.
 \!\!\!\!\!
\end{split}
\end{align}
The linearization of the element residual terms associated with axial tension results in:
\begin{align}
\label{linelestrongkirchhoff_residual_petrov7}
\begin{split}
\!\!\!\!\!\!
\Delta \bar{F}_{1}\!&=\!EA \Delta \bar{\epsilon} \!=\! EA \bar{\mbd{v}}_{\epsilon}^T \Delta \hat{\mbd{d}}, \,\,\,\, \Delta \bar{\mbd{v}}_{\epsilon}\!=\!\!\sum_{i=1}^{3} L^i(\xi) \Delta \mbd{v}_{\epsilon}(\xi^i), \,\,\,\, \Delta \mbd{v}_{\epsilon}\!=\!\frac{\mbd{H}^{\prime T}}{||\mb{r}^{\prime}||}(\mb{I}_3\!+\!\mb{g}_1\!\otimes\!\mb{g}_1^T)\mbd{H}^{\prime}\Delta \hat{\mbd{d}}.\!\!\!\!\!\!
\end{split}
\end{align}
Based on equation~\eqref{liegenalpha_finalvelacc}, the linearization of the inertia forces can be written as:
\begin{align}
\label{linelestrongkirchhoff_residual_petrov8}
\begin{split}
  \!\!\!\! -\mbd{H}^T \! \Delta \mb{f}_{\rho}\!=\! \rho A c_{\ddot{\mb{r}}1} \mbd{H}^T \mbd{H} \Delta \hat{\mbd{d}}, \quad c_{\ddot{\mb{r}}1}\!=\! \frac{1\!-\!\alpha_m}{(1\!-\!\alpha_m)\beta \Delta t^2}.
\end{split}
\end{align}
The time integration factor $c_{\ddot{\mb{r}}1}$ of the modified generalized-$\alpha$ scheme according to Section~\ref{sec:temporaldiscretization} slightly differs from the corresponding factor of the standard generalized-$\alpha$ scheme. The linearization of the inertia moments yields:
\begin{align}
\label{linelestrongkirchhoff_residual_petrov9}
\begin{split}
  \!\!\!\! -\Delta \mb{m}_{\rho}\!&=\! \mb{S}(\mb{m}_{\rho})\Delta \boldsymbol{\theta} + 
  [c_{\mb{W}}\{ \mb{S}(\mb{W})\mb{C}_{\rho} - \mb{S}(\mb{C}_{\rho}\mb{W}) \}+c_{\mb{A}}\mb{C}_{\rho}] \Delta \tilde{\boldsymbol{\Theta}}_{n+1},\\
 c_{\mb{A}}\!&=\! \frac{1\!-\!\alpha_m}{(1\!-\!\alpha_m)\beta \Delta t^2}, \quad
 c_{\mb{W}}\!=\! \frac{\gamma}{\beta \Delta t},\quad
 \Delta \tilde{\boldsymbol{\Theta}}_{n+1}\!=\!\boldsymbol{\Lambda}^T_{n} \Delta \tilde{\boldsymbol{\theta}}_{n+1}\!=\!\boldsymbol{\Lambda}^T_{n} \mb{T}(\tilde{\boldsymbol{\theta}}_{n+1})\Delta \boldsymbol{\theta}.
\end{split}
\end{align}
For clarity, the indices $n\!+\!1$ and $n$ of the current and previous time step have explicitly been noted for some of the quantities occurring in~\eqref{linelestrongkirchhoff_residual_petrov9}. All the other quantities are evaluated at $t_{n\!+\!1}$. As already introduced in Section~\ref{sec:temporaldiscretization}, the fields $\tilde{\boldsymbol{\Theta}}_{n+1}$ and $\tilde{\boldsymbol{\theta}}_{n+1}$ are the material and spatial multiplicative rotation increments relating the current configuration and the converged configuration of the previous time step $t_n$. The two vectors are related by the transformation
\begin{align}
\label{linelestrongkirchhoff_residual_petrov10}
\begin{split}
   \tilde{\boldsymbol{\Theta}}_{n+1}\!=\!\boldsymbol{\Lambda}^T_{n+1} \tilde{\boldsymbol{\theta}}_{n+1}\!=\!\boldsymbol{\Lambda}^T_{n} \tilde{\boldsymbol{\theta}}_{n+1} \rightarrow \Delta \tilde{\boldsymbol{\Theta}}_{n+1} \!=\!\boldsymbol{\Lambda}^T_{n} \Delta \tilde{\boldsymbol{\theta}}_{n+1}.
\end{split}
\end{align}
The second step in~\eqref{linelestrongkirchhoff_residual_petrov10} is valid since $\tilde{\boldsymbol{\theta}}_{n+1}$ is an eigenvector with eigenvalue one of the rotation tensor $\boldsymbol{\Lambda}_{n+1}\boldsymbol{\Lambda}^T_{n}$ between the configurations $n$ and $n\!+\!1$, thus $\boldsymbol{\Lambda}_{n+1}\boldsymbol{\Lambda}^T_{n}\tilde{\boldsymbol{\theta}}_{n+1}\!=\!\tilde{\boldsymbol{\theta}}_{n+1}$. Furthermore, $\Delta \tilde{\boldsymbol{\Theta}}_{n+1}$ and $\Delta \tilde{\boldsymbol{\theta}}_{n+1}$ represent the fields of additive increments of $\tilde{\boldsymbol{\Theta}}_{n+1}$ and $\tilde{\boldsymbol{\theta}}_{n+1}$ between two successive Newton iterations, whereas $\Delta \boldsymbol{\theta}$ as given by~\eqref{linelestrongkirchhoff_residual_petrov4} represents the field of multiplicative rotation increments between two successive Newton iterations.

%
\section{Linearization of WK-TAN element}
\label{anhang:linwktan}
%

The residual vector of the WK-TAN element is given in equation~\eqref{eleweakkirchhoff_residual} and repeated here:
\begin{align}
\label{eleweakkirchhoff_residual_petrov_anhang}
\begin{split}
 \!\!\!\!\!\! & \mbd{r}_{\hat{\mb{d}}} \!=\!\! \int \limits_{-1}^{1} \!\! \left( \bar{\mbd{v}}_{\theta_{\perp}}^{\prime} \! \mb{m} \!+\! \bar{\mbd{v}}_{\epsilon} \bar{F}_{1} \!-\! \mbd{H}^T \tilde{\mb{f}}_{\rho} \!-\! \bar{\mbd{v}}_{\theta_{\perp}} \! \tilde{\mb{m}}_{\rho} \right) \! J(\xi) d \xi \!-\! \Bigg[\mbd{H}^T \mb{f}_{\sigma} \!+\! \bar{\mbd{v}}_{\theta_{\perp}} \mb{m}_{\sigma} \Bigg]_{\varGamma_{\sigma}}\!\!\!\!\!\!, \!\!\!\!\! \\ \!\!\!\!\!\!
& \mbd{r}_{\mb{\hat{\Theta}}_1} \!=\!\!\int \limits_{-1}^{1} \!\! \left( \bar{\mbd{v}}_{\theta_{\parallel \Theta}}^{\prime} \mb{m}
-\bar{\mbd{v}}_{\theta_{\parallel \Theta}} \tilde{\mb{m}}_{\rho} \right) J(\xi) d \xi - \Bigg[ \bar{\mbd{v}}_{\theta_{\parallel \Theta}} \mb{m}_{\sigma} \Bigg]_{\varGamma_{\sigma}}\!\!\!\!\!\!.\!\!\!\!\!\!
\end{split}
\end{align}
The linearization of the element residual vector~\eqref{eleweakkirchhoff_residual_petrov_anhang} obeys the following general form:
\begin{align}
\label{lineleweakkirchhoff_residual_petrov1}
\begin{split}
 \!\!\!\!\!\! \Delta \mbd{r}_{\hat{\mb{d}}} \!&=\!\! \int \limits_{-1}^{1} \!\! \left( \Delta \bar{\mbd{v}}_{\theta_{\perp}}^{\prime} \! \mb{m} \!+\! \bar{\mbd{v}}_{\theta_{\perp}}^{\prime} \! \Delta \mb{m}  \!+\! \Delta \bar{\mbd{v}}_{\epsilon} \bar{F}_{1}
 \!+\! \bar{\mbd{v}}_{\epsilon} \Delta \bar{F}_{1} \right) \! J(\xi) d \xi \\
  \!\! \!&-\! \int \limits_{-1}^{1} \!\! \left( 
\mbd{H}^T \Delta \mb{f}_{\rho} 
 \!+\! \Delta \bar{\mbd{v}}_{\theta_{\perp}} \! \tilde{\mb{m}}_{\rho}
 \!+\! \bar{\mbd{v}}_{\theta_{\perp}} \! \Delta \mb{m}_{\rho} \right) \! J(\xi) d \xi\!-\! \Bigg[\Delta \bar{\mbd{v}}_{\theta_{\perp}} \! \mb{m}_{\sigma} \Bigg]_{\varGamma_{\sigma}}\!\!\!\!\!\!,\\
 \!\!\!\!\!
 \Delta \mbd{r}_{\mb{\hat{\Theta}}_1} \!&=\!\!\int \limits_{-1}^{1} \!\! \left( \Delta \bar{\mbd{v}}_{\theta_{\parallel \Theta}}^{\prime} \mb{m} \!+\! \bar{\mbd{v}}_{\theta_{\parallel \Theta}}^{\prime} \Delta \mb{m} -\Delta \bar{\mbd{v}}_{\theta_{\parallel \Theta}} \tilde{\mb{m}}_{\rho}
 - \bar{\mbd{v}}_{\theta_{\parallel \Theta}} \Delta \mb{m}_{\rho} \right) J(\xi) d \xi - \Bigg[ \Delta \bar{\mbd{v}}_{\theta_{\parallel \Theta}} \mb{m}_{\sigma} \Bigg]_{\varGamma_{\sigma}}\!\!\!\! .\!\!\!\!\!\!
\end{split}
\end{align}
In order to identify the element stiffness matrix $\mbd{k}_{WK-TAN}$,~\eqref{lineleweakkirchhoff_residual_petrov1} has to be brought in the form
\begin{align}
\label{linelestrongkirchhoff_residual_petrov1c}
\begin{split}
 \!\!\!\!\!\! \Delta \mbd{r}_{WK-TAN}=:\mbd{k}_{WK-TAN} \Delta \hat{\mbd{x}}_{TAN}.\!\!\!\!\!\!
\end{split}
\end{align}
The linearization of the vectors $\bar{\mbd{v}}_{...}$ and $\bar{\mbd{v}}_{...}^{\prime}$ originally defined in~\eqref{eleweakkirchhoff_residual} follows to:
\begin{align}
\label{lineleweakkirchhoff_residual_petrov1b}
\begin{split}
\Delta \bar{\mbd{v}}_{\theta_{\perp}}\!&=\!\! -\sum_{i=1}^{3} L^i(\xi) \Delta \mbd{v}_{\theta_{\perp}}(\xi^i), \quad
\Delta \bar{\mbd{v}}_{\epsilon}\!=\!\!\sum_{i=1}^{3} L^i(\xi) \Delta \mbd{v}_{\epsilon}(\xi^i), \\
\Delta \bar{\mbd{v}}_{\theta_{\parallel \Theta}}\!&=\!\! \sum_{i=1}^{3} L^i(\xi) \Delta \mbd{v}_{\theta_{\parallel \Theta} }(\xi^i) \quad \text{with} \quad
\Delta \bar{\mbd{v}}^{\prime}_{...}\!\!=\!\!\sum_{i=1}^{3} \frac{L^i{,\xi}(\xi)}{J(\xi)} \Delta \mbd{v}_{...} (\xi^i).
\end{split}
\end{align}
The linearization of the vectors $\mbd{v}_{...}$ and $\mbd{v}_{...}^{\prime}$ has already been stated in the last section. Also the linearization  of the moment stress resultant has the same form as in the last section:
\begin{align}
\label{lineleweakkirchhoff_residual_petrov2}
\begin{split}
\!\!\!\!\!\!
  \Delta  \mb{m} \!&=\! -\mb{S}(\mb{m}) \Delta \boldsymbol{\theta} \!+\! \mb{c}_m \Delta \boldsymbol{\theta}^{\prime}.
 \!\!\!\!\!\!
\end{split}
\end{align}
However, the fields $\Delta \boldsymbol{\theta}$ and $\Delta \boldsymbol{\theta}^{\prime}$ originally defined in Section~\ref{sec:spatialdiscretization_triadrelrot} are this time given by
\begin{align}
\label{lineleweakkirchhoff_residual_petrov2b}
\begin{split}
\!\!\!\!\!\!
 \Delta \boldsymbol{\theta} \!=\! \sum_{i=1}^{3} \tilde{\mb{I}}^i(\xi) \Delta {\boldsymbol{\theta}}(\xi^i), \,\,\,\,
 \Delta \boldsymbol{\theta}^{\prime} \!=\! \sum_{i=1}^{3} \frac{1}{J(\xi)}\tilde{\mb{I}}_{,\xi}^i(\xi) \Delta {\boldsymbol{\theta}}(\xi^i).
 \!\!\!\!\!\!
\end{split}
\end{align}
Due to the Kirchhoff constraint, the nodal increments $\Delta {\boldsymbol{\theta}}(\xi^i)$ can be expressed according to:
\begin{align}
\label{lineleweakkirchhoff_residual_petrov3}
\begin{split}
\!\!\!\!\!\!
 \Delta \boldsymbol{\theta}(\xi^i) \!&=\! \Delta \hat{\Theta}_1^i \mb{g}_1(\xi^i)\!+\! \mbd{v}_{\theta_{\perp}}^T(\xi^i) \Delta \hat{\mbd{d}},
\quad \Delta \hat{\Theta}_1^i = -\frac{ \bar{\mb{g} }_{1}^{i T} \mb{S}(\mb{g}_{1}^i)}{1+ \mb{g}_{1}^{i T} \bar{\mb{g}}_{1}^i}  \frac{\mbd{H}^{\prime}(\xi^i) \Delta \hat{\mbd{d}}}{|| \mb{t}^i ||} + \Delta \hat{\varphi}^i.
 \!\!\!\!\!\!
\end{split}
\end{align}
The linearization of the inertia forces is identical to the corresponding results of the last section:
\begin{align}
\label{lineleweakkirchhoff_residual_petrov8}
\begin{split}
  \!\!\!\! -\mbd{H}^T \! \Delta \mb{f}_{\rho}\!=\! \rho A c_{\ddot{\mb{r}}1} \mbd{H}^T \mbd{H} \Delta \hat{\mbd{d}}, \quad c_{\ddot{\mb{r}}1}\!=\! \frac{1\!-\!\alpha_m}{(1\!-\!\alpha_m)\beta \Delta t^2}.
\end{split}
\end{align}
This statement also holds for the linearization of the inertia moments, which reads:
\begin{align}
\label{lineleweakkirchhoff_residual_petrov9}
\begin{split}
  \!\!\!\! -\Delta \mb{m}_{\rho}\!&=\! \mb{S}(\mb{m}_{\rho})\Delta \boldsymbol{\theta} + 
  [c_{\mb{W}}\{ \mb{S}(\mb{W})\mb{C}_{\rho} - \mb{S}(\mb{C}_{\rho}\mb{W}) \}+c_{\mb{A}}\mb{C}_{\rho}] \Delta \tilde{\boldsymbol{\Theta}}_{n+1},\\
 c_{\mb{A}}\!&=\! \frac{1\!-\!\alpha_m}{(1\!-\!\alpha_m)\beta \Delta t^2}, \quad
 c_{\mb{W}}\!=\! \frac{\gamma}{\beta \Delta t},\quad
 \Delta \tilde{\boldsymbol{\Theta}}_{n+1}\!=\!\boldsymbol{\Lambda}^T_{n} \Delta \tilde{\boldsymbol{\theta}}_{n+1}\!=\!\boldsymbol{\Lambda}^T_{n} \mb{T}(\tilde{\boldsymbol{\theta}}_{n+1})\Delta \boldsymbol{\theta}.
\end{split}
\end{align}
However, for the WK-TAN element, the rotation increment field $\Delta \boldsymbol{\theta}$ is given by equation~\eqref{lineleweakkirchhoff_residual_petrov2b}.

%
\section{Linearization of SK-ROT and WK-ROT elements}
\label{anhang:linskwkrot}
%

The nodal primary variable variations of the SK/WK-TAN and the SK/WK-ROT elements read:
\begin{align}
\label{anhang_lintanrot1}
\begin{split}
\delta \hat{\mbd{x}}_{TAN}\!&:=\!(\delta \hat{\mb{d}}^{1T}\!, \delta \hat{\mb{t}}^{1T}\!, \delta \hat{\Theta}_{1}^1, \delta \hat{\mb{d}}^{2T}\!,\delta \hat{\mb{t}}^{2T}\!,\delta \hat{\Theta}_{1}^2, \delta \hat{\Theta}_{1}^3)^T, \\ \delta \hat{\mbd{x}}_{ROT}\!&:=\!(\delta \hat{\mb{d}}^{1T}\!, \delta \hat{\boldsymbol{\theta}}^{1T}\!, \delta \hat{t}^1, \delta \hat{\mb{d}}^{2T}\!,\delta \hat{\boldsymbol{\theta}}^{2T}\!,\delta \hat{t}_{1}, \delta \hat{\Theta}_{1}^3)^T.
\end{split}
\end{align}
In a similar manner, the set of iterative nodal primary variable increments have been defined as:
\begin{align}
\label{anhang_lintanrot2}
\begin{split}
\Delta \hat{\mbd{x}}_{TAN}\!&:=\!(\Delta \hat{\mb{d}}^{1T}\!, \Delta \hat{\mb{t}}^{1T}\!, \Delta \hat{\varphi}^1, \Delta \hat{\mb{d}}^{2T}\!,\Delta \hat{\mb{t}}^{2T}\!,\Delta \hat{\varphi}^2, \Delta \hat{\varphi}^3)^T, \\ \Delta \hat{\mbd{x}}_{ROT}\!&:=\!(\Delta \hat{\mb{d}}^{1T}\!, \Delta \hat{\boldsymbol{\theta}}^{1T}\!, \Delta \hat{t}^1, \Delta \hat{\mb{d}}^{2T}\!,\Delta \hat{\boldsymbol{\theta}}^{2T}\!,\Delta \hat{t}_{1}, \Delta \hat{\varphi}^3)^T.
\end{split}
\end{align}
The transformations between these primary variable variations and increments is given by:
\begin{align}
\label{anhang_lintanrot3}
\begin{split}
\delta \hat{\mbd{x}}_{TAN}\!=\!\tilde{\mbd{T}}_{\hat{\mbd{x}}} \delta \hat{\mb{x}}_{ROT} \quad \text{and} \quad
\Delta \hat{\mbd{x}}_{TAN}\!=\!\mbd{T}_{M\hat{\mbd{x}}} \Delta \hat{\mb{x}}_{ROT}.
\end{split}
\end{align}
The transformation matrices $\tilde{\mbd{T}}_{\hat{\mbd{x}}}$, originally defined in~\eqref{elestrongkirchhoff_trafovariations}, and $\mbd{T}_{M\hat{\mbd{x}}}$ have the following form:
\begin{align}
\label{anhang_lintanrot4}
   \tilde{\mbd{T}}_{\hat{\mb{x}}} \!=\!
   \left(
   \begin{array}{ccccc}
   \mb{I}_3 &&&&\\
   &\tilde{\mb{T}}^1&&&\\
   &&\mb{I}_3&&\\
   &&&\tilde{\mb{T}}^2&\\
   &&&&1
   \end{array}
   \right) \quad \text{and} \quad
   \mbd{T}_{M\hat{\mb{x}}} \!=\!
   \left(
   \begin{array}{ccccc}
   \mb{I}_3 &&&&\\
   &\mb{T}_M^1&&&\\
   &&\mb{I}_3&&\\
   &&&\mb{T}_M^2&\\
   &&&&1
   \end{array}
   \right).
\end{align}
These two different matrices are required, since the primary variable variations of the SK/WK-TAN elements are based on the multiplicative quantities $\delta \hat{\Theta}_{1}^i$, whereas
the corresponding iterative primary variable increments are based on the additive quantities $\Delta \hat{\varphi}^i$. The matrices $\tilde{\mb{T}}^i$ and $\mb{T}_M^i$ (\eqref{largerotations_tildetinvmatrixsr} and~\eqref{largerotations_tinvmatrixsr}) are evaluated at the element boundary nodes:
\begin{align}
\label{anhang_lintanrot5}
   \tilde{\mb{T}}^i := \tilde{\mb{T}}(\xi^i) \quad \text{and} \quad 
   \mb{T}_M^i := \mb{T}_M(\xi^i) \quad \text{for} \quad i=1,2.
\end{align}
In Section~\ref{sec:elestrongkirchhoff_variant1}, it has already been shown that the following residual transformation is valid:
\begin{align}
\label{anhang_lintanrot6}
   \mbd{r}_{ROT} \!=\! \tilde{\mbd{T}}_{\hat{\mb{x}}}^T \mbd{r}_{TAN}.
\end{align}
In a similar manner, also the linearized element residual vector can be transformed:
\begin{align}
\label{anhang_lintanrot7}
   \!\!\!\!\!\!
   \Delta \mbd{r}_{ROT} \!=\! \Delta \tilde{\mbd{T}}_{\hat{\mb{x}}}^T \mbd{r}_{TAN} \!+\! \tilde{\mbd{T}}_{\hat{\mb{x}}}^T \!\!\!\!\!\! \underbrace{\Delta \mbd{r}_{TAN}}_{=\mbd{k}_{TAN}\Delta \hat{\mbd{x}}_{TAN}} \!\!\!\!=:\!\!
   \underbrace{\left(\tilde{\mbd{H}}_{\hat{\mb{x}}} (\mbd{r}_{TAN})\!+\!\tilde{\mbd{T}}_{\hat{\mb{x}}}^T \mbd{k}_{TAN} \mbd{T}_{M\hat{\mb{x}}}\right)}_{:=\mbd{k}_{ROT}}    \!\Delta \hat{\mbd{x}}_{ROT}.\!\!\!\!\!\!
\end{align}
Here, the matrix $ \tilde{\mbd{H}}_{\hat{\mb{x}}} (\mbd{r}_{TAN})$ has been introduced, in order to represent the linearization of $\tilde{\mbd{T}}_{\hat{\mb{x}}}$:
\begin{align}
\label{anhang_lintanrot8}
    \tilde{\mbd{H}}_{\hat{\mb{x}}} (\mbd{r}_{TAN})  \Delta \hat{\mbd{x}}_{ROT} :=  \Delta \tilde{\mbd{T}}_{\hat{\mb{x}}}^T \mbd{r}_{TAN} \quad \text{with} \quad
    \tilde{\mbd{H}}_{\hat{\mb{x}}} (\mbd{r}_{TAN}) \!=\!
   \left(
   \begin{array}{ccccc}
   \mb{0} &&&&\\
   &\tilde{\mbd{H}}^1&&&\\
   &&\mb{0}&&\\
   &&&\tilde{\mbd{H}}^2&\\
   &&&&0
   \end{array}
   \right).
\end{align}
After calculating the derivative of $\tilde{\mbd{T}}_{\hat{\mb{x}}}$ and re-ordering the result, the submatrices $\tilde{\mbd{H}}^i$ can be stated:
\begin{align}
\label{anhang_lintanrot9}
   \tilde{\mbd{H}}^i=
   \left(
   \begin{array}{cc}
   \mb{S}(\mb{r}_{TAN,\hat{\mb{t}}^{i}}) \mb{S}(\mb{g}_1^{i}) - r_{TAN,\hat{\Theta}_1^{i}} \mb{S}(\mb{g}_1^{i})& \mb{S}(\mb{g}_1^{i})\mb{r}_{TAN,\hat{\mb{t}}^{i}}\\
   -\mb{r}_{TAN,\hat{\mb{t}}^{i}}^T\mb{S}(\mb{g}_1^{i}) & 0\!\!
   \end{array}
   \right), \quad i=1,2.
\end{align}
From~\eqref{anhang_lintanrot7}, the following transformation rule for the the element stiffness matrix can be stated:
\begin{align}
\label{anhang_lintanrot9b}
   \Delta \mbd{r}_{ROT}=\mbd{k}_{ROT} \Delta \hat{\mbd{x}}_{ROT} \quad \text{with} \quad
   \mbd{k}_{ROT}=\tilde{\mbd{H}}_{\hat{\mb{x}}} (\mbd{r}_{TAN})\!+\!\tilde{\mbd{T}}_{\hat{\mb{x}}}^T \mbd{k}_{TAN} \mbd{T}_{M\hat{\mb{x}}}.
\end{align}
In order to apply this transformation, the components of the element stiffness matrices $\mbd{k}_{TAN}$ and $\mbd{k}_{ROT}$ have to be arranged in the same order 
as the components of the element residual vectors:
\begin{align}
 \begin{split}
\label{anhang_lintanrot10}
\mbd{r}_{TAN}&:=(\mb{r}_{TAN,\hat{\mb{d}}^{1}}^T,\mb{r}_{TAN,\hat{\mb{t}}^{1}}^T,r_{TAN,\hat{\Theta}_1^{1}}, \mb{r}_{TAN,\hat{\mb{d}}^{2}}^T,\mb{r}_{TAN,\hat{\mb{t}}^{2}}^T,r_{TAN,\hat{\Theta}_1^{2}},r_{TAN,\hat{\Theta}_1^{3}})^T, \\
\mbd{r}_{ROT}&:=(\mbd{r}_{ROT,\hat{\mb{d}}^{1}}^T,\mbd{r}_{ROT,\hat{\boldsymbol{\theta}}^{1}}^T,r_{ROT,\hat{t}^{1}}, \mbd{r}_{ROT,\hat{\mb{d}}^{2}}^T,\mbd{r}_{ROT,\hat{\boldsymbol{\theta}^{2}}}^T,r_{ROT,\hat{t}^{2}},r_{ROT,\hat{\Theta}_1^{3}})^T.
\end{split}
\end{align}

%
\bibliographystyle{plain}
\bibliography{strongweakkirchhoff.bib}
%
%
\end{document}